\documentclass{egpubl}
\usepackage{eg2025}

\ConferencePaper         
\CGFStandardLicense

\usepackage[T1]{fontenc}
\usepackage{dfadobe}  

\biberVersion
\BibtexOrBiblatex
\usepackage[backend=biber,bibstyle=EG,citestyle=alphabetic,backref=true]{biblatex} 
\electronicVersion
\PrintedOrElectronic

\ifpdf \usepackage[pdftex]{graphicx} \pdfcompresslevel=9
\else \usepackage[dvips]{graphicx} \fi

\usepackage{egweblnk} 

\usepackage{amsmath}
\usepackage{booktabs} 
\usepackage{hyperref} 
\usepackage[ruled]{algorithm2e} 
\usepackage{pgfplots}
\usepackage{diagbox}
\usepackage{booktabs,siunitx}
\usepackage{pifont}
\usepackage{subcaption}
\usepackage{tabularx}
\usepackage{tabu}
\usepackage{dsfont}
\usepackage{graphicx}
\usepackage{pgffor}
\usepackage{xstring}
\usepackage{pgfplots}
\usepackage{pgfplotstable}
\usepackage{float}
\usepackage{placeins}
\usepackage{svg}
\usepackage{amssymb}
\usepackage{stmaryrd}

\hypersetup{
  colorlinks=true,
  linkcolor=blue,
  citecolor=blue,
  urlcolor=blue
}

\SetAlFnt{\small}
\SetAlCapFnt{\small}
\SetAlCapNameFnt{\small}
\SetAlCapHSkip{0pt}

\definecolor{purple}{rgb}{0.65,0,0.65}

\newcommand{\droff}{\color{black}}

\newcommand{\tcdot}{\!\cdot\!}
\newcommand{\best}{\color{Green}}
\newcommand{\secondbest}{\color{NavyBlue}}

\title[ReConForM]%
      {ReConForM : Real-time Contact-aware Motion Retargeting for more Diverse Character Morphologies}

\author[T. Cheynel \& T. Rossi \& B. Bellot-Gurlet \& D. Rohmer \& M.P. Cani]
{\parbox{\textwidth}{\centering T. Cheynel$^{1,2}$
        and T. Rossi$^{1}$
        and B. Bellot-Gurlet$^{1}$
        and D. Rohmer$^{2}$
        and M.P. Cani$^{2}$
        }
        \\
{\parbox{\textwidth}{\centering $^1$Kinetix \\
         $^2$LIX, École Polytechnique, CNRS, IP~Paris
       }
}
}

\begin{document}

\teaser{
 \includegraphics[width=\linewidth]{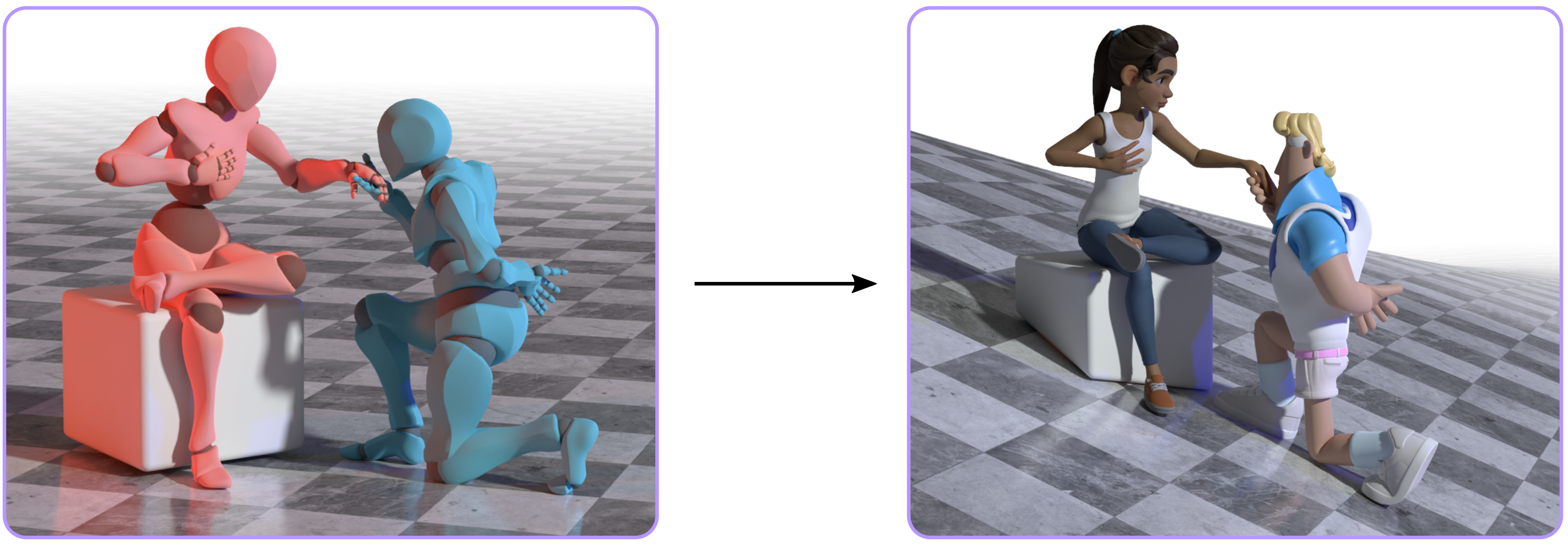}
 \centering
  \caption{Output of our retargeting method, showcasing several of our contributions. The source pose (left scene) shows complex contacts carrying semantic information : self-contacts, foot-ground contacts, and inter-character contacts). This pose is retargeted onto very different characters from the animated movie industry, evolving on a non-flat terrain.}
  \vspace*{4\baselineskip}
\label{fig:teaser}
}

\maketitle




\begin{abstract}
Preserving semantics, in particular in terms of contacts, is a key challenge when retargeting motion between characters of different morphologies. Our solution relies on a low-dimensional embedding of the character's mesh, based on rigged key vertices that are automatically transferred from the source to the target. Motion descriptors are extracted from the trajectories of these key vertices, providing an embedding that contains combined semantic information about both shape and pose. A novel, adaptive algorithm is then used to automatically select and weight the most relevant features over time, enabling us to efficiently optimize the target motion until it conforms to these constraints, so as to preserve the semantics of the source motion. Our solution allows extensions to several novel use-cases where morphology and mesh contacts were previously overlooked, such as multi-character retargeting and motion transfer on uneven terrains. As our results show, our method is able to achieve real-time retargeting onto 
a wide variety of characters. Extensive experiments and comparison with 
state-of-the-art methods using several relevant metrics demonstrate improved results, both in terms of motion smoothness and contact accuracy. 

\ccsdesc[500]{Computing methodologies~Animation}
\ccsdesc[300]{Computing methodologies~Motion processing}

\printccsdesc   
\end{abstract}

%
%

%


\sloppy

\section{Introduction}

Motion transfer between animated characters,
also called motion retargeting, is a crucial aspect of character animation with major applications in the fields of 
cinema, video games and virtual reality.
The complexity of this task arises from the variety of 3D character models.
While different skeletal topologies and skin meshes may make them difficult to compare,
%
retargeting remains an ill-posed problem even when a correspondence exists between the source and target models.
Differences in morphological proportions may cause 
self-penetrations, inaccurate contacts, and a loss of the overall sense of the pose. 
Finding the right compromises to mitigate these issues while maintaining the expected fidelity to the source motion is a complex task.
In particular, well-thought-out metrics, capturing at least in part the semantics of motion, need to be designed. 
Alternatively, machine learning shows promise in enabling pose semantics to be discovered and retargeted without any explicit definition. 
However, the lack of paired retargeting data makes the use of supervised learning difficult.


Our work was greatly inspired by the evidence that the preservation of contacts between different body parts is of utmost 
importance when humans assess the quality of a retargeted motion~\cite{basset22}. We argue that contacts with the ground, present in almost every motion, are equally important.
It would, however, be very time-consuming to manually identify the most relevant contact interactions of a specific movement, which may change over time, and then find a compromise between their influences in order to preserve the relevant mesh contacts.
Instead, we introduce specific motion features to describe the relative positioning of body parts and propose an automatic, adaptive solution that dynamically selects and weights these features over time, focusing on periods around key events such as collisions and contacts.
The selected weighted features are then used to compute lightweight objective functions, optimized in real time to generate the target motion.
In short, our main contributions are:
\begin{itemize}
    \item A new joint representation
    for
    the character's morphology and motion, used to compute relevant motion features;
    \item An adaptive proximity-based method to select and weight motion features over time;
    \item Extensions to multi-character retargeting and non-flat grounds, 
    enabled by the adaptability of our framework.
\end{itemize}
The evaluation, both qualitative via a user study and quantitative thanks to various metrics, shows that our method achieves better results than the state of the art, both in terms of temporal fluidity of the target animation and semantic similarity with the source.
Furthermore, adaptively selected constraints being sparse in space and time, our method (named ReConForM) achieves retargeting tasks several orders of magnitude faster than former optimization-based methods, while accommodating arbitrary morphologies and being easily extensible to new use cases.

\section{Related Work}

\begin{figure*}[!t]
    \centering
    \begin{subfigure}{0.24\textwidth}
        \centering
        \includegraphics[trim={300px 0 480px 120px},clip,width=.92\textwidth]{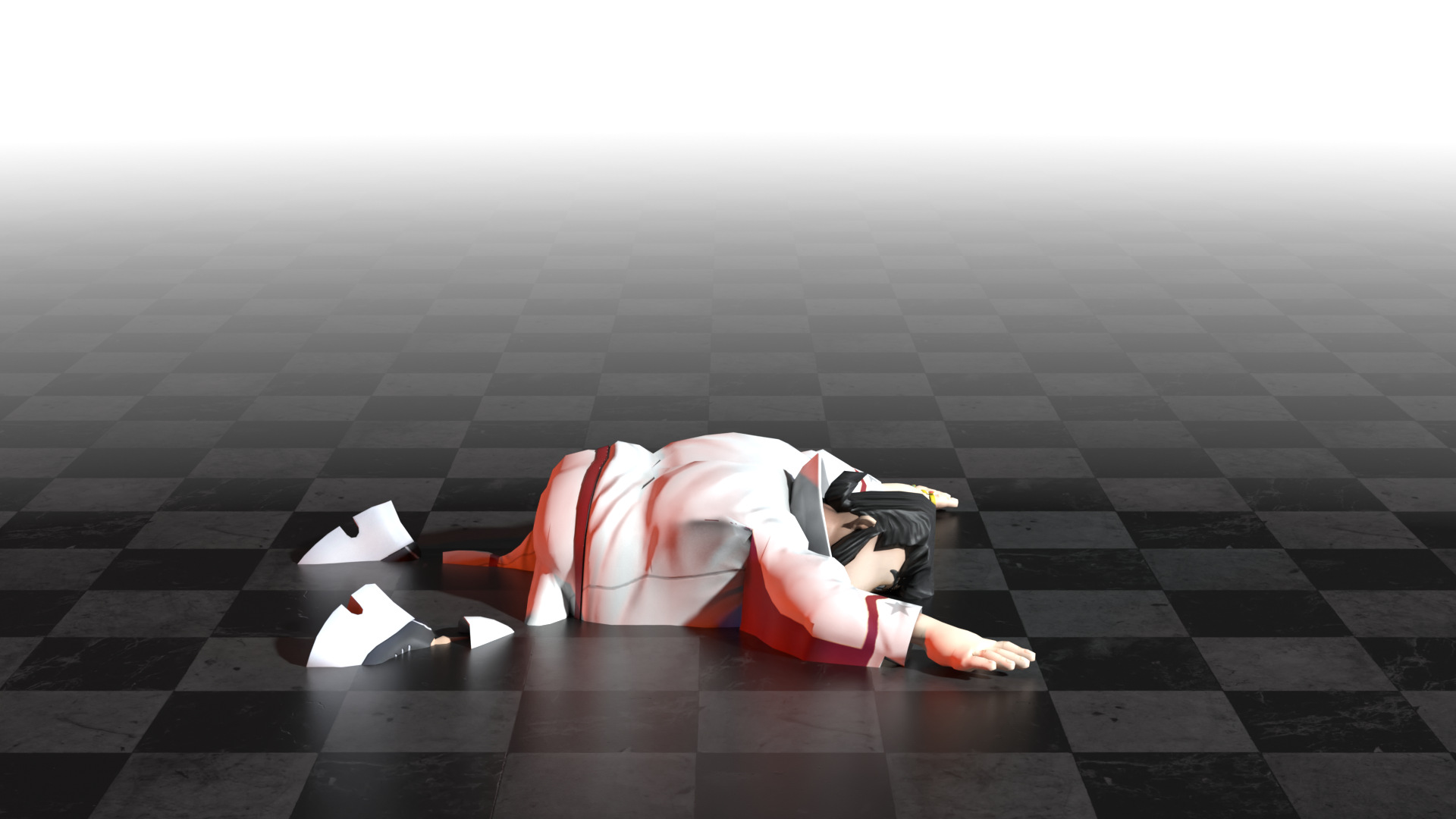}
        \caption{Penetration with the floor (legs, belly), hands floating above ground}
        \label{subfig:BigVegas}
    \end{subfigure}%
    \hfill%
    \begin{subfigure}{0.24\textwidth}
        \centering
        \includegraphics[trim={430px 90px 430px 60px},clip,width=.92\textwidth]{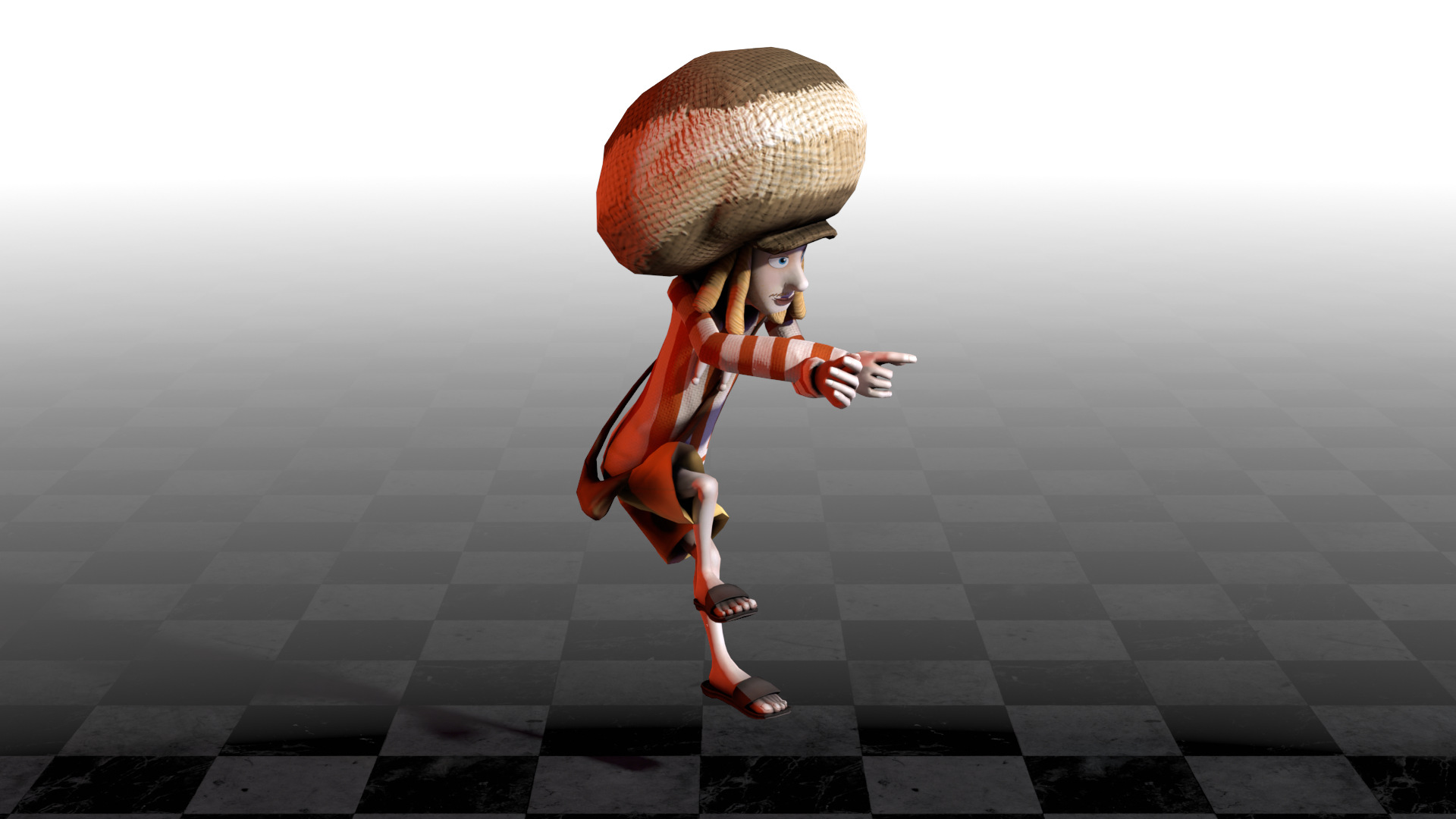}
        \caption{Colliding limbs (arms and legs) \\ $ $}
        \label{subfig:Kaya}
    \end{subfigure}%
    \hfill%
    \begin{subfigure}{0.24\textwidth}
        \centering
        \includegraphics[trim={530px 60px 360px 100px},clip,width=.92\textwidth]{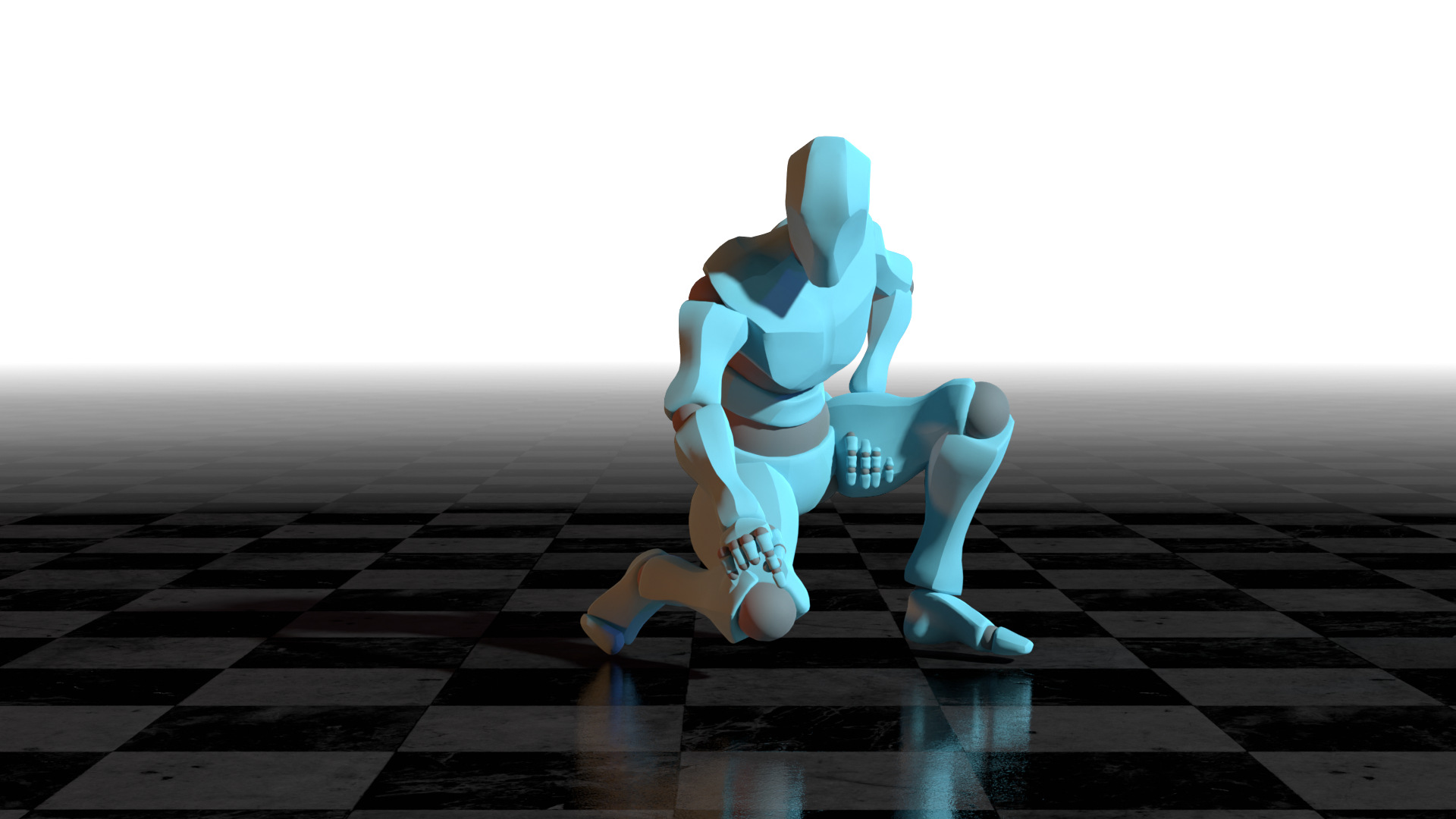}
        \caption{Self-collision on the thigh, feet floating above ground}
        \label{subfig:XBot}
    \end{subfigure}%
    \hfill%
    \begin{subfigure}{0.24\textwidth}
        \centering
        \includegraphics[trim={300px 60px 600px 90px},clip,width=.92\textwidth]{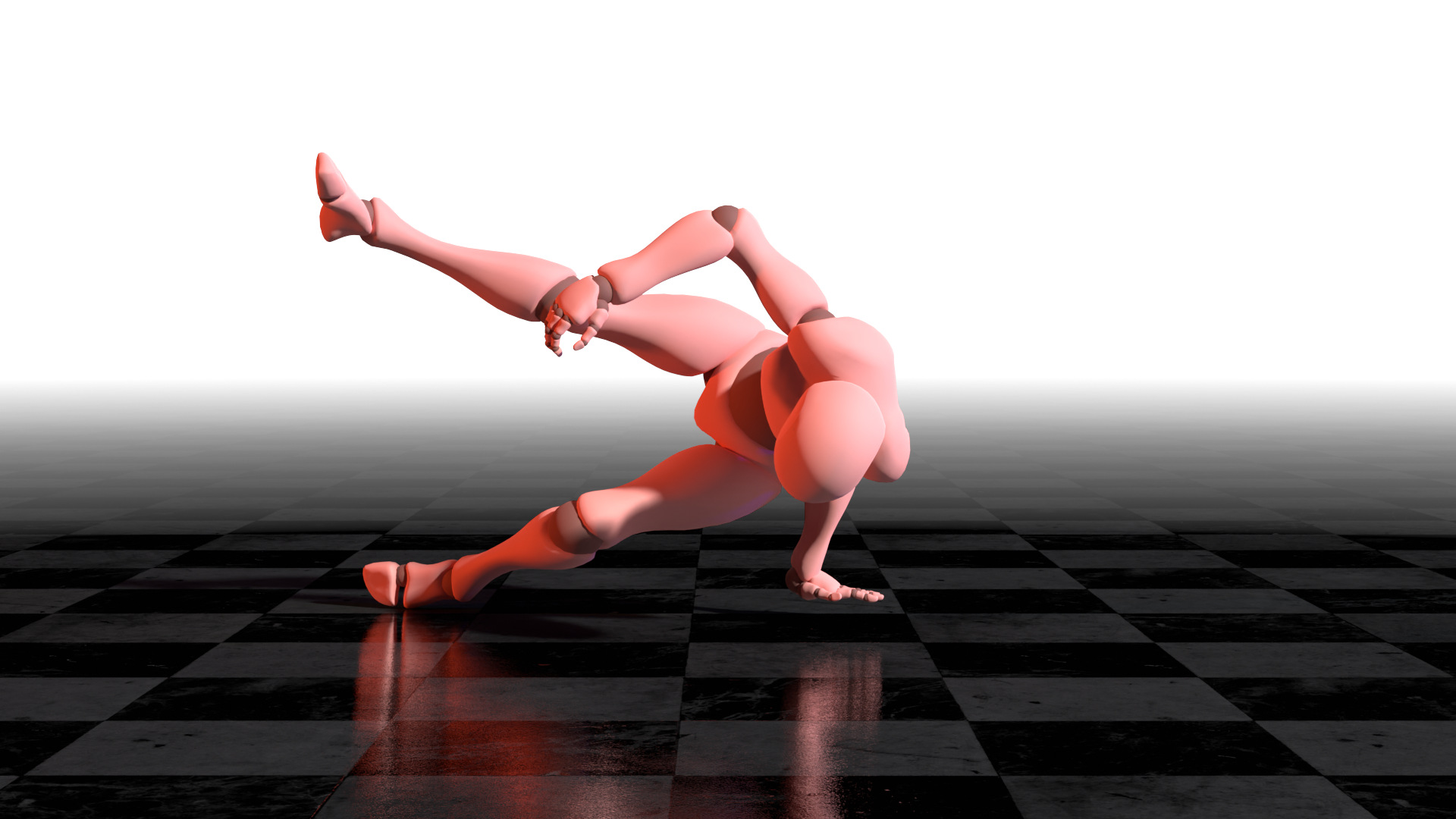}
        \caption{Hand floating above ground\\ $  $}
        \label{subfig:YBot}
    \end{subfigure}
    \vspace{-.6\baselineskip}
    \caption{Examples of issues found in NKN's dataset~\cite{nkn}. Figures \ref{subfig:BigVegas} and \ref{subfig:Kaya} are called ``ground-truth'' although they were retargeted using Mixamo, while figures~\ref{subfig:XBot} and~\ref{subfig:YBot} show issues without having undergone any retargeting (original source characters for those motions).}
    \label{fig:mixamo_shortcomings}
\end{figure*}

\subsection{Optimization-based retargeting}
A first category of methods formulate motion retargeting tasks as an optimization problem. 
Gleicher~\cite{gleicher} pioneered this approach 
with a solution requiring the manual identification of a number of spacetime constraints. Still only considering skeletal motion, the method was improved thanks to inverse kinematics (IK) solvers~\cite{lee},
the addition of constraints to the IK framework~\cite{choi}, and 
of dynamic constraints 
to improve the respect of physical laws~\cite{tak}.

Noting that character morphology should have a significant impact on motion, recent optimization-based methods 
addressed motion retargeting for skinned characters.
Jin et al.~\cite{taeil17} introduced the concept of Aura Mesh to retarget two characters whose meshes are interacting, and Basset et al.~\cite{basset20} 
optimized energy functions based on volume preservation and collision management. However, both methods are limited to parametric skin mesh models such as SMPL~\cite{smpl}, used to directly pair the vertices of the source and target meshes.
Ho et al.~\cite{ho2010} 
performed multi-character motion retargeting using an optimization method based on interacting joints and hard constraints to prevent collisions of bone-attached primitives (serving as character shape proxies).
However, this method cannot retarget motion to characters of arbitrary morphologies, as the iterative motion morphing step requires the source and target characters to have identical skeleton topology.


\subsection{Learning-based retargeting}
Other methods leveraged machine learning techniques 
for motion retargeting. Shon et al.~\cite{shon} modeled common latent representation of motion using Gaussian process regression, but this method required datasets of paired motion data for both characters, which are usually unavailable.
To alleviate this issue, Villegas et al.~\cite{nkn} introduced a cycle consistency loss for an adversarial unsupervised training framework. Lim et al.~\cite{pmnet} learned to disentangle pose (joints' local coordinates) from movement (the root bone's global trajectory). By encoding motion in the shared latent space of a common primal skeleton~\cite{san, Aberman_2019, Hu_2023, yan2024imitationnet}, retargeting was performed in deep feature space, thus increasing generalization to various skeletons. 
Yet, as these methods only considered the skeleton and not the mesh, 
they failed to capture self-contacts, despite them being a key component of motion.

In contrast, Villegas et al.~\cite{villegas} detected 
self-contacts and used geometry-conditioned RNN to preserve them while optimizing motion in a latent space.
Despite yielding good results, this method may not be adapted to industrial use cases as the model cannot handle arbitrary skeleton topology and is not suitable for real-time applications. Besides, comparison with this method is difficult as no implementation is available online.
More recently, R$^2$ET \cite{r2et} introduced a shape-aware module that improves self-contacts as a post-processing step on the target mesh, and avoids unwanted penetrations, in real-time. Their method can handle very diverse characters, but collisions on the source mesh are not taken into account, meaning that semantic information can be lost during retargeting.

Recent works \cite{zhang2024semanticsawaremotionretargetingvisionlanguage} focus on motion semantics by performing a differentiable rendering of the animated character's mesh, and encoding the video through a vision-language model. However, this complex setup has only been shown to perform on simple motions that are easily transcribed into a textual description (e.g. ``shrugging", ```waving", ``praying while standing up"), with little to no interaction between the limbs and minimal foot motion ; and requires the same skeleton between source and target characters.

Lastly, some recent works~\cite{reda23, yunbo23} used a combination of deep reinforcement learning and physically-based simulation to allow for a physically accurate retargeting, taking mesh contacts into account. The main issue is that of generalization: unlike previously mentioned methods, the model has to be retrained for each new target character, making it impossible to scale up to the diversity of virtual characters used in most 3D computer graphics applications. 


\subsection{Mesh correspondence}
Some methods \cite{zhou20unsupervised, wang2020neural, liao2022pose} aim to disentangle the pose and shape of a mesh to transfer poses to a target with a different shape. Because we have skeletal animations and a common standard pose amongst our characters, this disentanglement is trivial. 
However, we still need a way to compare shapes by computing a correspondence between the source and target meshes
(see~\cite{van2011survey} and ~\cite{recent_advances} for surveys on this topic). 
A recent family of solutions relied on optimal transport. Solomon et al.~\cite{solomon} used a Gromov-Wasserstein distance with an entropy term to find a dense mapping between 3D meshes. Mandad et al.~\cite{desbrun} increased efficiency by using a coarse-to-fine approach relying on diffusion geometry, while Schmidt et al.~\cite{schmidt2023surface} 
used an automatic triangulation to approximate the source and target surfaces. 
In this work, we take inspiration from these techniques to compute a correspondence between the source and target models.




\subsection{Retargeting datasets}
Recent works in the field of motion retargeting~\cite{pmnet, r2et, san} made use of the dataset introduced in NKN~\cite{nkn}, with 2400 motion sequences sampled from Mixamo~\cite{mixamo}. 
We visually assessed its quality, and expose two major shortcomings.
First, the variety of characters was obtained using Mixamo's in-house retargeting algorithm, despite many of these retargeted animations showing degraded motion semantics (where the intention behind a movement is completely lost), severe self-collisions, and ground collisions. Examples of such issues are shown in Figures~\ref{subfig:BigVegas} and \ref{subfig:Kaya}. Second, some of the original motions from Mixamo show severe flaws on their original character (before any retargeting), as illustrated by Figures~\ref{subfig:XBot} and ~\ref{subfig:YBot}.
Therefore, we claim that using this dataset as ground truth, with metrics such as the MSE of joint positions, is not an accurate way to estimate the quality of motion retargeting methods. It may even have impacted the quality of previous work: while it was used for unsupervised training by Villegas et al.~\cite{nkn}, some other methods used this dataset as evaluation data, training data \cite{pmnet, r2et} and even as ground-truth for supervised learning \cite{san}.


In contrast, we use a few quality metrics such as self-penetration, foot sliding, and jerk, to assess the quality of the animations we generate, while referring to
human feedback, through a user-study, to assess the overall semantic-preserving quality of our method.




\section{Shape 
and motion descriptors}
\label{section:motionsemantic}
Let us consider a \emph{source} and a \emph{target} humanoid characters, both with a skin-mesh rigged to a skeleton. Neither the skeletons or the meshes need to have the same topologies.
The skeletons may have different numbers of joints so long as a \emph{bone mapping}, i.e. a list of paired bones on both characters, is predefined; and the meshes can exhibit different numbers of vertices and large shape differences. 
In addition, an input, kinematic animation to be transferred is provided for the source character. 

To compute relevant, semantic-based motion transfer from source to target, some common representation for the characters' shapes and a set of motion descriptors for the input animation are required. Tracking the distance between relevant pairs of points was already used as a simple and robust way to encode the semantic of motion in the fields of robotics~\cite{handa2019dexpilot} and motion retargeting~\cite{ho2010, yunbo23}. 
Using points located on the skin mesh instead of skeletal joints captures finer semantic information and helps preventing collisions \cite{1013569}. 
Taking inspiration from these works, we propose a light morphological representation based on 
key-vertices, described in section \ref{section:shape_encoding}, and then use it to build a set of time-varying descriptors encoding the successive poses in the input animation. 


\subsection{Sparse shape encoding and correspondence}
\label{section:shape_encoding}

\begin{figure}[b!]
    \centering
    \begin{subfigure}{.23\textwidth}
        \centering
        \includegraphics[width=.95\textwidth]{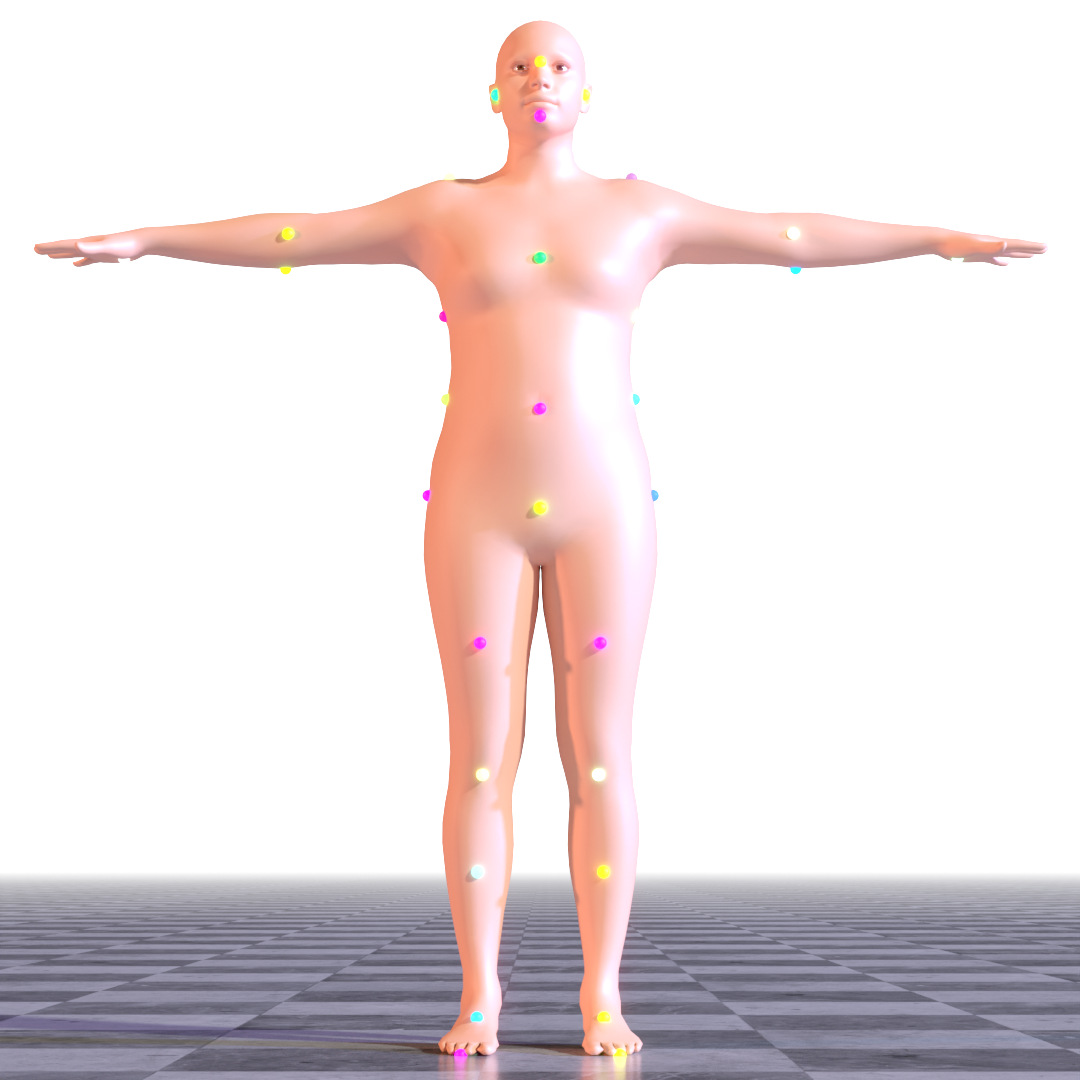}
    \end{subfigure}%
    \hfill%
    \begin{subfigure}{.23\textwidth}
        \centering
        \includegraphics[width=.95\textwidth]{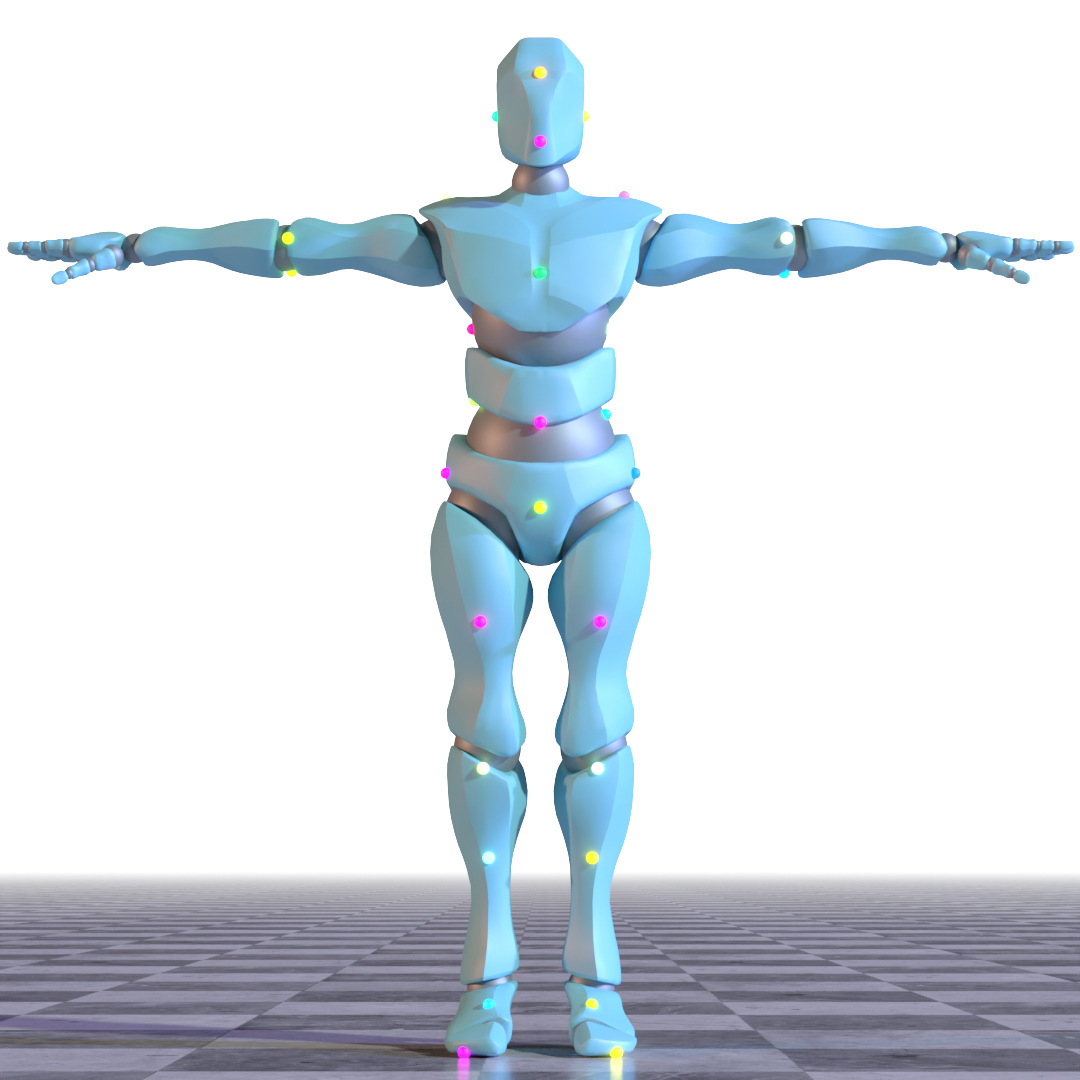}
    \end{subfigure}
    \begin{subfigure}{.23\textwidth}
        \centering
        \includegraphics[width=.95\textwidth]{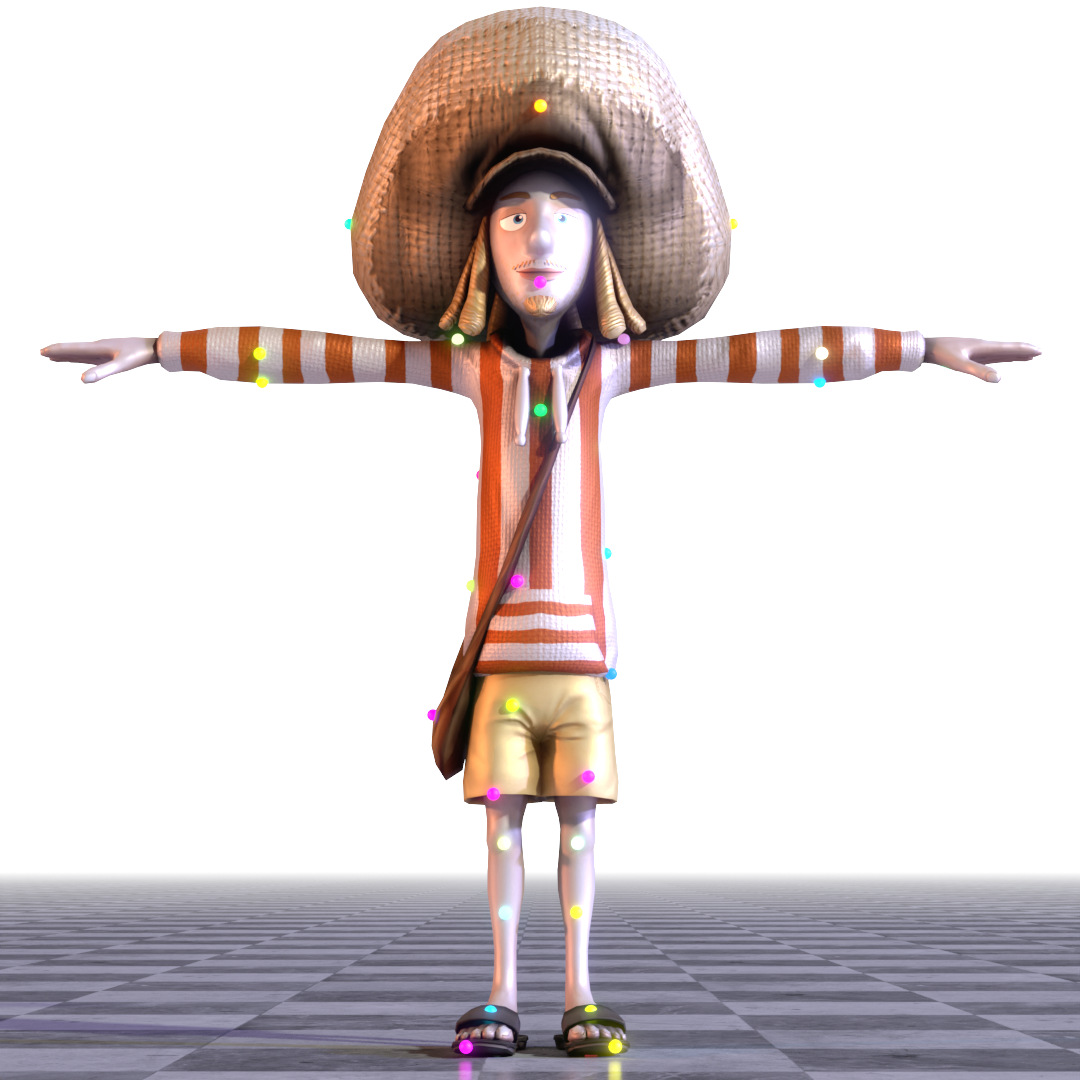}
    \end{subfigure}%
    \hfill%
    \begin{subfigure}{.23\textwidth}
        \centering
        \includegraphics[width=.95\textwidth]{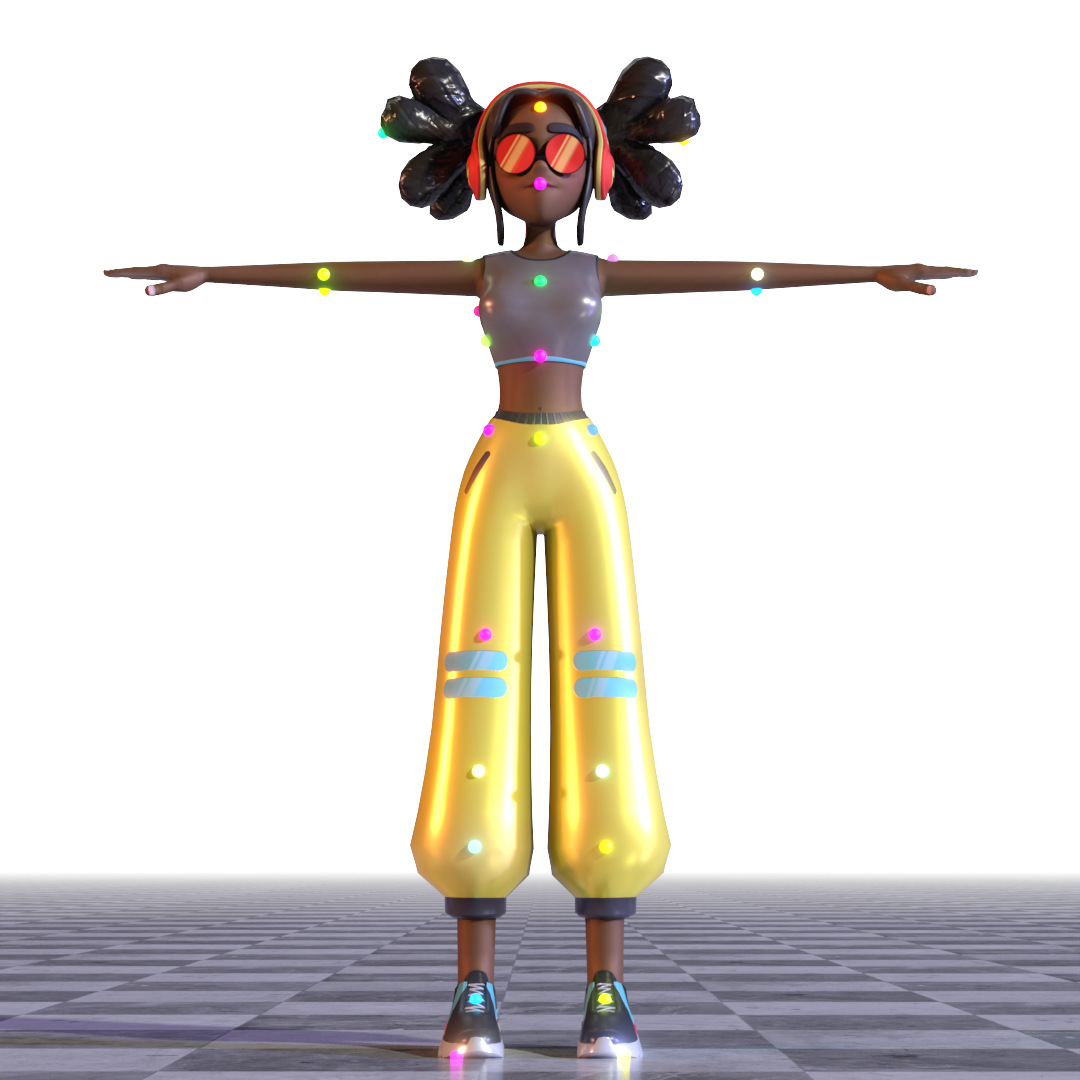}
    \end{subfigure}

    \begin{subfigure}{.23\textwidth}
        \centering
        \includegraphics[width=.95\textwidth]{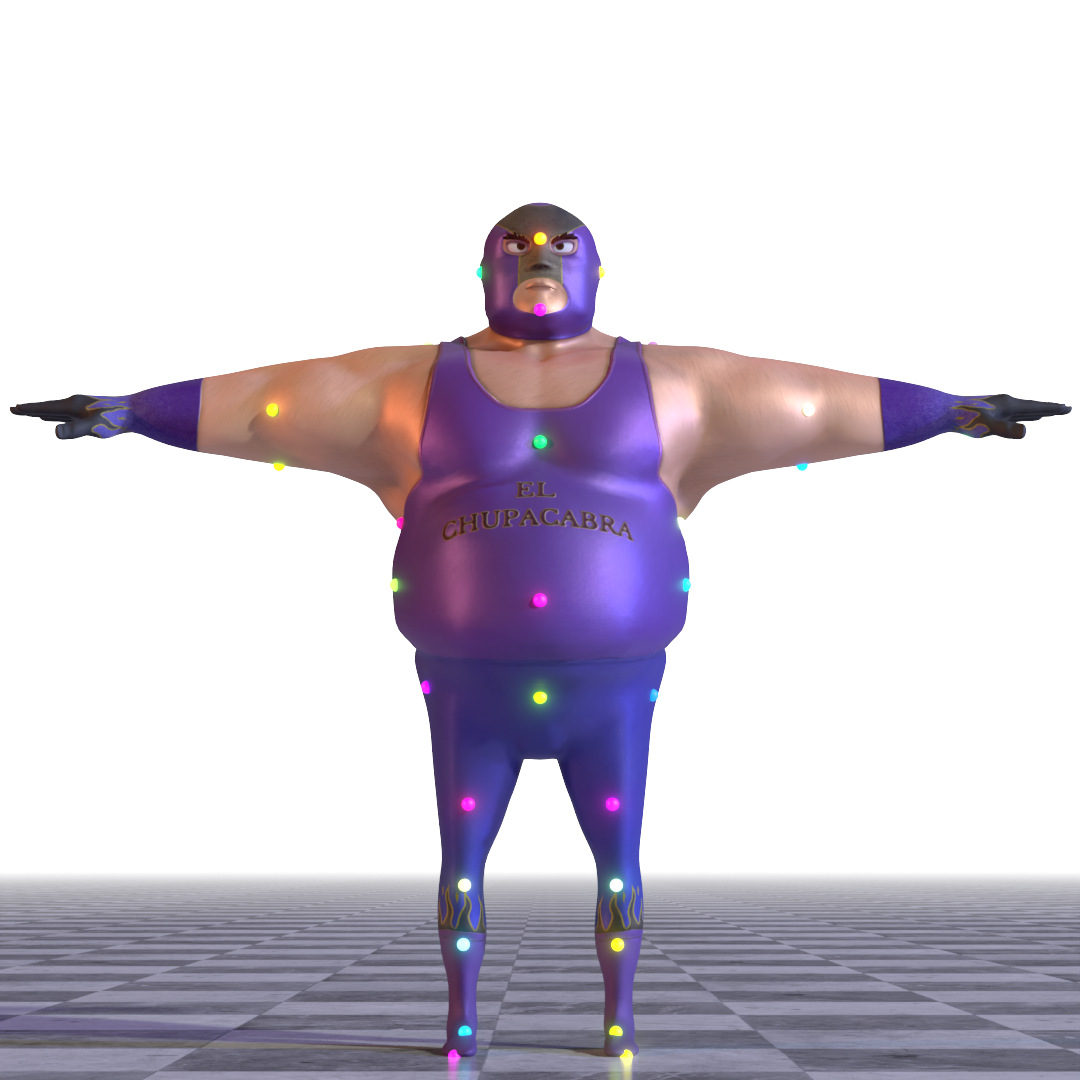}
    \end{subfigure}%
    \hfill%
    \begin{subfigure}{.23\textwidth}
        \centering
        \includegraphics[width=.95\textwidth]{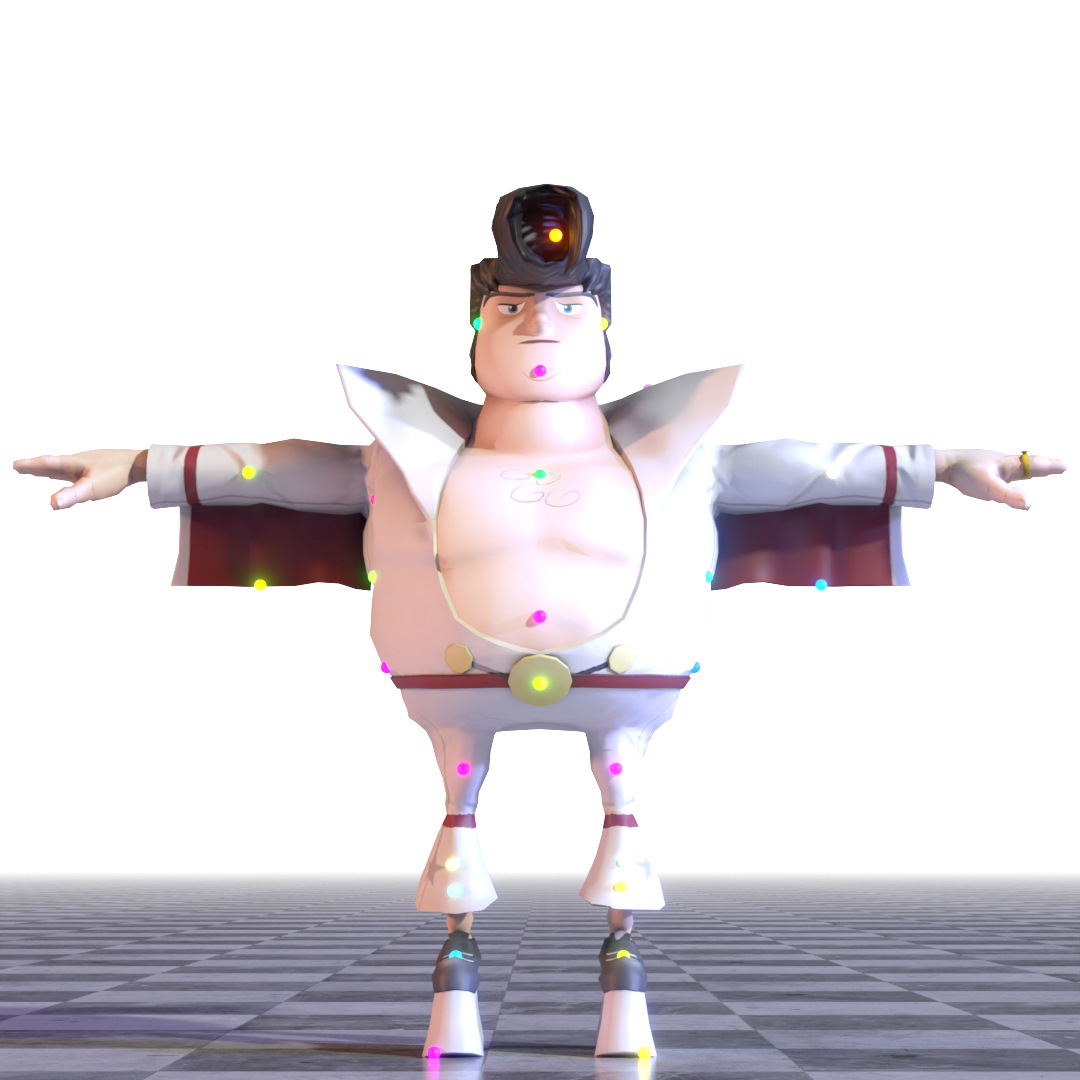}
        \label{fig:keypoints_results}
    \end{subfigure}

    \caption{Chosen location of 
    key-vertices on the template mesh (top-left), and results of the key vertices transfer to several character from Mixamo \cite{mixamo}. Key vertices are shown with corresponding colored
    spheres to show the automatic transfer to the different meshes.}
    \label{fig:keypoints}
\end{figure}

The ReConForM method uses a generic humanoid template mesh
(based on the SMPL model~\cite{smpl}), on which we pre-selected $N$ specific vertices, called 
\emph{key-vertices}
($N=41$ unless mentioned otherwise, see Figure~\ref{fig:keypoints}).
 The latter are chosen as to provide (i) a sparse yet comprehensive coverage of the character's surface, 
 and (ii) a good sampling of typical areas prone to contact, such as hands and feet.
 While our method was developed for humanoid characters, its principle could easily be extended to different categories of characters, just by changing this template model.

To provide a sparse one-to-one correspondence between the source and target model, we automatically transfer the template's keypoints to their two skin meshes, as follows.
Taking inspiration from optimal transport 
between arbitrary meshes~\cite{solomon}, we view the task as an optimization problem.
We input both the template and destination (i.e., source or target) meshes in their T-pose. Using the skinning weights, we first split the mesh into its various limbs (arms, legs, torso, head, feet and hands). For each limb, we consider the position of all $N^l$ vertices, and normalize them to have zero mean and unit variance.
For each limb $l$, we convert the template and destination point-clouds into two distributions : 
\begin{equation*}
    \mathcal{T} = \frac{1}{N^l_t} \sum\limits_{i=1}^{N^l_t} \delta_{v^l_{i, t}} \ \ \text{   and   } \ \ \mathcal{D} = \frac{1}{N^l_d} \sum\limits_{j=1}^{N^l_d} \delta_{v^l_{j, d}} 
\end{equation*}
where $v^l_{i, t}$ and $v^l_{j, d}$ are, respectively, the normalized position of the $i$-th and $j-th$ vertices of limb $l$ for the template and destination meshes, and $\delta$ is the Dirac distribution.

The constraints of our setup (namely, that the meshes are in T-pose and normalized) allow us to use a simple euclidian distance as criterion, instead of relying on the Gromov-Wasserstein distance used in Solomon et al.~\cite{solomon}. Our goal is to find an optimal transport plan, which is a matrix $P^l \in \mathbb{R}_+^{N^l_t \times N^l_d}$ defining a coupling between the two distributions $\mathcal{T}$ and $\mathcal{D}$.

In order to account for an uneven distribution of vertices, we weight each vertex by the inverse of the local vertex density : 
\begin{equation*}
w^l_i \ = \ \sum\limits_{f \in \mathcal{F}_i} \frac{A(f)}{n(f)}
\end{equation*}
where $\mathcal{F}_i$ is the set of faces containing vertex $i$, $A(f)$ the area of face $f$, and $n(f)$ its number of vertices. 

We solve the following optimization problem:
\begin{align*}
    {P^l}^* \ \ = \ \ & \underset{P^l}{\operatorname{argmin}} \ \langle P^l \, , \, C^l \rangle\ \  = \ \ \underset{P^l}{\operatorname{argmin}} \ \sum\limits_{i = 1}^{N^l_t} \sum\limits_{j = 1}^{N^l_d} P^l_{i,j} C^l_{i,j} \\[.2\baselineskip]
    &\text{s.t. } \ \ \mathds{1} \cdot P^l \ =\ \left(w^l_{0, d} \ ; \ \cdots \ ; \ w^l_{N^l_d, d}\right) \\[.0\baselineskip]
        & \hspace{.63cm} P^l \cdot \mathds{1}^T \ =\ \left(w^l_{0, t} \ ; \  \cdots  \ ; \ w^l_{N^l_t, t}\right)^T
\end{align*}
where $C^l_{i,j} = \left\lVert v^l_{i, t} - v^l_{j, d} \right\rVert^2$, and $\mathds{1}$ is a vector filled with ones.

Similarly to Feydy et al.~\cite{feydy2019interpolating}, we approximate this optimal transport plan by solving the entropy regularized version of the problem, which is faster to compute using Sinkhorn's algorithm \cite{cuturi2013sinkhorndistanceslightspeedcomputation}. Thus, for each key-vertex on the template, we are thus able to get a corresponding vertex on the destination mesh, which effectively gives us the position of the key-vertices for both the source and target characters.

Examples of key-vertex transfer is shown in Figure~\ref{fig:keypoints}, and results on additional characters are shown in the Supp. Mat, along with results for our extended template with $N=96$ keypoints. Note that the output of this automatic transfer can be overridden, or further refined if needed via interactive authoring. 
This enables the user to fine-tune key-vertex positions on the source, target, or even on the template shape, to account for very large morphological differences, or to highlight specific areas where careful sampling needs to be considered.

\subsection{Motion encoding via time-varying pose descriptors}
\label{section:pose_descriptors}


Using the key-vertices defined before, time-varying pose descriptors can be computed on the source animation to extract the semantic elements, in order to transfer them to the target character. Yet, not all descriptors convey the same amount of semantic information: it is necessary to adaptively focus on the pairs of key-vertices interacting with each other or with the ground.


Given an animated character, we call $p_{i}$ (or $p_{i,t}$ explicitly when necessary) the 3D position of the $i^{th}\in\llbracket1,N\rrbracket$ key-vertex at time $t$, and $n_i$ its normal to the mesh surface. 
We compute three time-varying descriptors $\mathcal{M}_\text{dist}$, $\mathcal{M}_\text{dir}$, and $\mathcal{M}_\text{pen}$, 
each stored in a matrix-like structure representing
the pairwise relationship between all possible pairs of key-vertices $(i,j)\in\llbracket1,N\rrbracket^2$ of a character. 

These descriptors respectively encode:
\begin{itemize}
    \item  The distances between all pairs of key-vertices,
    $$\mathcal{M}_\text{dist} \in \mathbb{R}^{N\times N} \text{,   where } \mathcal{M}_\text{dist}(i,j) = \left\lVert p_j - p_i\right\rVert;$$
    \item  The vector,
    and therefore the offset direction, between pairs of key-vertices, 
    $$\mathcal{M}_\text{dir} \in \mathbb{R}^{N\times N \times 3} \text{,   where } \mathcal{M}_\text{dir}(i,j) = p_j - p_i;$$
    \item 
    The signed distance between two key vertices $i$ and $j$, computed along the mesh normal at vertex $i$, which locally serves as a measure of penetration,
    $$\mathcal{M}_\text{pen} \in \mathbb{R}^{N\times N} \text{,   where } \mathcal{M}_\text{pen}(i,j) = n_i \cdot (p_j - p_i).$$
\end{itemize}
Two additional descriptors, $\mathcal{M}_\text{height}$ and $\mathcal{M}_\text{sliding}$, are used to represent the relationship between individual key-vertices and the environment, considered to be an horizontal ground floor at this stage.
Calling $\overrightarrow{up}$ the unit vertical direction, these two descriptors are stored in vector-like containers encoding:
\begin{itemize}
    \item The height of each key-vertex with respect to the ground,
    $$\mathcal{M}_\text{height} \in \mathbb{R}^{N\times 1} \text{,   where } \mathcal{M}_\text{height}(i) = \overrightarrow{\text{up}} \cdot p_i;$$
    \item The horizontal velocity of each key-vertex,
    \begin{gather*}
    \mathcal{M}_\text{sliding} \in \mathbb{R}^{N\times 2} \text{,   where }  
    \mathcal{M}_\text{sliding}(i) = \frac{H(p_{i,t+\Delta t})-H(p_{i,t})}{\Delta t} \\
    \mbox{and }\; \text{H}(p_i)=p_i-(\overrightarrow{\text{up}} \cdot p_i)\, \overrightarrow{\text{up.}}
    \end{gather*}
\end{itemize}
\droff{}

\section{Adaptive 
weighting of constraints}

\begin{figure*}[t]
    \centering 
    \includegraphics[width=\linewidth, trim={0 0 0 0}, clip]{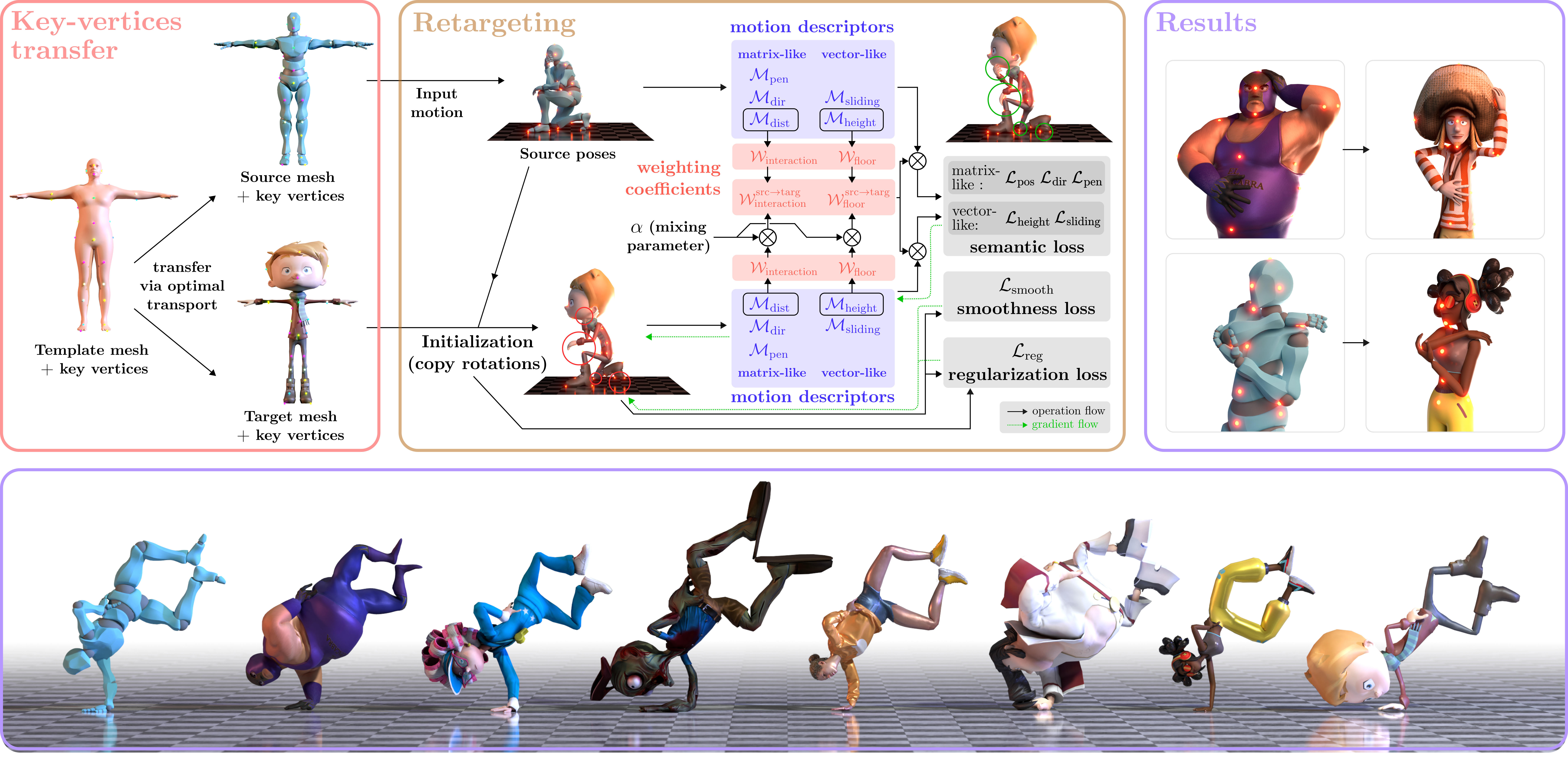}
    \centering
    \caption{Top: graphical representation of our method (left: key-vertices transfer presented in Section \ref{section:shape_encoding}; middle: motion retargeting process presented in Sections \ref{section:pose_descriptors} through \ref{section:sparse_objective}). Top-right shows two poses being retargeted onto different characters, with key-vertices shown in red. Bottom shows a complex pose showcasing collisions with the floor as well as self-collisions (the leftmost character is the source character, all the others are target characters from Mixamo).}
    \label{fig:method}
\end{figure*}

%
%
The motion descriptors defined above are computed at each time on the source animation and used as constraints to be matched on the target animation. 
Trying to maintain all these constraints at once might however be unfeasible when the source and target morphologies are different, as they would conflict with one another. 
Instead, our insight is that, depending on the input animation, only a few of these descriptors are significant at each time step to the perceived semantics of motion. We therefore introduce a new, dynamic weighting 
formulation to select only the relevant ones over time and weight them accordingly.

\subsection{Weighting coefficients}
\label{section:weightingcoeffs}


Our descriptor weighting method exploits the notion of spatial proximity between pairwise key-vertices (indicating an interaction between limbs), and of height relative to the ground (indicating a likely occurrence of floor contact).
We therefore make use of two weighting characteristics, also stored in matrix/vector-like structures with values in \([0,1]\), obtained via a clamped normalized representation of the distances between, and heights of key-vertices:
\begin{equation*}
\begin{array}{rl}
\mathcal{W}_\text{interaction}(i,j) &\displaystyle = \text{clamp}\left(1-\frac{\mathcal{M_\text{dist}}(i,j) - d_\text{min}}{d_\text{max} - d_\text{min}}\right)
\\
\mathcal{W}_\text{floor}(i) &\displaystyle = \text{clamp}\left(1 - \frac{\mathcal{M_\text{height}}(i) - h_\text{min}}{h_\text{max} - h_\text{min}}\right)
\end{array}
\end{equation*}
where \(\text{clamp}(x)=\min(1,\max(0,x))\) and $d_\text{min}$, $d_\text{max}$, and $h_\text{min}$, $h_\text{max}$ are per-character constant thresholds characterizing typical distances 
for which nearby limb interaction or ground contact should be considered. For a character of height $h_c$, we typically set $d_{\min}=h_{\min}=5\%\,h_c$, and $d_{\max}=h_{\max}=15\%\,h_c$.

These weights provide a measure of the importance of each constraint, at each animation frame. 
More precisely,
\(\mathcal{W}_\text{interaction}(i,j)\) (resp. \(\mathcal{W}_\text{floor}(i)\)) will be close to 1 when a constraint between two interacting limbs (resp. between a limb and the floor) must be taken into account, and be 0 for unimportant relationships to be discarded. Furthermore, even though a pair of key-vertices may interact at a specific animation frame, the associated weight decreases back to 0 as soon as they spread apart. The resulting sparsity of constraints, in space and time, is a key-advantage for the efficiency of our solution.

\pgfmathdeclarefunction{customf}{1}{%
  \pgfmathparse{min(1,max(0,1-(#1-0.09)/(0.27-0.09)))}%
}



\subsection{Sparse objective to be optimized}
\label{section:sparse_objective}

We formulate the retargeting task as an optimization problem over the skeletal poses variables defined by the local joint rotation $\textbf{q}$ and root joint position in world space $p$, which are related to the key-vertex positions $p_i$ and normal $n_i$ through linear-blend skinning.

The optimization starts with a rough initialization of $\textbf{q}$ and $p$ through a naive retargeting, obtained by copying the rotation of the source bones onto their equivalent target bones using the \textit{bone mapping} mentioned in Section ~\ref{section:motionsemantic}.

We then optimize ($\textbf{q}$, $p$) over the entire animation at once to minimize the combination of the following three losses :
\begin{equation}
\begin{split}
  \min\limits_{\textbf{q},p} \quad & w_\text{reg} \mathcal{L}_\text{reg} + w_\text{smooth} \mathcal{L}_\text{smooth} + w_\text{sem} \mathcal{L}_\text{sem}
\end{split}
\end{equation}

The regularization loss $\mathcal{L}_\text{reg}=\sum_{i \in \llbracket1,N\rrbracket} \sum_{t} \|p_{i,t}-p_{i,t}^{\text{init}}\|^2$ is a simple Mean Squared Error (MSE) comparing the key-vertices' positions to that of the naive retargeting $p_{i,t}^{\text{init}}$, ensuring that the solution stays in the neighborhood of the initialization.

The smoothness loss $\mathcal{L}_\text{smooth} = \sum_{i \in \llbracket1,N\rrbracket} \sum_t \lVert \dddot{p}_{i,t} \rVert$ ensures that the change in acceleration, or jerk, is as small as possible.

The semantic loss $\mathcal{L}_\text{sem}$ 
is defined as the weighted sum of the following terms (where $\circ$ denotes the element-wise product):
\newcommand{\stot}{\text{src}\to\text{targ}}
\begin{itemize}
    \item A term ensuring that the distance between pairs of key-vertices is preserved:
    $$ \mathcal{L}_\text{dist} = \sum_t \left\lVert \mathcal{W}_\text{interaction}^{\stot} \circ \left(\mathcal{M}^\text{source}_\text{dist} - \mathcal{M}^\text{target}_\text{dist} \right)\right\rVert^2$$

    \item A term ensuring that the direction between pairs of key-vertices is preserved (with $S_\text{cosine}$ being the element-wise cosine similarity):
    $$ \mathcal{L}_\text{dir} = \sum_t \left\lVert \mathcal{W}_\text{interaction}^{\stot} \circ S_\text{cosine}\left(\mathcal{M}^\text{source}_\text{dir}, \mathcal{M}^\text{target}_\text{dir} \right)\right\rVert^2$$

    \item A term matching the amount of penetration of pairs of interacting key vertices, between source and target:
    $$ \mathcal{L}_\text{pen} = \sum_t \left\lVert \mathcal{W}_\text{interaction}^{\stot} \circ \left(\mathcal{M}^\text{source}_\text{pen} - \mathcal{M}^\text{target}_\text{pen} \right)\right\rVert^2$$
    
    \item A term penalizing key-vertices penetrating the floor, and encouraging similar heights on source and target for key-vertices close to the ground:
    $$ 
    \begin{array}{ll}
    \mathcal{L}_\text{height} = & \displaystyle \sum_t \left\lVert \text{min}\left(0, \mathcal{M}^\text{target}_\text{height} \right)\right\rVert^2 + \\ & \displaystyle
    \sum_t
    \left\lVert \mathcal{W}_\text{floor}^{\stot} \circ \left(\mathcal{M}^\text{source}_\text{height} - \mathcal{M}^\text{target}_\text{height} \right)\right\rVert^2
    \end{array}
    $$
    
    \item A term preventing key-vertices that are in contact with the ground from sliding across consecutive frames:
    $$\mathcal{L}_\text{sliding} = \sum_t \left\lVert \mathcal{W}_\text{floor}^{\stot} \circ \left(\mathcal{M}^\text{source}_\text{sliding} - \mathcal{M}^\text{target}_\text{sliding} \right)\right\rVert^2$$\\[-1.2\baselineskip]
\end{itemize}

The matrices $\mathcal{W}_\text{interaction}^{\stot}$ and  $\mathcal{W}_\text{floor}^{\stot}$ are a combination of the weighting matrices of the source and target characters:
\begin{equation}
\begin{array}{ll}
    \mathcal{W}_\text{interaction}^{\stot} &\displaystyle = \mathcal{W}^\text{source}_\text{interaction} + \alpha \;\text{Sg}\left(\mathcal{W}^\text{target}_\text{interaction}\right)\;, \\ \\
    \mathcal{W}_\text{floor}^{\stot} &\displaystyle = \mathcal{W}^\text{source}_\text{floor} + \alpha \, \text{Sg}\left(\mathcal{W}^\text{target}_\text{floor}\right)\;,
\end{array}
\end{equation}
where the mixing parameter $\alpha$ increases linearly from 0 to 1 during the optimization process. Both $\mathcal{W}^\text{target}_\text{interaction}$ and $\mathcal{W}^\text{target}_\text{floor}$ 
are weighting matrices computed from the target's motion descriptors at the current optimization step.
This evolution from source to target weighting helps to guide the numerical optimization toward a correct solution during the first steps, and prevent false positive interactions (i.e., ending up with contacts that were not present in the source motion) for later steps. $\text{Sg}()$ is the stop-gradient operation, through which gradient flow is prevented during the optimization to reduce computational time and selectively prevent the optimization process from attempting to minimize unrelated parameters.

The terms of $\mathcal{L}_\text{sem}$ were weighted manually to give each loss a roughly equal contribution at the start of the optimization. The values of the hyperparameters, which yield satisfying results regardless of the animation and characters, are provided in supplementary material.

\section{Results}

\subsection{Implementation}

We implemented our ReConForM method using the Pytorch framework to leverage the differentiability of the losses relative to the variables $(\textbf{q},p)$.
We optimized the target poses using Adam \cite{adam}.

\subsection{Evaluation data}
\label{sec:evaluation}

We tested our method using animations and characters from Mixamo \cite{mixamo}. 
The evaluation sets provided by previous works \cite{r2et, nkn, san} lack quality even on the source character.
Instead, we manually selected 45 animations from Mixamo \cite{mixamo} with accurate contacts and asked a 3D animation expert to annotate their estimated difficulty to be retargeted based on the amount of floor contacts and self-collisions present in them: 24 of them were quantified as \emph{Hard}, 21 as \emph{Easy}.

\subsection{Quantitative results}

\subsubsection{Inference time}
\begin{figure}[!hb]
    \begin{minipage}{.48\columnwidth}
        \centering
        \scalebox{.75}{\begin{tikzpicture}
\begin{axis}[
    title={},
    width=1.45\columnwidth,
    height=5.5cm,
    xlabel={\shortstack{Duration of source animation (s) \\ \phantom{0}}},
    xlabel style={align=center},
    ylabel={\shortstack{Duration of retargeting (s) \\ \phantom{0}}},
    legend pos=south east,
    xmin=0, xmax=85,
    ymin=0, ymax=32,
    grid=both,
    legend style={nodes={scale=0.68, transform shape}}, 
    y label style={at={(axis description cs:0.125,.5)}}
]

\addplot[
    only marks,
    mark=*,
    mark options={draw=blue,fill=blue}
]
coordinates{
(4.04,3.87200331687927)
(7.56,4.79238843917846)
(10.68,6.02954316139221)
(4.04,4.05491375923156)
(6.2,4.38879418373107)
(4.04,3.86439490318298)
(7.04,4.68316221237182)
(14.88,7.22319650650024)
(3.36,3.81598758697509)
(23.04,10.1149051189422)
(7.56,4.5926194190979)
(10.68,5.71096491813659)
(6.2,4.26563096046447)
(7.04,4.5237283706665)
(14.88,7.11200785636901)
(4.16,3.89491581916809)
(15.44,7.74832272529602)
(49.64,18.4637961387634)
(30,12.390343427658)
(78.92,29.2441575527191)
(62.8,24.236147403717)
(36.88,14.9805083274841)
(24.32,10.7317581176757)
(63.16,23.5052063465118)
(16.72,8.29764723777771)
(65.16,24.3001554012298)
(59.84,22.3504650592803)
(52.92,19.8861153125762)
(64.84,24.0680136680603)
(63.76,23.6077966690063)
(59.44,22.1923902034759)
(52.92,19.8603155612945)
(72.04,26.4954731464386)
(56.72,21.2447750568389)
(27.52,11.4237840175628)
(9.36,5.58571887016296)
(9.04,5.37753796577453)
(4.76,4.09439373016357)
(12.04,6.26829361915588)
(5,4.24791145324707)
(7.16,4.85827350616455)
(4.28,4.18254041671752)
(9.8,5.64722752571106)
(4.92,4.30693697929382)
(12.68,6.64022660255432)
(10.68,6.0418803691864)
(12.2,6.4834713935852)
(9.6,5.75861668586731)
(9.76,5.799649477005)
(10.04,5.91278028488159)
(11.24,6.37373256683349)
(8.4,5.56686353683471)
(12.64,7.05557751655578)
(10,5.97793769836425)
(46,18.2509372234344)
(47.96,19.4702570438385)
(42.04,16.7281250953674)
(16.08,7.98792934417724)
(44.04,17.2894029617309)
(35.96,14.6223828792572)
(20.4,9.27856469154357)
(42.56,16.9192836284637)
(31.32,12.8573677539825)
(26.6,11.0249783992767)
(37.68,14.6139779090881)
(22.2,9.65031576156616)
(45.8,17.3298373222351)
};

\addlegendentry{Measurements}

\addplot[
    domain=0:90, 
    samples=2, 
    color=green,
    thick,
]
{x};

\addlegendentry{Real-time}

\addplot[
    domain=0:90, 
    samples=2, 
    color=red,
    thick,
]
{0.335*x + 2.45};

\addlegendentry{Trend line {\footnotesize ($r^2 = 0.999$)}}


\end{axis}
\end{tikzpicture}\hspace{.1cm}}
    \end{minipage}%
    \hfill%
    \begin{minipage}{.48\columnwidth}
        \centering
        \scalebox{.75}{\begin{tikzpicture}
\begin{axis}[
    title={},
    width=1.45\columnwidth,
    height=5.5cm,
    xlabel={\shortstack{Duration of source animation \\ ($\times 10^3$ frames)}},
    xlabel style={align=center},
    ylabel={\shortstack{Framerate (Hz) \\ \phantom{0}}},
    legend pos=south east,
    xmin=0, xmax=2000,
    ymin=0, ymax=80,
    grid=both,
    legend style={nodes={scale=0.68, transform shape}}, 
    y label style={at={(axis description cs:0.125,.5)}},
    xtick={0,500,1000,1500,2000},    
    xticklabels={0,0.5,1,1.5,2},     
    ytick={0,20,40,60}    
]

\addplot[
    only marks,
    mark=*,
    mark options={draw=blue,fill=blue}
]
coordinates{
(101,26.0846883988217)
(189,39.4375377535965)
(267,44.2819618092509)
(101,24.9080513167659)
(155,35.3172177849154)
(101,26.1360452361661)
(176,37.5814443358484)
(372,51.5007448108649)
(84,22.0126502210628)
(576,56.9456651571871)
(189,41.152985421362)
(267,46.7521695242909)
(155,36.3369455624737)
(176,38.9059610964372)
(372,52.3059039743415)
(104,26.7014756745662)
(386,49.8172331851669)
(1241,67.2126138456767)
(750,60.5310098447984)
(1973,67.4664673257634)
(1570,64.7792726231404)
(922,61.5466431341616)
(608,56.6542772706176)
(1579,67.1766066088727)
(418,50.3757255546995)
(1629,67.036608330396)
(1496,66.9337302840074)
(1323,66.5288307547588)
(1621,67.3508010406012)
(1594,67.5200664572267)
(1486,66.9598896908029)
(1323,66.6152557302955)
(1801,67.973875765343)
(1418,66.745823206235)
(688,60.2252282555654)
(234,41.8925487370927)
(226,42.0266674895431)
(119,29.0641320406785)
(301,48.0194480807576)
(125,29.4262254229548)
(179,36.8443645202089)
(107,25.5825382038924)
(245,43.384120594495)
(123,28.5585790066906)
(317,47.7393346603651)
(267,44.1915403293485)
(305,47.0427000421055)
(240,41.6766756830687)
(244,42.0715081087977)
(251,42.4504189072919)
(281,44.0871964823592)
(210,37.7232167827496)
(316,44.7872621707453)
(250,41.8204425362961)
(1150,63.0104627461755)
(1199,61.5811079073264)
(1051,62.8283202097203)
(402,50.325933377595)
(1101,63.6806257819891)
(899,61.4810874139597)
(510,54.9653978771966)
(1064,62.8868233055688)
(783,60.8989347572694)
(665,60.3175784946325)
(942,64.4588356339441)
(555,57.5110715247652)
(1145,66.0710183661623)
};

\addlegendentry{Measurements}

\addplot[
    domain=0:2000, 
    samples=2, 
    color=green,
    thick,
]
{25};

\addlegendentry{Real-time}

\addlegendentry{%
    $\pgfmathprintnumber{\pgfplotstableregressiona} \cdot x
    \pgfmathprintnumber[print sign]{\pgfplotstableregressionb}$}

\end{axis}
\end{tikzpicture}}
    \end{minipage}
    \vspace{-.7\baselineskip}
    \caption{Speed and framerate of our retargeting method}
    \label{fig:speed}
\end{figure}
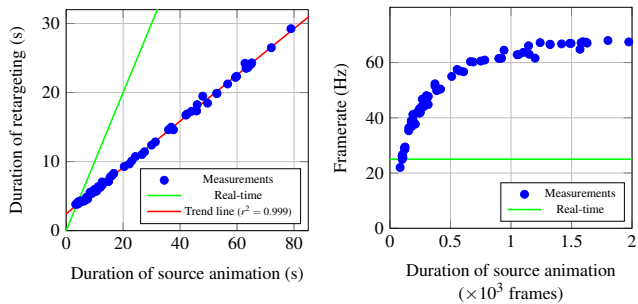

Our optimization objectives being very sparse, our method is able to consistently run in real-time (for motions longer than 3 seconds) on a laptop with a Nvidia RTX 3060 GPU, with an asymptotic speed of 67 frames per second.
This makes it suitable for use in video games or motion capture pre-visualization. Since all frames are processed at once in a batched fashion, our method reaches higher framerates for longer animations. A three-second batch (around 75 frames) gives a good compromise between speed and delay for use cases where an online, on-the-fly retargeting is necessary, such as motion capture visualization.
Figure \ref{fig:speed} shows the inference speed for animations of various durations.

\begin{table*}[t!]
\centering

\sisetup{detect-all}
\definecolor{Green}{HTML}{00A64F}
\definecolor{NavyBlue}{HTML}{006EB8}

\small
\begin{tabular}{
  @{\extracolsep{\fill}}
  l{c}
  c{c}
  c{c}
  c{c}
  c{c}
  c{c}
  c{c}
  c{c}
  c{c}
  c{c}
  c
  @{}
}

\toprule
 & \multicolumn{2}{c}{Jerk (m.s$^{-3}$) $\downarrow$} & \multicolumn{2}{c}{Self-penetration (\%) $\downarrow$} & \multicolumn{2}{c}{Floor penetration (\%) $\downarrow$}  & \multicolumn{2}{c}{Grounded feet} & \multicolumn{2}{c}{Foot sliding} \\
\cmidrule(lr){2-3} \cmidrule(lr){4-5} \cmidrule(l){6-7}  \cmidrule(l){8-9}  \cmidrule(l){10-11}
& {mean} & {max} & {mean} & {max} & {mean} & {max}  & {F1 $\uparrow$} & {ROC AUC $\uparrow$}  & {F1 $\uparrow$} & {ROC AUC $\uparrow$} \\
\midrule
\midrule

 Source              &   254.9    & 3980.2   & \(1.68\tcdot 10^{-8}\) & \(1.31\tcdot 10^{-7}\) & \(1.54\tcdot 10^{-3}\) & \(9.08 \tcdot 10^{-3}\) & --& -- & -- & -- \\
\midrule
\midrule

Mixamo & 281.2 & 6495.44 & \(5.65\tcdot 10^{-8}\) &  \(2.84\tcdot 10^{-7}\) & \(6.56\tcdot 10^{-3}\) & \(3.31\tcdot 10^{-2}\) &  0.944 & 0.908 & 0.923 & 0.900 \\

Unreal Engine 5 & 261.2 & 4020.3 & \(9.50\tcdot 10^{-8}\) & \(4.74\tcdot 10^{-7}\) & \(1.76\tcdot 10^{-2}\) & \(7.14\tcdot 10^{-2}\) & 0.573 & 0.515 & 0.283 & 0.541 \\

Maya + MotionBuilder & 275.1 & 4364.8 & \(6.43\tcdot 10^{-8}\) & \(3.33\tcdot 10^{-7}\) & \(7.10\tcdot 10^{-3}\) & \(3.45\tcdot 10^{-2}\) & 0.945 & 0.922 & 0.928 & 0.927 \\
\midrule
\midrule

Copy Rotations      &  \secondbest 258.6  & 3854.9   & \(1.05\tcdot 10^{-7}\)  & \(4.82\tcdot 10^{-7}\)  & \(9.01\tcdot 10^{-3}\)  & \(3.92 \tcdot 10^{-2}\) & 0.826 & 0.787 & \secondbest 0.670 & \secondbest 0.701\\

SAN             &  293.9 & \secondbest 3803.1 & \(6.78 \tcdot 10^{-8}\) & \(3.91 \tcdot 10^{-7}\) & \(\secondbest 7.18  \tcdot 10^{-3}\) & \secondbest \(2.26 \tcdot 10^{-2}\) & 0.302 & 0.508 & 0.139 & 0.510 \\

R$^2$ET             & 1540.1   & 17249.9   & \( \secondbest 5.75\tcdot 10^{-8}\) &  \(\secondbest 3.52 \tcdot 10^{-7}\) & \(9.85\tcdot 10^{-3}\) & \(4.41 \tcdot 10^{-2}\) & \secondbest 0.828 & \secondbest 0.790 & 
0.505 & 0.641\\

R$^2$ET + 1€ filter & 398.5     & 4628.6   & \(5.96\tcdot 10^{-8}\)  & \best \(3.27\tcdot 10^{-7}\) & \(9.82\tcdot 10^{-3}\) & \(5.53 \tcdot 10^{-2}\) & 0.773 & 0.708 & 0.379 & 0.586\\
\midrule

Ours ($N$ = 41)               & \best 212.8 & \best 2978.8 & \best \(4.65\tcdot 10^{-8}\) & \(3.60\tcdot 10^{-7}\) & \best \(3.22\tcdot 10^{-3}\) & \best \(1.72\tcdot 10^{-2}\) & \best 0.925 & \best 0.901 & \best 0.835 &  \best 0.804 \\

\midrule\midrule

Ours ($N$ = 96)               & 238.8 & 3358.4 & \(3.62\tcdot 10^{-8}\) & \(3.55\tcdot 10^{-7}\) & \(2.76 \tcdot 10^{-3}\) & \(1.46\tcdot 10^{-2}\) & 0.908 & 0.888 & 0.820 & 0.795 \\

Ours  (w/o k-v transfer)  & 200.1    & 2560.5   & \(4.09\tcdot 10^{-8}\) &  \(3.34\tcdot 10^{-7}\) & \(3.99\tcdot 10^{-3}\) & \(2.02 \tcdot 10^{-2}\) & 0.926 & 0.900 & 0.836 & 0.805\\


\bottomrule

\end{tabular}

\caption{Metrics on our evaluation set. {\best Green} : best, {\secondbest blue} : second best. Bottom two lines show the results of the ablation study mentioned in section \ref{ablation}.}
\label{table:results}
\vspace*{-1\baselineskip}
\end{table*}

\subsubsection{Metrics}

We evaluated, and report in Table~\ref{table:results}, the performance of ReConForM through several metrics, comparing it to the naive baseline (``Copy Rotations'' -- see Supp. Mat. for implementation details), to SAN \cite{san}, as well as R$^2$ET \cite{r2et} with and without the addition of a 1€ filter \cite{1eurofilter} to 
reduce its noise. We also provide comparison with industrial retargeting algorithms of Mixamo~\cite{mixamo}, Unreal Engine 5~\cite{ue5}, and Autodesk Maya/MotionBuilder~\cite{humanIK}. Overall, our approach provides the best compromise to avoid self-penetration and foot skating, while preserving a smooth animation.

We evaluated the motion smoothness using the average and maximum magnitude of the jerk of joint positions throughout each animation sequence. Thanks to the integration of our smoothness loss within the optimization, our approach achieves remarkably low jerk (same order of magnitude as the source), leading to very smooth motion.


Quantifying the self-penetration by computing the collision volume between the limbs, normalized by the total volume of the character, we found that our method surpasses the current state-of-the-art while remaining temporally smooth. On the other hand, R$^2$ET, being solely dedicated to reducing self-collisions, shows higher temporal noise. Using a post-process filter~\cite{1eurofilter} helps reduce the noise, but spoils, in return, the amount of self-penetration.  

%
%
Similarly, we evaluated the penetration with the floor by computing the relative proportion of the volume of the character that is below the ground at each frame.
Our method shows a 70\% reduction in floor penetration compared to the state-of-the-art.
%
%

We finally evaluated the quality of the feet-to-ground contacts along two quantitative measures. 
First, we call a foot to be \emph{grounded} when its 
distance to the floor is between -1\% and 1\% 
of the character's height. 
Second,
we consider that the foot is \emph{locked} on its position if its horizontal speed is lower than 0.1\% of the character's height per second. We computed the F1 and ROC AUC scores to quantify how much the semantic information of feet-to-ground contacts is kept by the retargeting method for these two metrics, with the source motion serving as ground truth. In both cases, our scores indicates that our approach 
generates more accurate feet-to-ground contact than the academic state-of-the-art, much closer to the quality of IK-based methods such as Mixamo and MotionBuilder. These methods indeed allow for a very precise placement of the feet, however they tend to do so without consideration for other semantic aspects of motion, leading for example to higher self-collisions and more frequent loss of semantic contacts.

%

\subsubsection{Ablation study}
\label{ablation}
We replaced the vertex transfer process described in section \ref{section:shape_encoding} by key-vertices placed manually by a human expert, to evaluate its impact on the overall retargeting method. We also evaluate with different amounts of key-vertices. These results (bottom lines of Table \ref{table:results}) confirm the robustness of our method with respect to the definition and placement of key-vertices. Additional experiments demonstrating the usefulness of each loss can be found in the supplementary material.

\subsection{Qualitative results}

\subsubsection{Visual output}
\begin{figure*}[p]
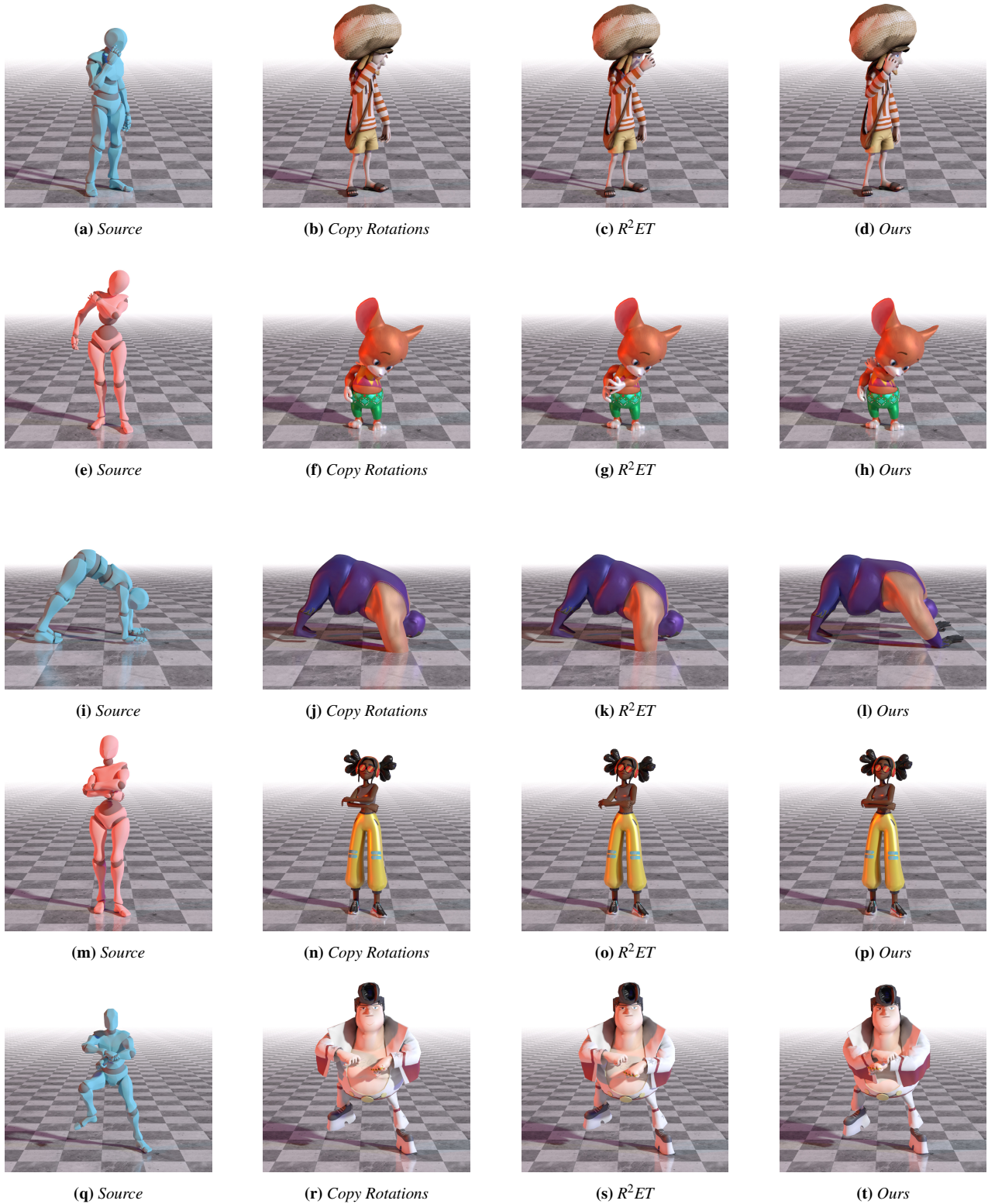

    \centering
    \foreach \folder in {Phone_Call, Shoulder_Rubbing, Pike_Walk, Angry, Gangnam_Style} {
        \begin{subfigure}[b]{\textwidth}
            \centering
            \hfill%
            \begin{subfigure}[b]{.2\textwidth}
                \centering
                \includegraphics[width=\textwidth]{contents/images/results/\folder/source.jpg}
                \caption{Source}
            \end{subfigure}%
            \hfill%
            \begin{subfigure}[b]{.2\textwidth}
                \centering
                \includegraphics[width=\textwidth]{contents/images/results/\folder/copy_rotations.jpg}
                \caption{Copy Rotations}
            \end{subfigure}%
            \hfill%
            \begin{subfigure}[b]{.2\textwidth}
                \centering
                \includegraphics[width=\textwidth]{contents/images/results/\folder/r2et.jpg}
                \caption{R$^2$ET}
            \end{subfigure}%
            \hfill%
            \begin{subfigure}[b]{.2\textwidth}
                \centering
                \includegraphics[width=\textwidth]{contents/images/results/\folder/ours.jpg}
                \caption{Ours}
            \end{subfigure}%
            \hfill
        \end{subfigure}
        \newline
    }
\vspace{-1\baselineskip}
\caption{Results of our method on several animations taken from our validation dataset. Animation references from top to bottom : Phone Call, Shoulder Rubbing, Pike Walk, Angry, Gangnam Style. From left to right : Source, Copy Rotations, R$^2$ET \cite{r2et}, Ours.}
\label{fig:results}
\end{figure*}

Figure \ref{fig:teaser} and Figure \ref{fig:method} show the results of our retargeting on several poses, and Figure \ref{fig:results} provides a visual comparison to R$^2$ET. Additional results are displayed in the Supp. Mat.


\begin{figure}[!b]
    \centering
    
    \pgfplotsset{
    compat=1.9,
    compat/bar nodes=1.8,
}
\pgfplotstableread{
    X Label ours tie r2et
    1 \textbf{Overall}    0.5878205128   0.2705128205  0.1416666667 
    2 Easy 0.5150645624	0.32313   0.1635581062  
    3 Hard 0.6465816918	0.2294322132 0.123986095
    4.5 \textbf{Overall} 0.531694696	0.3421733506	0.1261319534
    5.5 Easy 0.4515195369	0.3777134588	0.1707670043
    6.5 Hard 0.5964912281	0.3134502924	0.09005847953 
}\testdata

\begin{tikzpicture}
    \begin{axis}[
        width=1.02\columnwidth, 
        height=4.6cm, 
        ybar stacked,
        bar width=22pt,
        ymin=0,
        ymax=1,
        enlarge x limits=0.15,
        xtick=data,
        legend style={
            at={(0.5,1.2)}, 
            anchor=north,
            legend columns=-1,
            /tikz/every even column/.append style={column sep=10mm} 
        },
        ylabel={Proportion of answers},
        xticklabels from table={\testdata}{Label},
        xticklabel style={text width=2cm,align=center},
        reverse legend=true,
        clip=false,
    ]
        \addplot [
            fill=green!45,
            nodes near coords={\pgfmathprintnumber[fixed, fixed zerofill, precision=1]\pgfplotspointmeta\%},
            point meta={100*\thisrow{ours}}, 
            every node near coord/.append style={anchor=north, font=\scriptsize}
        ]
            table [y=ours, x=X]
                {\testdata};
                    \addlegendentry{Ours}
        \addplot [
            fill=blue!50,
            nodes near coords={\pgfmathprintnumber[fixed, fixed zerofill, precision=1]\pgfplotspointmeta\%},
            point meta={100*\thisrow{tie}}, 
            every node near coord/.append style={anchor=north, font=\scriptsize}
        ]
            table [y=tie, x=X]
                {\testdata};
                    \addlegendentry{Tie}
        \addplot [
            fill=red!45,
            nodes near coords={\pgfmathprintnumber[fixed, fixed zerofill, precision=1]\pgfplotspointmeta\%},
            point meta={100*\thisrow{r2et}}, 
            every node near coord/.append style={anchor=north, font=\scriptsize}
        ]
            table [y=r2et, x=X]
                {\testdata};
                    \addlegendentry{R$^2$ET}

    \end{axis}
    \node at (1.8,-.7) {Fidelity w.r.t. source}; 
    \node at (5.3,-.7) {Pleasantness}; 

\end{tikzpicture}
    \vspace*{-1.8\baselineskip}
    \captionof{figure}{User preference across all answers, split by difficulty}
    \label{fig:userpref1}

    \vspace*{.5\baselineskip}
        
    \pgfplotsset{
    compat=1.9,
    compat/bar nodes=1.8,
}
\pgfplotstableread{
    X Label ours tie r2et
    1 \footnotesize{None\ \ }    0.5774240232	0.2850940666	0.1374819103
    2 \footnotesize{Mod-\ }\\[-.5\baselineskip]\footnotesize{erate} 0.6833333333	0.2333333333	0.08333333333
    3 \footnotesize{\ Expert} 0.8648648649	0.09459459459	0.04054054054
    4.5 \footnotesize{None\ \ } 0.5065406977	0.3582848837	0.1351744186
    5.5 \footnotesize{Mod-\ }\\[-.5\baselineskip]\footnotesize{erate} 0.619047619	0.3174603175	0.06349206349
    6.5 \footnotesize{\ Expert} 0.8441558442	0.09090909091	0.06493506494
}\testdata

\begin{tikzpicture}
    \begin{axis}[
        width=1.02\columnwidth, 
        height=4.6cm, 
        ybar stacked,
        bar width=22pt,
        ymin=0,
        ymax=1,
        enlarge x limits=0.15,
        xtick=data,
        ylabel={Proportion of answers},
        xticklabels from table={\testdata}{Label},
        xticklabel style={text width=2cm,align=center},
        clip=false,
    ]
        \addplot [
            fill=green!45,
            nodes near coords={\pgfmathprintnumber[fixed, fixed zerofill, precision=1]\pgfplotspointmeta\%},
            point meta={100*\thisrow{ours}}, 
            every node near coord/.append style={anchor=north, font=\tiny}
        ]
            table [y=ours, x=X]
                {\testdata};
        \addplot [
            fill=blue!50,
            nodes near coords={\pgfmathprintnumber[fixed, fixed zerofill, precision=1]\pgfplotspointmeta\%},
            point meta={100*\thisrow{tie}}, 
            every node near coord/.append style={anchor=north, font=\tiny}
        ]
            table [y=tie, x=X]
                {\testdata};
        \addplot [
            fill=red!45,
            nodes near coords={\pgfmathprintnumber[fixed, fixed zerofill, precision=1]\pgfplotspointmeta\%},
            point meta={100*\thisrow{r2et}}, 
            every node near coord/.append style={anchor=north, font=\tiny}
        ]
            table [y=r2et, x=X]
                {\testdata};

    \end{axis}
    \node at (1.8,-.7) {Fidelity w.r.t. source}; 
    \node at (5.3,-.7) {Pleasantness}; 

\end{tikzpicture}
    \vspace*{-1.8\baselineskip}
    \captionof{figure}{User preference across all answers, split by expertise level}
    \label{fig:userpref2}
    
\end{figure}

\subsubsection{User study}

We designed a preferential blind user study to evaluate the quality of ReConForM compared to the state-of-the-art R$^2$ET.
We gathered a panel of 133 online respondents including both experts and non-experts.
Each participant evaluated twelve randomly selected animations displayed as videos containing the source animation, the result of our retargeting, and the one obtained using R$^2$ET~\cite{r2et} (in random order) on one of eight different characters taken from Mixamo~\cite{mixamo}. Participants were asked to choose a preferred output along two questions: first with respect to \emph{fidelity} with the source, and second, with respect to \emph{pleasantness} of the motion. 


Figure \ref{fig:userpref1} summarizes the reported preferences of the users over all the animations (Overall) covering a total of 1596 answers, as well as on both \emph{Easy} and \emph{Hard} subsets.
We observe a clear
preference for our results, with our outputs obtaining the majority of the votes on 41 out of 45 animations. 
Moreover, we can note that the average score on animations from the \emph{Hard} subset exceeds that on the \emph{Easy} subset.
This makes us hypothesize that the benefits of using our approach 
increase with the complexity of the retargeting. 
A
$\chi^2$ test 
confirms
this claim with $p<0.01$.

Figure \ref{fig:userpref2} shows another interesting result: the more respondents are familiar with the task of retargeting, the more they perceive our outputs to be best.
A $\chi^2$ test indicates a statistically significant distribution shift between the answers of experts and neophytes, with $p<0.01$.
All study details are in the supplementary material.

\subsection{Additional characters}

\begin{figure*}[!t]
    \centering
    \begin{subfigure}[b]{.16\textwidth}
        \includegraphics[width=\textwidth]{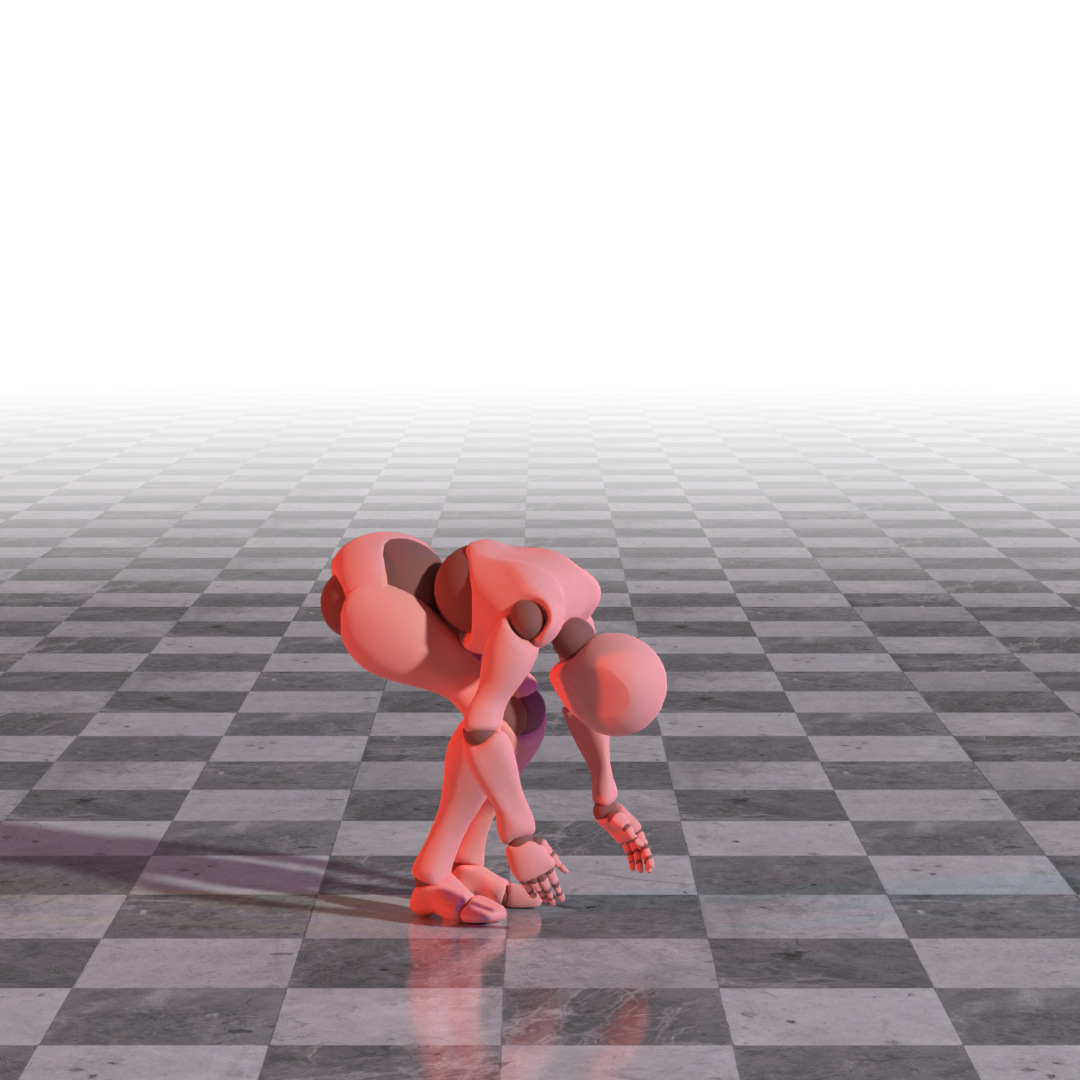}
        \centering
    \end{subfigure}%
    \hfill%
    \begin{subfigure}[b]{.16\textwidth}
        \includegraphics[width=\textwidth]{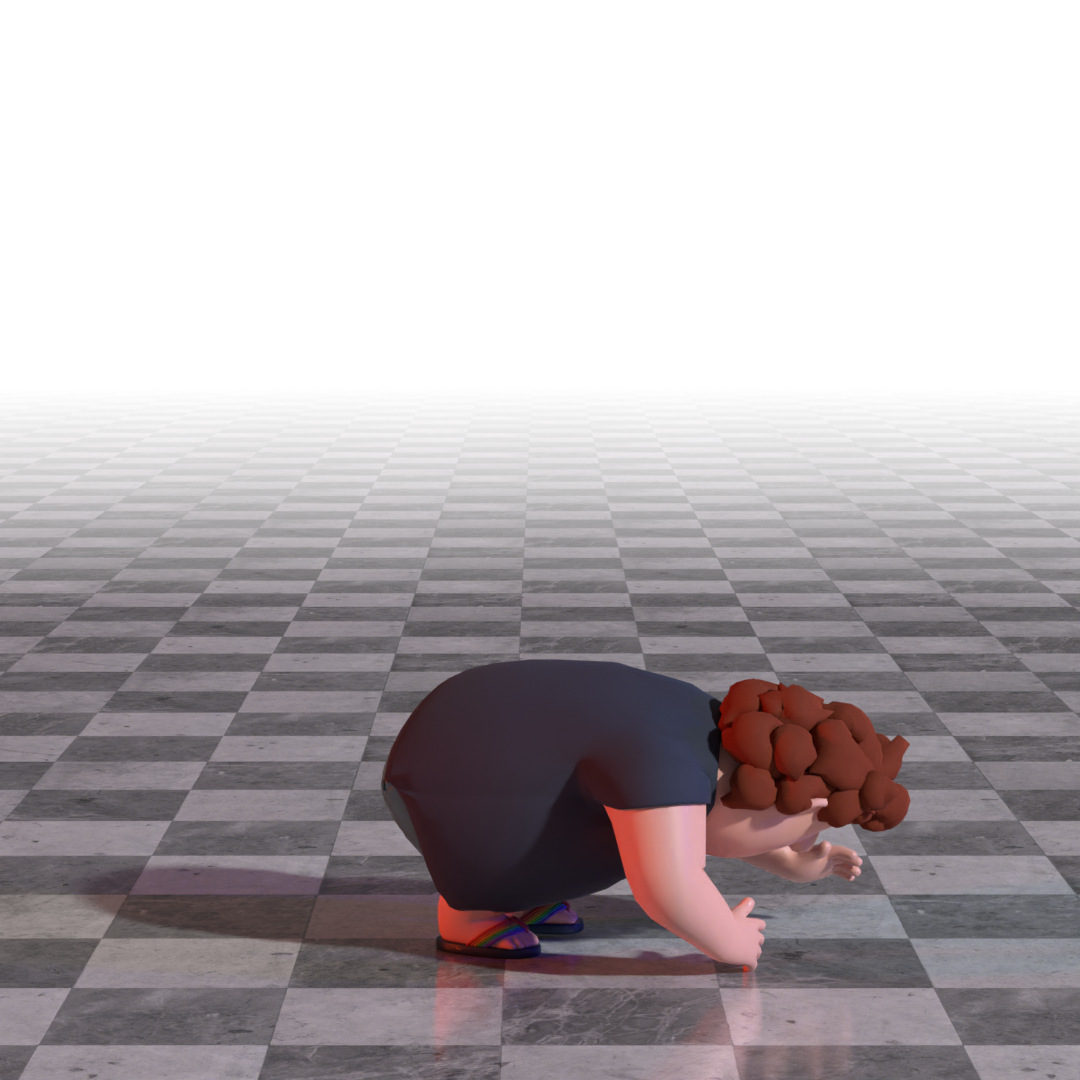}
        \centering
    \end{subfigure}%
    \hfill%
    \begin{subfigure}[b]{.16\textwidth}
        \includegraphics[width=\textwidth]{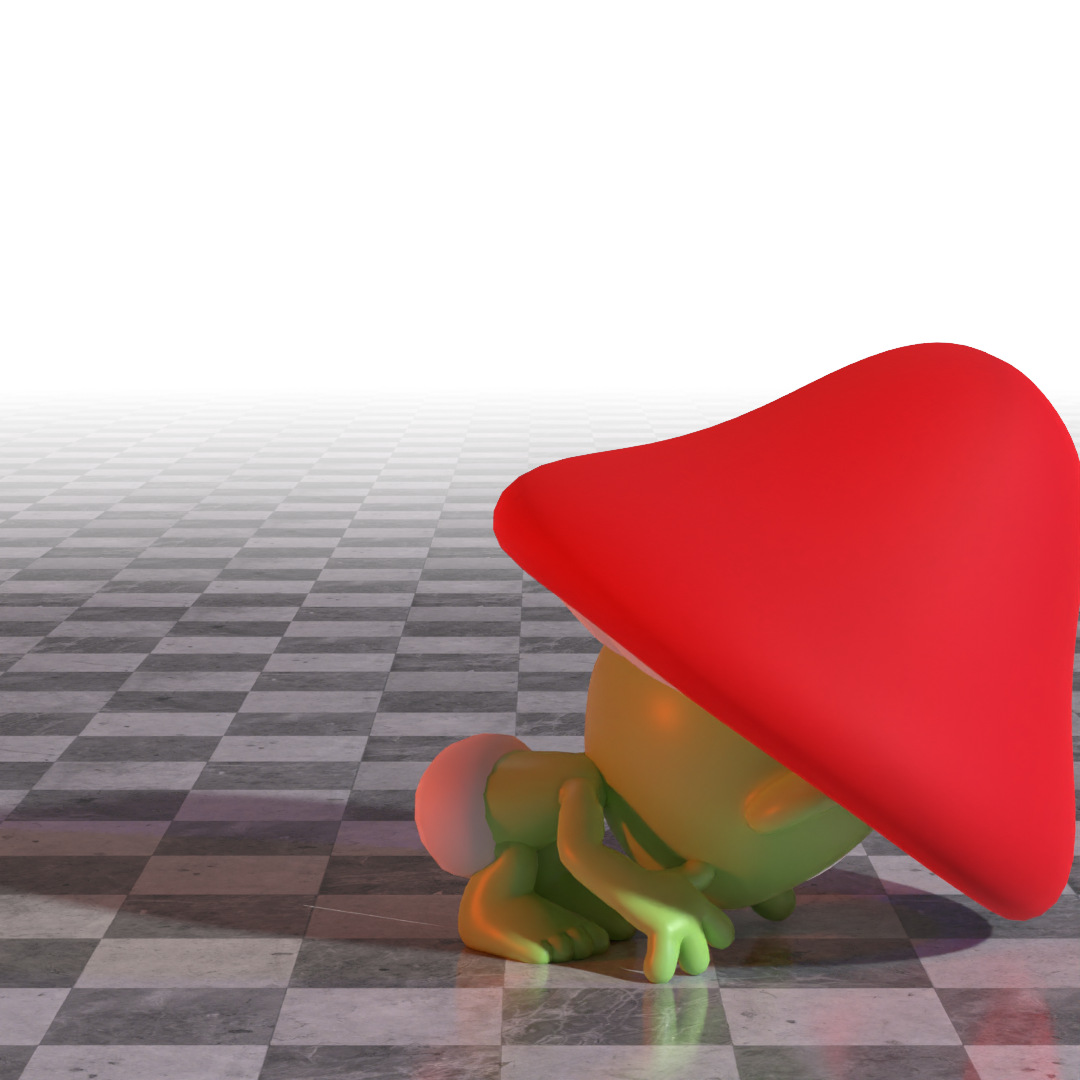}
        \centering
    \end{subfigure}%
    \hfill%
    \begin{subfigure}[b]{.16\textwidth}
        \includegraphics[width=\textwidth]{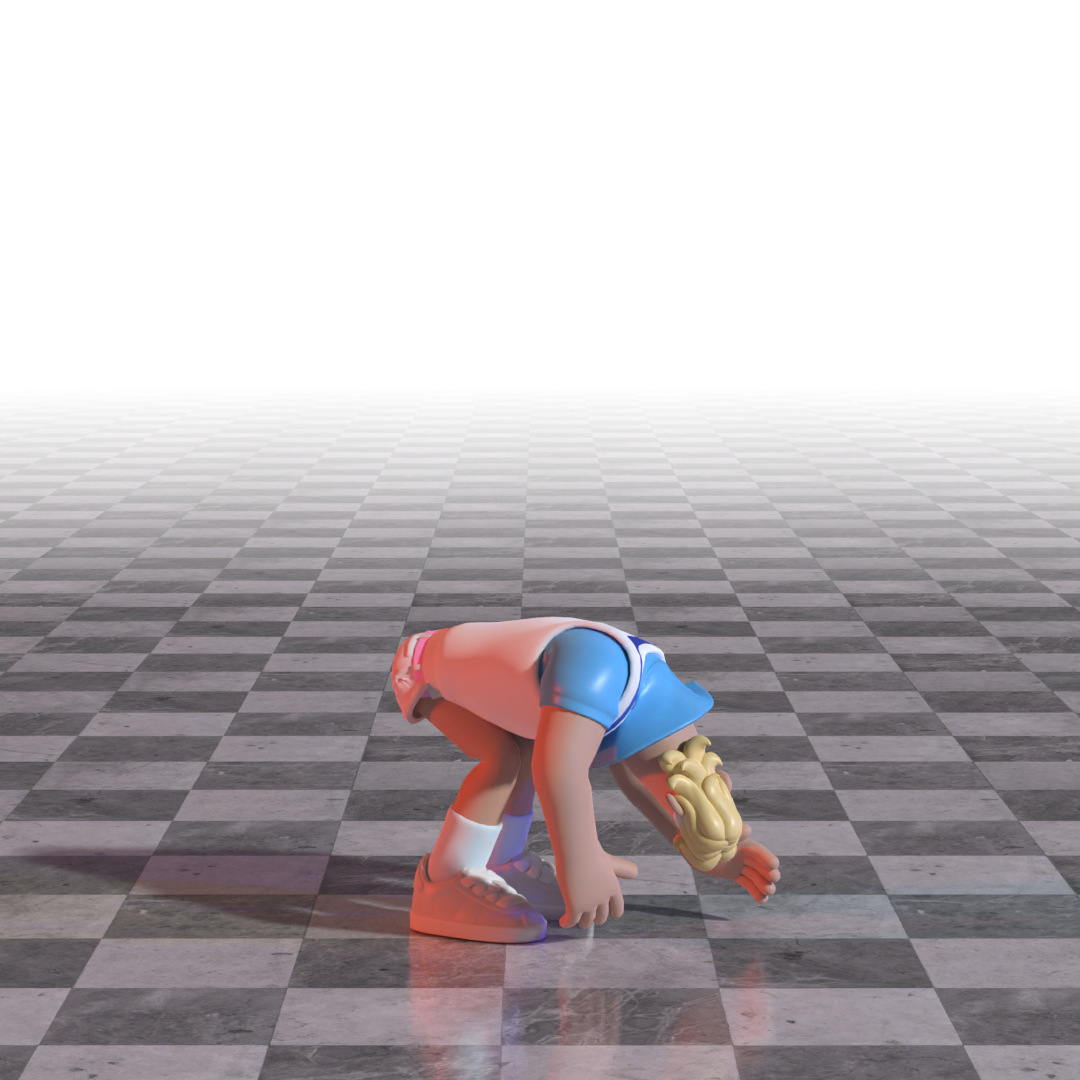}
        \centering
    \end{subfigure}%
    \hfill%
    \begin{subfigure}[b]{.16\textwidth}
        \includegraphics[width=\textwidth]{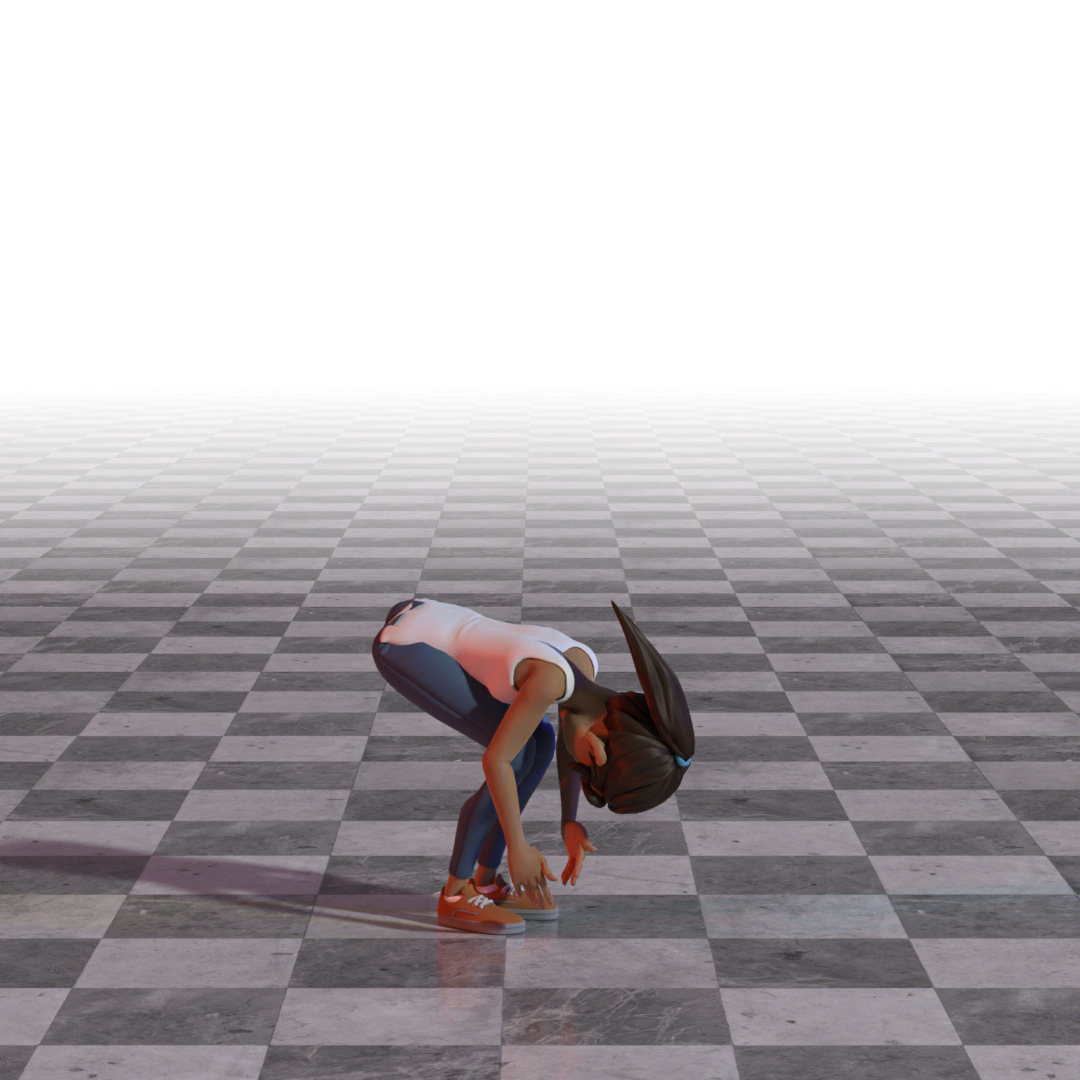}
        \centering
    \end{subfigure}%
    \hfill%
    \begin{subfigure}[b]{.16\textwidth}
        \includegraphics[width=\textwidth]{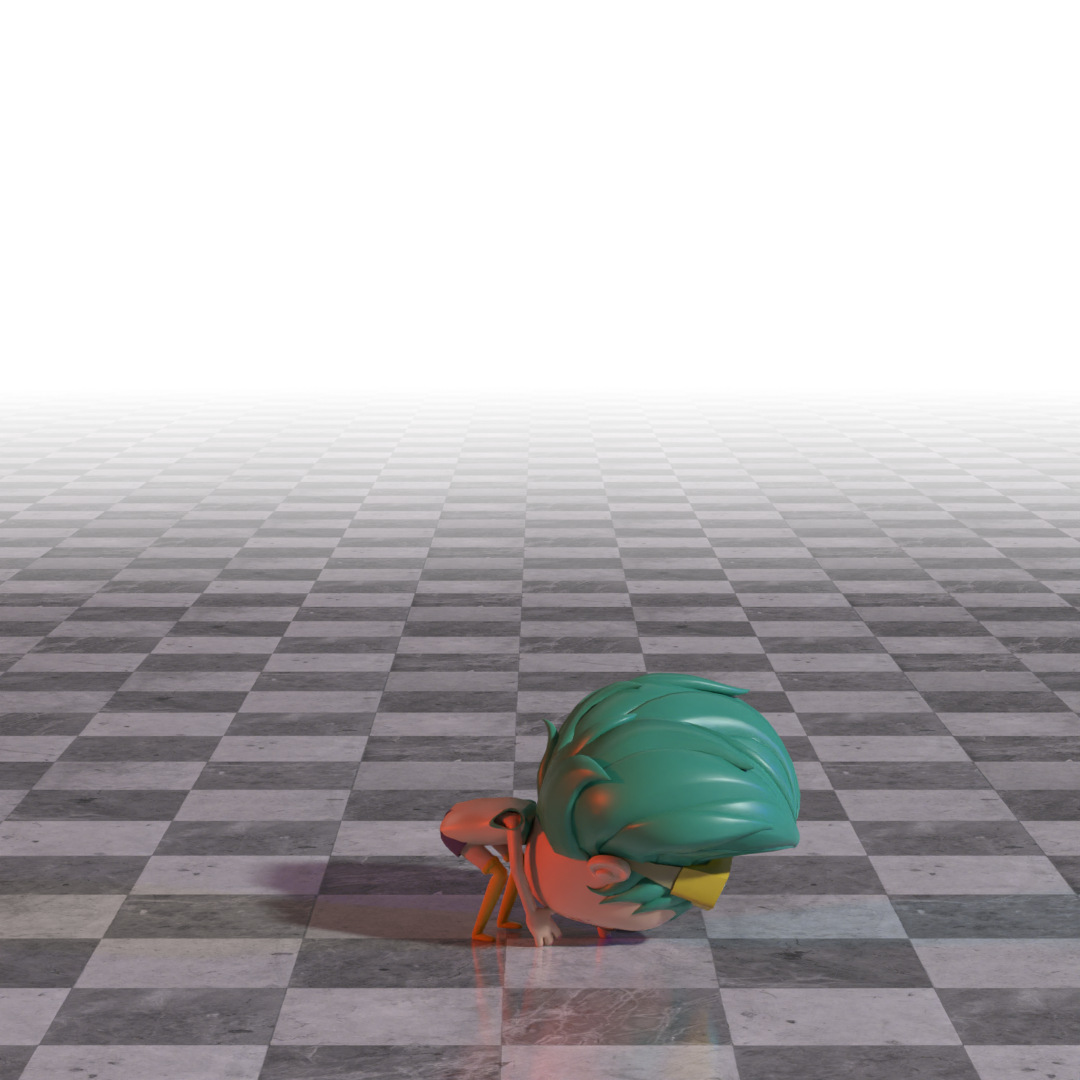}
        \centering
    \end{subfigure}

    \vspace*{1.5\baselineskip}
    \centering
    \begin{subfigure}[b]{.16\textwidth}
        \includegraphics[width=\textwidth]{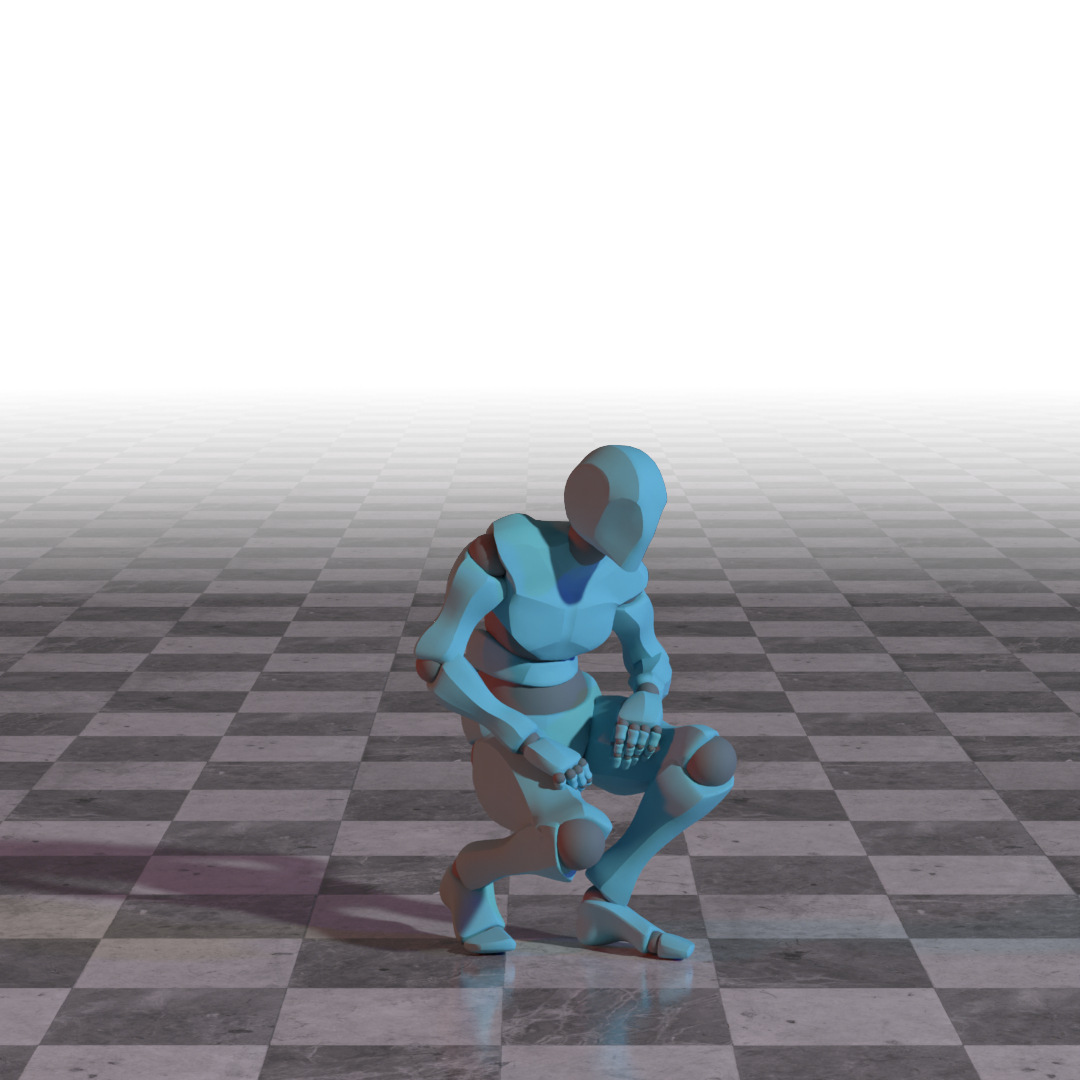}
        \centering
    \end{subfigure}%
    \hfill%
    \begin{subfigure}[b]{.16\textwidth}
        \includegraphics[width=\textwidth]{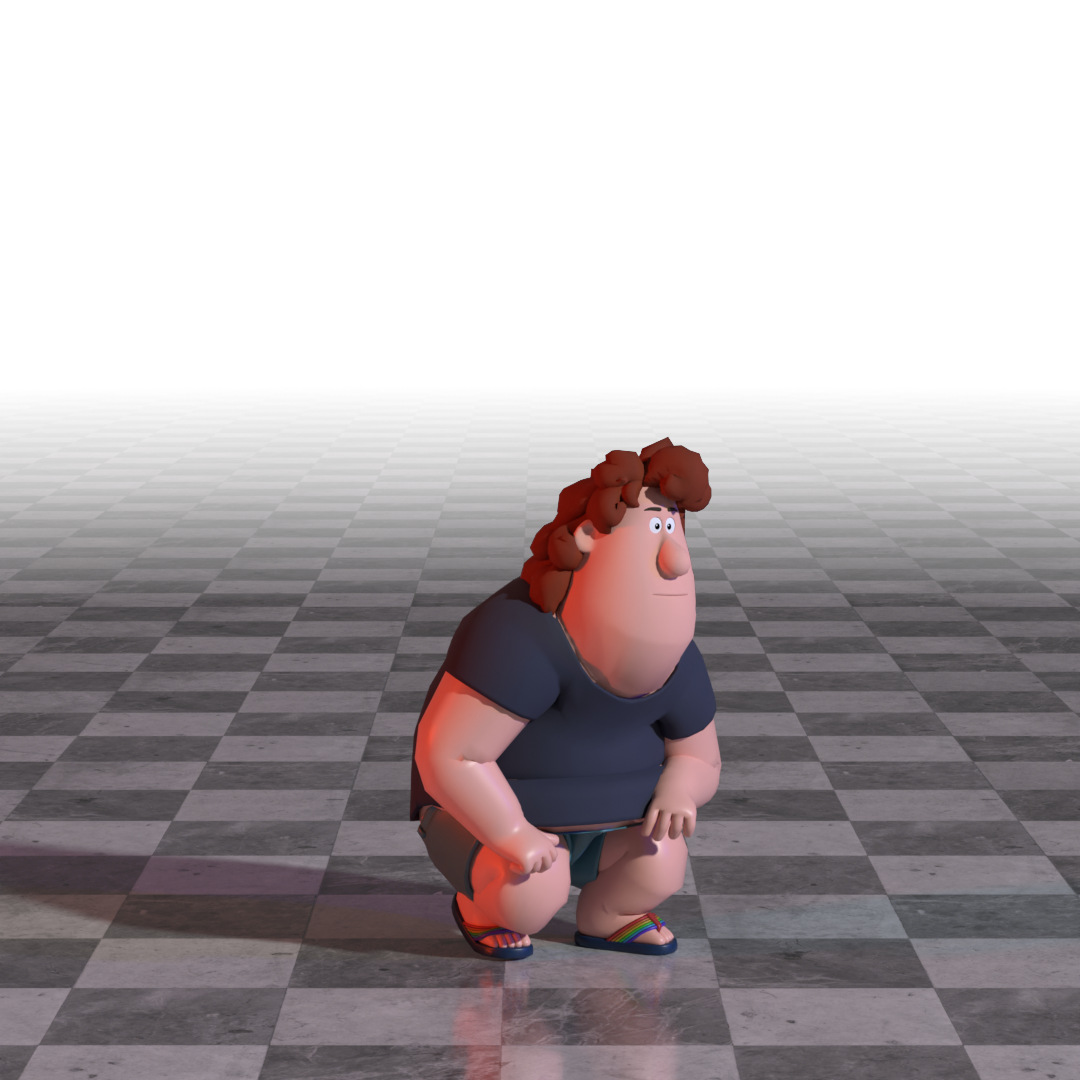}
        \centering
    \end{subfigure}%
    \hfill%
    \begin{subfigure}[b]{.16\textwidth}
        \includegraphics[width=\textwidth]{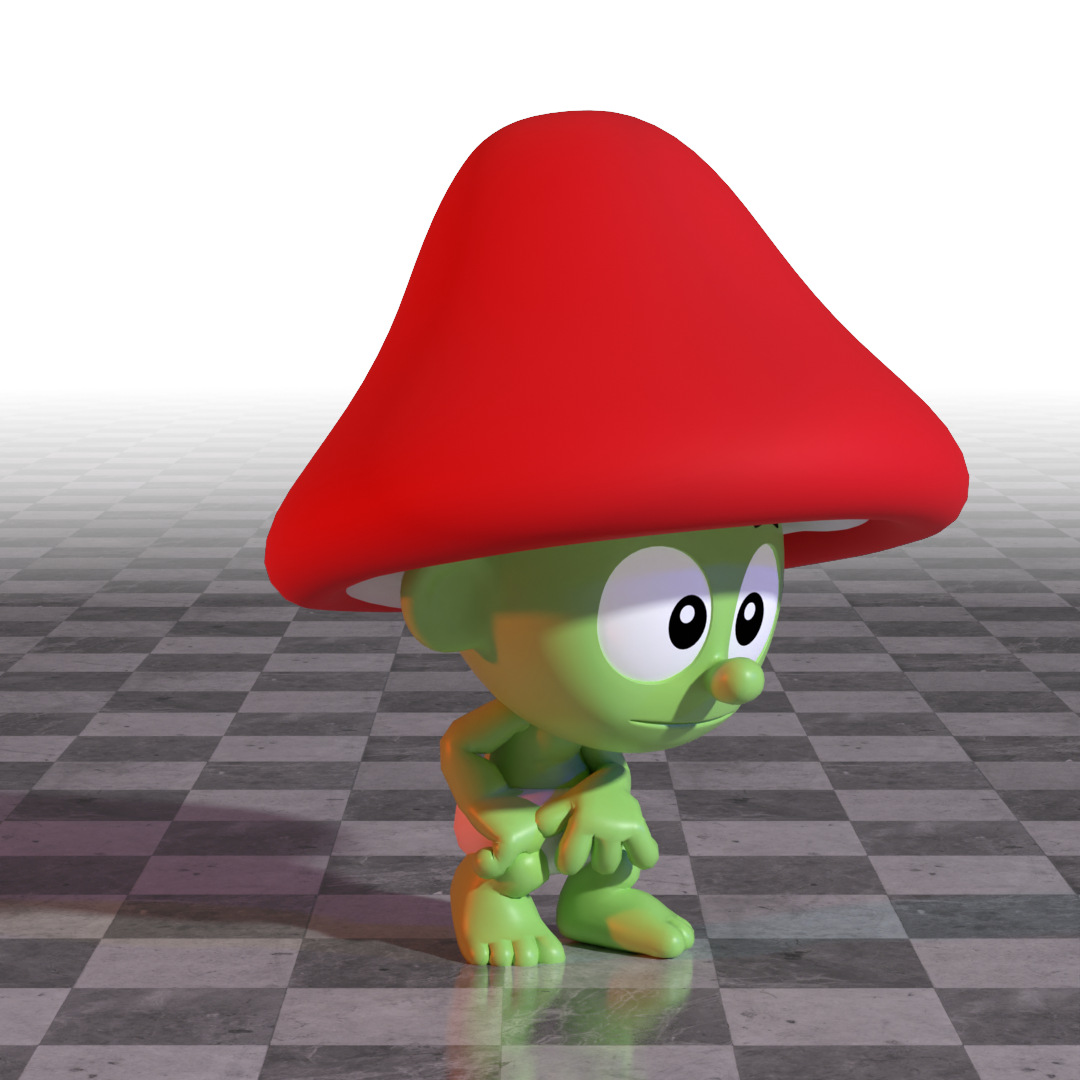}
        \centering
    \end{subfigure}%
    \hfill%
    \begin{subfigure}[b]{.16\textwidth}
        \includegraphics[width=\textwidth]{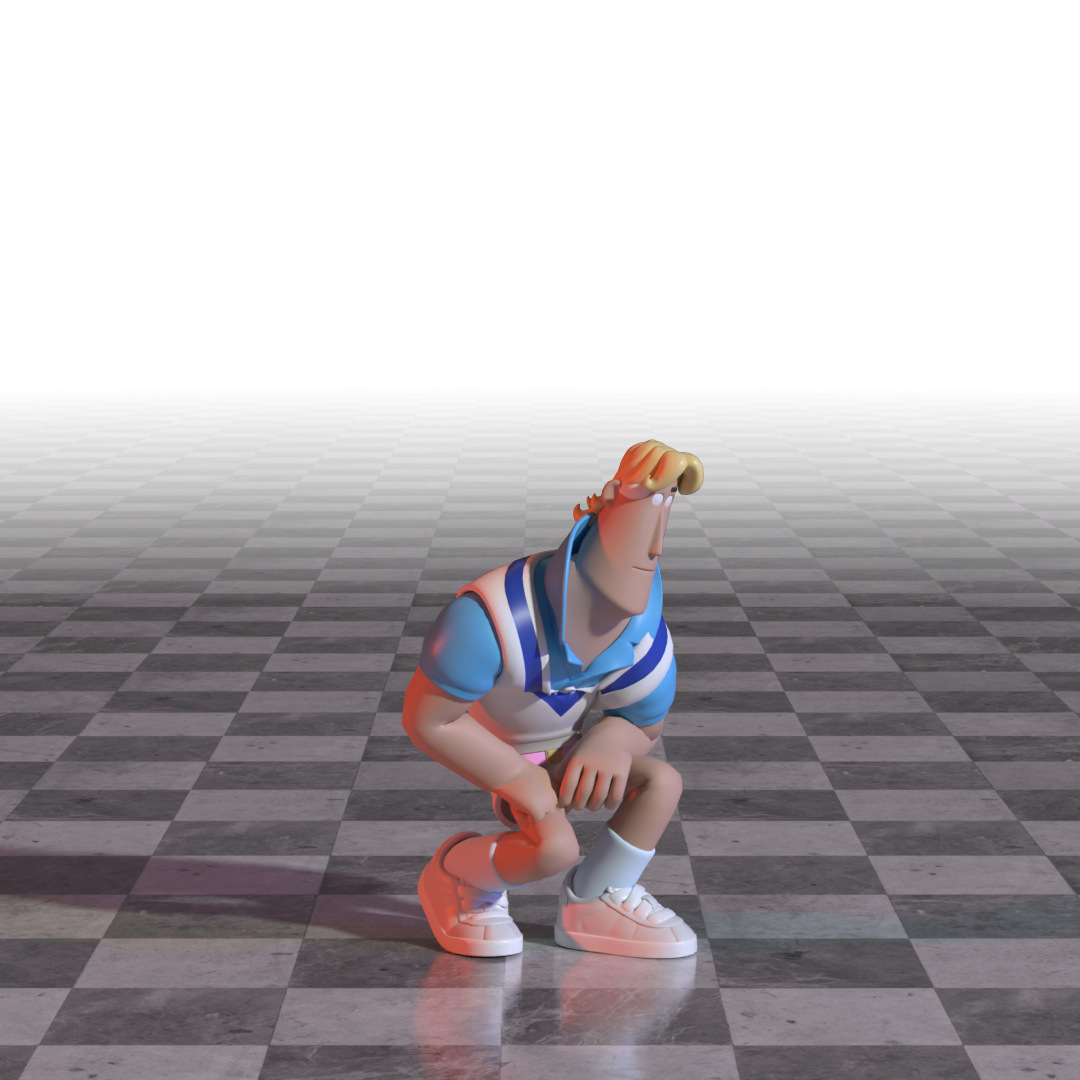}
        \centering
    \end{subfigure}%
    \hfill%
    \begin{subfigure}[b]{.16\textwidth}
        \includegraphics[width=\textwidth]{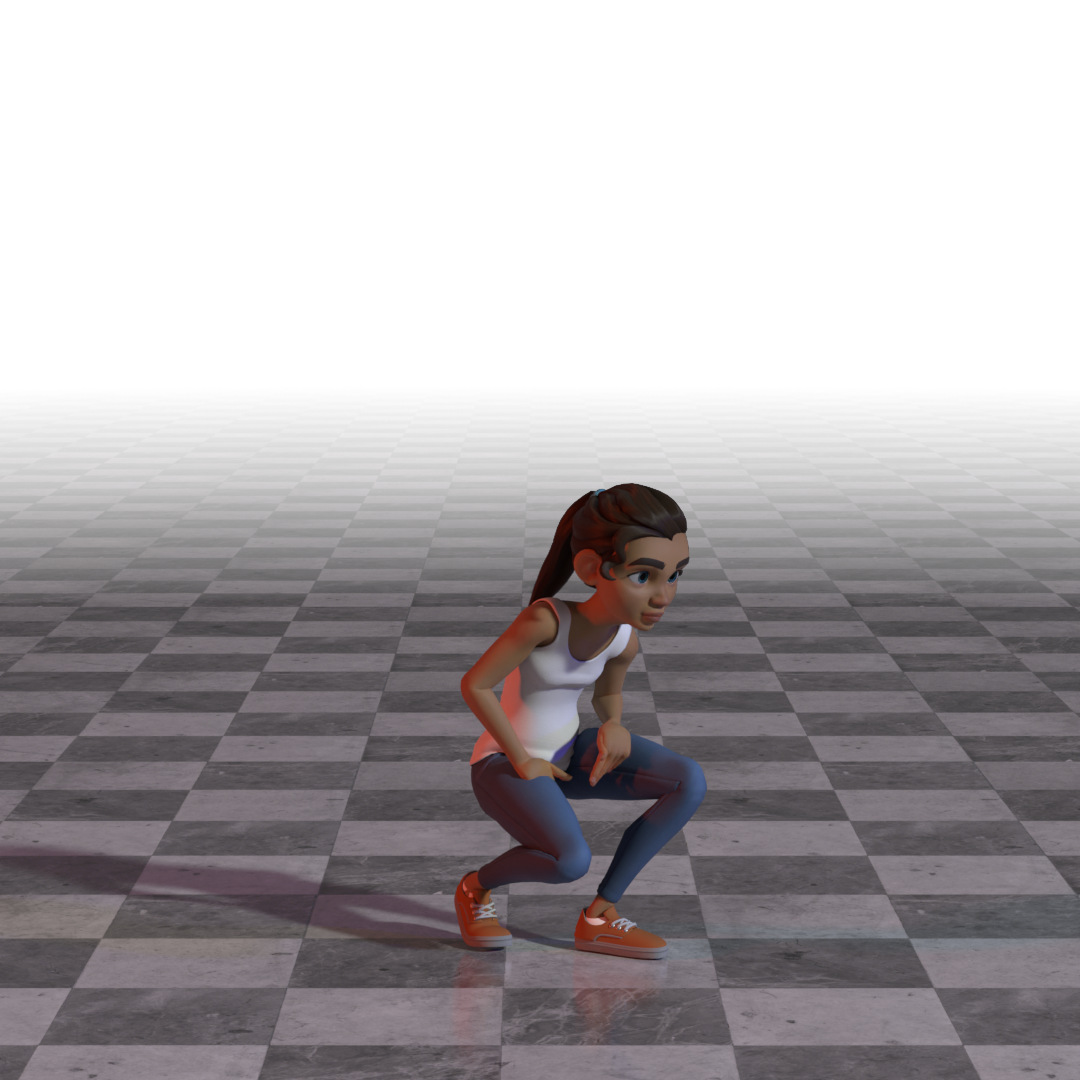}
        \centering
    \end{subfigure}%
    \hfill%
    \begin{subfigure}[b]{.16\textwidth}
        \includegraphics[width=\textwidth]{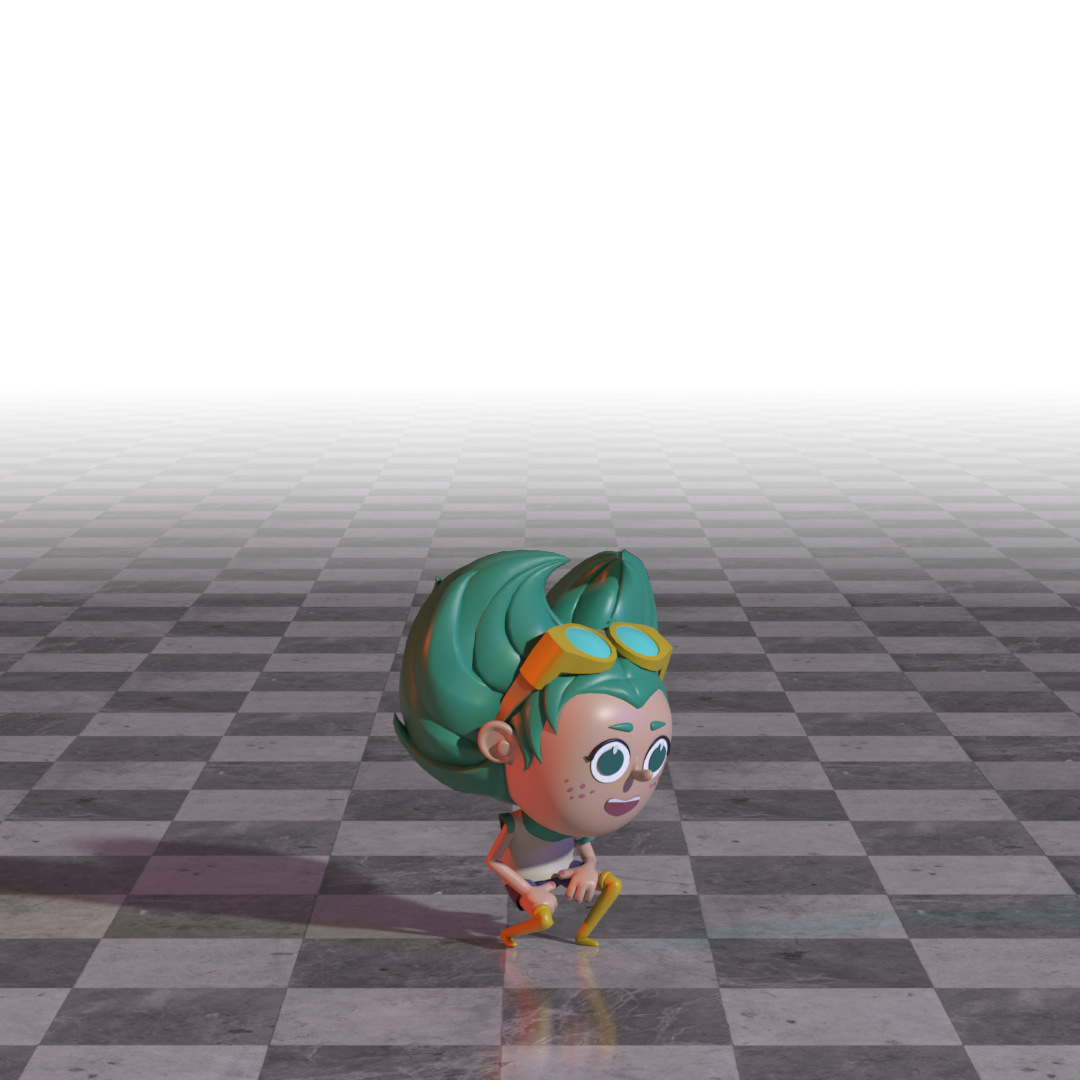}
        \centering
    \end{subfigure}

    \centering
    \begin{subfigure}[b]{.16\textwidth}
        \includegraphics[width=\textwidth]{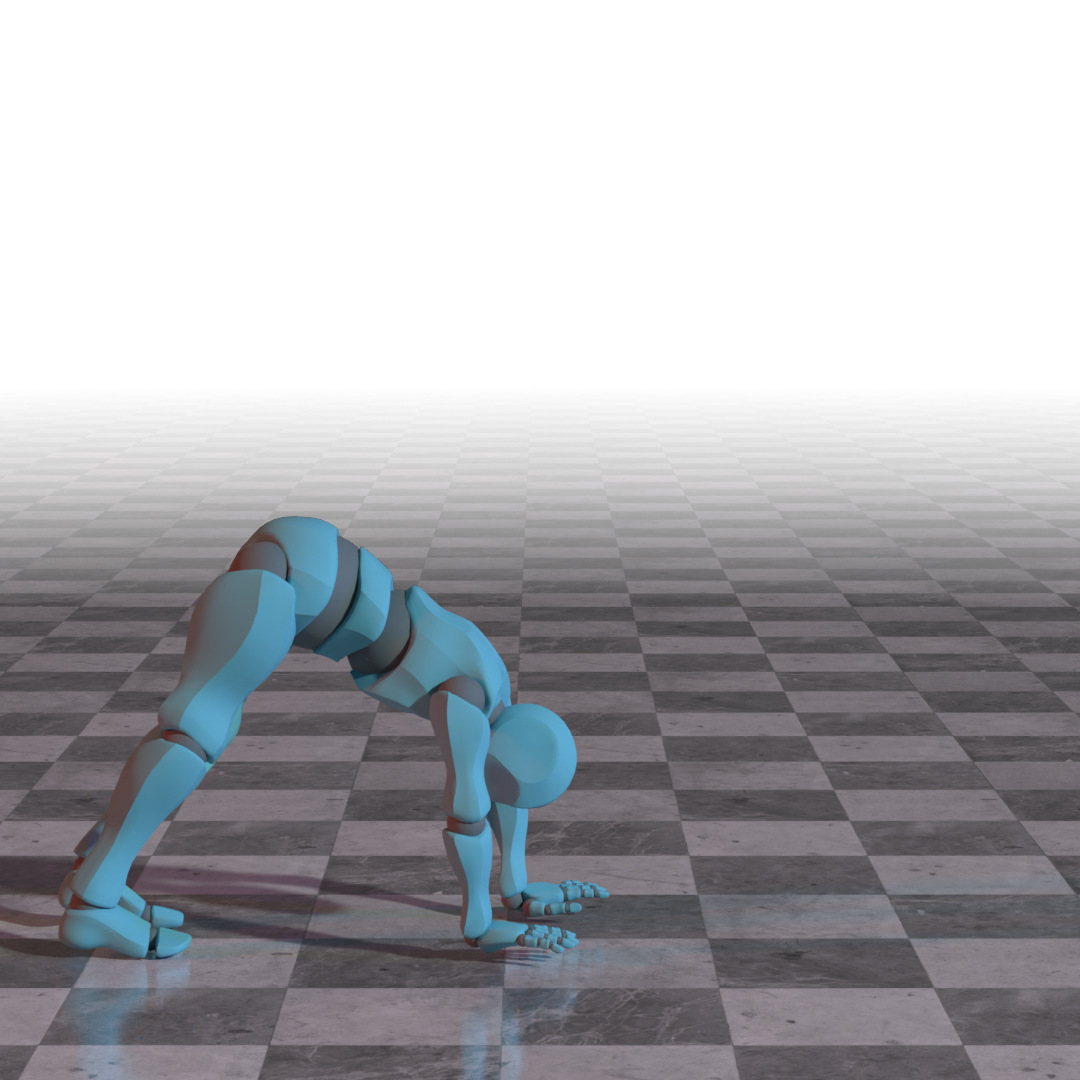}
        \centering
    \end{subfigure}%
    \hfill%
    \begin{subfigure}[b]{.16\textwidth}
        \includegraphics[width=\textwidth]{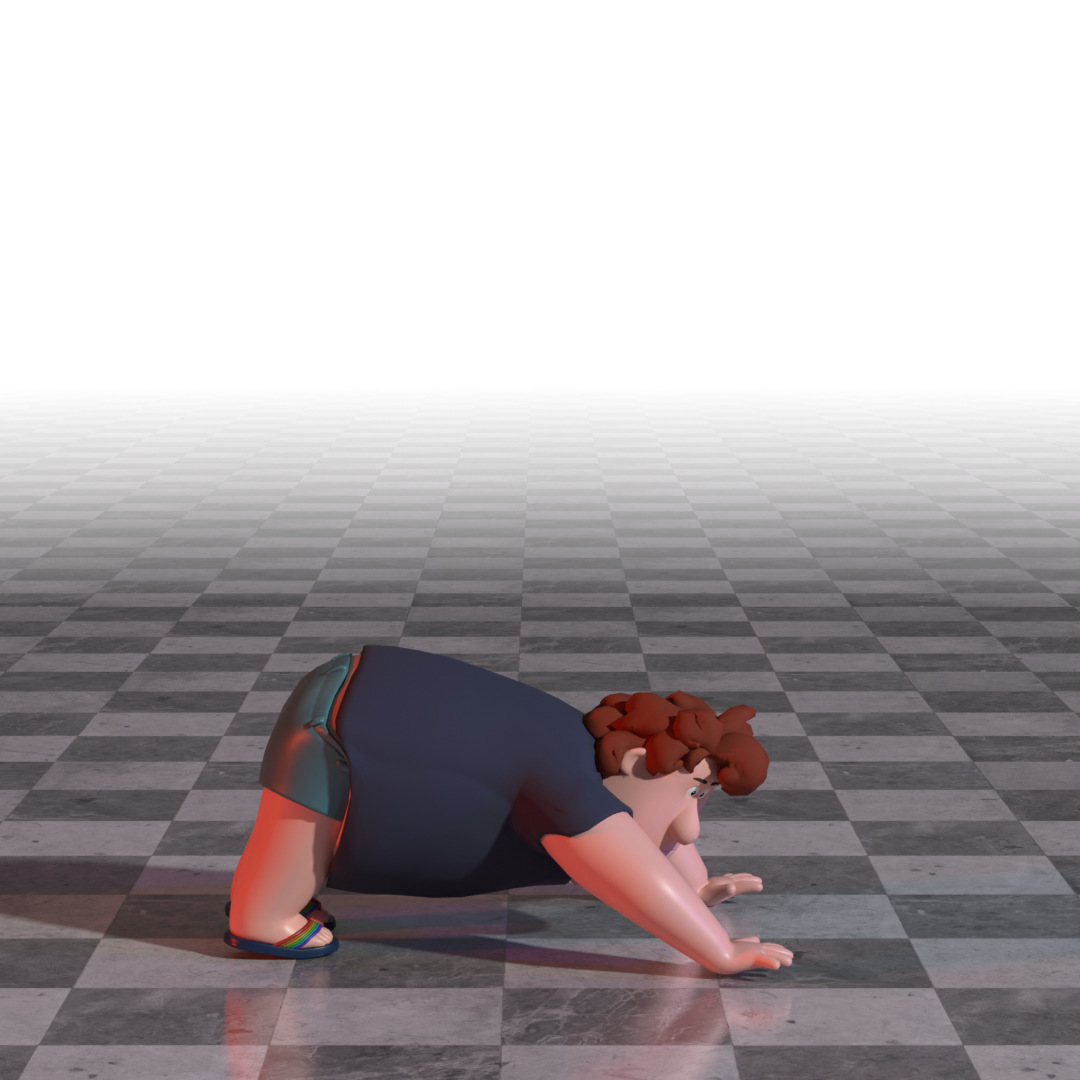}
        \centering
    \end{subfigure}%
    \hfill%
    \begin{subfigure}[b]{.16\textwidth}
        \includegraphics[width=\textwidth]{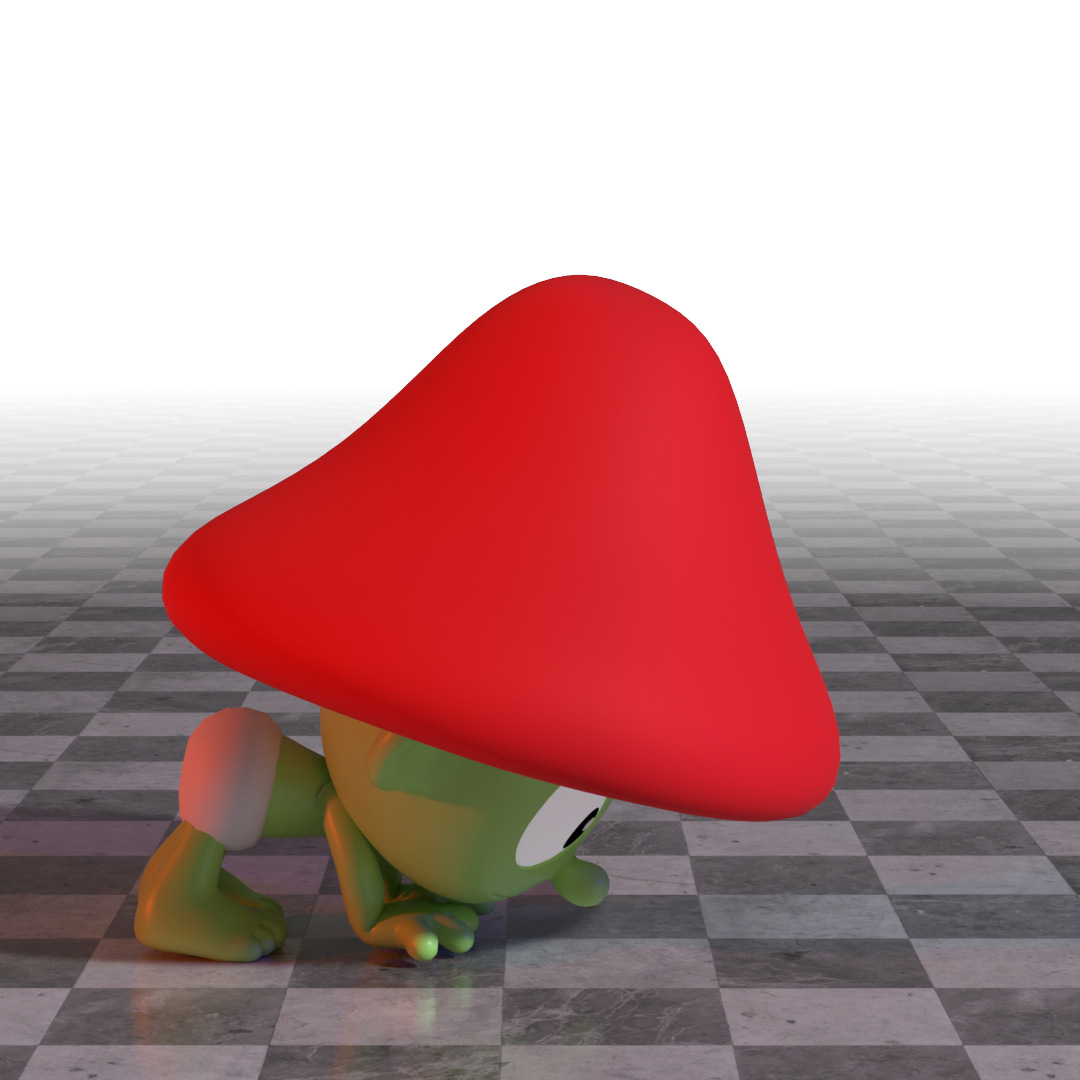}
        \centering
    \end{subfigure}%
    \hfill%
    \begin{subfigure}[b]{.16\textwidth}
        \includegraphics[width=\textwidth]{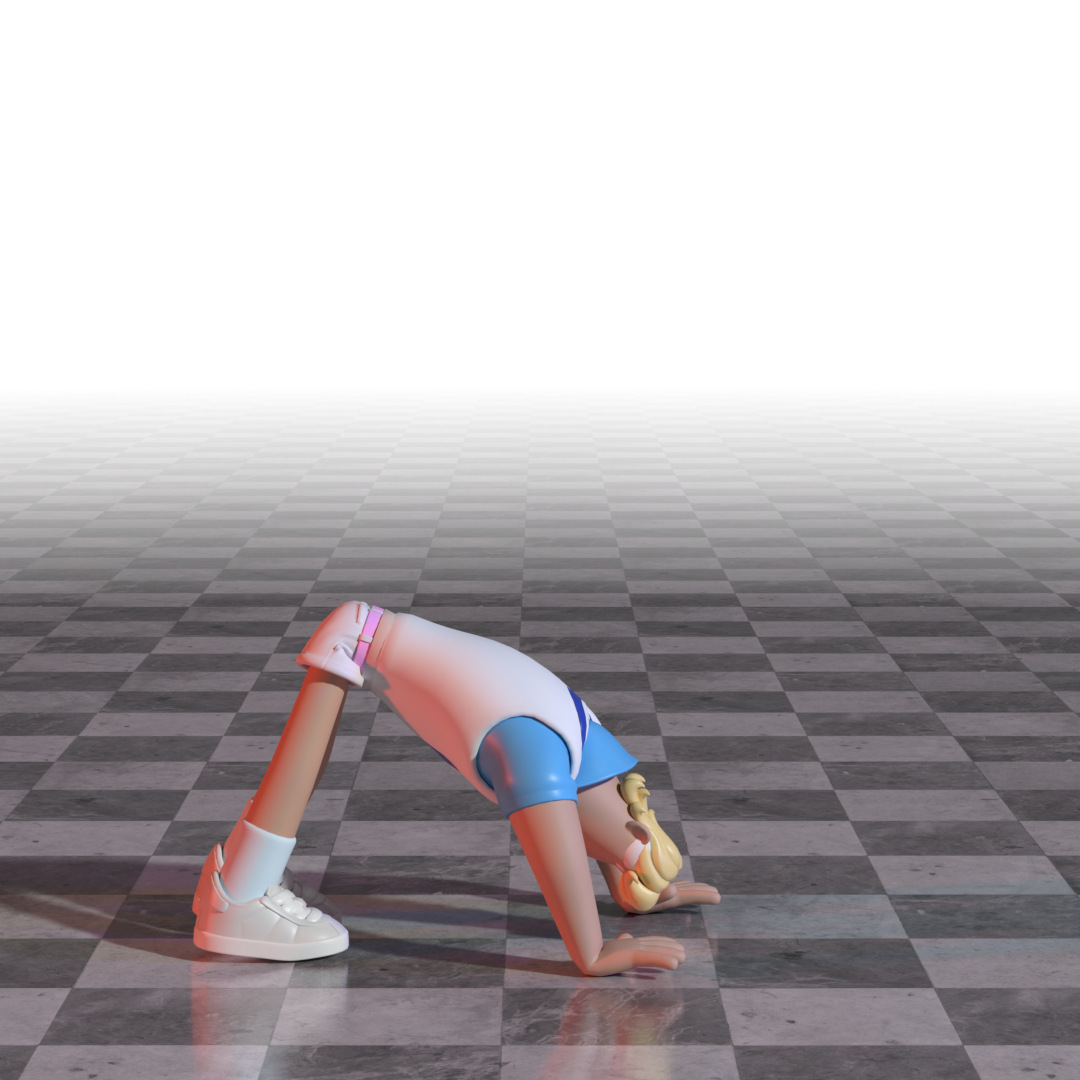}
        \centering
    \end{subfigure}%
    \hfill%
    \begin{subfigure}[b]{.16\textwidth}
        \includegraphics[width=\textwidth]{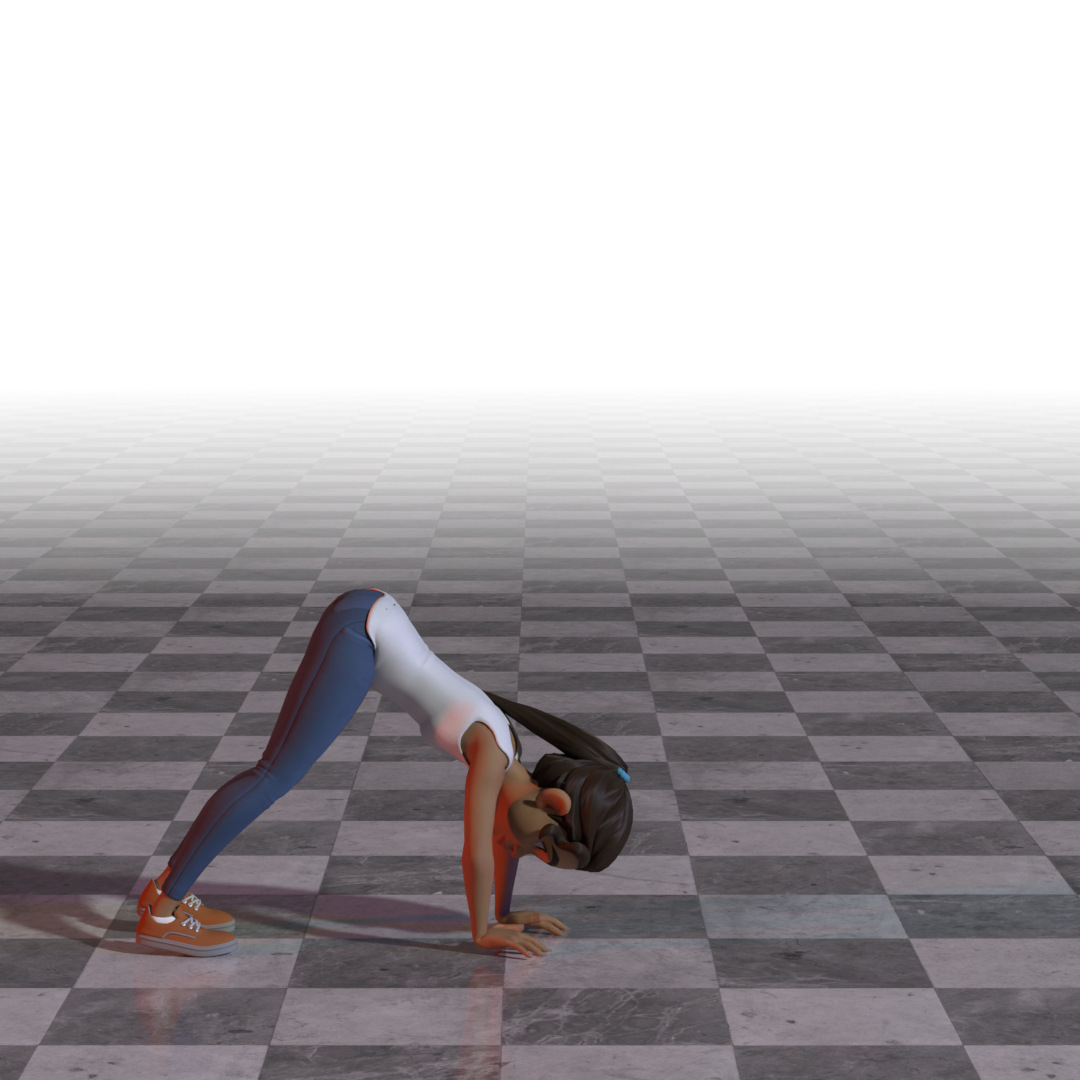}
        \centering
    \end{subfigure}%
    \hfill%
    \begin{subfigure}[b]{.16\textwidth}
        \includegraphics[width=\textwidth]{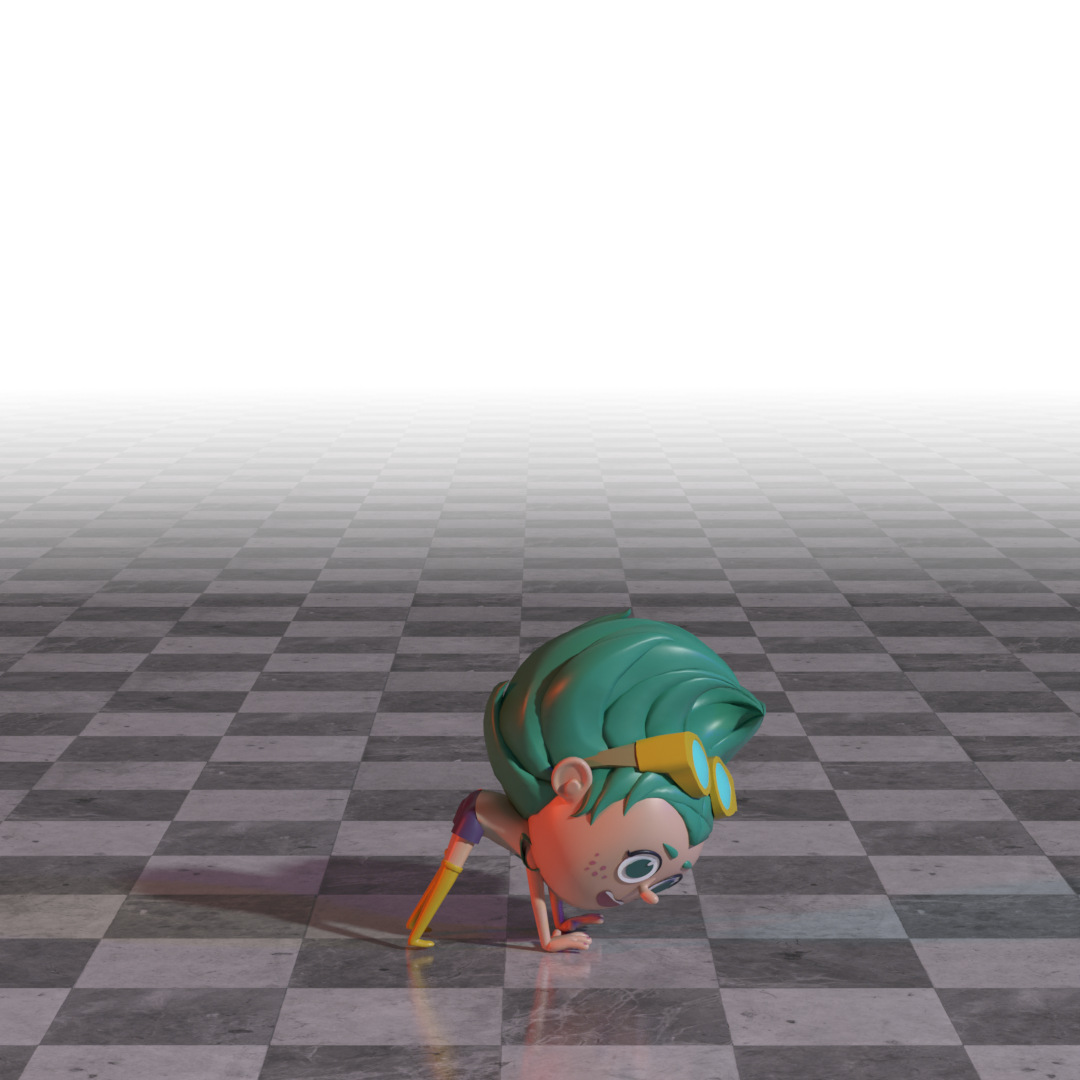}
        \centering
    \end{subfigure}

    \centering
    \begin{subfigure}[b]{.16\textwidth}
        \includegraphics[width=\textwidth]{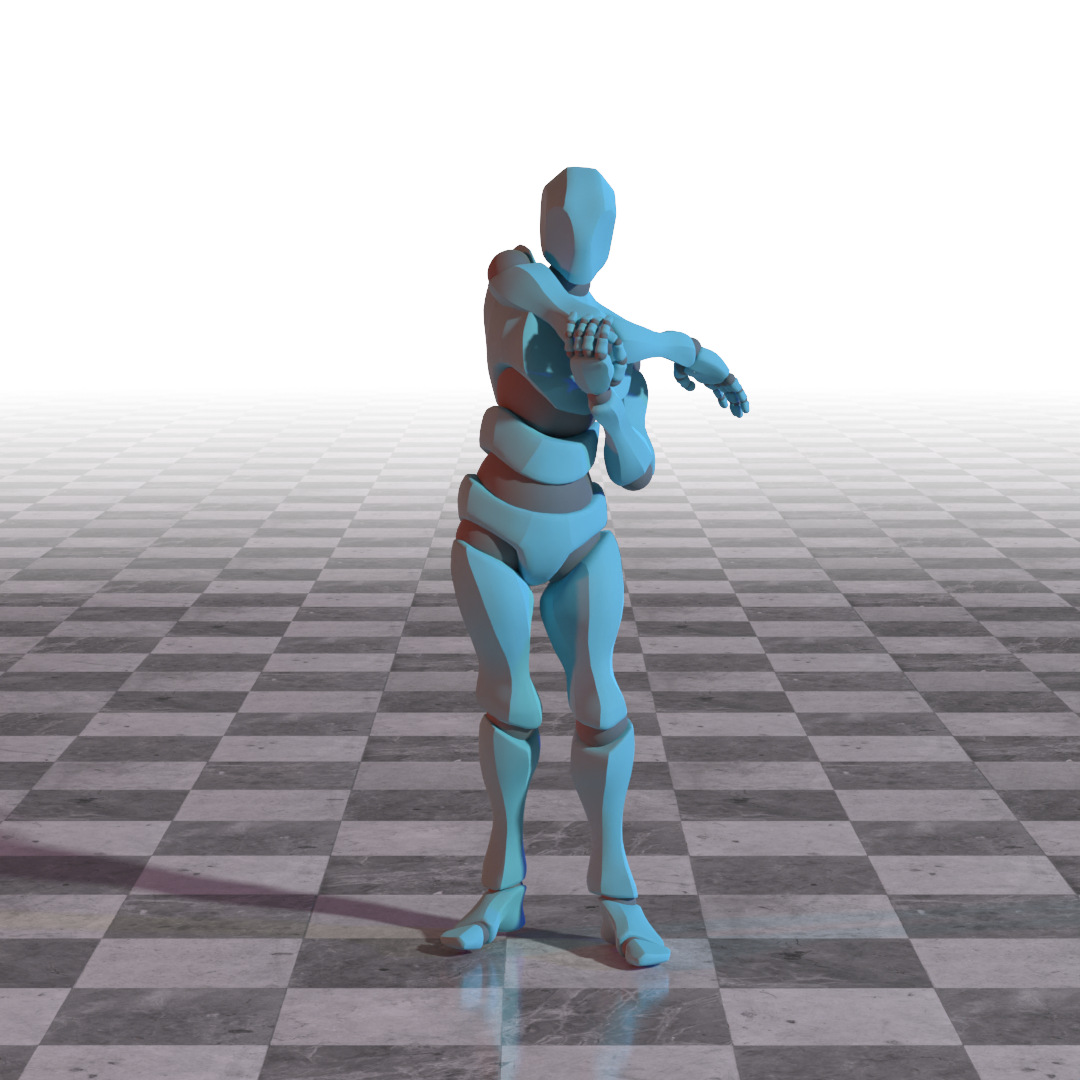}
        \centering
    \end{subfigure}%
    \hfill%
    \begin{subfigure}[b]{.16\textwidth}
        \includegraphics[width=\textwidth]{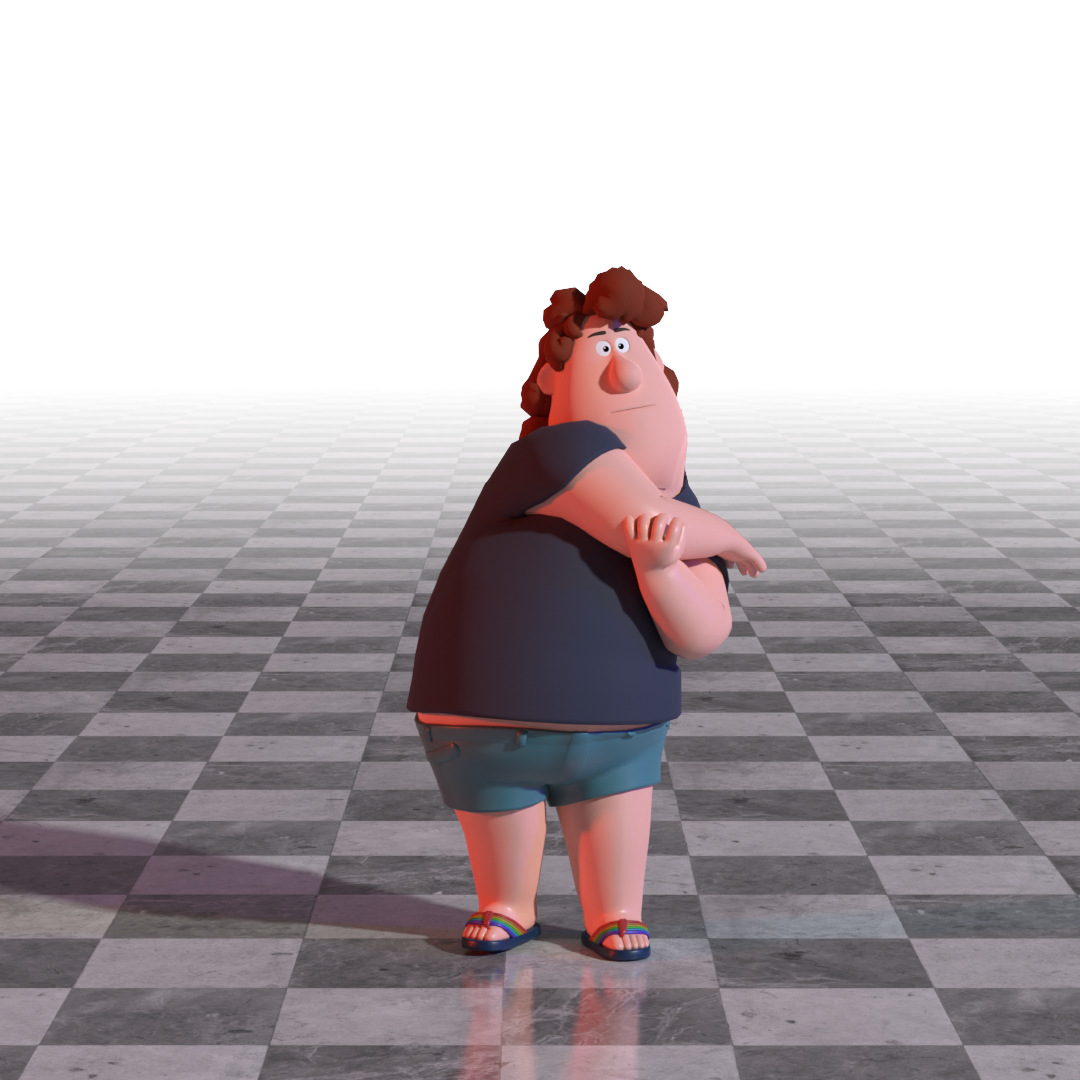}
        \centering
    \end{subfigure}%
    \hfill%
    \begin{subfigure}[b]{.16\textwidth}
        \includegraphics[width=\textwidth]{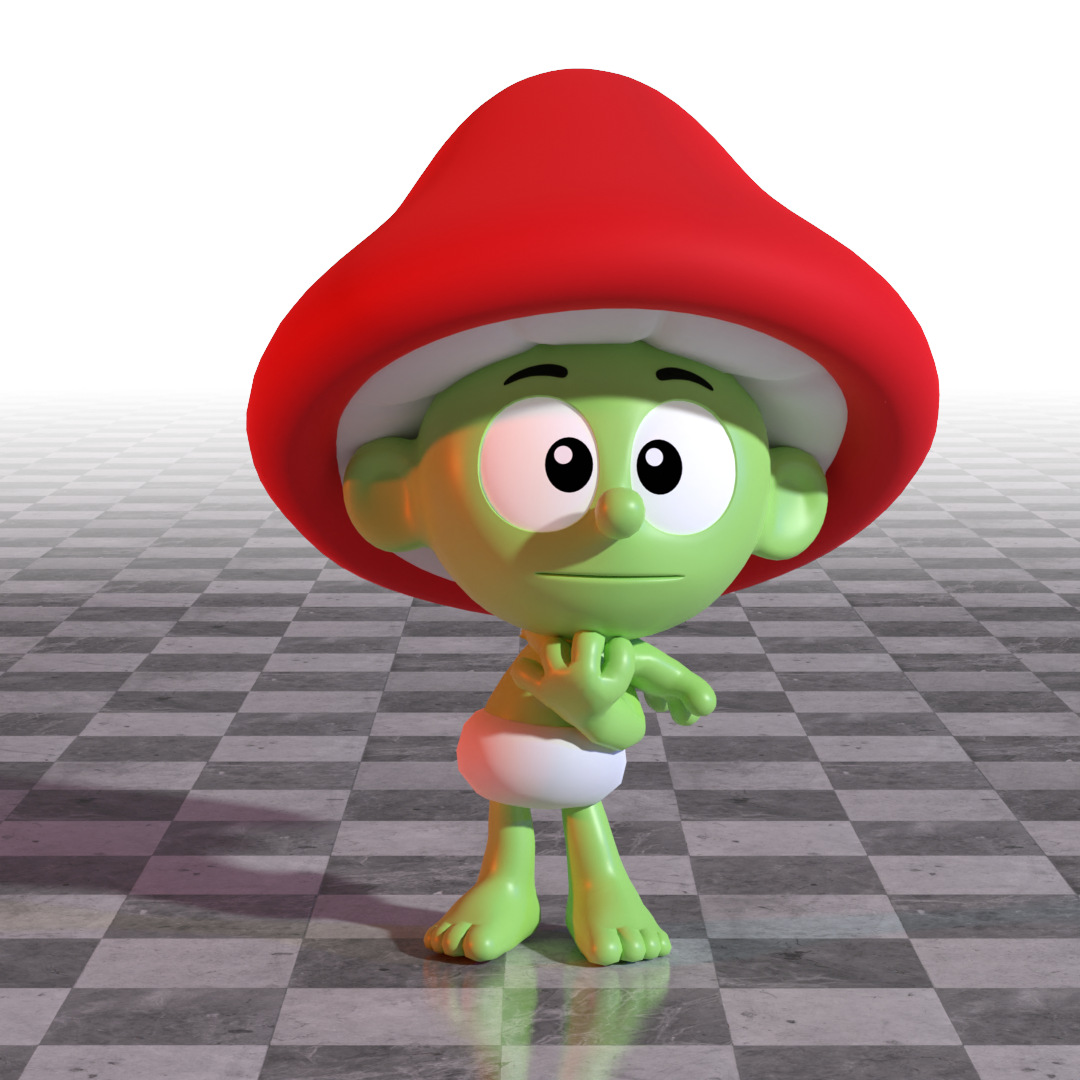}
        \centering
    \end{subfigure}%
    \hfill%
    \begin{subfigure}[b]{.16\textwidth}
        \includegraphics[width=\textwidth]{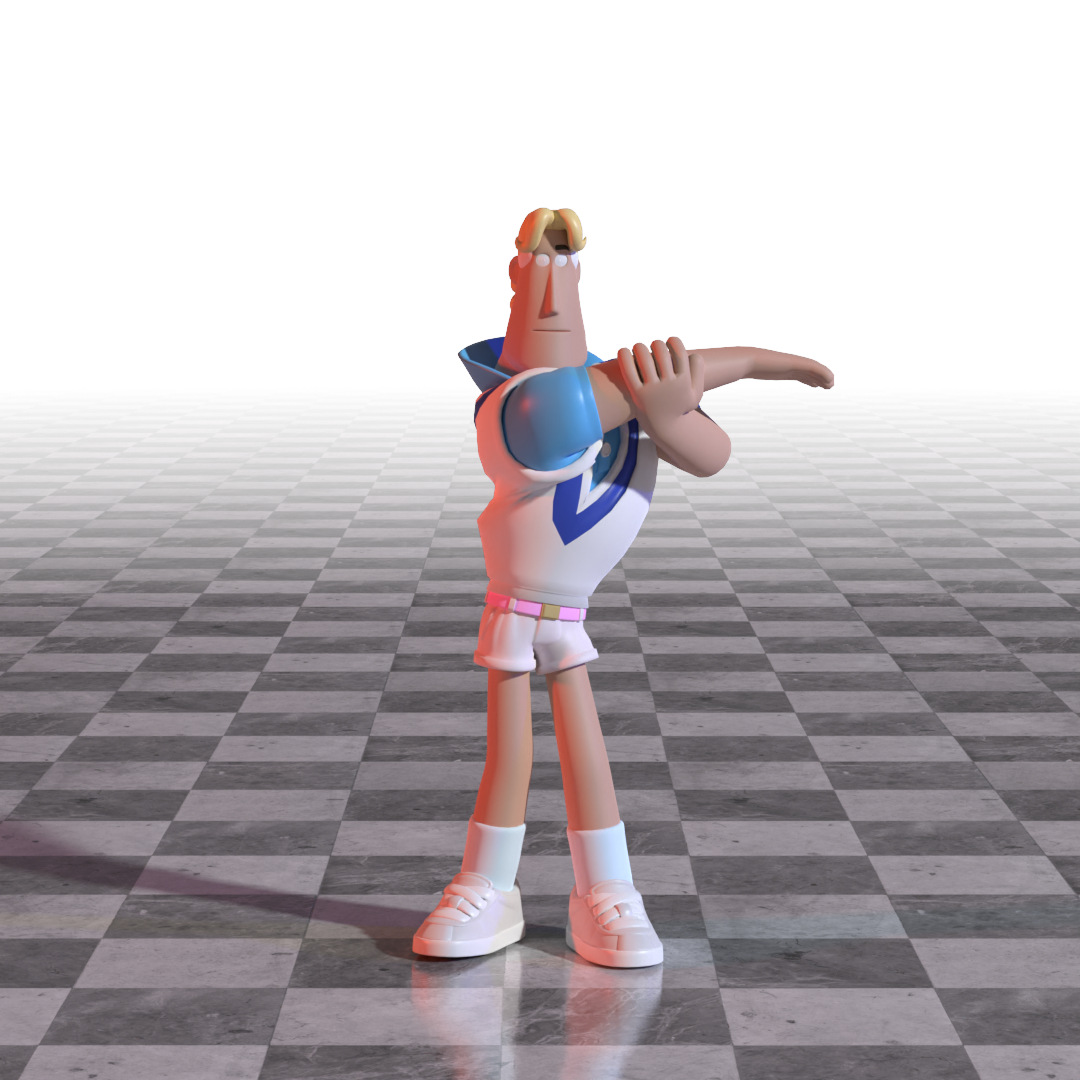}
        \centering
    \end{subfigure}%
    \hfill%
    \begin{subfigure}[b]{.16\textwidth}
        \includegraphics[width=\textwidth]{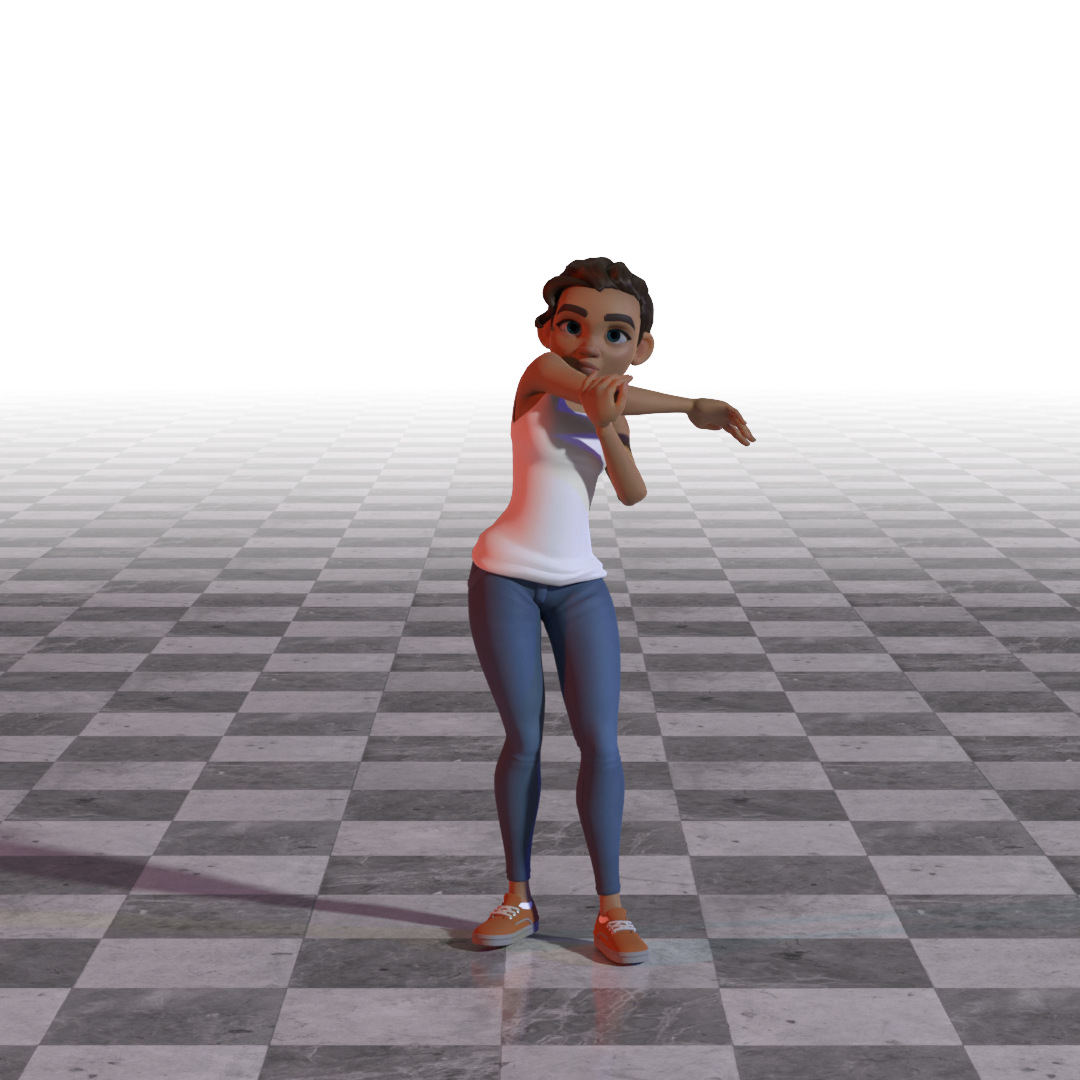}
        \centering
    \end{subfigure}%
    \hfill%
    \begin{subfigure}[b]{.16\textwidth}
        \includegraphics[width=\textwidth]{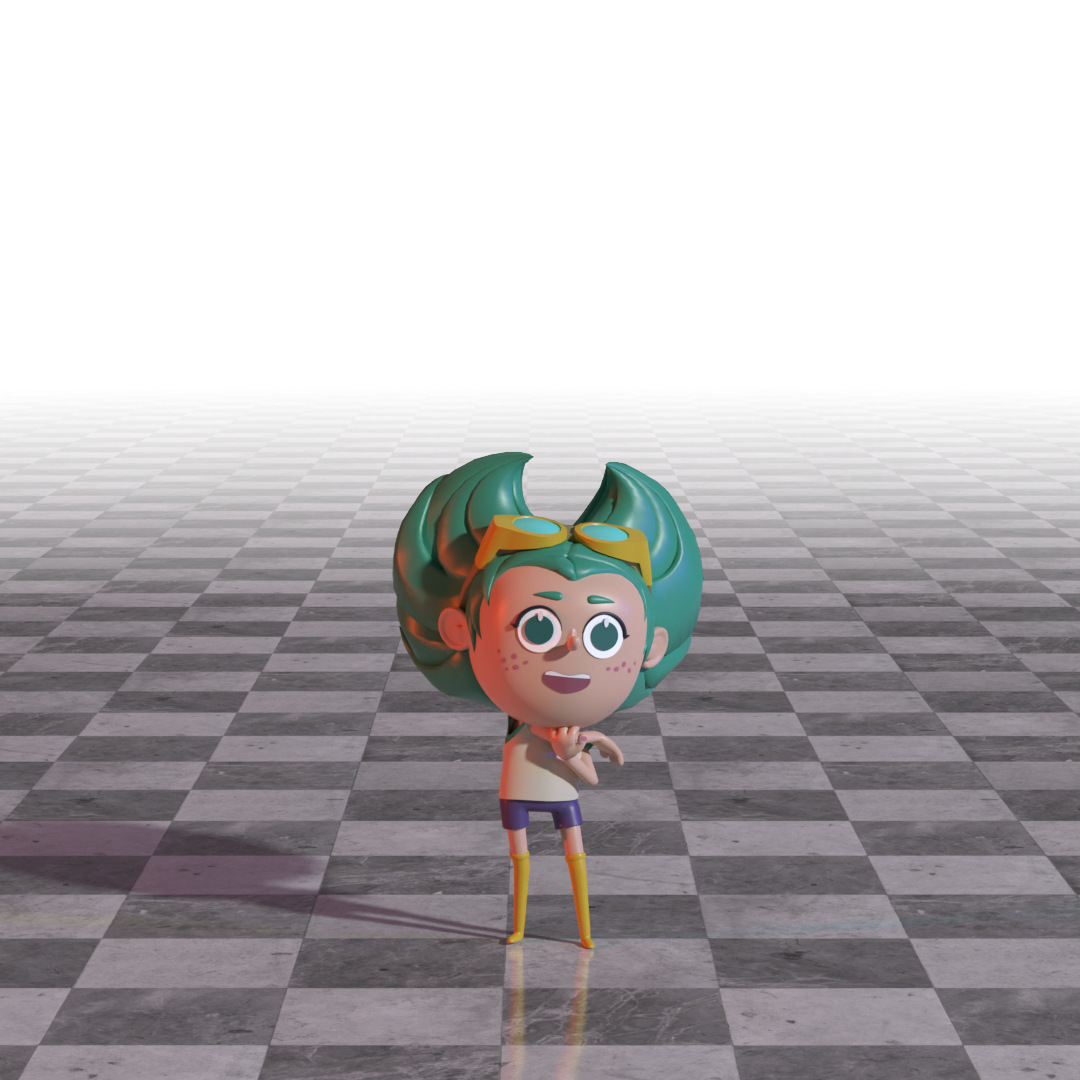}
        \centering
    \end{subfigure}
    \caption{Results of retargeting on characters from Blender Studio. Source poses on the left, followed by the results on several characters (from left to right : Phil, Sprite, Rex, Rain, Gabby)}
    \label{figure:more-characters}
\end{figure*}

To demonstrate the robustness of our method on various characters, we provide additional results on production-ready characters provided by Blender Studio (CC-BY license). Despite their extreme proportions, and their different skeleton structure (fewer spine bones, different hierarchy, presence of additional ``roll bones''...), our key-vertex transfer method provides a remarkably robust output. Several retargeted poses are shown in Figure \ref{figure:more-characters}. Whenever a good target pose is reachable, our optimization yields good results, and when the characters' morphology prevents it from mimicking a complex pose, the outputs stay visually pleasing, without any catastrophic failure.

\subsection{Failure cases}

\subsubsection{Key-vertices transfer}

We identified several failure cases which might cause suboptimal results: 
\begin{itemize}
    \item \textbf{Lack of geometry}: low-poly character with few vertices can cause poor results because of a low number of vertices to choose from.
    \item \textbf{Asymmetry}: while usually insignificant, some extreme asymmetries in the mesh can ``pull'' the key-vertices of some limbs toward one side, causing an asymmetry in the key-vertices that might not always be preferred.
    \item \textbf{Accessories / Props}: character modeled with props (i.e. bags, hats, large haircuts...) can cause local errors on the placement of key vertices (directly on or close to the prop).
    \item \textbf{Overlapping meshes}: when meshes overlap, the transfer might place key-vertices on a ``hidden'' part of the mesh, which could prevent the retargeting process from considering the outer part of the mesh.
\end{itemize}
A more complete description of the failure cases, along with examples can be found in the Supplementary Material. We also propose several straightforward solutions to alleviate these specific issues.

\subsubsection{Motion Retargeting}

We also identified some failure cases of the retargeting process:

\begin{itemize}
    \item \textbf{Complex poses}: poses with several points of contact (such as yoga poses) are harder to retarget, and finding a compromise between all semantic aspects of the pose can prove challenging when limb proportions are very different.
    \item \textbf{Unrealistic motion trajectories}: preventing floor collisions can cause a modification of the trajectory of the character, without enforcing the laws of dynamics, might create motions that are not physically accurate, although this effect is only visible on some cartoon characters with extreme proportions.
    \item \textbf{Fine-grained contacts on undersampled parts of the mesh}: using few keypoints constrains us to a simplified representation of the mesh, which might fail to capture contacts on some parts of the mesh.
\end{itemize}

A detailed analysis of those failure cases can be found in the Supplementary Material, with examples of each type of failure case.

\section{Extensions}


Thanks to the modular nature of our method, the optimization terms can be easily extended to account for specific and challenging scenarios previously unaddressed in the literature, 
as well as providing some degree of authoring and interactive control.

\subsection{Multi-character retargeting}

Handling multi-character interaction can be 
achieved
by extending 
the pose descriptors to account for all interactions between the characters.
These interactions may encompass not only physical contact 
but also 
some long-distance interactions such as gaze direction. In particular,
handling eye contact
is key in modeling plausible human interaction.

To this end, we first expand $\mathcal{W}_\text{interaction}$ and $\mathcal{W}_\text{height}$ to account for all possible distances 
between pairs of key-vertices of all the characters. 
The respective dimensions of these matrices become, respectively, $N\,N_\text{char}\times N\,N_\text{char}$ and $N\,N_\text{char}\times 1$, where $N_\text{char}$ designates the number of characters.
Second, we propose a modified $\mathcal{W}_\text{interaction}$ descriptor formulation accounting for an additional gaze direction term. Let us call $\mathcal{I}^{\text{eye}}=(i_c^{\text{eye}})_{c\in \llbracket1,N_\text{char}\rrbracket}$ the set of indices of the key-vertices associated with the eyes of a character $c$, and typically positioned at the center of the character's face. We further call $\overrightarrow{\text{gaze}_c}$ the averaged gaze direction of character $c$ at a given time, 
and define the following sparse matrix:
\begin{equation*}
\mathcal{M}_\text{gaze} \in \mathbb{R}^{N_\text{char} N \times 3} \text{,   where } \mathcal{M}_\text{gaze}(i) =
\left\{
\begin{array}{ll}
\overrightarrow{\text{gaze}_c} & \text{for } i\in \mathcal{I}^{\text{eye}} \\
0 & \text{otherwise.} \\
\end{array}
\right.
\end{equation*}
We finally 
define the extended $\mathcal{W}_\text{interaction}$ as :
\begin{equation*}
\mathcal{W}_\text{interaction} \mathrel{+}= \text{clamp}\left( \frac{S_\text{cosine}(\mathcal{M}_\text{dir}, \mathcal{M}_{\text{gaze}}) - \cos(a_\text{min})}{\cos(a_\text{min}) - \cos(a_\text{max})} \right)
\end{equation*}

\begin{figure}[b!]
    \centering
    \begin{subfigure}[b]{.23\textwidth}
        \includegraphics[width=\textwidth]{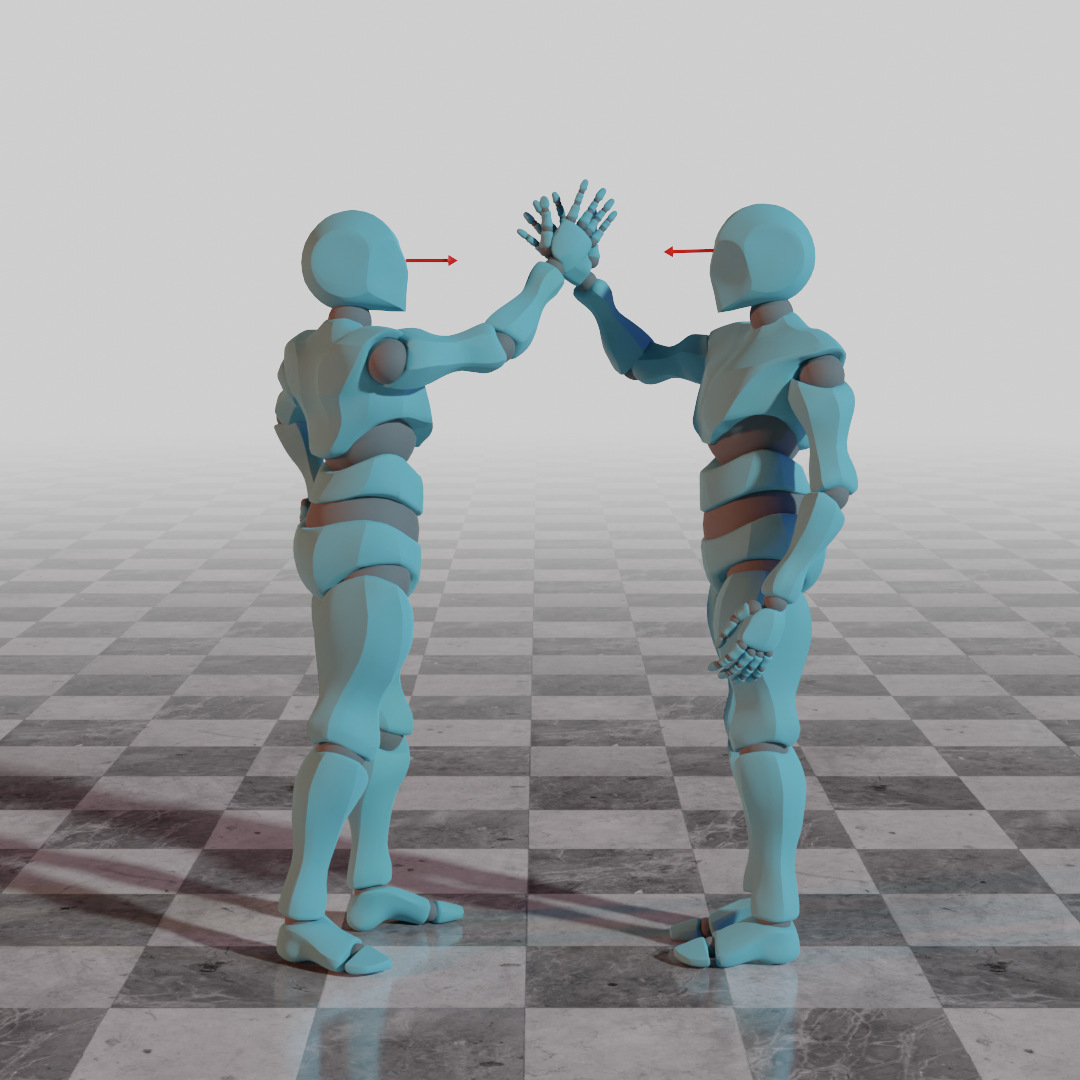}
        \caption{Source characters\\ $ $}
    \end{subfigure}%
    \hfill%
    \begin{subfigure}[b]{.23\textwidth}
        \includegraphics[width=\textwidth]{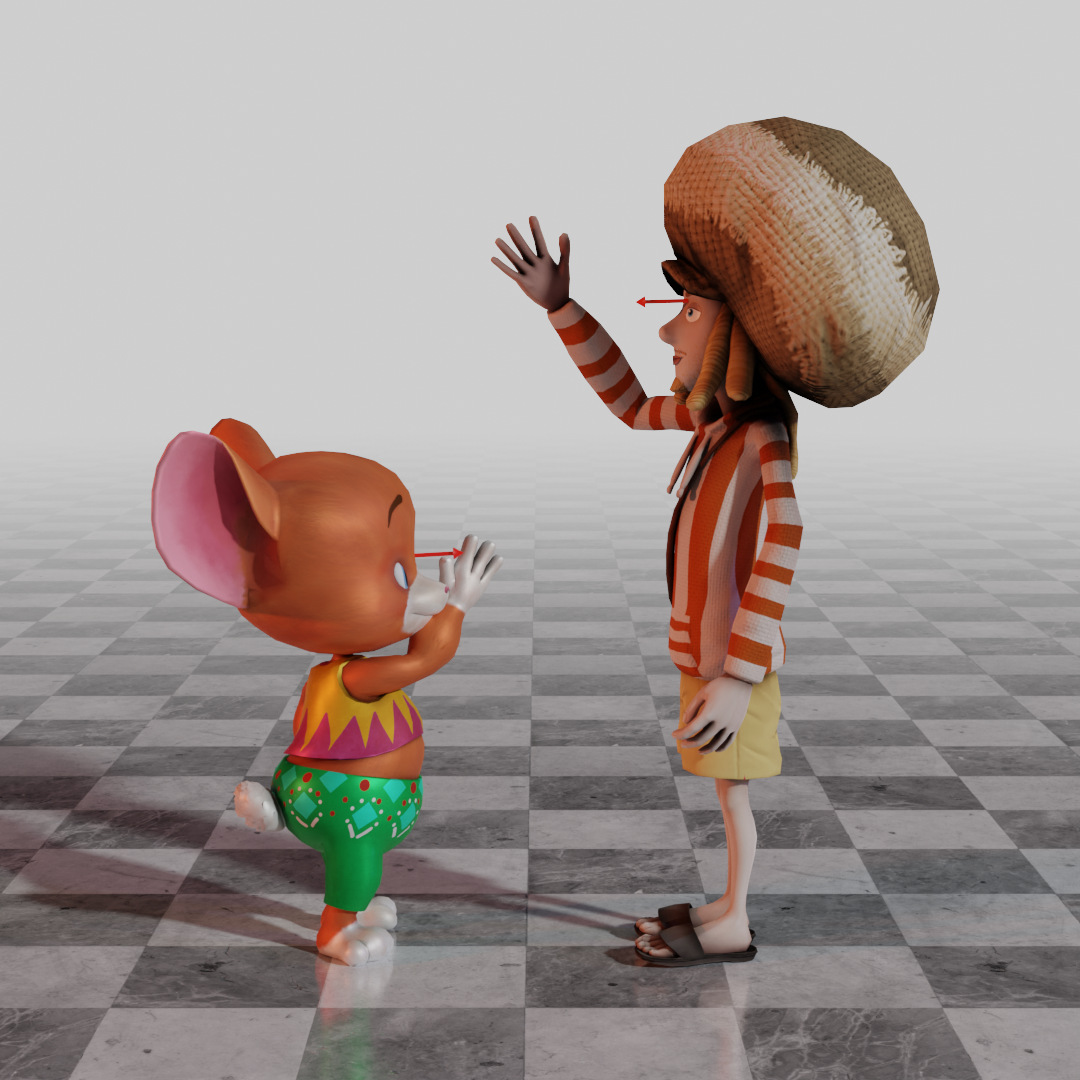}
        \caption{With no extra pose descriptor. \\ $ $
        }
    \end{subfigure}%
    
    \begin{subfigure}[b]{.23\textwidth}
        \includegraphics[width=\textwidth]{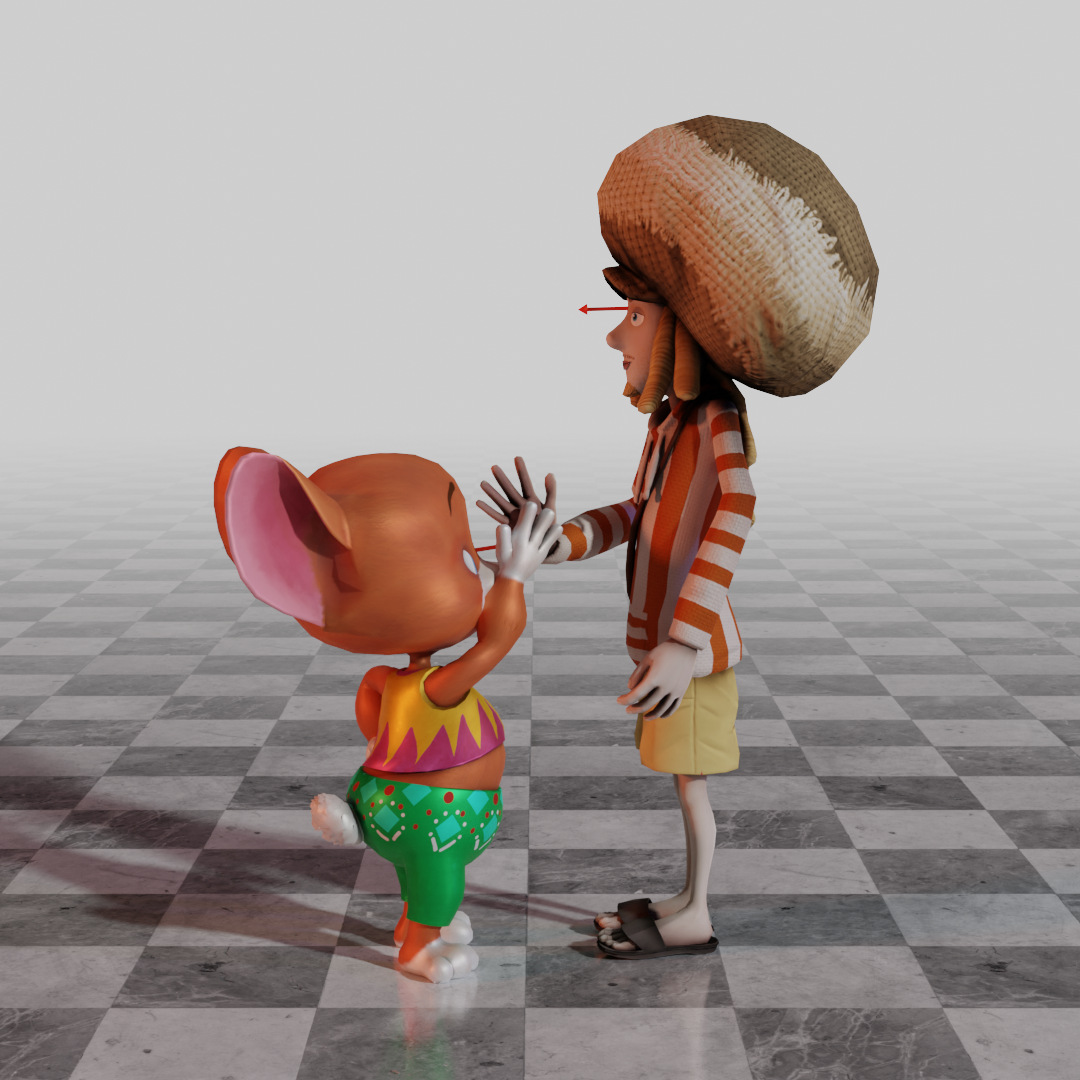}
        \caption{With inter-character contact features, but no gaze direction feature}
    \end{subfigure}%
    \hfill%
    \begin{subfigure}[b]{.23\textwidth}
        \includegraphics[width=\textwidth]{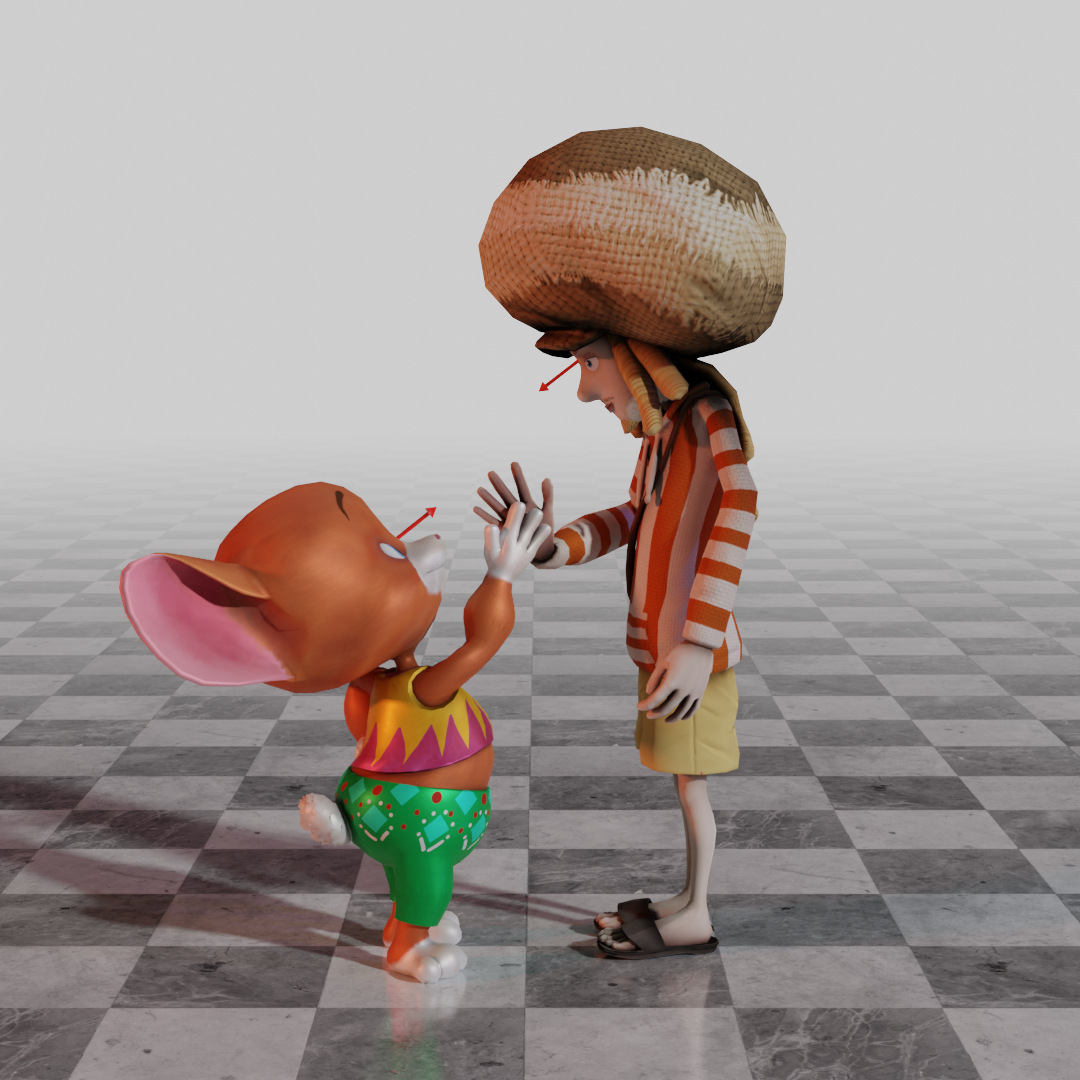}
        \caption{With both inter-character contact and gaze direction features.}
    \end{subfigure}
    \vspace{-0.2\baselineskip}
\caption{Results of the multi-character extension on a high-five pose. Arrows denote the gaze direction.}
\label{fig:multichar}
\end{figure}

The thresholds $a_{\min}$ and $a_{\max}$ are characteristic angles, set to 2\textdegree \ and 5\textdegree.
Figure \ref{fig:multichar} illustrates our results on multi-character interactions and shows the separate impacts of adding the inter-character physical contact and the gaze-based extension. Note how the optimization automatically changes the 
relative position of the characters, as well as their poses, to account for their height difference. 

\begin{figure}[b!]
    \centering
    \begin{subfigure}[b]{.23\textwidth}
        \includegraphics[width=\textwidth, trim=420 0 420 0, clip]{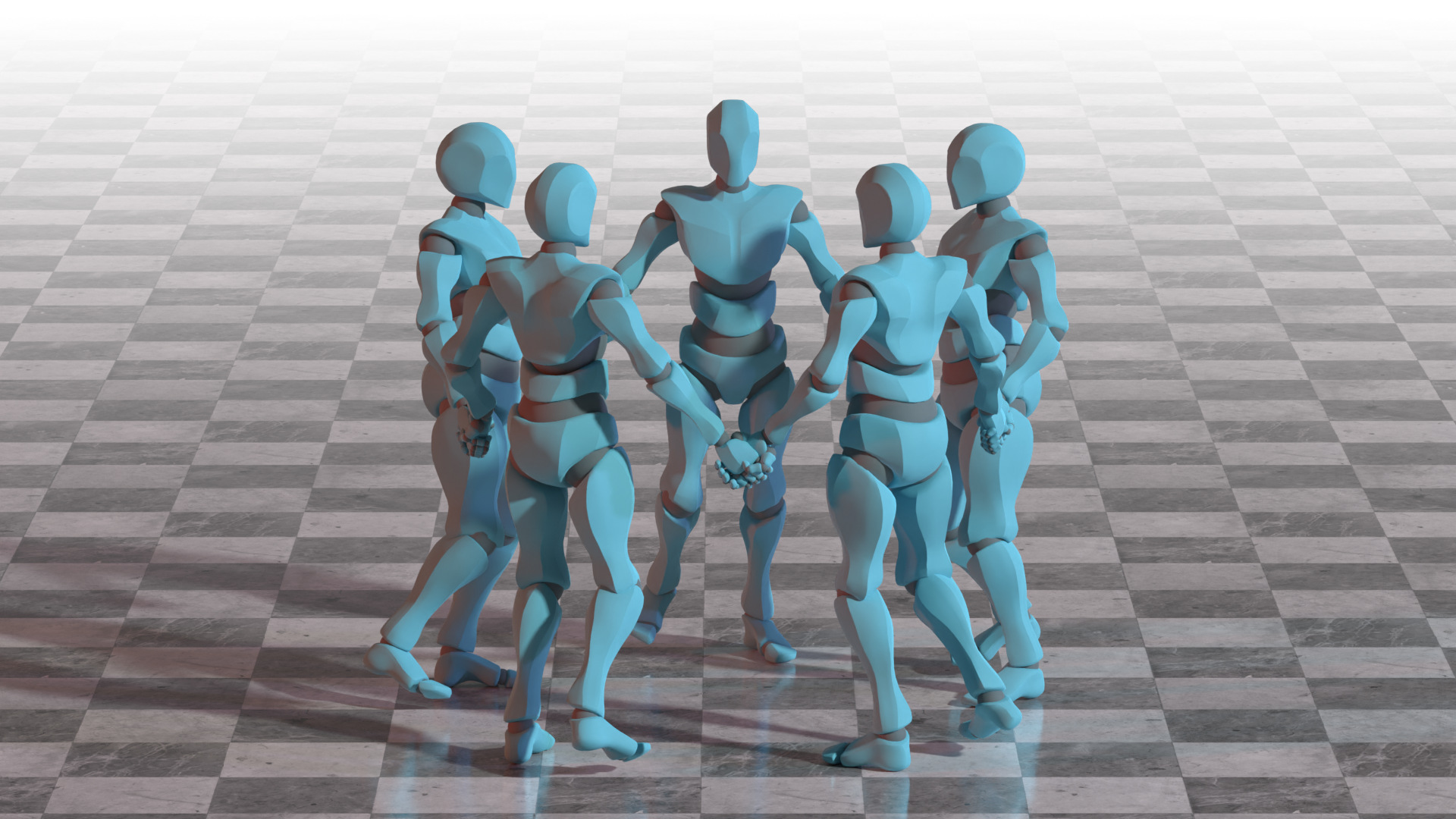}
        \caption{Source (side view)}
    \end{subfigure}%
    \hfill%
    \begin{subfigure}[b]{.23\textwidth}
        \includegraphics[width=\textwidth, trim=420 0 420 0, clip]{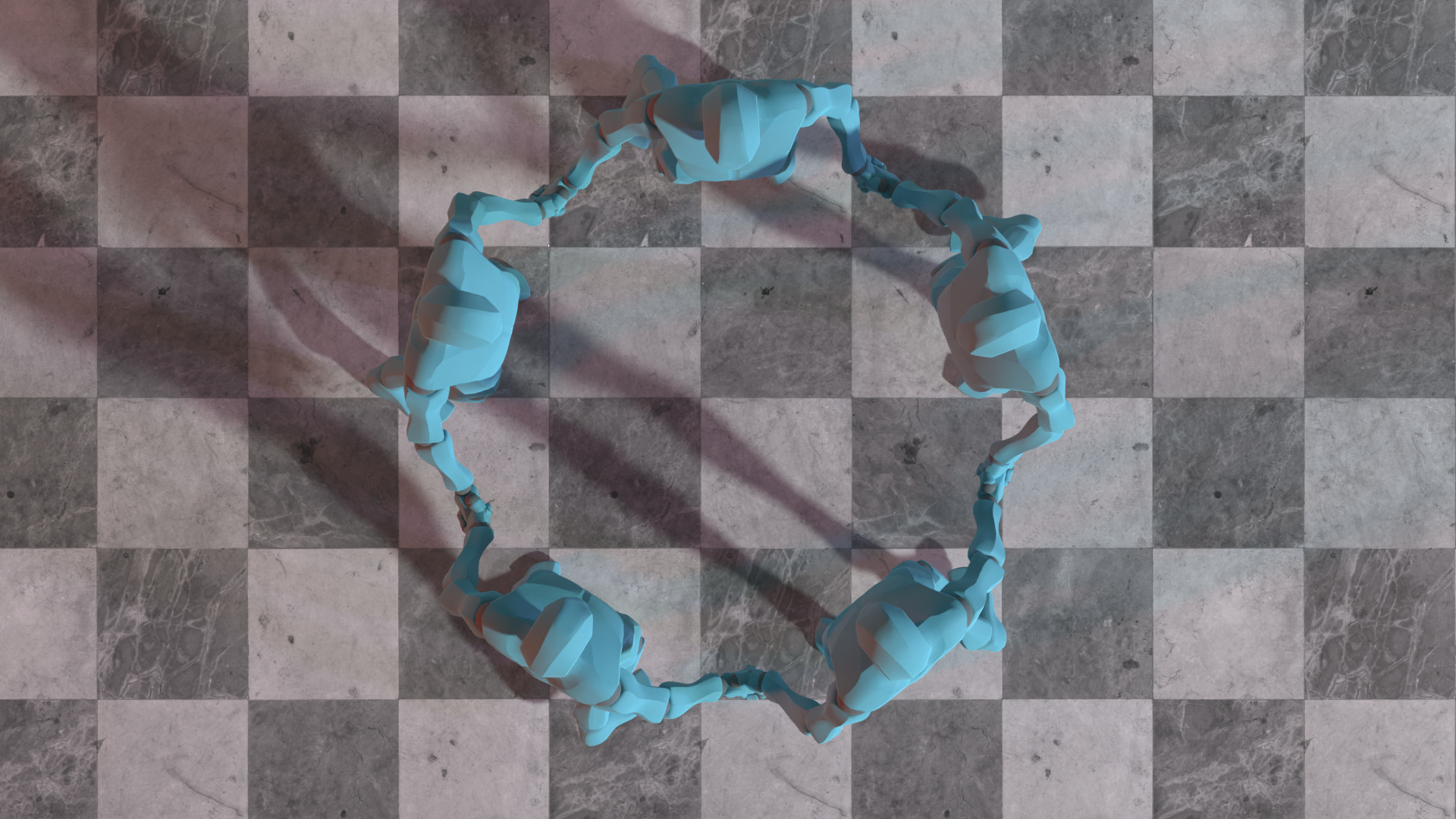}
        \caption{Source (top view)}
    \end{subfigure}%
    
    \begin{subfigure}[b]{.23\textwidth}
        \includegraphics[width=\textwidth, trim=420 0 420 0, clip]{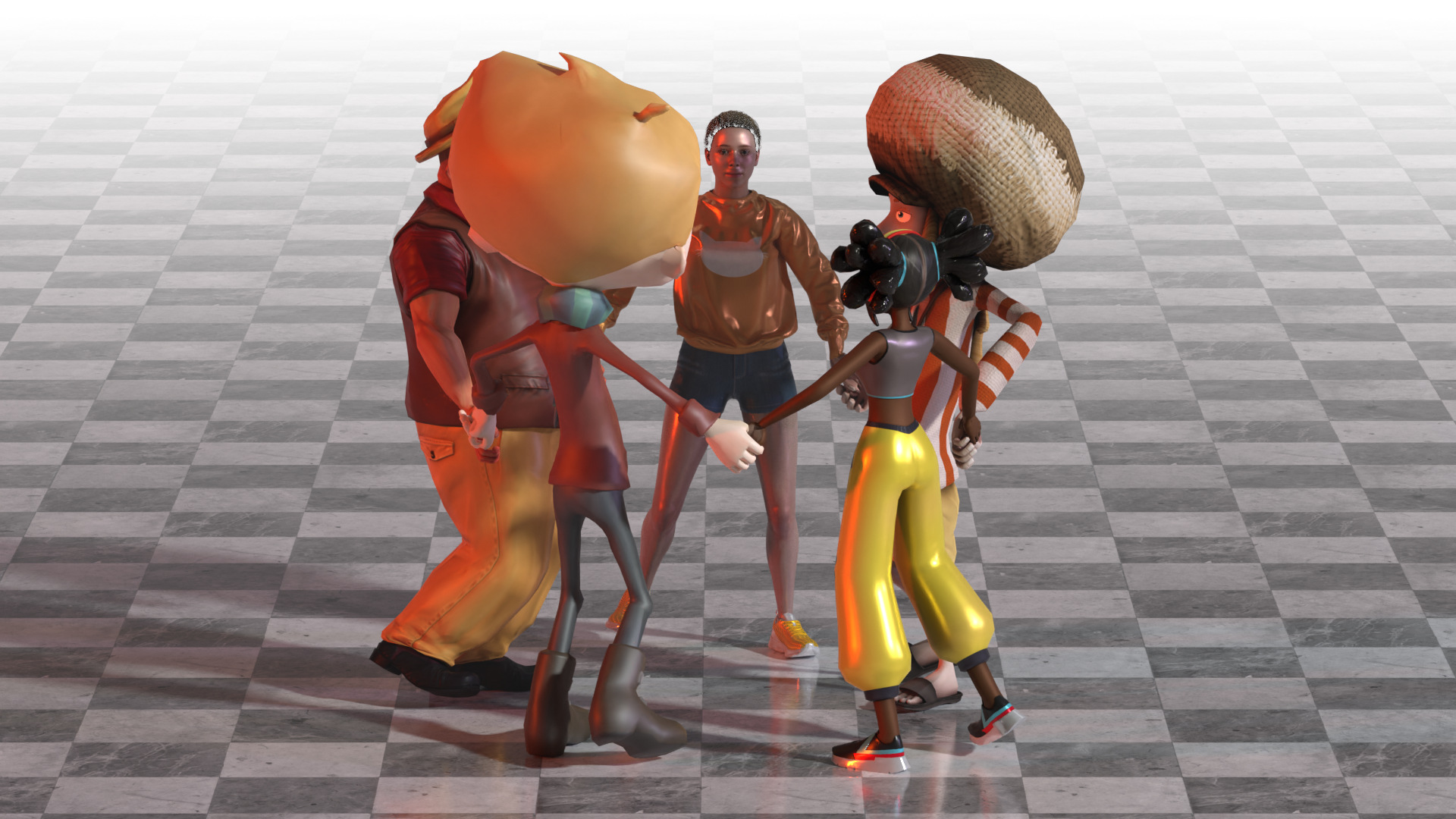}
        \caption{Retargeted (side view)}
    \end{subfigure}%
    \hfill%
    \begin{subfigure}[b]{.23\textwidth}
        \includegraphics[width=\textwidth, trim=420 0 420 0, clip]{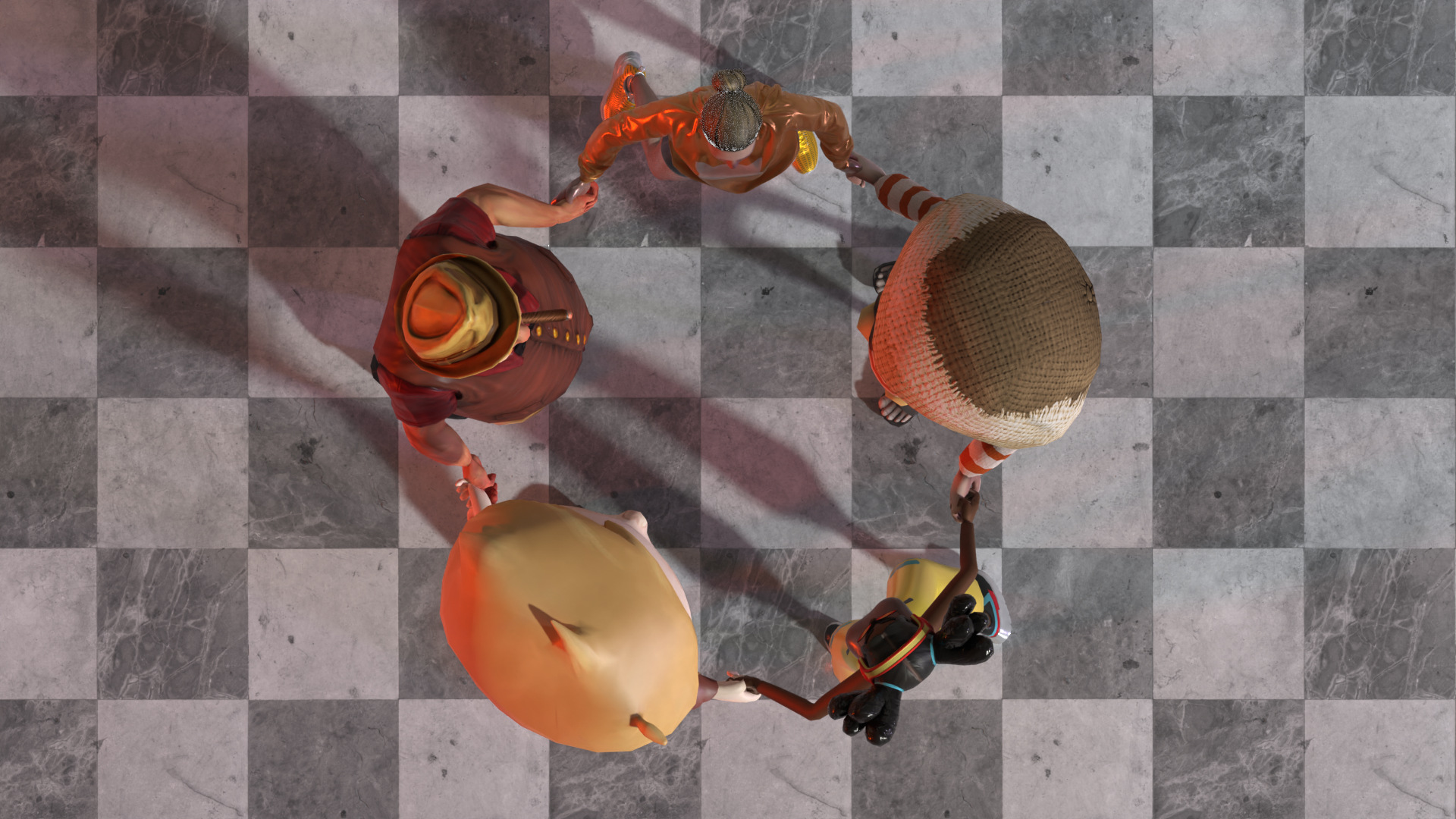}
        \caption{Retargeted (top view)}
    \end{subfigure}
    \vspace{-0.2\baselineskip}
\caption{Results of the multi-character extension on a five-character motion.}
\label{fig:multichar2}
\end{figure}

We also experiment with multi-character retargeting with strictly more than two characters. We design and retarget a five-character motion inspired from a traditional Breton dance, shown in Figure \ref{fig:multichar2}. We found that the multi-character retargeting only runs 19\% slower than five independent retargeting processes, due to the sparse nature of the interactions in the method. The speed and memory footprints could likely be optimized further using specific sparse computations libraries.

\subsection{Non-flat grounds}

\begin{figure}[!b]
    \centering

    \begin{subfigure}[b]{.32\columnwidth}
        \includegraphics[width=\textwidth, clip, trim=0 0 0 150]{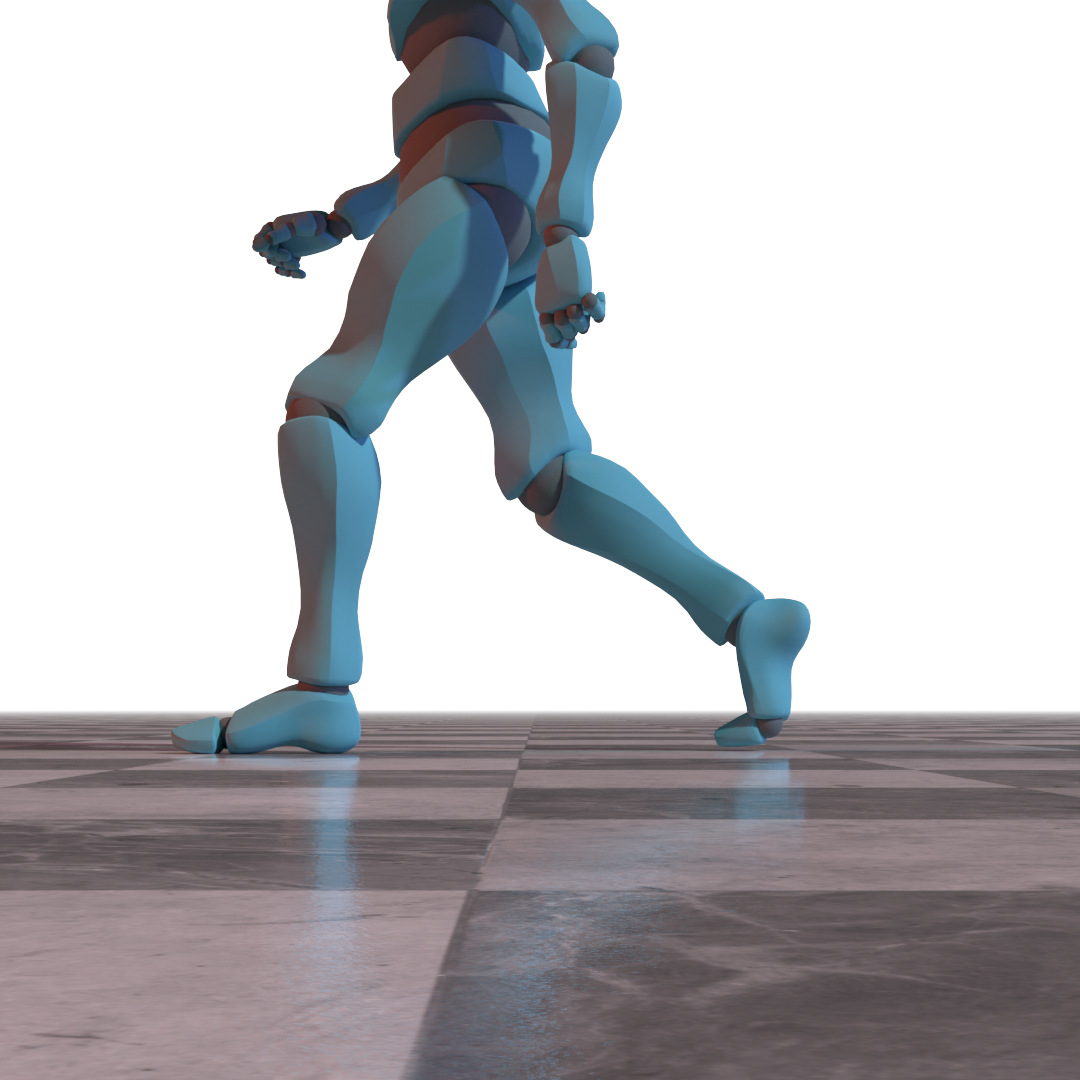}
        \caption{Source, flat ground}
    \end{subfigure}%
    \hfill%
    \begin{subfigure}[b]{.32\columnwidth}
        \includegraphics[width=\textwidth, clip, trim=0 0 0 150]{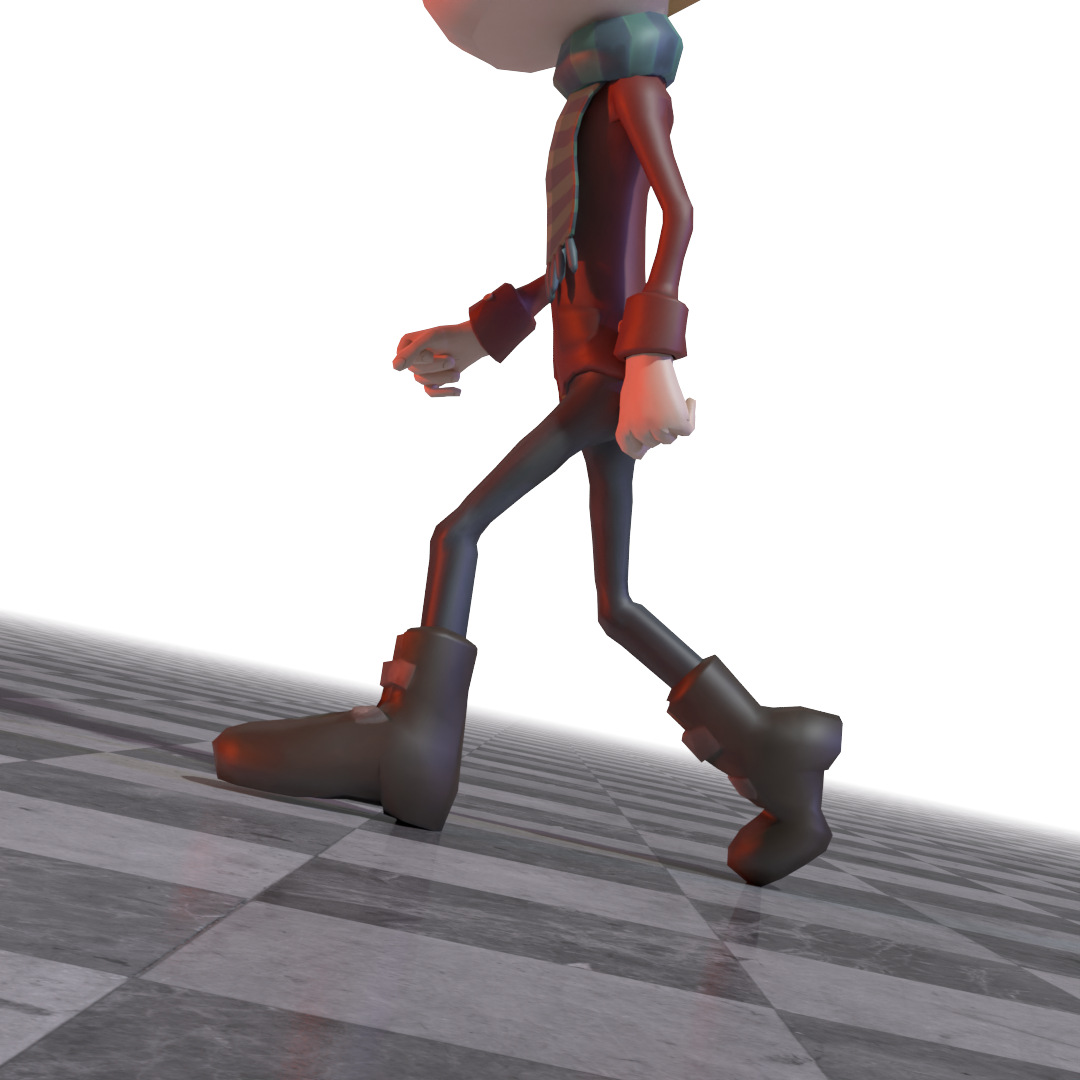}
        \caption{Target on slope 
        }
    \end{subfigure}%
    \hfill%
    \begin{subfigure}[b]{.32\columnwidth}
        \includegraphics[width=\textwidth, clip, trim=0 0 0 150]{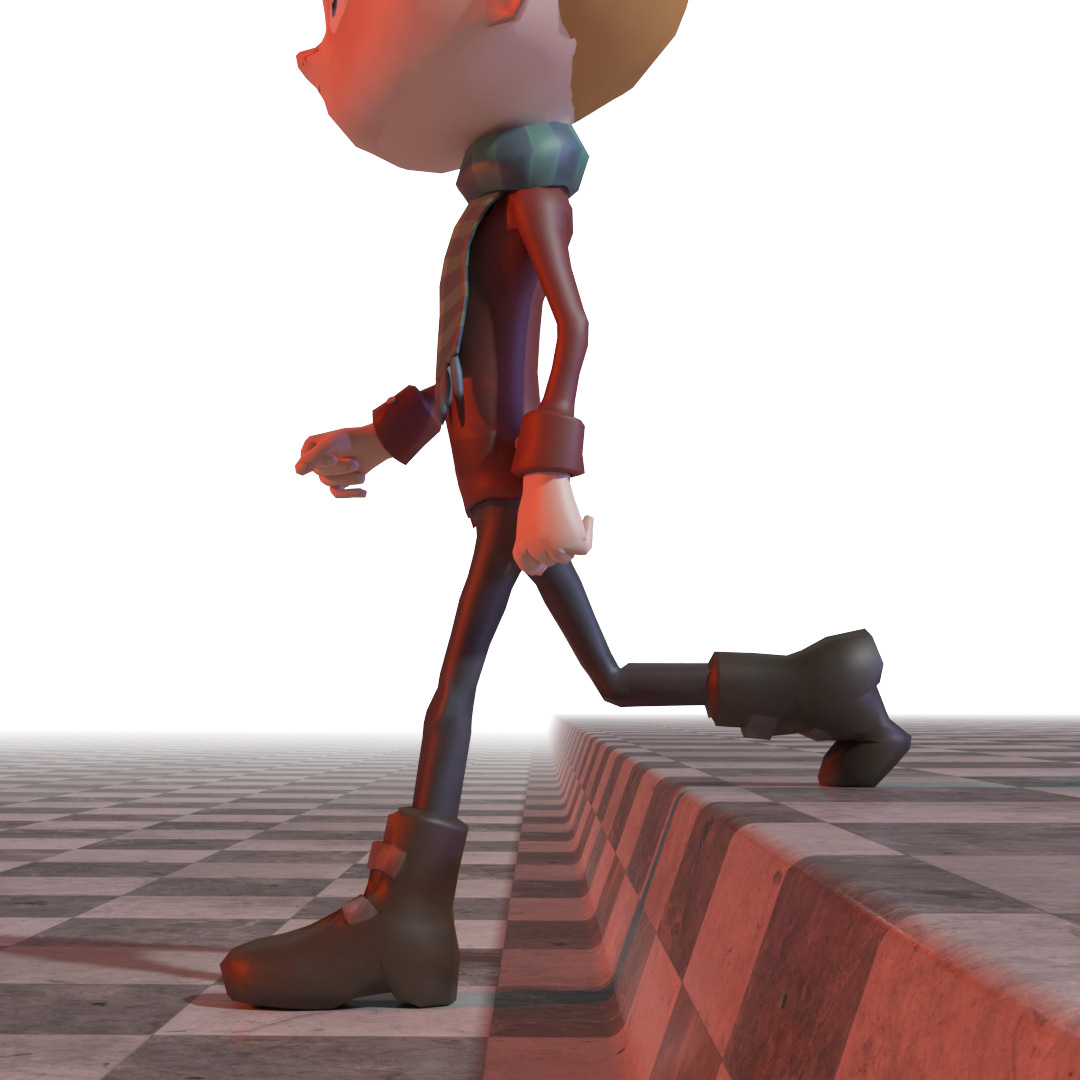}
        \caption{Target on step}
    \end{subfigure}
    \vspace{-0.6\baselineskip}
    \caption{Results of the non-flat terrain extension. }
    \label{fig:ground}
\end{figure}

A second extension of our approach is to generalize the retargeting to non-flat grounds. This can be straightforwardly implemented by adapting the height evaluation of the key-vertices to become coordinates-dependent. Considering a terrain defined as a differentiable height-field $y=f(x,z)$ and key-vertex position coordinates as $p_i=(x_i,y_i,z_i)$, we can replace $p_i\leftarrow p_i-f(x_i,z_i)$ in the computation of $\mathcal{M}_\text{height}$ and $\mathcal{M}_\text{sliding}$ descriptors.
Illustrations of results obtained using this approach are provided in Figure~\ref{fig:ground}.

\subsection{Identifying and solving conflicting constraints}



\begin{figure}[!t]
    \centering
    \hfill
    \begin{minipage}{.5\columnwidth}
        \centering
        \includegraphics[width=.9\textwidth, clip, trim=0 100 0 160]{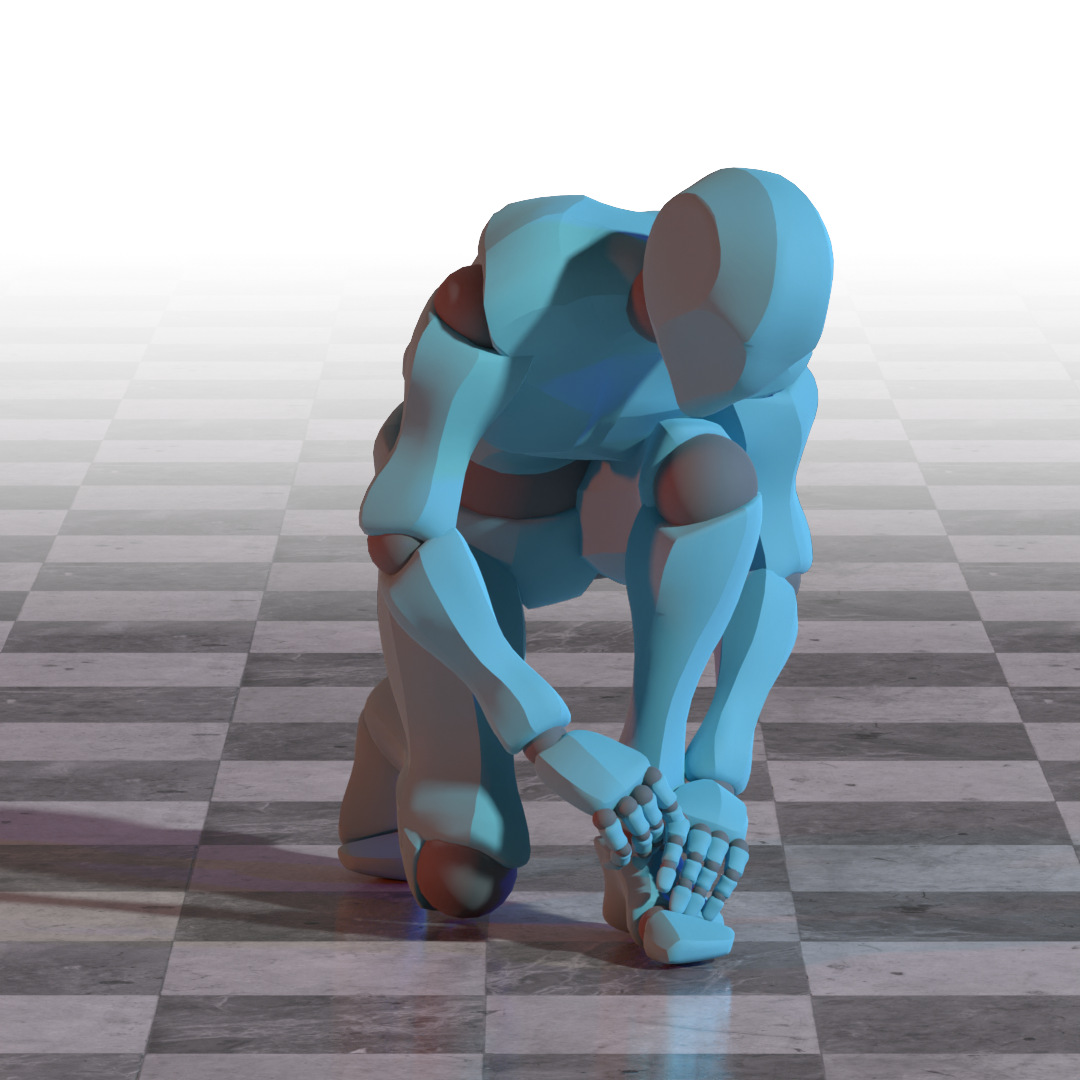}
        \subcaption{Source pose}
    \end{minipage}%
    \begin{minipage}{.5\columnwidth}
        \centering
        \includegraphics[width=.9\textwidth, clip, trim=0 100 0 160]{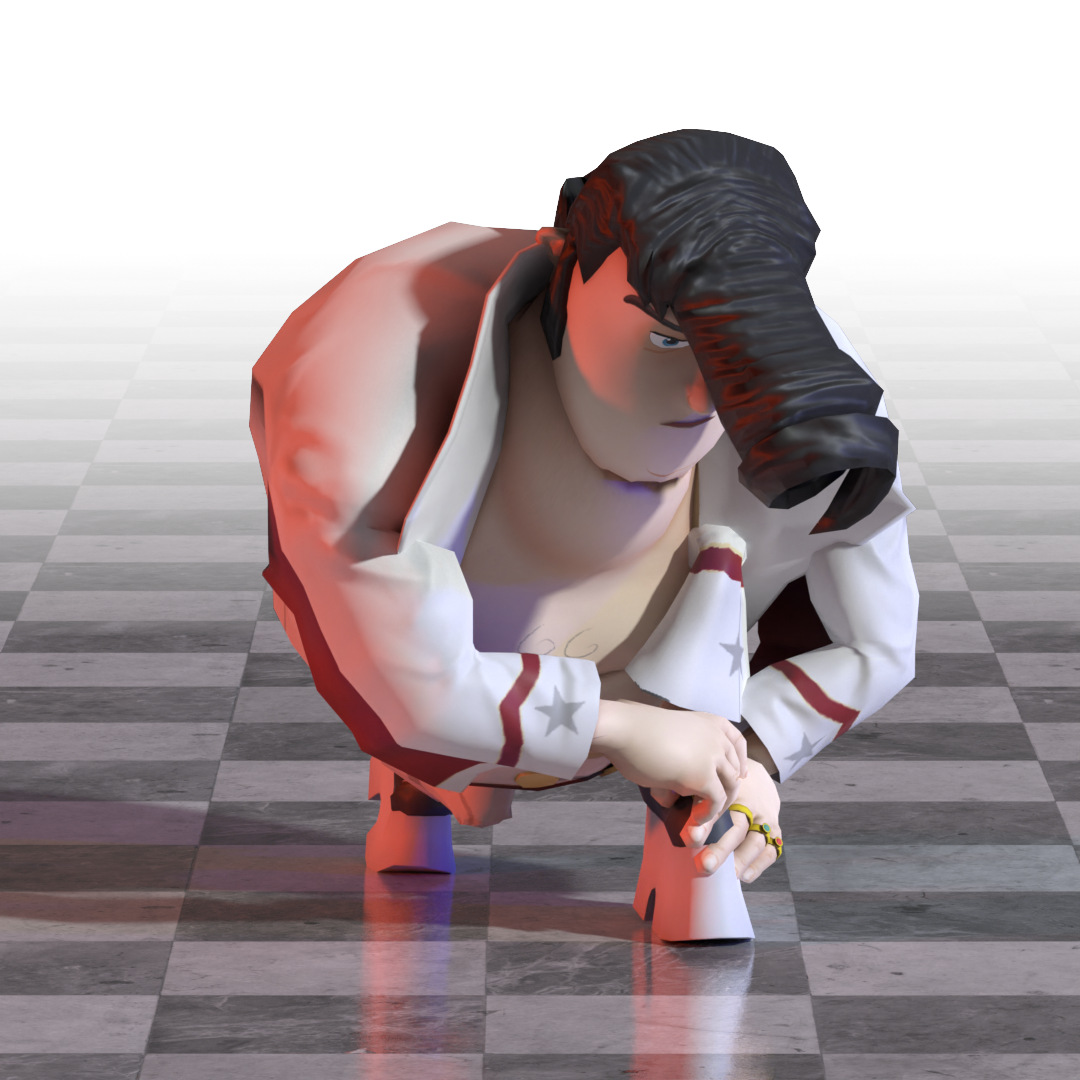}
        \subcaption{Conflict b/w $\mathcal{L}_\text{pen}$ and $\mathcal{L}_\text{dist}$}
    \end{minipage}
    \begin{minipage}{.5\columnwidth}
        \centering
        \includegraphics[width=.9\textwidth, clip, trim=0 100 0 160]{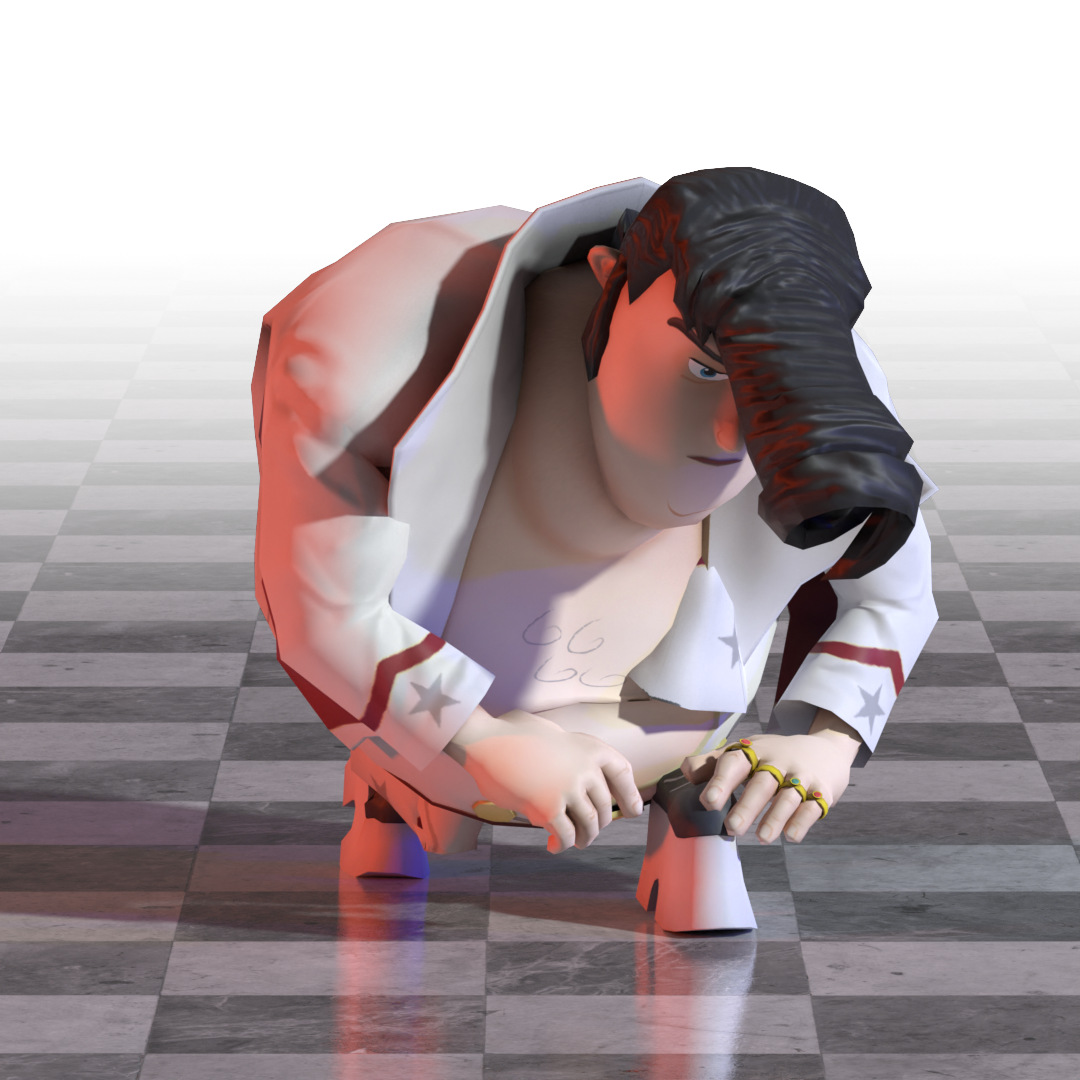}
        \subcaption{User gives priority to $\mathcal{L}_\text{dist}$}
        \label{fig:conflicting_losses1}
    \end{minipage}%
    \begin{minipage}{.5\columnwidth}
        \centering
        \includegraphics[width=.9\textwidth, clip, trim=0 100 0 160]{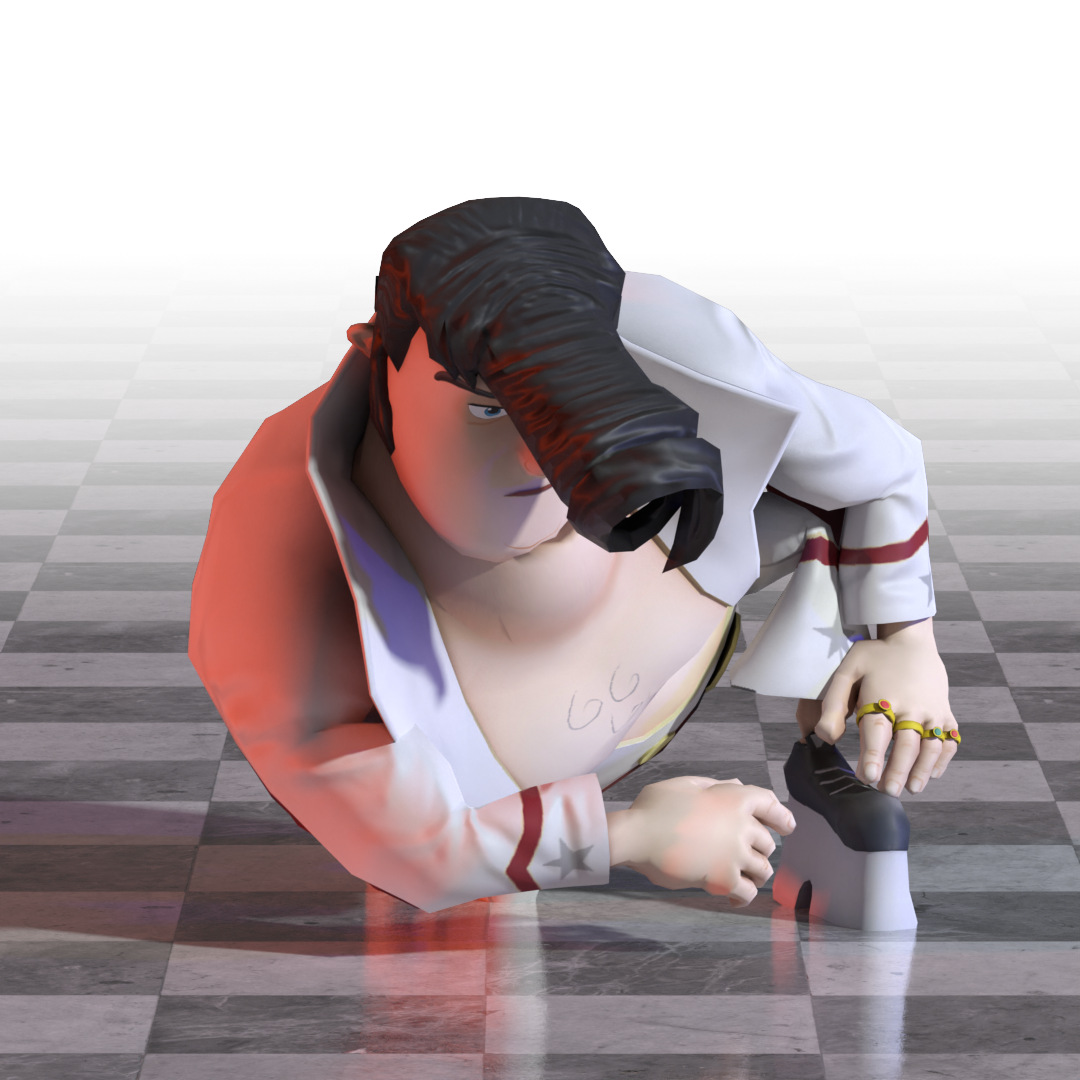}
        \subcaption{User gives priority to $\mathcal{L}_\text{pen}$}
        \label{fig:conflicting_losses2}
    \end{minipage}
    \vspace*{-.5\baselineskip}
    \caption{Example of conflicting losses and conflict resolution}
    \label{fig:conflicting_losses}
\end{figure}

An advantage of our formulation is its ability to automatically detect and identify conflicting semantic objectives during the optimization process. This identification can be coupled with an interactive adjustment of the weights assigned to the associated losses, thus offering an efficient authoring tool to handle complex retargeting scenarios.
Let us consider the loss component as a scalar function $\mathcal{L}_k(\mathbf{q_b})$, where $k$ is the type of loss (among dist, dir, pen, height, and sliding), and $b$ is a body part. The cosine similarity 
\begin{equation}
S_{\text{cosine}}\big(\nabla\mathcal{L}_{k_1}(\mathbf{q}_b), \nabla\mathcal{L}_{k_2}(\mathbf{q}_b)\big)
\end{equation}
is an indicator of the conflict between the loss $k_1$ and $k_2$ for the body part $b$. More precisely, a negative value close to -1 indicates strong conflicting goals between two losses.
Figure~\ref{fig:conflicting_losses} shows an example such use case where a conflict is identified between $\mathcal{L}_\text{pen}$ and $\mathcal{L}_\text{dist}$ on the torso, as the character is unable to reach its feet because of its large belly and short limbs. 
The user can then use an interactive cursor to adapt a weight between the two losses, so as to find the most pertinent output for their use case. In one case (Figure \ref{fig:conflicting_losses1}), collision is neglected to account for more accurate contacts; in the other (Figure \ref{fig:conflicting_losses2}), the output is free of collisions, but contacts are lost.

\section{Conclusion and future work}

We presented ReConForM, a novel motion retargeting method for characters of arbitrary shape and skeleton structure, focusing on self-collisions and ground contacts.
Our solution efficiently encodes shape and motion by creating time-varying pose descriptors based on the trajectory of selected key vertices of the character's meshes. Thanks to a proximity-based criterion, we designed a way to select and weight the descriptors that carry the most importance in each frame. By optimizing the target poses to conform as well as possible to the source's motion descriptions, we are able to retarget motion in real time. 
Compared to state-of-the-art methods, our algorithm achieves more accurate contacts with the ground, smoother motion, and effectively avoids self-collisions.

Our method's simplicity and flexibility allow easy adaptation to new use cases. We presented three extensions: accounting for inter-character contacts and gaze alignment to achieve multi-character retargeting; adapting the height of the ground for retargeting motion on non-flat terrains; and identifying conflicting constraints, allowing for interactive feedback and authoring, as prioritizing one of the constraints is often more desirable than getting high errors for them all. 

\subsection{Limitations and future work}

Our method, while achieving real-time results,
 has a few limitations, which we would like to address in future work. 
Firstly, it does not account for how a character's morphology (e.g., weight) affects its motion style (e.g., inertia).
Addressing this may request additional modifications beyond pure pose-based analysis, paving the way to include dynamic speed-based or
acceleration-based descriptors. 
Secondly, while our method can handle 
different skeleton topologies
and large geometrical variations from source to target, it cannot transfer motions between drastically different  
articulated structures that lack a common template, e.g. a target character with additional limbs.
To handle this case, additional work 
would be required to 
extend the method to non-bijective mapping between key-vertices. 
%
%
In addition, our method would highly benefit from an adaptive set of key-vertices, generated in collision areas of the source motion, and automatically transferred to the target character. This could enable the use of an even sparser set, but of more precise of motion descriptors over time.  
%
Lastly, combining our semantic losses with deep learning methods could accelerate inference, while reinforcement learning methods could be used to ensure physically plausible motions.
The latter may be particularly useful when retargeting to non-flat grounds, where equilibrium might be lost.
Finally, since assessing the quality of retargeting strongly benefits from human 
feedback, human-in-the-loop reinforcement learning might give new insights on how to move beyond manually-designed motion descriptors.

\printbibliography 

@String{Computing = "Computing" }

@String{Computer = "{IEEE} Computer" }

@String{Springer = "Springer-Verlag" }

@inproceedings{basset22,
author = {Basset, Jean and Ouannas, Badr and Hoyet, Ludovic and Multon, Franck and Wuhrer, Stefanie},
title = {Impact of Self-Contacts on Perceived Pose Equivalences},
year = {2022},
isbn = {9781450398886},
url = {https://doi.org/10.1145/3561975.3562946},
doi = {10.1145/3561975.3562946},
booktitle = {Motion, Interaction and Games (MIG)},
articleno = {6},
numpages = {10},
}

@inproceedings{villegas,
  title={Contact-aware retargeting of skinned motion},
  author={Villegas, Ruben and Ceylan, Duygu and Hertzmann, Aaron and Yang, Jimei and Saito, Jun},
  booktitle={Proceedings of the IEEE/CVF International Conference on Computer Vision},
  pages={9720--9729},
  year={2021}
}

@article{basset20,
title = {Contact preserving shape transfer: Retargeting motion from one shape to another},
journal = {Computers \& Graphics},
volume = {89},
pages = {11-23},
year = {2020},
issn = {0097-8493},
doi = {https://doi.org/10.1016/j.cag.2020.04.002},
url = {https://www.sciencedirect.com/science/article/pii/S0097849320300406},
author = {Jean Basset and Stefanie Wuhrer and Edmond Boyer and Franck Multon},
}

@inproceedings{taeil17,
author = {Jin, Taeil and Kim, Meekyung and Lee, Sung-Hee},
title = {Motion Retargeting to Preserve Spatial Relationship between Skinned Characters},
year = {2017},
isbn = {9781450350914},
url = {https://doi.org/10.1145/3099564.3106647},
doi = {10.1145/3099564.3106647},
booktitle = {Symposium on Computer Animation (SCA)},
articleno = {25},
numpages = {2},
}

@InProceedings{nkn,
  author = {Villegas, Ruben and Yang, Jimei and Ceylan, Duygu and Lee, Honglak},
  title = {Neural Kinematic Networks for Unsupervised Motion Retargetting},
  booktitle = {The IEEE Conference on Computer Vision and Pattern Recognition (CVPR)},
  month = {June},
  year = {2018}
}

@InProceedings{pmnet,
  author = {Lim, Jongin and Chang, Hyung Jin and Choi, Jin Young},
  title = {PMnet: Learning of Disentangled Pose and Movement for Unsupervised Motion Retargeting},
  booktitle = {British Machine Vision Conference (BMVC)},
  year = {2019}
}

@article{san,
  author = {Aberman, Kfir and Li, Peizhuo and Lischinski, Dani and Sorkine-Hornung, Olga and Cohen-Or, Daniel and Chen, Baoquan},
  title = {Skeleton-Aware Networks for Deep Motion Retargeting},
  journal = {ACM Transactions on Graphics (TOG), Proc. SIGGRAPH},
  volume = {39},
  number = {4},
  pages = {62},
  year = {2020},
  publisher = {ACM}
}

@InProceedings{r2et,
    author    = {Zhang, Jiaxu and Weng, Junwu and Kang, Di and Zhao, Fang and Huang, Shaoli and Zhe, Xuefei and Bao, Linchao and Shan, Ying and Wang, Jue and Tu, Zhigang},
    title     = {Skinned Motion Retargeting With Residual Perception of Motion Semantics \& Geometry},
    booktitle = {Proceedings of the IEEE/CVF Conference on Computer Vision and Pattern Recognition (CVPR)},
    month     = {June},
    year      = {2023},
    pages     = {13864-13872}
}

@article{reda23,
author = {Reda, Daniele and Won, Jungdam and Ye, Yuting and Panne, Michiel and Winkler, Alexander},
year = {2023},
month = {08},
pages = {1-19},
title = {Physics-based Motion Retargeting from Sparse Inputs},
volume = {6},
journal = {SCA, Proceedings of the ACM on Computer Graphics and Interactive Techniques},
doi = {10.1145/3606928}
}

@inproceedings{yunbo23,
author = {Zhang, Yunbo and Gopinath, Deepak and Ye, Yuting and Hodgins, Jessica and Turk, Greg and Won, Jungdam},
title = {Simulation and Retargeting of Complex Multi-Character Interactions},
year = {2023},
isbn = {9798400701597},
url = {https://doi.org/10.1145/3588432.3591491},
doi = {10.1145/3588432.3591491},
booktitle = {ACM SIGGRAPH 2023 Conference Proceedings},
articleno = {65},
numpages = {11},
}

@article{solomon,
author = {Solomon, Justin and Peyr\'{e}, Gabriel and Kim, Vladimir G. and Sra, Suvrit},
title = {Entropic Metric Alignment for Correspondence Problems},
year = {2016},
issue_date = {July 2016},
publisher = {Association for Computing Machinery},
address = {New York, NY, USA},
volume = {35},
number = {4},
issn = {0730-0301},
url = {https://doi.org/10.1145/2897824.2925903},
doi = {10.1145/2897824.2925903},
journal = {ACM Transactions on Graphics},
month = {jul},
articleno = {72},
numpages = {13}
}

@misc{mixamo,
  author = {Adobe},
  title = {Mixamo},
  howpublished = {\url{https://www.mixamo.com/}},
  note = {Accessed: 2023-09-12},
  year={{2024}},
}

@misc{prolific,
  author = {Prolific},
  howpublished = {\url{https://www.prolific.com}},
  note = {Accessed: 2023-12},
  year={2023},
}

@article{shon,
  title={Learning shared latent structure for image synthesis and robotic imitation},
  author={Shon, Aaron and Grochow, Keith and Hertzmann, Aaron and Rao, Rajesh P},
  journal={Advances in neural information processing systems},
  volume={18},
  year={2005}
}

@inproceedings{gleicher,
  title={Retargetting motion to new characters},
  author={Gleicher, Michael},
  booktitle={Proc. ACM SIGGRAPH},
  pages={33--42},
  year={1998}
}

@article{choi,
  title={Online motion retargetting},
  author={Choi, Kwang-Jin and Ko, Hyeong-Seok},
  journal={The Journal of Visualization and Computer Animation},
  volume={11},
  number={5},
  pages={223--235},
  year={2000},
  publisher={Wiley Online Library}
}

@article{tak,
  title={A physically-based motion retargeting filter},
  author={Tak, Seyoon and Ko, Hyeong-Seok},
  journal={ACM Transactions on Graphics},
  volume={24},
  number={1},
  pages={98--117},
  year={2005},
  publisher={ACM New York, NY, USA}
}

@inproceedings{lee,
  title={A hierarchical approach to interactive motion editing for human-like figures},
  author={Lee, Jehee and Shin, Sung Yong},
  booktitle={Proc. ACM SIGGRAPH},
  pages={39--48},
  year={1999}
}

@article{smpl,
      author = {Loper, Matthew and Mahmood, Naureen and Romero, Javier and Pons-Moll, Gerard and Black, Michael J.},
      title = {{SMPL}: A Skinned Multi-Person Linear Model},
      journal = {ACM Transactions on Graphics (Proc. SIGGRAPH Asia)},
      number = {6},
      pages = {248:1--248:16},
      publisher = {ACM},
      volume = {34},
      year = {2015}
    }

@article{Hu_2023,
   title={Pose-Aware Attention Network for Flexible Motion Retargeting by Body Part},
   ISSN={2160-9306},
   url={http://dx.doi.org/10.1109/TVCG.2023.3277918},
   DOI={10.1109/tvcg.2023.3277918},
   journal={IEEE Transactions on Visualization and Computer Graphics},
   publisher={Institute of Electrical and Electronics Engineers (IEEE)},
   author={Hu, Lei and Zhang, Zihao and Zhong, Chongyang and Jiang, Boyuan and Xia, Shihong},
   year={2023},
   pages={1–17} }

@inproceedings{van2011survey,
  title={A survey on shape correspondence},
  author={Van Kaick, Oliver and Zhang, Hao and Hamarneh, Ghassan and Cohen-Or, Daniel},
  booktitle={Computer graphics forum},
  volume={30},
  number={6},
  pages={1681--1707},
  year={2011},
  organization={Wiley Online Library}
}

@article{recent_advances,
author = {Sahillio\u{g}lu, Yusuf},
title = {Recent Advances in Shape Correspondence},
year = {2020},
issue_date = {Aug 2020},
publisher = {Springer-Verlag},
address = {Berlin, Heidelberg},
volume = {36},
number = {8},
issn = {0178-2789},
url = {https://doi.org/10.1007/s00371-019-01760-0},
doi = {10.1007/s00371-019-01760-0},
journal = {Vis. Comput.},
month = {aug},
pages = {1705–1721},
numpages = {17}
}

@article{adam,
  title={Adam: A Method for Stochastic Optimization},
  author={Diederik P. Kingma and Jimmy Ba},
  journal={CoRR},
  year={2014},
  volume={abs/1412.6980},
  url={https://api.semanticscholar.org/CorpusID:6628106}
}

@article{schmidt2023surface,
  title={Surface Maps via Adaptive Triangulations},
  author={Schmidt, Patrick and Pieper, D\"orte and Kobbelt, Leif},
  year={2023},
  journal={Computer Graphics Forum},
  volume={42},
  number={2},
}

@inproceedings{feydy2019interpolating,
    title={Interpolating between Optimal Transport and MMD using Sinkhorn Divergences},
    author={Feydy, Jean and S{\'e}journ{\'e}, Thibault and Vialard, Fran{\c{c}}ois-Xavier and Amari, Shun-ichi and Trouve, Alain and Peyr{\'e}, Gabriel},
    booktitle={The 22nd International Conference on Artificial Intelligence and Statistics},
    pages={2681--2690},
    year={2019}
}

@inproceedings{1eurofilter,
  title={1€ filter: a simple speed-based low-pass filter for noisy input in interactive systems},
  author={Casiez, G{\'e}ry and Roussel, Nicolas and Vogel, Daniel},
  booktitle={Proceedings of the SIGCHI Conference on Human Factors in Computing Systems},
  pages={2527--2530},
  year={2012}
}

@article{desbrun,
    TITLE = {{Variance-Minimizing Transport Plans for Inter-surface Mapping}},
    AUTHOR = {Mandad, Manish and Cohen-Steiner, David and Kobbelt, Leif and Alliez, Pierre and Desbrun, Mathieu},
    URL = {https://inria.hal.science/hal-01519006},
    JOURNAL = {{ACM Transactions on Graphics, Proc. SIGGRAPH}},
    PUBLISHER = {{Association for Computing Machinery}},
    VOLUME = {36},
    PAGES = {14},
    YEAR = {2017},
    DOI = {10.1145/3072959.3073671},
    HAL_ID = {hal-01519006},
    HAL_VERSION = {v1},
}

@article{ho2010,
    author = {Ho, Edmond and Komura, Taku and Tai, Chiew-Lan},
    year = {2010},
    month = {07},
    pages = {},
    title = {Spatial Relationship Preserving Character Motion Adaptation},
    volume = {29},
    journal = {ACM Trans. Graph.},
    doi = {10.1145/1833351.1778770}
}

@misc{minecraft,
    title={Minecraft}, 
    author={Mojang},
    year={2024},
    url={http://minecraft.net},
}

@inproceedings{liao2022pose,
    title = {Skeleton-free Pose Transfer for Stylized 3D Characters},
    author = {Liao, Zhouyingcheng and Yang, Jimei and Saito, Jun and Pons-Moll, Gerard and Zhou, Yang},
    booktitle = {European Conference on Computer Vision ({ECCV})},
    month = {October},
    organization = {{Springer}},
    year = {2022},
}

@inproceedings{wang2020neural,
  title={Neural Pose Transfer by Spatially Adaptive Instance Normalization},
  author={Wang, Jiashun and Wen, Chao and Fu, Yanwei and Lin, Haitao and Zou, Tianyun and Xue, Xiangyang and Zhang, Yinda},
  booktitle={Proceedings of the IEEE/CVF Conference on Computer Vision and Pattern Recognition},
  pages={5831--5839},
  year={2020}
}

@inproceedings{zhou20unsupervised,
    title = {Unsupervised Shape and Pose Disentanglement for 3D Meshes},
    author = {Zhou, Keyang and Bhatnagar, Bharat Lal and Pons-Moll, Gerard},
    booktitle = {European Conference on Computer Vision (ECCV)},
    month = {August},
    year = {2020},
}

@INPROCEEDINGS{1013569,
  author={Kuffner, J. and Nishiwaki, K. and Kagami, S. and Kuniyoshi, Y. and Inaba, M. and Inoue, H.},
  booktitle={Proceedings 2002 IEEE International Conference on Robotics and Automation (Cat. No.02CH37292)}, 
  title={Self-collision detection and prevention for humanoid robots}, 
  year={2002},
  volume={3},
  number={},
  pages={2265-2270 vol.3},
  doi={10.1109/ROBOT.2002.1013569}}

@INPROCEEDINGS{handa2019dexpilot,
  author={Handa, Ankur and Van Wyk, Karl and Yang, Wei and Liang, Jacky and Chao, Yu-Wei and Wan, Qian and Birchfield, Stan and Ratliff, Nathan and Fox, Dieter},
  booktitle={2020 IEEE International Conference on Robotics and Automation (ICRA)}, 
  title={DexPilot: Vision-Based Teleoperation of Dexterous Robotic Hand-Arm System}, 
  year={2020},
  volume={},
  number={},
  pages={9164-9170},
  doi={10.1109/ICRA40945.2020.9197124}
}

@INPROCEEDINGS{yan2024imitationnet,
  author={Yan, Yashuai and Mascaro, Esteve Valls and Lee, Dongheui},
  booktitle={2023 IEEE-RAS 22nd International Conference on Humanoid Robots (Humanoids)}, 
  title={ImitationNet: Unsupervised Human-to-Robot Motion Retargeting via Shared Latent Space}, 
  year={2023},
  volume={},
  number={},
  pages={1-8},
  doi={10.1109/Humanoids57100.2023.10375150}}

@article{Aberman_2019,
   title={Learning character-agnostic motion for motion retargeting in 2D},
   volume={38},
   ISSN={1557-7368},
   url={http://dx.doi.org/10.1145/3306346.3322999},
   DOI={10.1145/3306346.3322999},
   number={4},
   journal={ACM Transactions on Graphics},
   publisher={Association for Computing Machinery (ACM)},
   author={Aberman, Kfir and Wu, Rundi and Lischinski, Dani and Chen, Baoquan and Cohen-Or, Daniel},
   year={2019},
   month=jul, pages={1–14} }

@software{ue5,
  author = {{Epic Games}},
  title = {Unreal Engine},
  url = {https://www.unrealengine.com},
  version = {4.22.1},
  year={2020}
}

@software{humanIK,
  author = {{Autodesk, Inc.}},
  title = {Maya (MotionBuilder)},
  url = {https://www.autodesk.com/fr/support/technical/product/motionbuilder},
  version = {2025.2},
  year={2025}
}

@misc{blenderstudio,
      title={Characters}, 
      author={Blender Studio},
      date = {2024-04-30},
      url={https://studio.blender.org/characters/},
}

@inproceedings{cuturi2013sinkhorndistanceslightspeedcomputation,
 author = {Cuturi, Marco},
 booktitle = {Advances in Neural Information Processing Systems},
 editor = {C.J. Burges and L. Bottou and M. Welling and Z. Ghahramani and K.Q. Weinberger},
 pages = {},
 publisher = {Curran Associates, Inc.},
 title = {Sinkhorn Distances: Lightspeed Computation of Optimal Transport},
 url = {https://proceedings.neurips.cc/paper_files/paper/2013/file/af21d0c97db2e27e13572cbf59eb343d-Paper.pdf},
 volume = {26},
 year = {2013}
}

@article{zhang2024semanticsawaremotionretargetingvisionlanguage,
  title={Semantics-aware Motion Retargeting with Vision-Language Models},
  author={Haodong Zhang and ZhiKe Chen and Haocheng Xu and Lei Hao and Xiaofei Wu and Songcen Xu and Zhensong Zhang and Yue Wang and Rong Xiong},
  booktitle={Proceedings of the IEEE/CVF Conference on Computer Vision and Pattern Recognition (CVPR)},
  year={2024}
}

\end{document}


\maketitle

\sloppy

\section{Related work}

We propose in Table~\ref{tab:method_comparison} a comparative summary of the features provided by the previous works regarding the constraints (avoiding mesh penetration, preserving foot contacts, handling multiple characters, adaptation to diverse and unseen morphologies), and their computational efficiency.


\begin{figure*}[!h]
    \centering
    \vspace*{1\baselineskip}
    \hfill%
    \begin{subfigure}[b]{.25\textwidth}
        \centering
        \includegraphics[trim=150 80 150 100, clip,width=\textwidth]{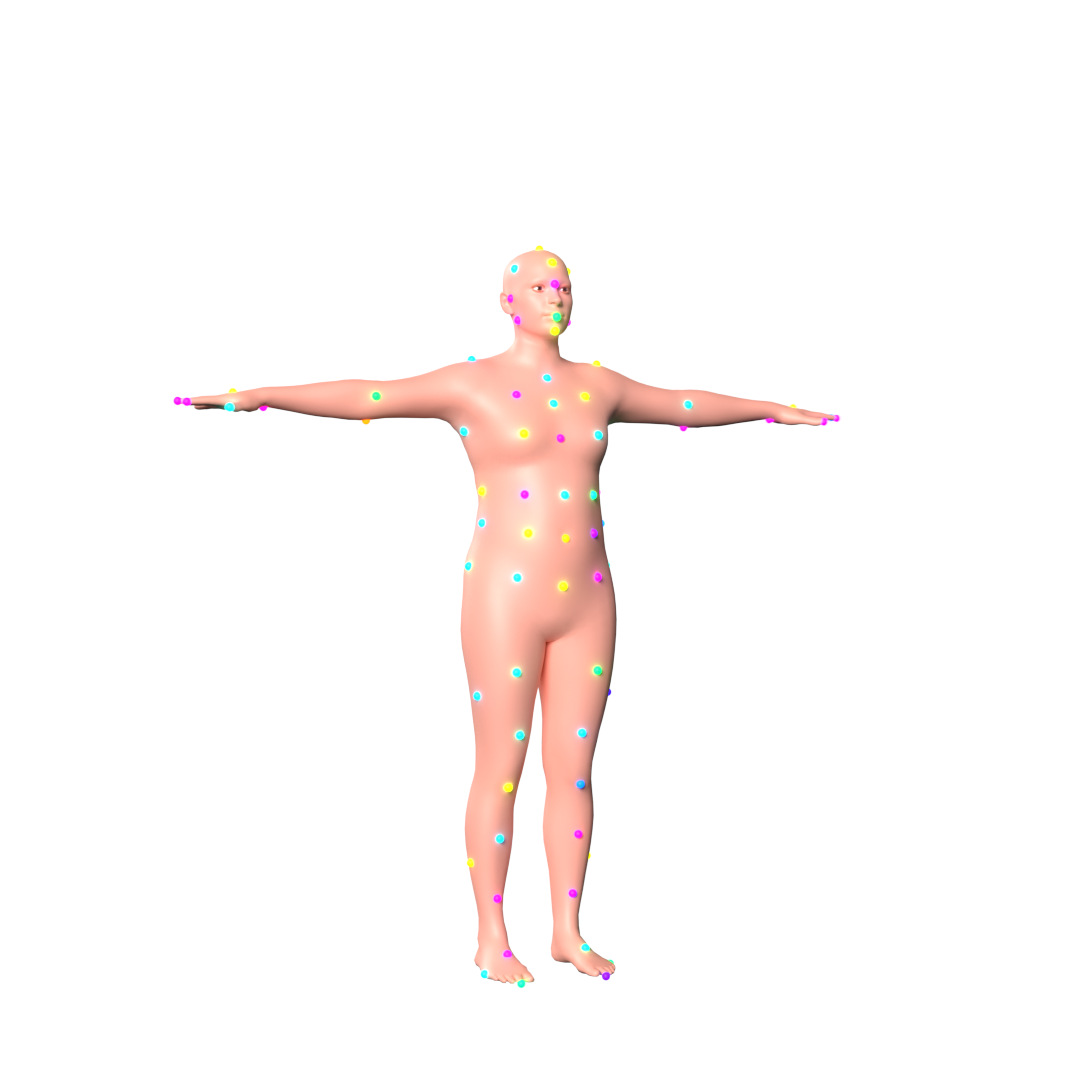}
    \end{subfigure}%
    \hfill%
    \begin{subfigure}[b]{.25\textwidth}
        \centering
        \includegraphics[trim=50 90 130 140, clip, width=\textwidth]{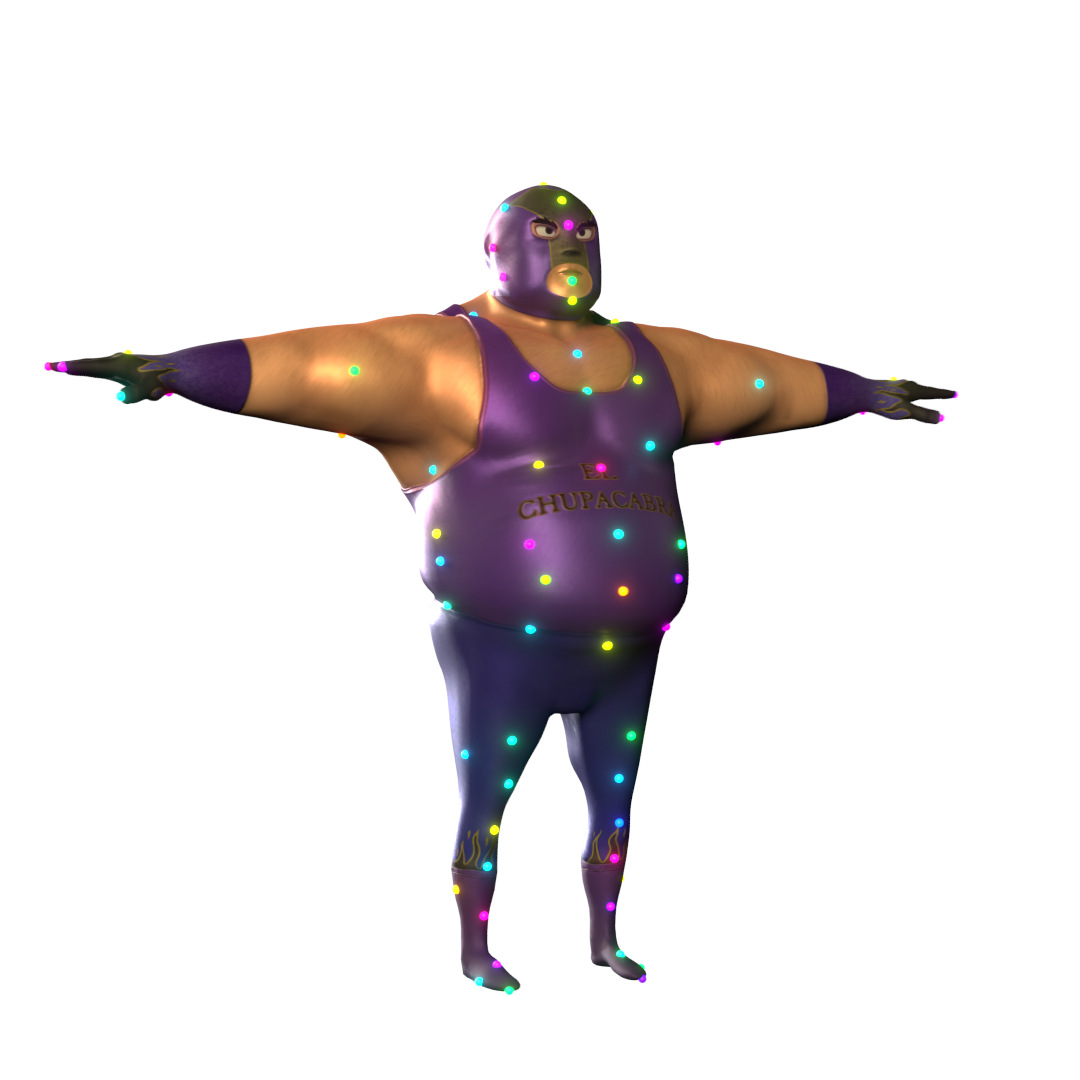}
    \end{subfigure}%
    \hfill%
    \begin{subfigure}[b]{.25\textwidth}
        \centering
        \includegraphics[trim=200 85 200 200, clip, width=\textwidth]{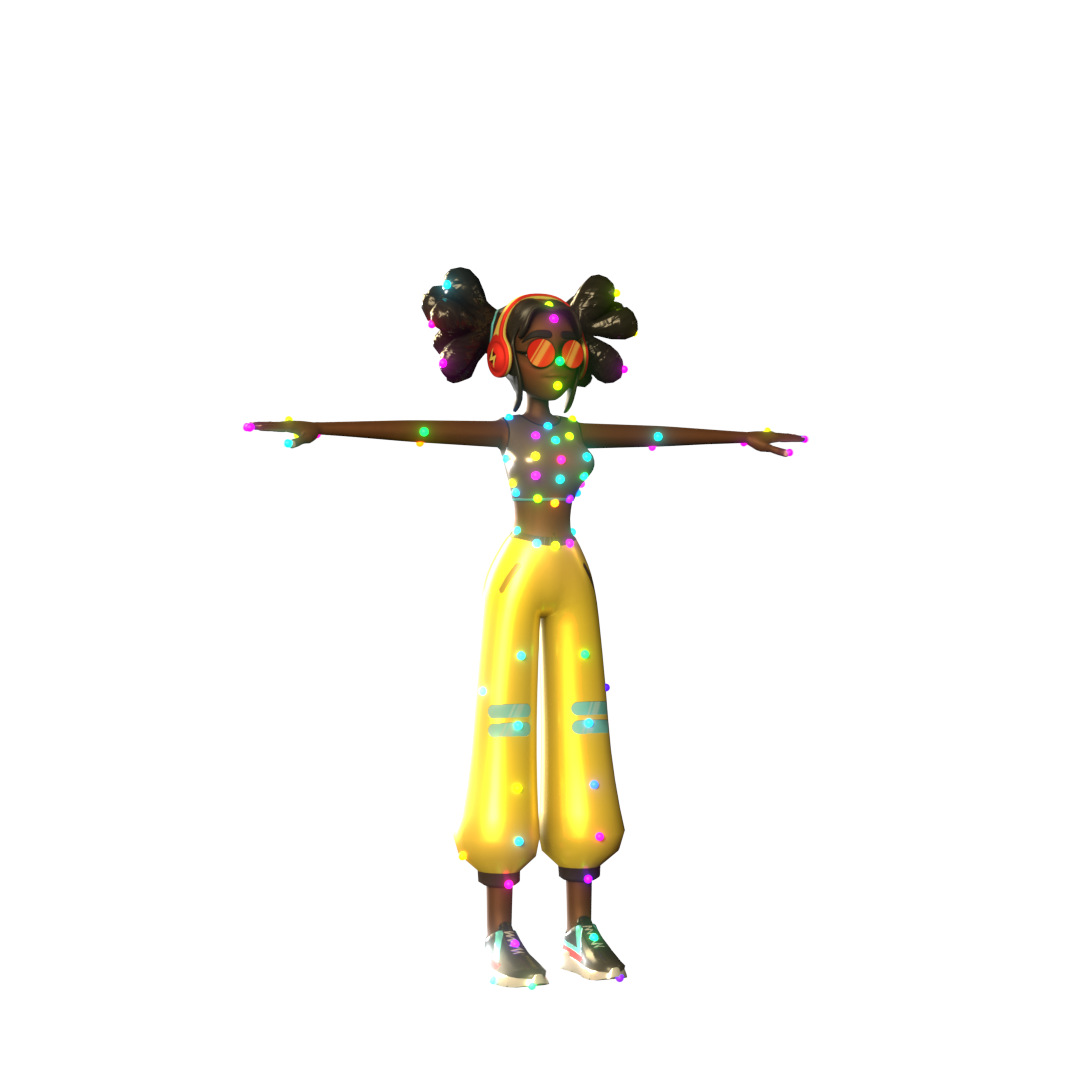}
    \end{subfigure}%
    \hfill%
    \begin{subfigure}[b]{.25\textwidth}
        \centering
        \includegraphics[trim=130 80 150 100, clip, width=\textwidth]{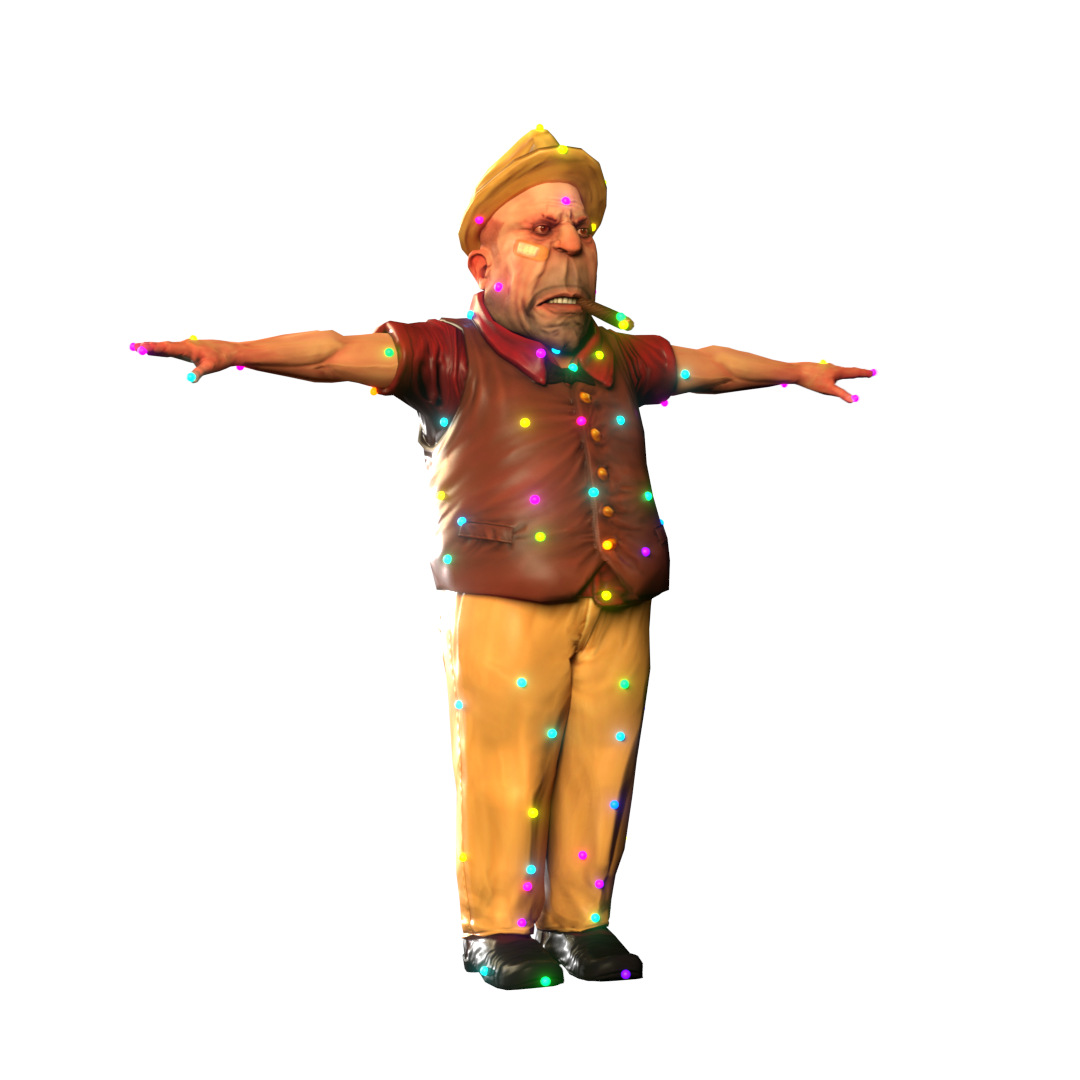}
    \end{subfigure}%
    \hfill
    \hfill%
    \begin{subfigure}[b]{.25\textwidth}
        \centering
        \includegraphics[trim=50 30 50 20, clip,width=\textwidth]{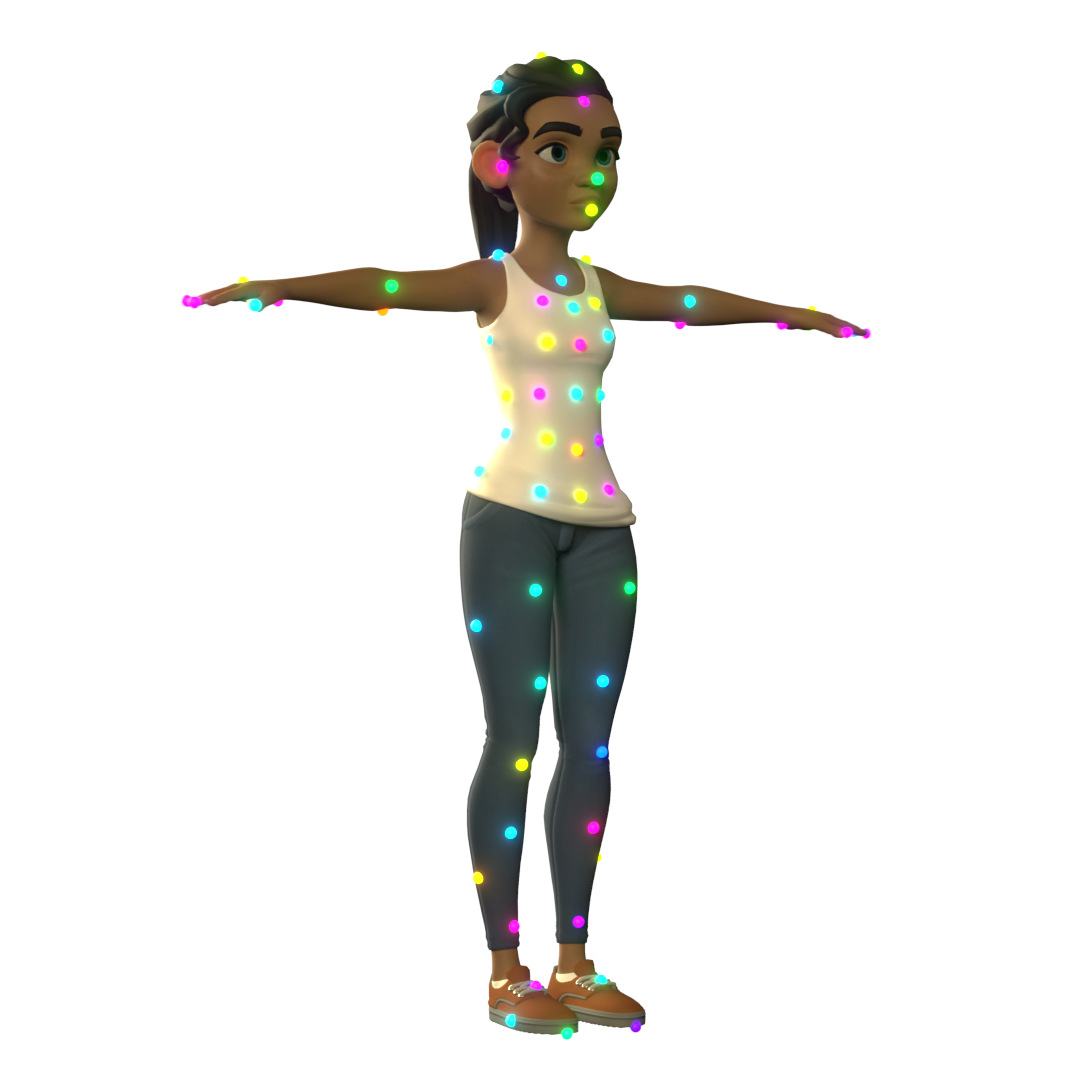}
    \end{subfigure}%
    \hfill%
    \begin{subfigure}[b]{.25\textwidth}
        \centering
        \includegraphics[trim=0 30 80 10, clip, width=\textwidth]{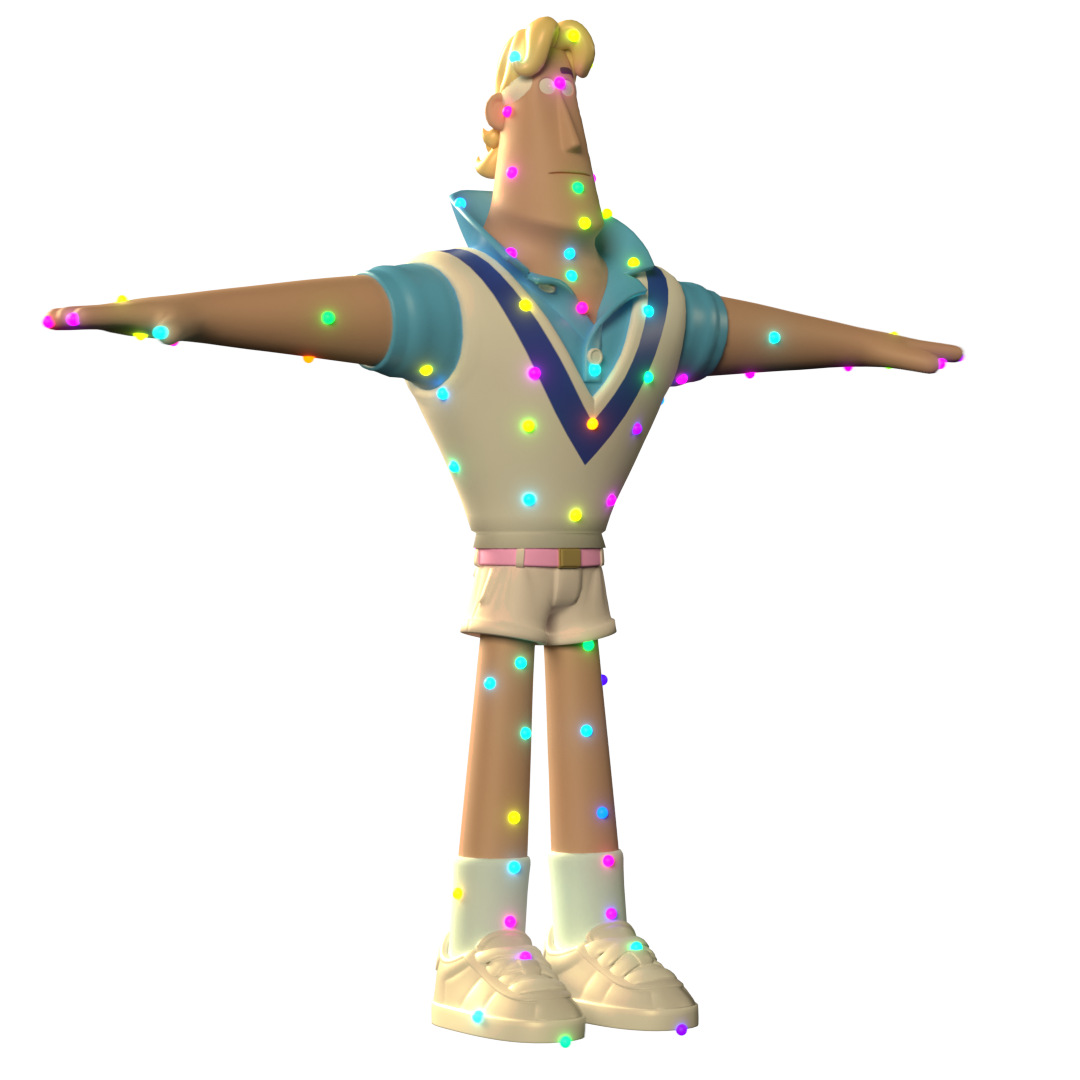}
    \end{subfigure}%
    \hfill%
    \begin{subfigure}[b]{.25\textwidth}
        \centering
        \includegraphics[trim=200 55 200 200, clip, width=\textwidth]{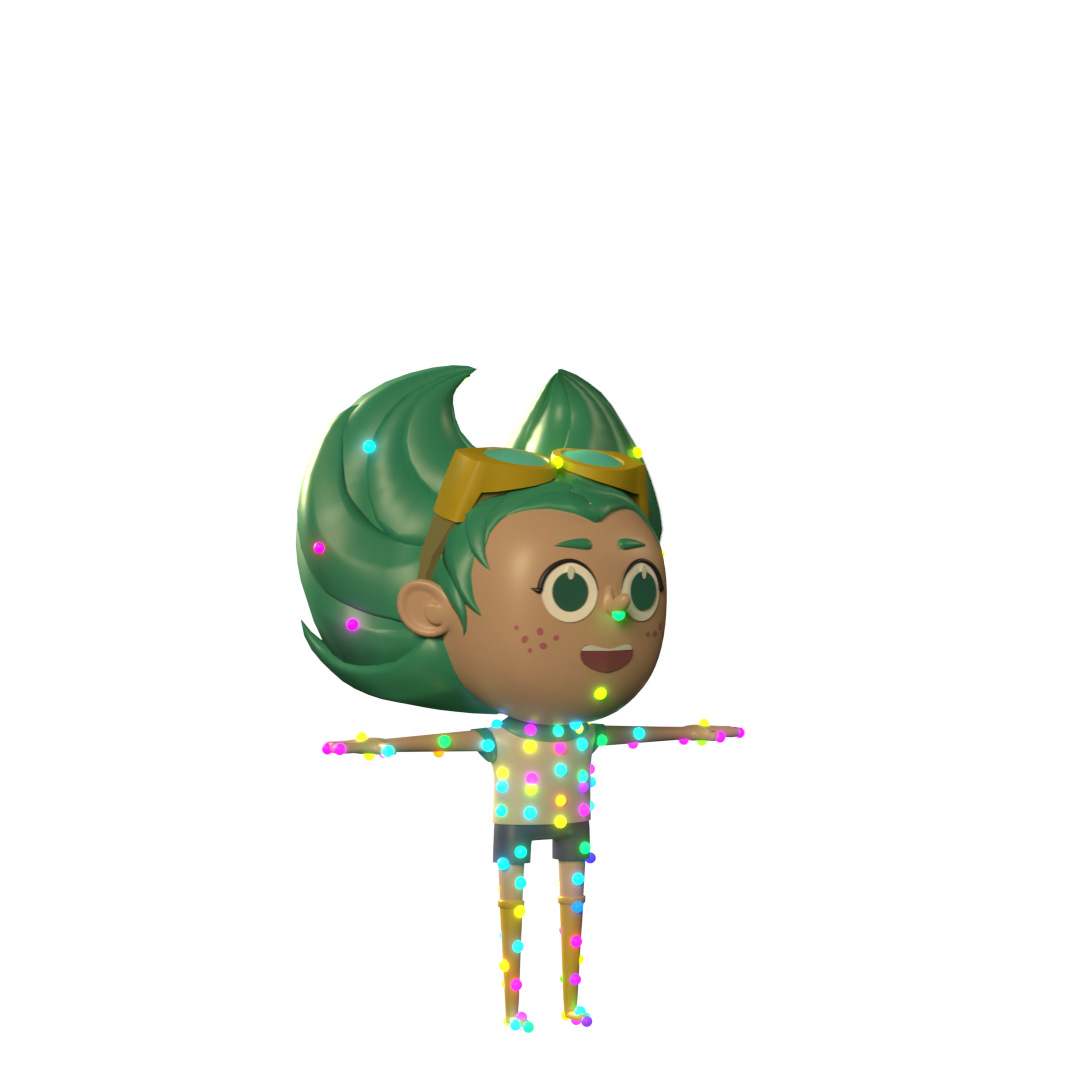}
    \end{subfigure}%
    \hfill%
    \begin{subfigure}[b]{.25\textwidth}
        \centering
        \includegraphics[trim=130 80 150 100, clip, width=\textwidth]{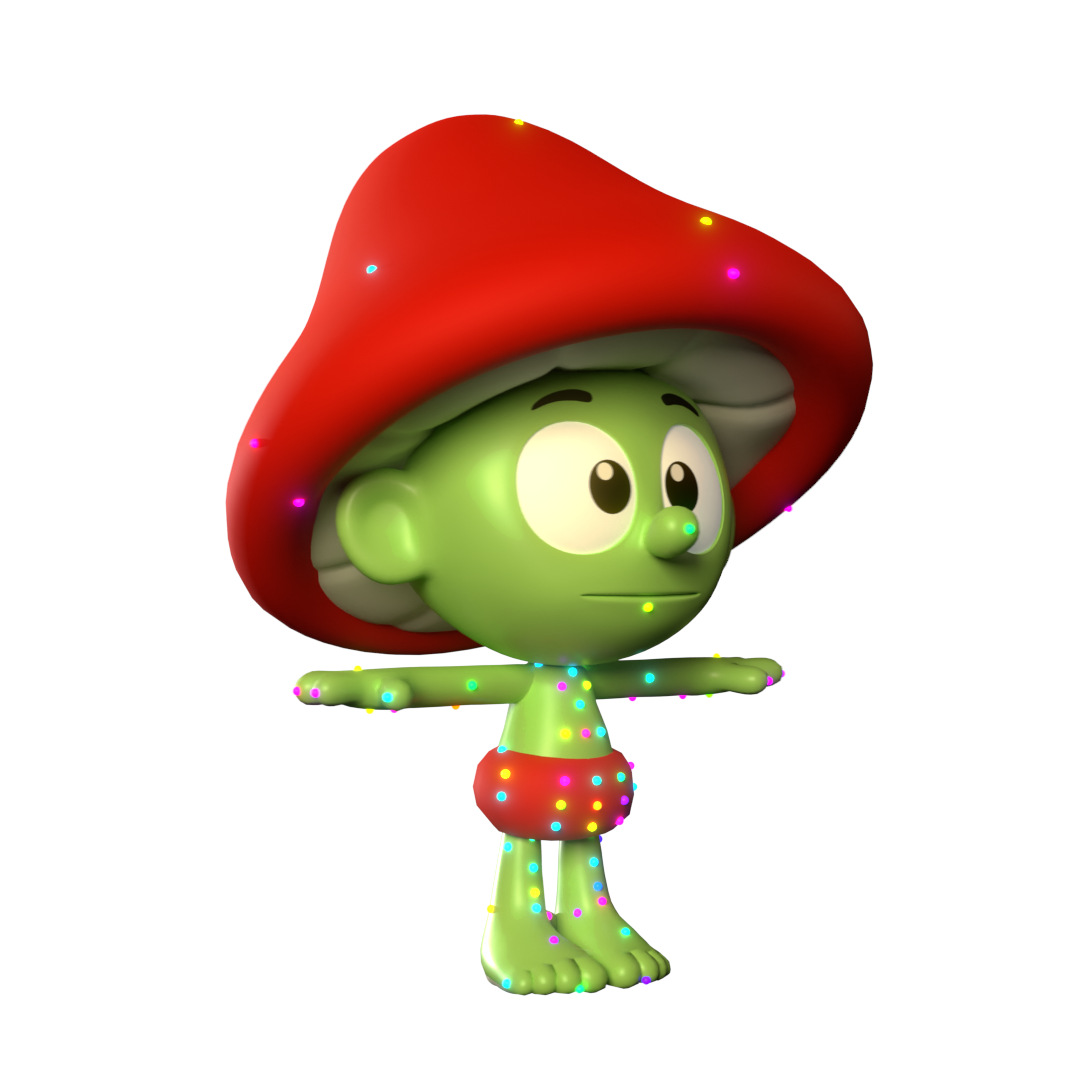}
    \end{subfigure}%
    \hfill
    \caption{Results of the key vertices transfer algorithm ($N=96$). Top-left character shows the vertices locations on the template mesh \cite{smpl}, top row shows characters from Mixamo \cite{mixamo}, while bottom row shows additional, production-ready characters from Blender Studio~\cite{blenderstudio}. Key vertices are shown with corresponding colored spheres to show the automatic transfer to the different meshes.}
    \label{fig:keyvertices_transfer_big}
    \vspace*{.5\baselineskip}
\end{figure*}

\section{Validation set}

Table~\ref{table:dataset} itemizes the contents of our validation set, with all animations taken from Mixamo \cite{mixamo}. The source character is either ``XBot'' (the red character with a feminine morphology) or ``YBot'' (the blue character with a masculine morphology), as those seem to be the original characters on which the motion was designed, and thus are the best available quality for those motions. The column ``Difficulty'' corresponds to the difficulty of the motion considering mesh contacts, as explained in the main article (Sec.~5.2.1).

\section{Key-vertices transfer}

We propose a quantitative evaluation of the accuracy of our key-vertices transfer algorithm. For each of our 10 Mixamo characters, we asked a 3D character expert to manually transfer the key-vertices, and we measured the average distance between those key-vertices and the ones automatically produced by our method. On average, we report an error of 6.46 cm, while the average edge length of these meshes is 3.72 cm.

As mentioned in the main article, we performed an ablation study with regards to the number of key-vertices. For this, we used $N=96$ vertices, in order to represent the mesh of the characters in finer detail. The location of these vertices, as well as the results of the transfer algorithm on several characters, are illustrated in Figure~\ref{fig:keyvertices_transfer_big}.

\section{Hyperparameters} 
We report hyperparameters used in the fine-tuned version of our method. We use a learning rate of 0.01 for the optimization process, and set the number $N$ of keypoints to 41 on the template, source and target meshes. 
\autoref{table:hyper} reports the weights of main loss terms $w_{reg}, w_{smooth}$ and $ w_{sem}$, along with additional ponderations for each term within the semantic loss (resp. distance, direction, penetration, height, and sliding) $w_{dist}, w_{dir}, w_{pen}, w_{height}$, and $ w_{sliding}$. 
Finally, we provide values for characteristic distances used in the computation of the weighting matrices (see section 4.1): $d_{min}, d_{max}, h_{min}, h_{max}$. We empirically found that this set of hyperparameters works well in most of the cases, although they can easily be refined thanks to their reduced amount and explicit purpose in the method.

\begin{table*}[p!]
\small
\centering
\scalebox{1.0}{
\begin{tabular}{|l|l|c|c|c|c|c|c|c|}

\hline
Method & Approach & \begin{tabular}[c]{@{}c@{}}\small{Focus on}\\ \small{mesh}\\ \small{penetration}\end{tabular} & \begin{tabular}[c]{@{}c@{}}\small{Focus}\\ \small{on foot}\\ \small{contacts}\end{tabular} & \begin{tabular}[c]{@{}c@{}}\small{Temporal}\\ \small{continuity}\end{tabular}& \begin{tabular}[c]{@{}c@{}}\small{Multi}\\ \small{char.}\end{tabular} & \begin{tabular}[c]{@{}c@{}}\small{Can adapt}\\ \small{to diverse}\\ \small{characters}\end{tabular} & \begin{tabular}[c]{@{}c@{}}\small{Generalizes}\\ \small{to unseen}\\ \small{morphologies}\end{tabular} & \small{Speed}
\\ \hline\hline

\begin{tabular}[l]{@{}l@{}}NKN\\\cite{nkn}\end{tabular} & Self-supervised & \xmark & \xmark & \cmark  & \xmark & \begin{tabular}[c]{@{}c@{}}\xmark \\[-.2em] \scriptsize{(fixed number}\\[-.4em]\scriptsize{of joints)}\end{tabular} & \cmark & unreported \\ \hline

\begin{tabular}[l]{@{}l@{}}PMNet\\\cite{pmnet}\end{tabular} & Self-supervised & \xmark & \xmark & \cmark & \xmark & \begin{tabular}[c]{@{}c@{}}\xmark \\[-.2em] \scriptsize{(fixed number}\\[-.4em]\scriptsize{of joints)}\end{tabular} & \cmark & unreported \\ \hline

\begin{tabular}[l]{@{}l@{}}SAN\\\cite{san}\end{tabular} & Self-supervised & \xmark & \cmark & \cmark & \xmark & \begin{tabular}[c]{@{}c@{}}\cmark \\[-.2em] \scriptsize{(common primal}\\[-.4em]\scriptsize{skeleton)}\end{tabular} & \cmark & unreported\\\hline

\begin{tabular}[l]{@{}l@{}}R$^2$ET\\\cite{r2et}\end{tabular} & Self-supervised & \cmark & \xmark & \cmark & \xmark & \begin{tabular}[c]{@{}c@{}}\xmark \\[-.2em] \scriptsize{(fixed number}\\[-.4em]\scriptsize{of joints)}\end{tabular} & \cmark & $\sim$120 fps\\\hline

\begin{tabular}[l]{@{}l@{}}Villegas et al.\\\cite{villegas}\end{tabular} & \begin{tabular}[c]{@{}c@{}}Supervised \\ + Optimization\end{tabular} & \cmark & \cmark & \cmark & \xmark & \cmark & \begin{tabular}[c]{@{}c@{}}\xmark \\[-.2em] \scriptsize{(needs retraining}\\[-.4em]\scriptsize{for each new char.)}\end{tabular} & unreported\\\hline

\begin{tabular}[l]{@{}l@{}}Gleicher\\\cite{gleicher}\end{tabular}& Optimization & \xmark & \cmark  & \cmark & \cmark & \begin{tabular}[c]{@{}c@{}} \xmark\\[-.2em] \scriptsize{(only tested on}\\[-.4em]\scriptsize{one character)}\end{tabular} & \cmark & \begin{tabular}[c]{@{}c@{}} $\sim$5 fps \\[-.2em] \scriptsize{(need to manually}\\[-.4em]\scriptsize{identify constraints)}\end{tabular}\\\hline

\begin{tabular}[l]{@{}l@{}}AuraMesh\\\cite{taeil17}\end{tabular}& Optimization & \cmark & \cmark & \xmark & \cmark & \begin{tabular}[c]{@{}c@{}}\xmark \\[-.2em] \scriptsize{(only for SMPL}\\[-.4em]\scriptsize{\cite{smpl})}\end{tabular} & \cmark & $\sim$1 fps\\\hline

\begin{tabular}[l]{@{}l@{}}Basset et al.\\\cite{basset20}\end{tabular} & Optimization & \cmark & \cmark & \xmark & \xmark & \begin{tabular}[c]{@{}c@{}}\xmark \\[-.2em] \scriptsize{(only for SMPL}\\[-.4em]\scriptsize{\cite{smpl})}\end{tabular} & \cmark & $\sim$0.003 fps\\\hline

\begin{tabular}[l]{@{}l@{}}Ho et al.\\\cite{ho2010}\end{tabular} & Optimization & \begin{tabular}[c]{@{}c@{}}\xmark \\[-.2em] \scriptsize{(only to avoid}\\[-.4em]\scriptsize{collisions)}\end{tabular} & \cmark & \cmark & \cmark & \begin{tabular}[c]{@{}c@{}}\xmark \\[-.2em] \scriptsize{(same skeleton}\\[-.4em]\scriptsize{structure)}\end{tabular} & \cmark & $\sim$1.7 fps\\\hline

\begin{tabular}[l]{@{}l@{}}PAN\\\cite{Hu_2023}\end{tabular} & \begin{tabular}[c]{@{}c@{}}Unsupervised \\ Learning\end{tabular} & \xmark & \cmark & \cmark & \cmark & \begin{tabular}[c]{@{}c@{}}\xmark \\[-.2em] \scriptsize{(common primal}\\[-.4em]\scriptsize{skeleton)}\end{tabular} & \begin{tabular}[c]{@{}c@{}}\xmark \\[-.2em] \scriptsize{(needs retraining}\\[-.4em]\scriptsize{for each new char.)}\end{tabular} & unreported \\\hline

\begin{tabular}[l]{@{}l@{}}Zhang et al.\\\cite{yunbo23}\end{tabular} & \begin{tabular}[c]{@{}c@{}}Reinforcement \\ Learning\end{tabular} & \cmark & \cmark & \cmark & \cmark & \begin{tabular}[c]{@{}c@{}}\xmark \\[-.2em] \scriptsize{(only one type}\\[-.4em]\scriptsize{of character)}\end{tabular} & \begin{tabular}[c]{@{}c@{}}\xmark \\[-.2em] \scriptsize{(needs retraining}\\[-.4em]\scriptsize{for each new char.)}\end{tabular} & \begin{tabular}[c]{@{}c@{}}\scriptsize{(needs retraining}\\[-.4em]\scriptsize{for each new motion)}\end{tabular}\\\hline

\begin{tabular}[l]{@{}l@{}}Reda et al.\\\cite{reda23}\end{tabular} & \begin{tabular}[c]{@{}c@{}}Reinforcement \\ Learning\end{tabular} & \cmark & \cmark & \cmark & \cmark & \begin{tabular}[c]{@{}c@{}}\xmark \\[-.2em] \scriptsize{(only one type}\\[-.4em]\scriptsize{of character)}\end{tabular} & \begin{tabular}[c]{@{}c@{}}\xmark \\[-.2em] \scriptsize{(needs retraining}\\[-.4em]\scriptsize{for each new char.)}\end{tabular} & $\sim$36 fps

\\ \hline\hline
\textbf{Ours} & Optimization & \cmark & \cmark & \cmark & \cmark & \cmark & \cmark & $\sim$67 fps\\\hline
\end{tabular}}
\caption{Comparison 
of existing retargeting methods}
\label{tab:method_comparison}
\end{table*}

\begin{table}[htbp!]
\vspace{1\baselineskip}
\centering
\caption{Hyperparameters used in our method}
\label{table:hyper} 
\begin{tabular}{ p{2.2cm} p{3.9cm} }
  \toprule
    Hyperparameter           & \textbf{Value}                       \\ 
  \midrule
    $d_{\min}$        &  5\% of character height                      \\
    $d_{\max}$        &  15\% of character height                     \\
    $h_{\min}$        &  5\% of character height                      \\
    $h_{\max}$        &  15\% of character height                     \\
    $w_{reg}$       & $1\cdot10^{-2}$\\
    $w_{smooth}$    & $1\cdot10^{-4}$\\
    $w_{sem}$       & $1$\\ 
    $w_{dist}$      & $1$\\
    $w_{dir}$       & $0.5$ \\
    $w_{pen}$       & $10.0$ \\
    $w_{height}$    & $1$ \\
    $w_{sliding}$   & $0.5$ \\
    $lr$            & $1\cdot10^{-2}$ \\
    $N$             & $41$ {\small (or $96$ when explicitly mentioned)}\\
  \bottomrule
\end{tabular} 
\vspace{1\baselineskip}
\end{table}

\begin{table}[htbp!]
\vspace{1\baselineskip}
\centering
\label{table:initialization} 
\begin{tabular}{ p{3.5cm} p{4.2cm} }
  \toprule
    Range of uniform noise & Average joint position difference   \\
  \midrule
    $\pm$ 5 degrees           & $1.2\cdot10^{-2}$ m \\
    $\pm$ 20 degrees          & $3.2\cdot10^{-2}$ m \\
  \bottomrule
\end{tabular} 
\caption{Impact of random noise after initialization}
\end{table}

\section{Metrics}

Figure~\ref{fig:violinplots} shows the distribution of the values of each metric across all animations of our validation dataset. This allows us to obtain finer information than the values reported in the table of the main article.

\section{Robustness}

We provide several additional results that demonstrate the robustness of our method.


    

\subsection{Initialization}

As explained in the main article, we initialize the target character with the ``copy rotations'' method, before optimizing the poses. To evaluate the robustness of our method with respect to the initialization, we perform the following experiment: we add uniform random noise to the poses obtained from the ``copy rotations'' method, and measure the average difference in joint positions. Table \hyperref[table:initialization]{3} shows the results of our experiment (performed on the motions of character ``Sophie''). While this experiment shows some degree of robustness against initialization, it is important to mention that a proper initialization makes the results less prone to the loss of semantic information during the optimization process. It also helps the optimization process converge faster.

\subsection{Individual impact of the losses}

We demonstrate the usefulness of each component of the semantic loss $\mathcal{L}_\text{sem}$ by performing an ablation study. For each component, we show its impact on a selected case that highlights specific aspects of a pose or motion.

\subsubsection{Distance and direction losses}
\begin{figure*}[!p]
    \label{fig:distance_and_direction_loss}
    \centering
    \begin{subfigure}[b]{.25\textwidth}
        \centering
        \includegraphics[width=\linewidth]{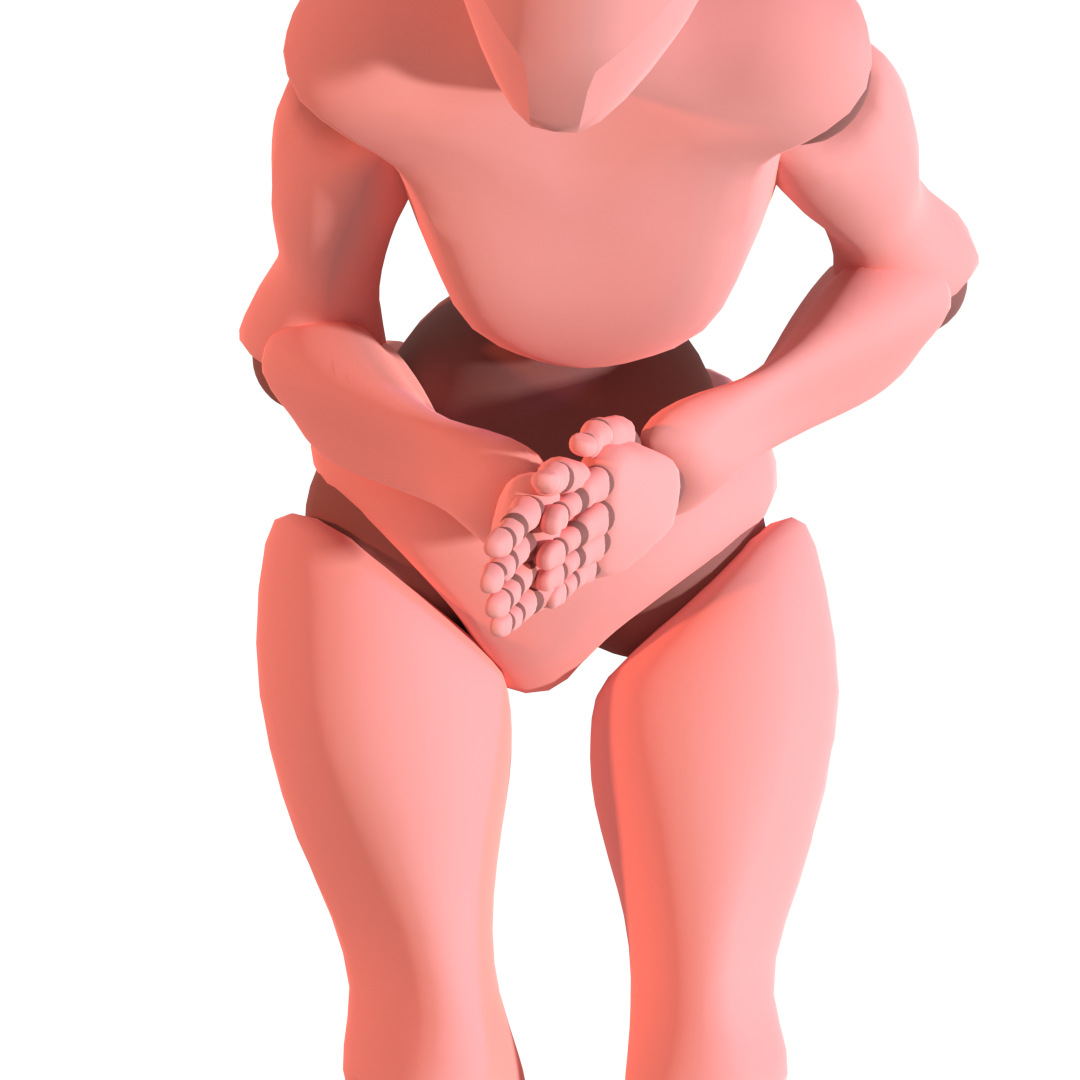}    
        \caption{Source pose (XBot - Plotting)}
        \label{fig:distance_and_direction_loss1}
    \end{subfigure}%
    \begin{subfigure}[b]{.25\textwidth}
        \centering
        \includegraphics[width=\linewidth]{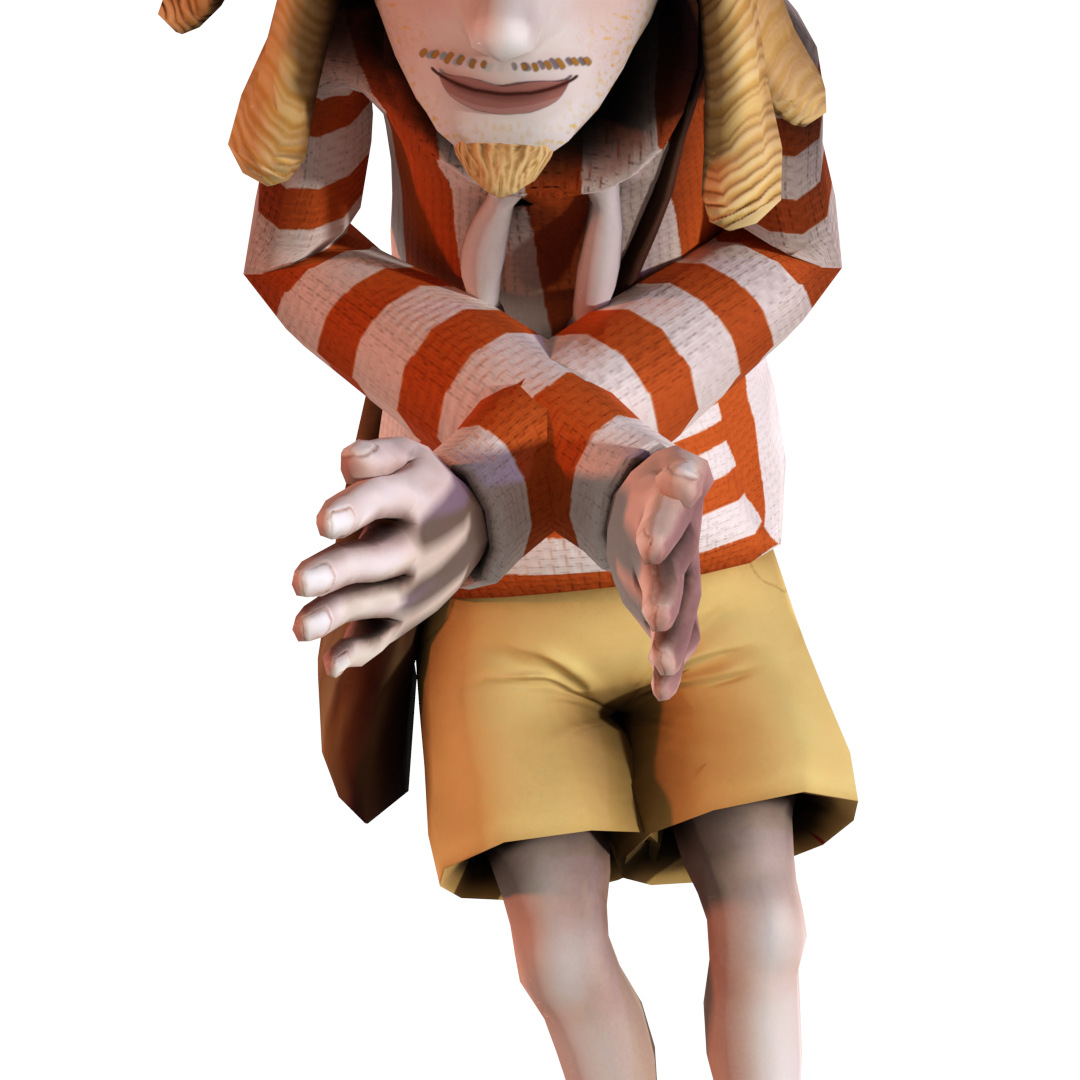}    
        \caption{No distance loss, no direction loss}
        \label{fig:distance_and_direction_loss2}
    \end{subfigure}%
    \begin{subfigure}[b]{.25\textwidth}
        \centering
        \includegraphics[width=\linewidth]{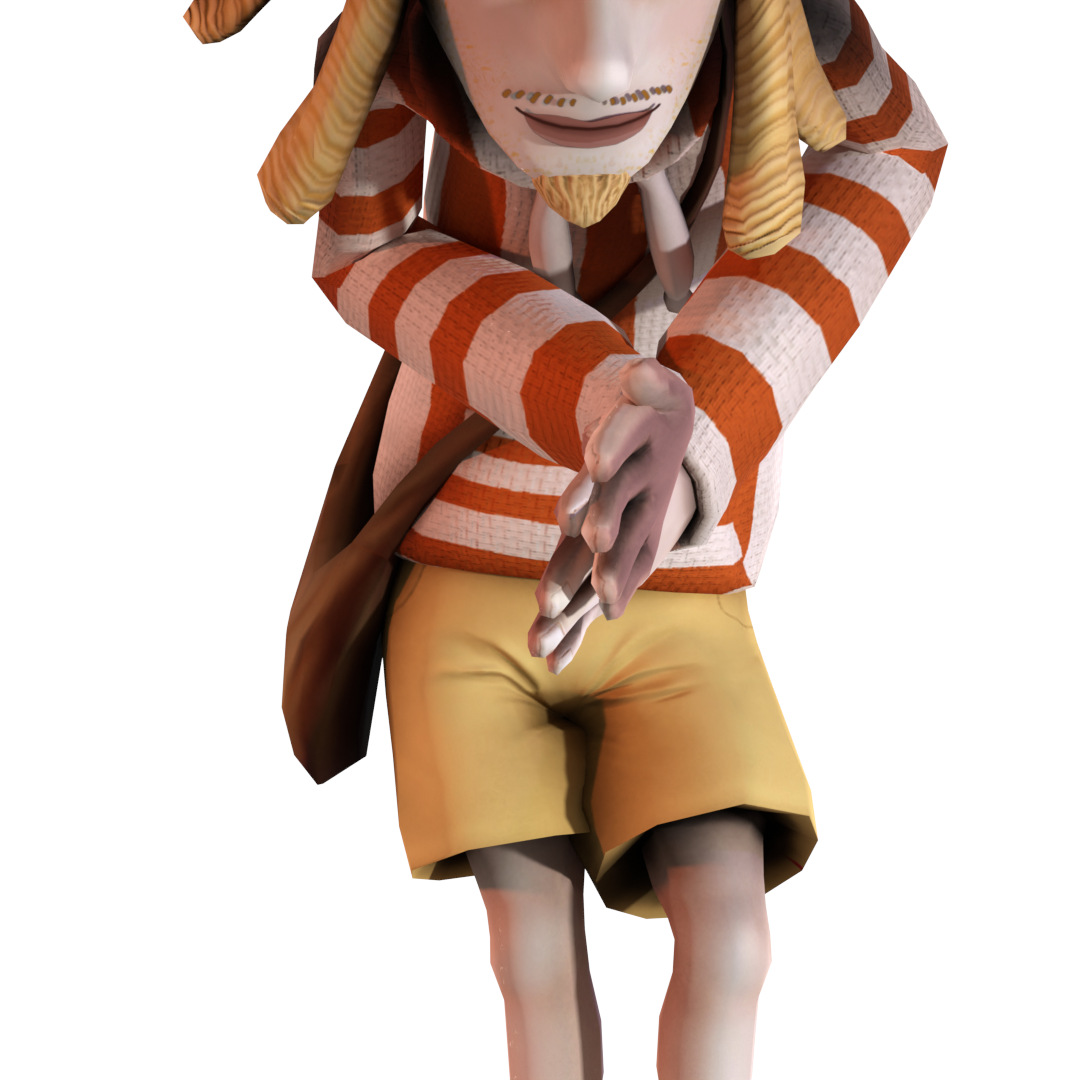}    
        \caption{Distance loss, no direction loss}
        \label{fig:distance_and_direction_loss3}
    \end{subfigure}%
    \begin{subfigure}[b]{.25\textwidth}
        \centering
        \includegraphics[width=\linewidth]{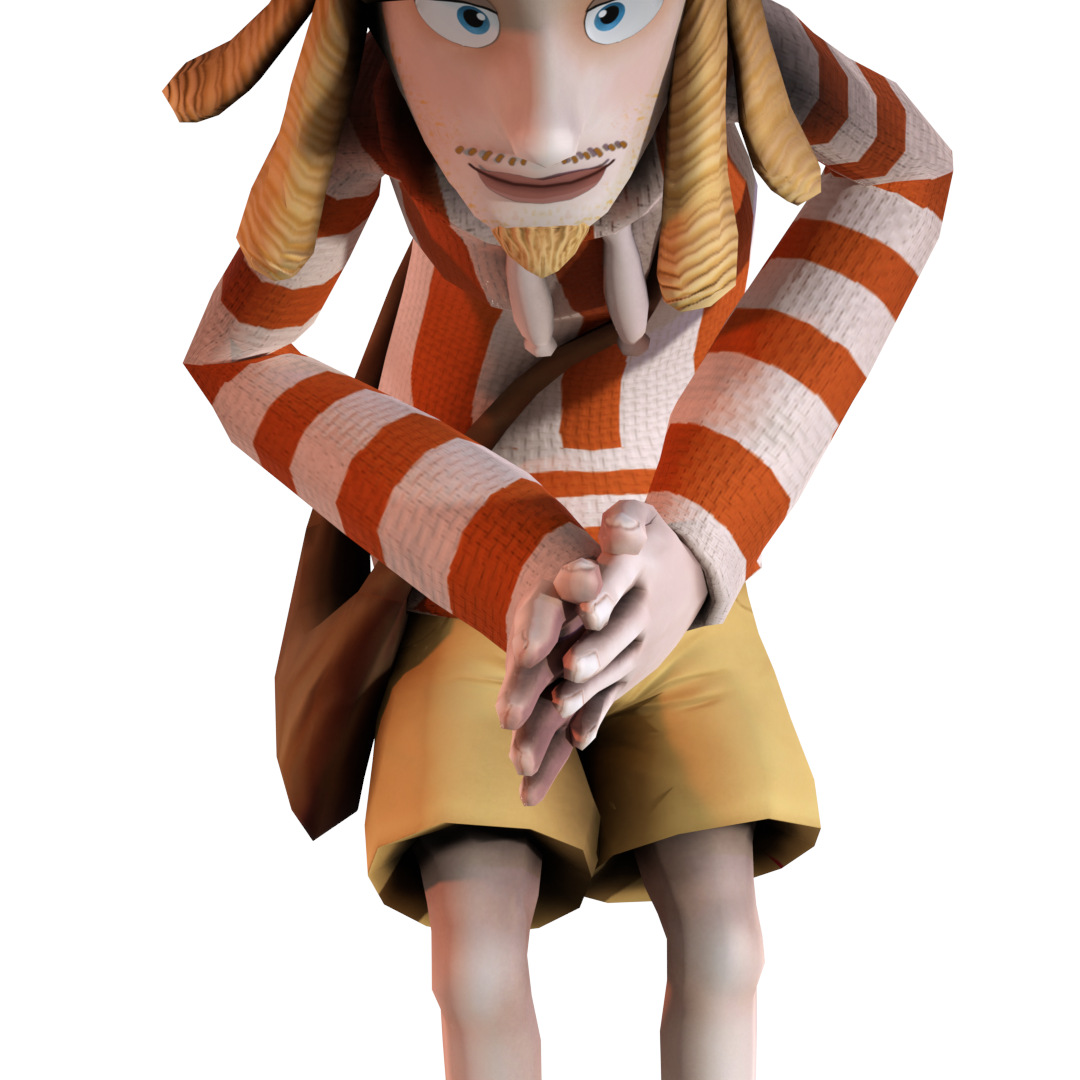}    
        \caption{Distance loss, direction loss}
        \label{fig:distance_and_direction_loss4}
    \end{subfigure}
\caption{Visualizations of the impact of $\mathcal{L}_\text{dist}$ and $\mathcal{L}_\text{dir}$.}
\end{figure*}

Figure \hyperref[fig:distance_and_direction_loss]{5} shows the impact of $\mathcal{L}_\text{dist}$ and $\mathcal{L}_\text{dir}$, and explains the need for both in our optimization:
\begin{itemize}
    \item In Figure~\ref{fig:distance_and_direction_loss2}, neither $\mathcal{L}_\text{dist}$ nor $\mathcal{L}_\text{dir}$ are active. Because the target character has longer forearms, the hands do not meet in the middle, leading to a considerable loss of motion semantics.
    \item In Figure~\ref{fig:distance_and_direction_loss3}, only $\mathcal{L}_\text{dist}$ is active. The distance between the two palms is respected (a few centimeters), but the hands are not in a similar pose, because of their respective position at the initialization of our optimization.
    \item In Figure~\ref{fig:distance_and_direction_loss4}, both $\mathcal{L}_\text{dist}$ and $\mathcal{L}_\text{dir}$ are active. This time, the hands of the target character are placed in order to respect both the distance (a few centimeters) and direction (right hand forward) present on the source pose. The arms are open wider, to compensate for the target character having longer forearms.
\end{itemize}

\subsubsection{Penetration loss}
\begin{figure*}
    \label{fig:pen_loss}
    \centering
    \begin{subfigure}[b]{.33\textwidth}
        \centering
        \includegraphics[width=.95\linewidth]{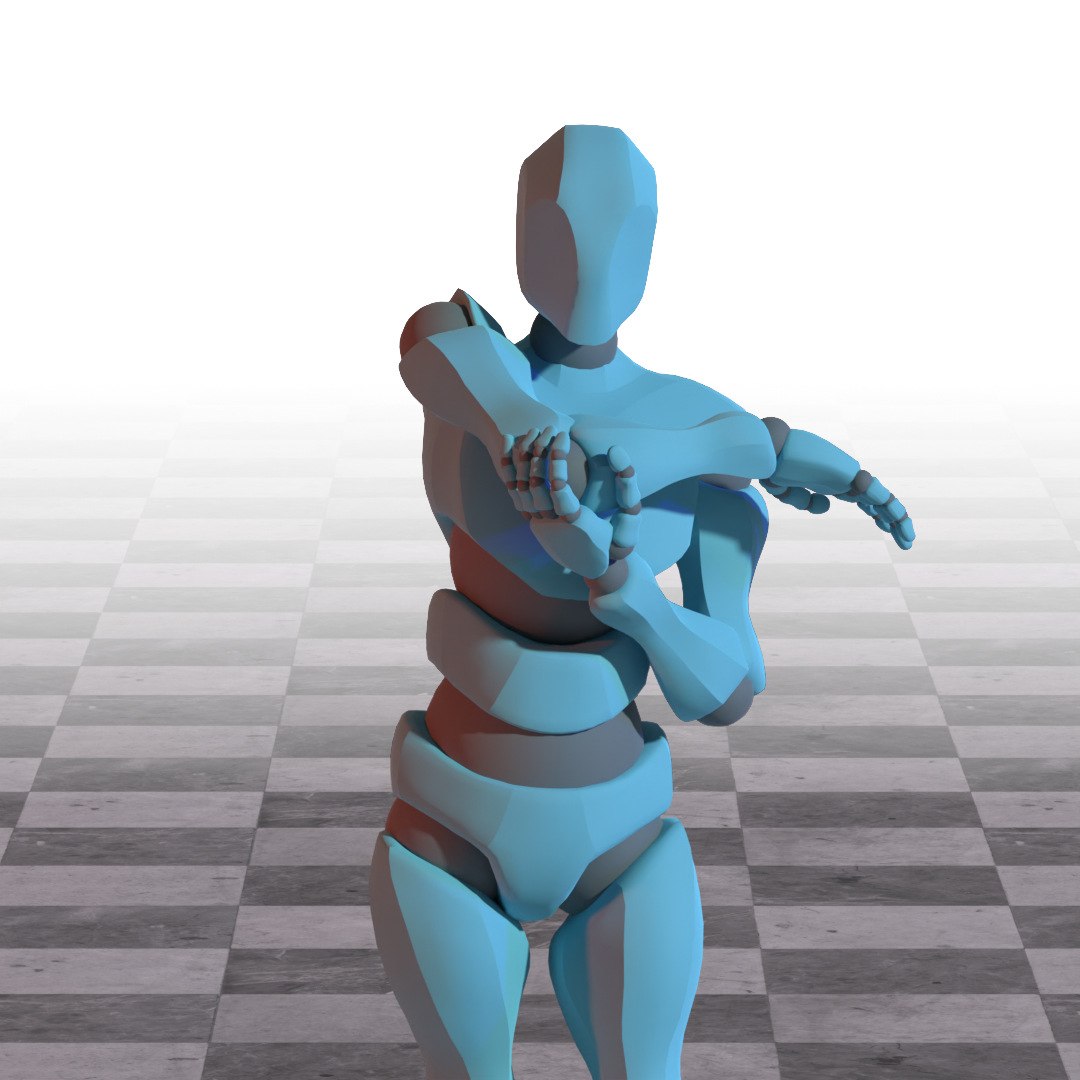}    
        \caption{Source pose (YBot - Arm Stretching)}
        \label{fig:pen_loss1}
    \end{subfigure}%
    \hfill%
    \begin{subfigure}[b]{.33\textwidth}
        \centering
        \includegraphics[width=.95\linewidth]{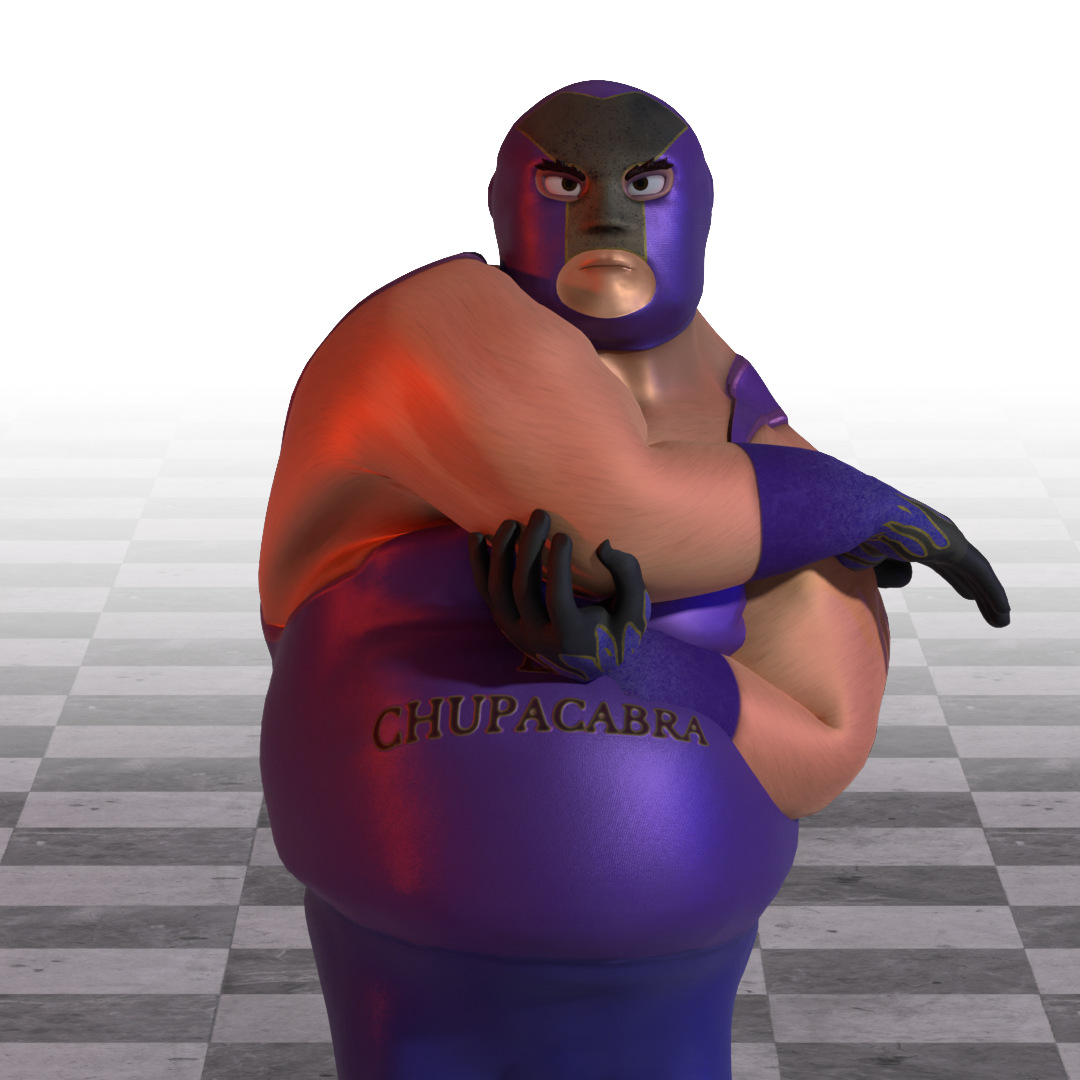}    
        \caption{All losses active, except $\mathcal{L}_\text{pen}$}
        \label{fig:pen_loss2}
    \end{subfigure}%
    \hfill%
    \begin{subfigure}[b]{.33\textwidth}
        \centering
        \includegraphics[width=.95\linewidth]{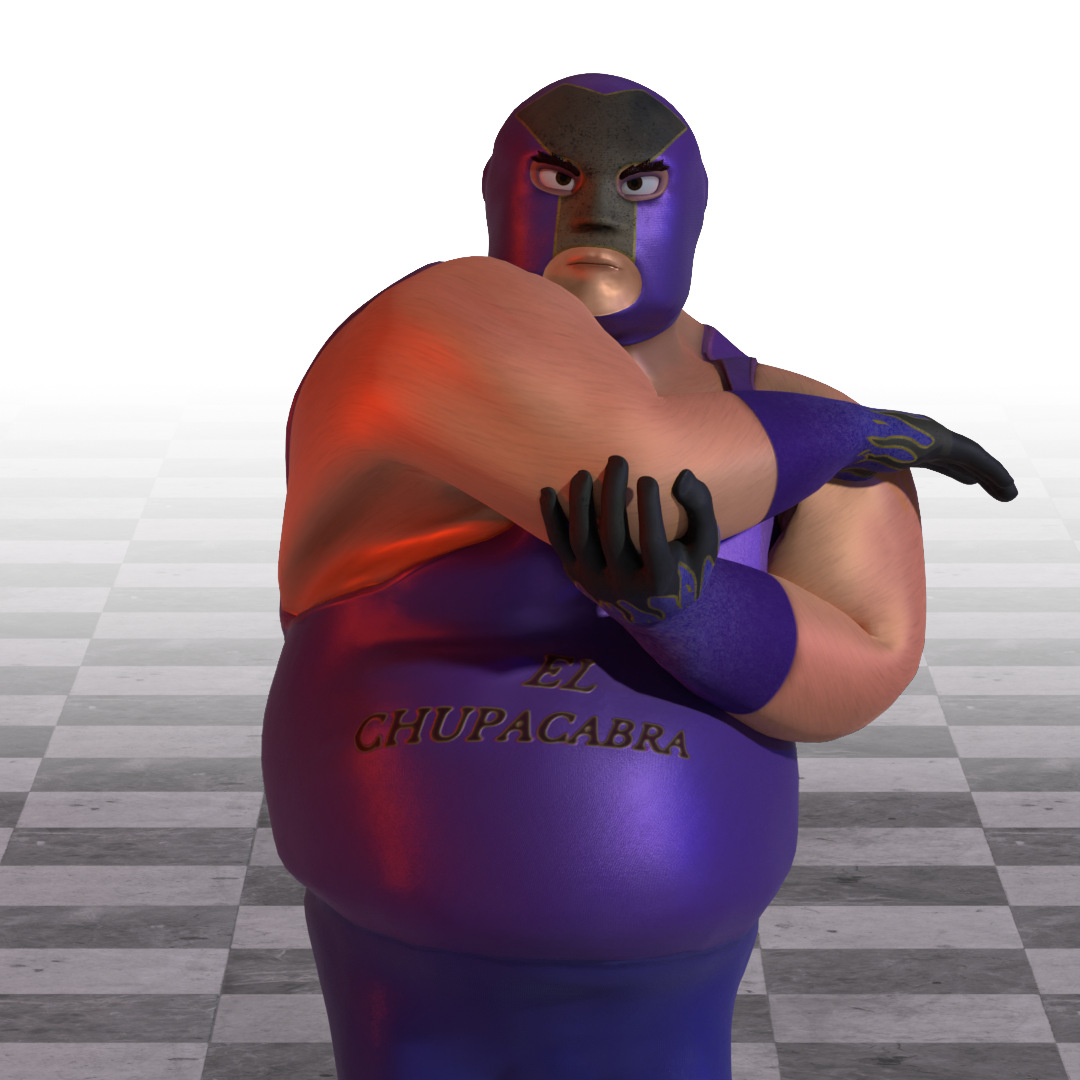}    
        \caption{All losses active, including $\mathcal{L}_\text{pen}$}
        \label{fig:pen_loss3}
    \end{subfigure}
\caption{Visualizations of the impact of $\mathcal{L}_\text{pen}$}
\end{figure*}

Figure \hyperref[fig:pen_loss]{6} shows the impact of the penetration loss. With all other losses active (Figure~\ref{fig:pen_loss2}), the left elbow is colliding with the body. Adding the penetration loss (Figure~\ref{fig:pen_loss3}) greatly reduces collisions.

\subsubsection{Height and foot sliding losses}
\begin{figure}
    \label{fig:foot_traj}
    \centering
    \begin{subfigure}[b]{\columnwidth}
        \centering
        \scalebox{-1}[1]{\includegraphics[width=\linewidth]{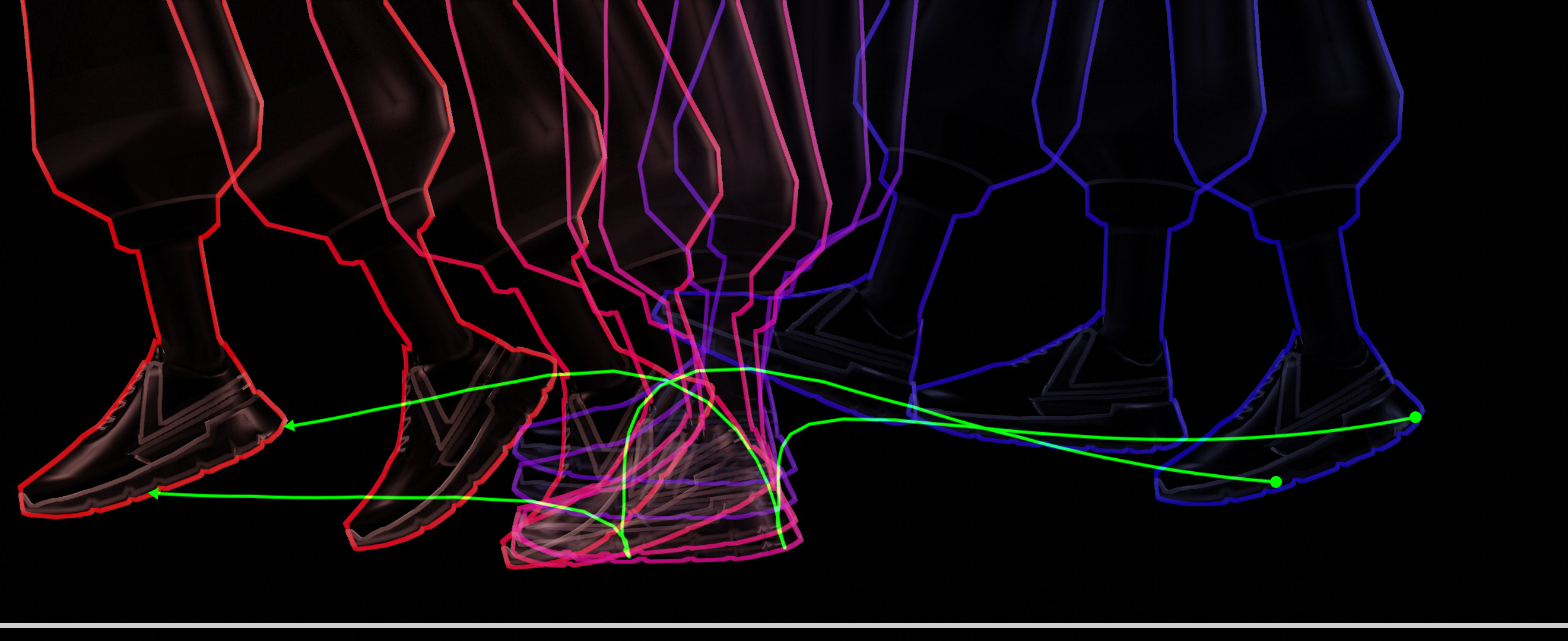}}
        \caption{No height loss, no foot sliding loss}
        \label{fig:foot_traj3}
    \end{subfigure}
    \begin{subfigure}[b]{\columnwidth}
        \centering
        \scalebox{-1}[1]{\includegraphics[width=\linewidth]{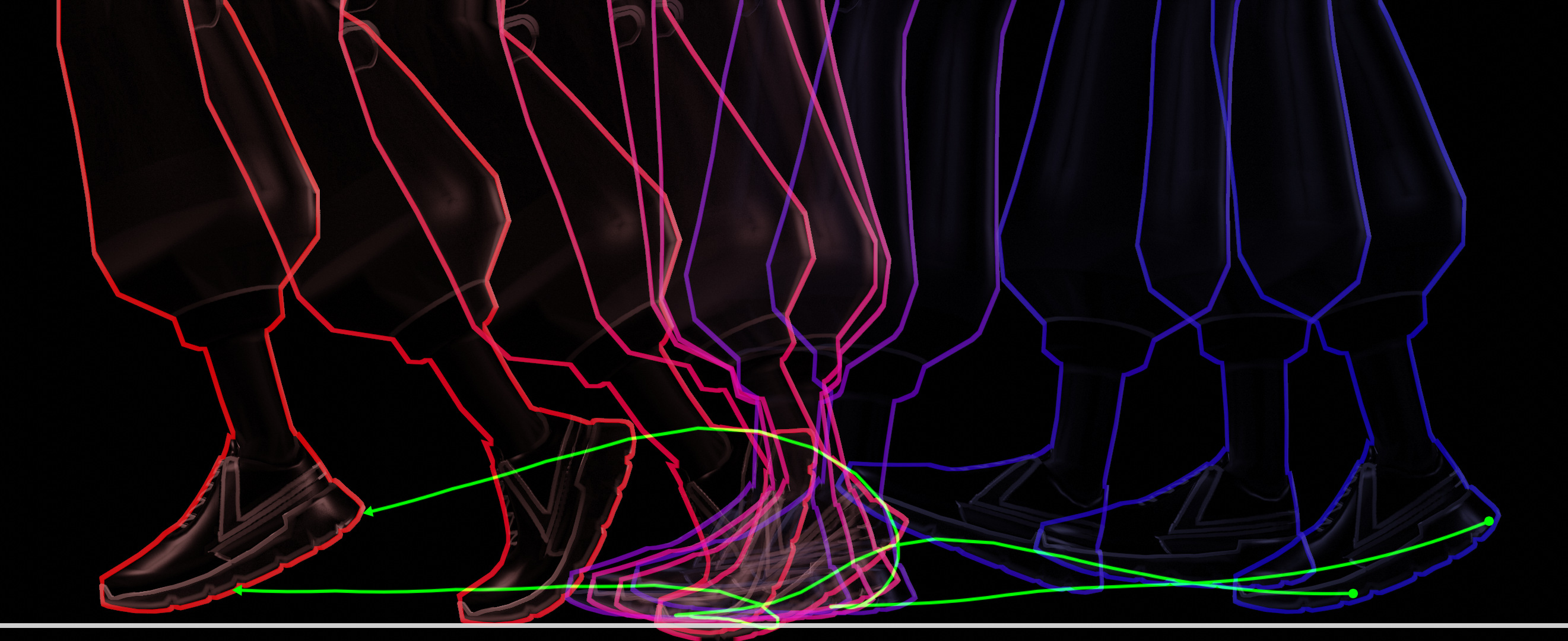}}
        \caption{Height loss only}
        \label{fig:foot_traj2}
    \end{subfigure}
    \begin{subfigure}[b]{\columnwidth}
        \centering
        \scalebox{-1}[1]{\includegraphics[width=\linewidth]{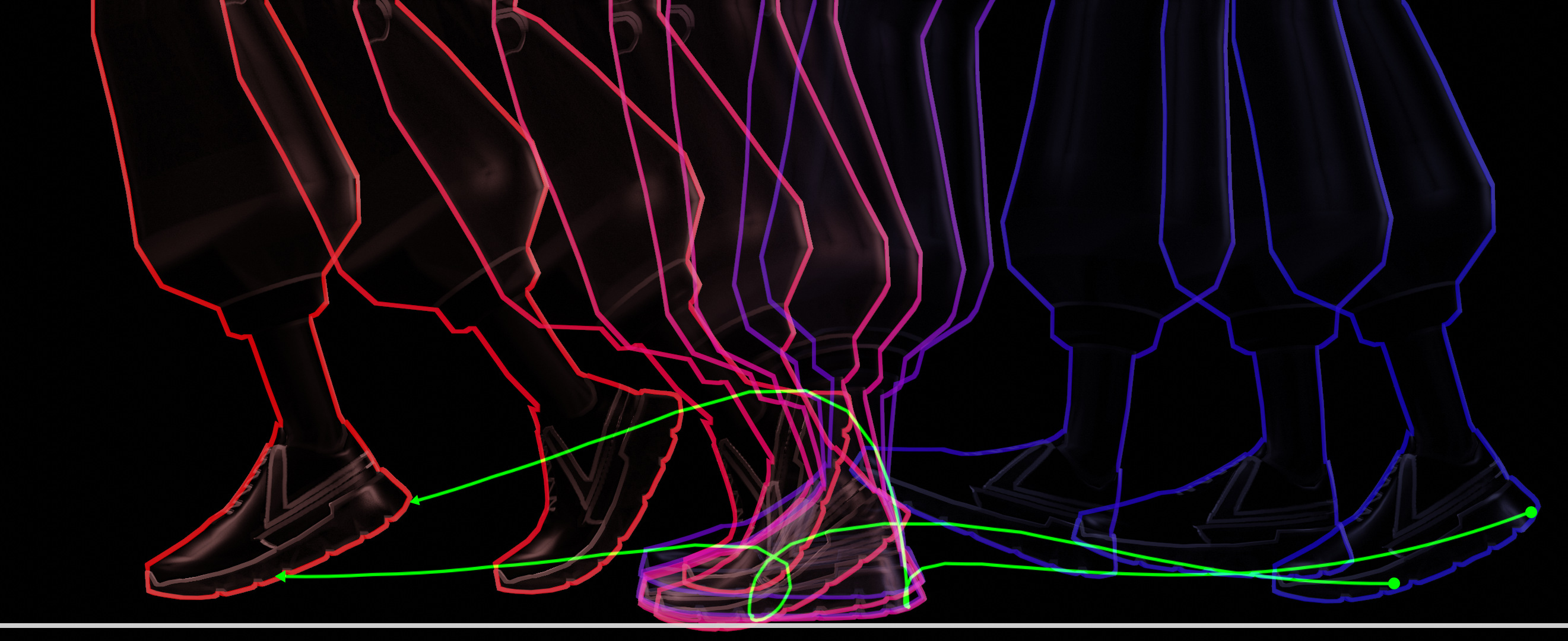}}  
        \caption{Height loss and foot sliding loss}
        \label{fig:foot_traj1}
    \end{subfigure}
\caption{Visualizations of the impact of the losses on foot contact. Overlay of the position of the foot at every third frame (blue: past, red: future). The trajectory of two vertices from the sole and heel plotted in green. Ground floor is represented by the white horizontal line.}
\end{figure}

Figure \hyperref[fig:foot_traj]{7} depicts our experiment to evaluate the impact of $\mathcal{L}_\text{height}$ and $\mathcal{L}_\text{sliding}$ on the foot contacts:
\begin{itemize}
    \item Without those losses (Figure~\ref{fig:foot_traj3}) we observe an offset to the ground, of an amount that varies during the contact. It is due to the difference of bone lengths along the kinematic chain.
    \item In response to this, the height loss (Figure~\ref{fig:foot_traj2}) forces the foot to touch the ground during the entire duration of the contact. However, this causes horizontal sliding during the contact period.
    \item The foot sliding loss (Figure~\ref{fig:foot_traj3}) mitigates this issue by finding a compromise between the two constraints, producing cleaner contacts with very little foot skating. 
\end{itemize}

\clearpage

\section{Failure cases}

In the following paragraphs, we discuss several failure cases regarding the key-vertices transfer and regarding the retargeting process, and identify possible solutions to fix them.

\subsection{Key-vertices transfer}


\subsubsection{Lack of geometry}
\begin{figure}
    \centering
    
    \hfill%
    \begin{subfigure}[b]{\columnwidth}
        \begin{subfigure}[b]{.45\columnwidth}
            \centering
            \includegraphics[trim={140 50 150 50},clip,width=\textwidth]{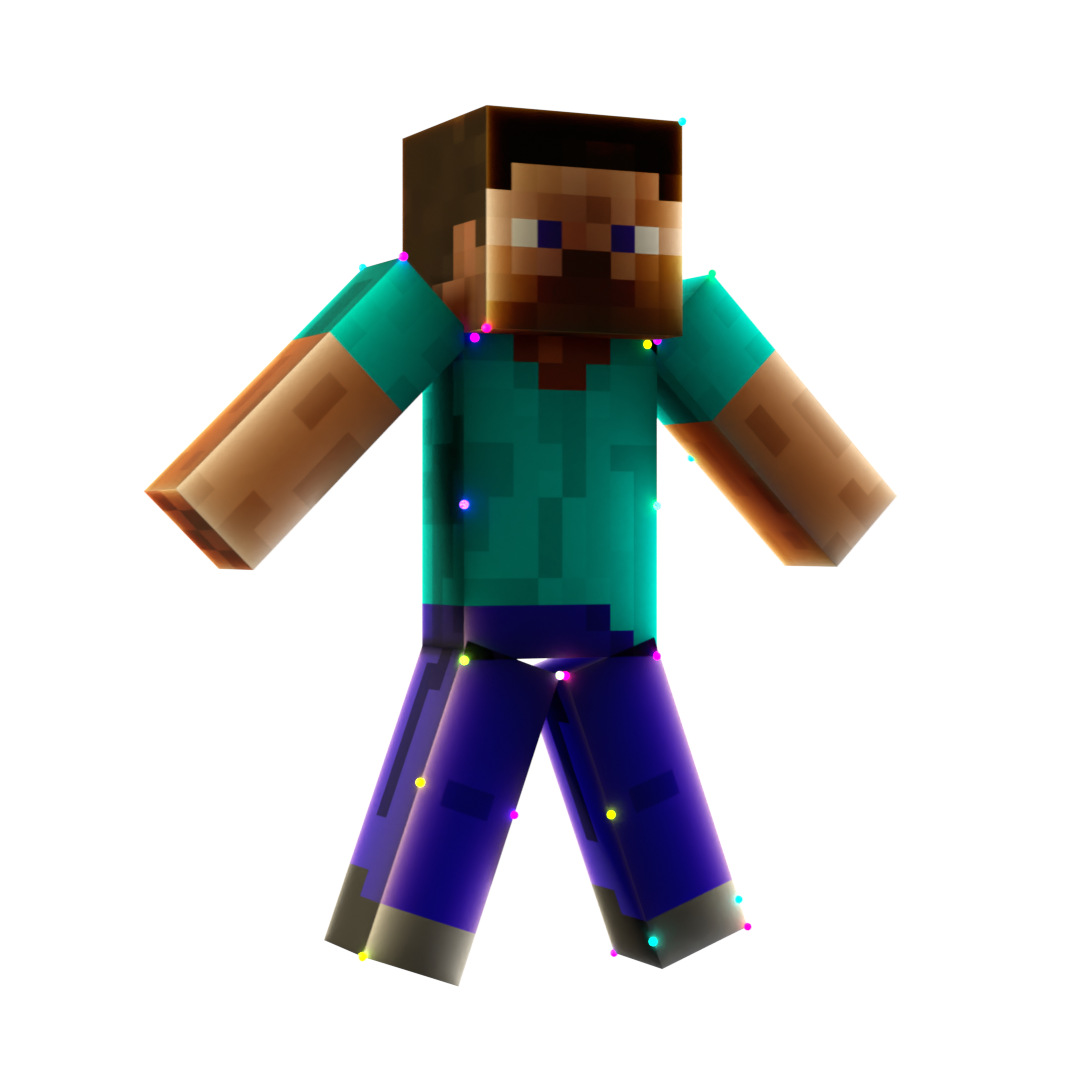}
        \end{subfigure}%
        \hfill%
        \begin{subfigure}[b]{.45\columnwidth}
            \centering
            \includegraphics[trim={140 50 150 50},clip,width=\textwidth]{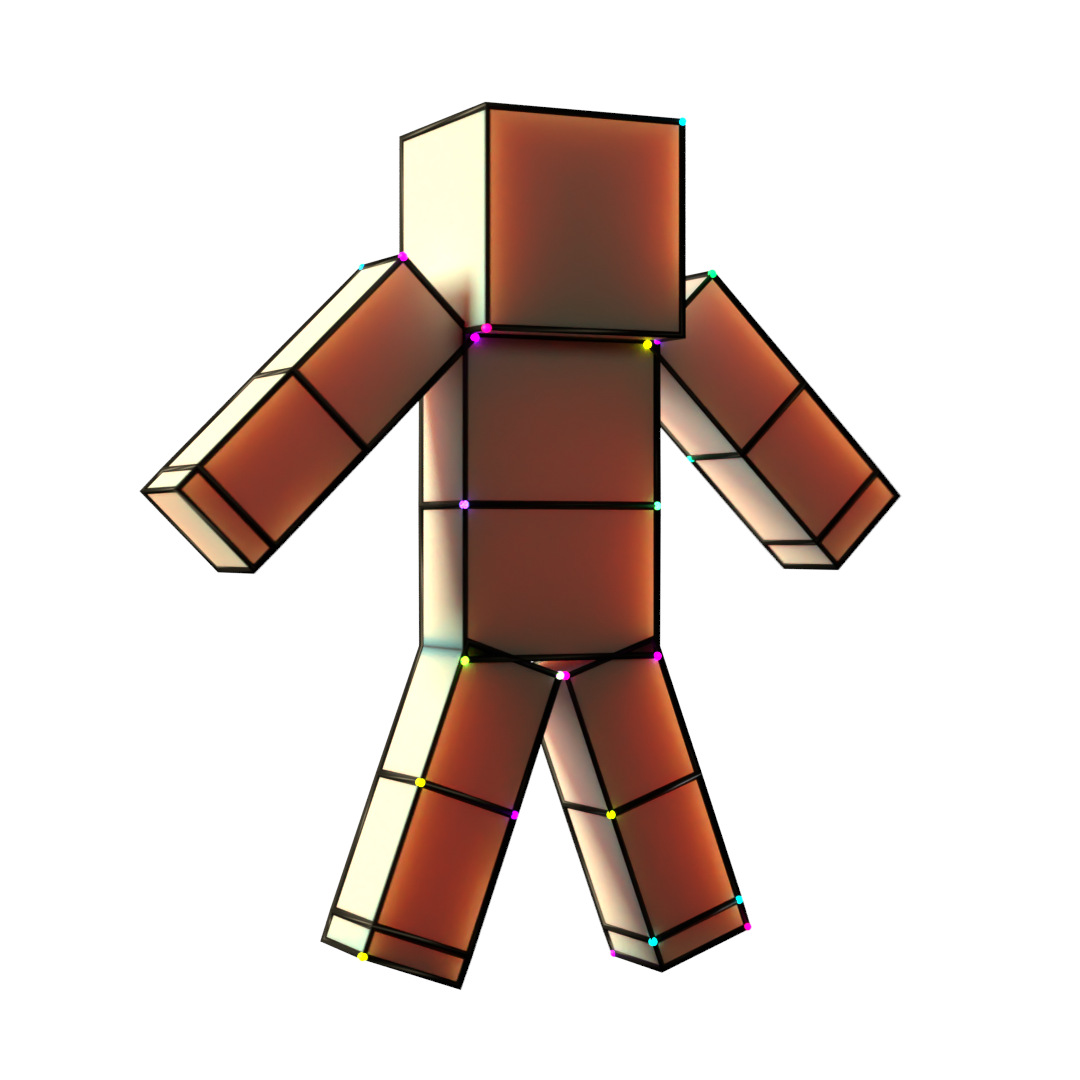}
        \end{subfigure}%
        \hfill
        \subcaption[]{A low-res character from Minecraft~\cite{minecraft} with only 84 vertices.}
        \label{fig:minecraft-low}
    \end{subfigure}

    \hfill%
    \begin{subfigure}[b]{\columnwidth}
    
        \begin{subfigure}[b]{.45\columnwidth}
            \centering
            \includegraphics[trim={140 50 150 50},clip,width=\textwidth]{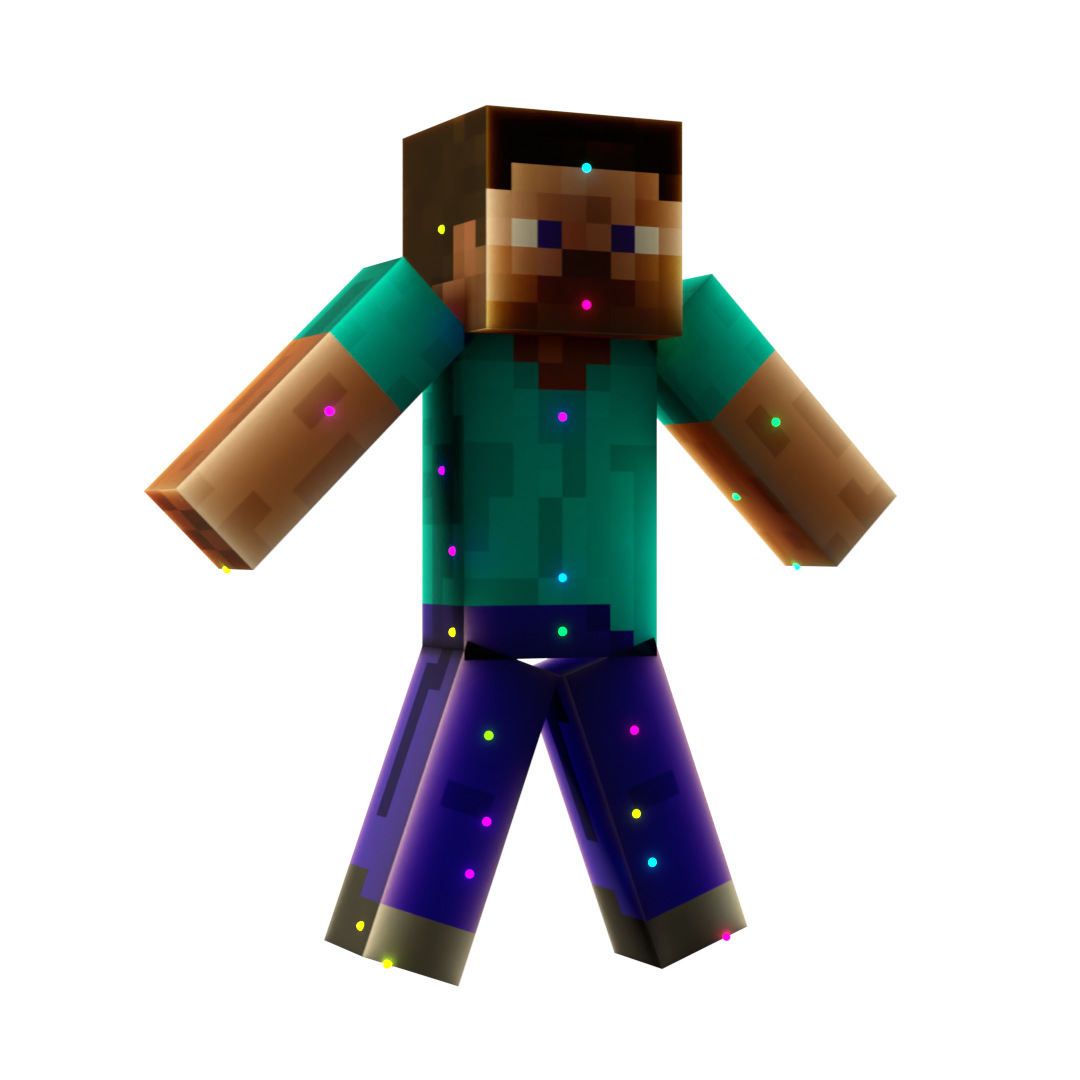}
        \end{subfigure}%
        \hfill%
        \begin{subfigure}[b]{.45\columnwidth}
            \centering
            \includegraphics[trim={140 50 150 50},clip,width=\textwidth]{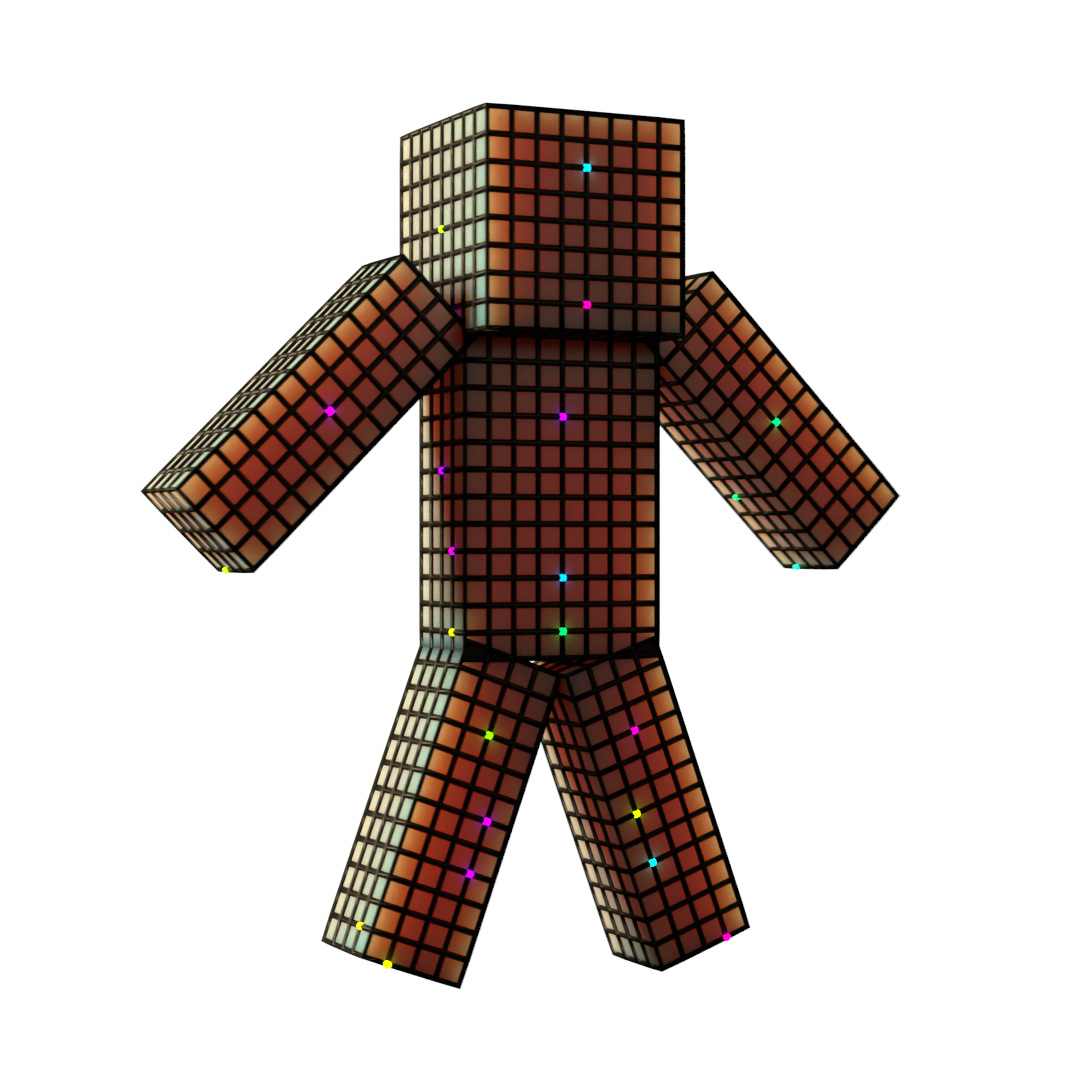}
        \end{subfigure}%
        \hfill
        \subcaption[]{Higher-resolution version of the same character, with 1644 vertices.}
        \label{fig:minecraft-high}
    \end{subfigure}

    \begin{subfigure}[b]{\columnwidth}
        \begin{subfigure}[b]{.45\columnwidth}
            \centering
            \includegraphics[trim={145 60 180 60},clip,width=\textwidth]{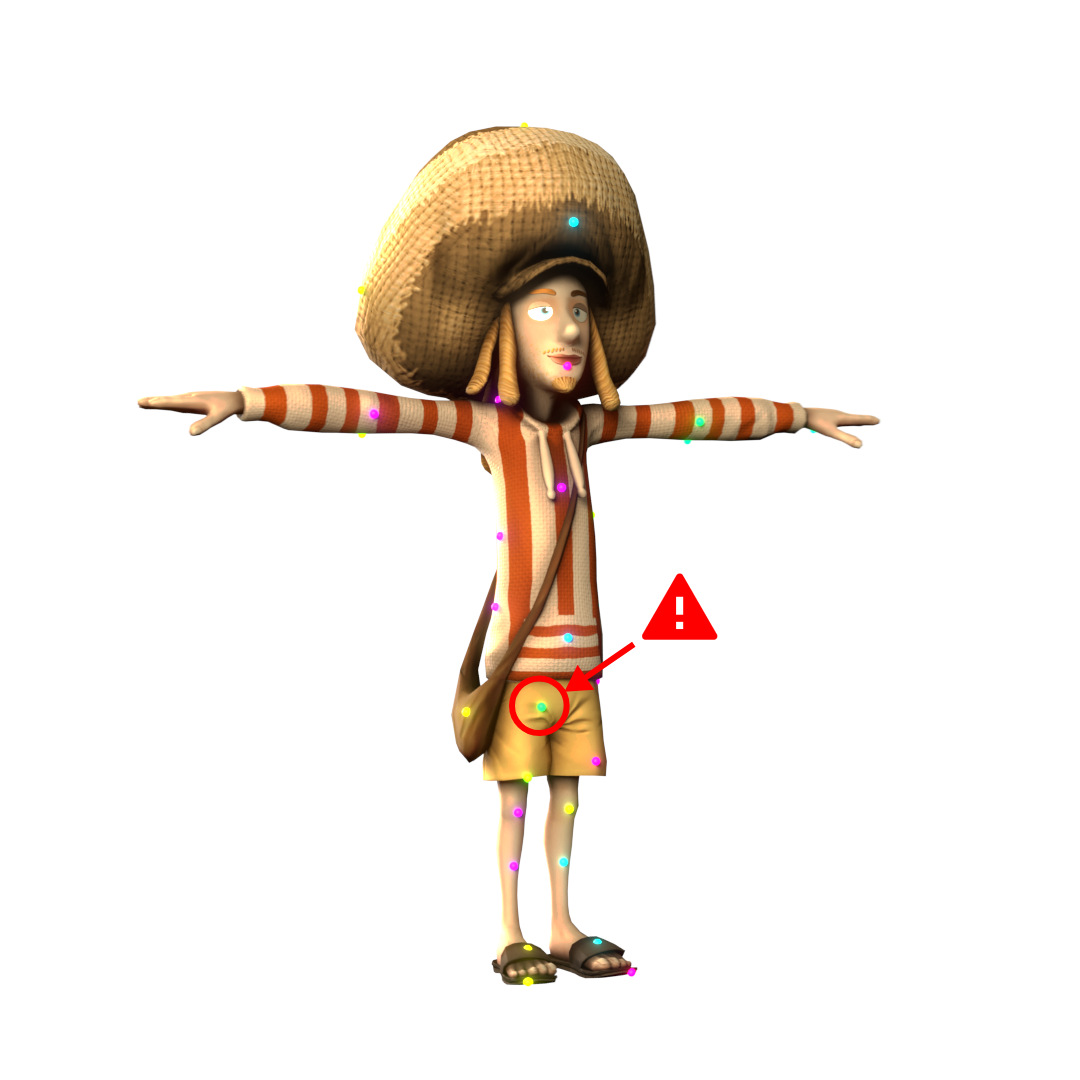}
            \caption{Asymmetric character}
            \label{fig:asymmetric}
        \end{subfigure}%
        \hfill%
        \begin{subfigure}[b]{.45\columnwidth}
            \centering
            \includegraphics[trim={145 60 180 60},clip,width=\textwidth]{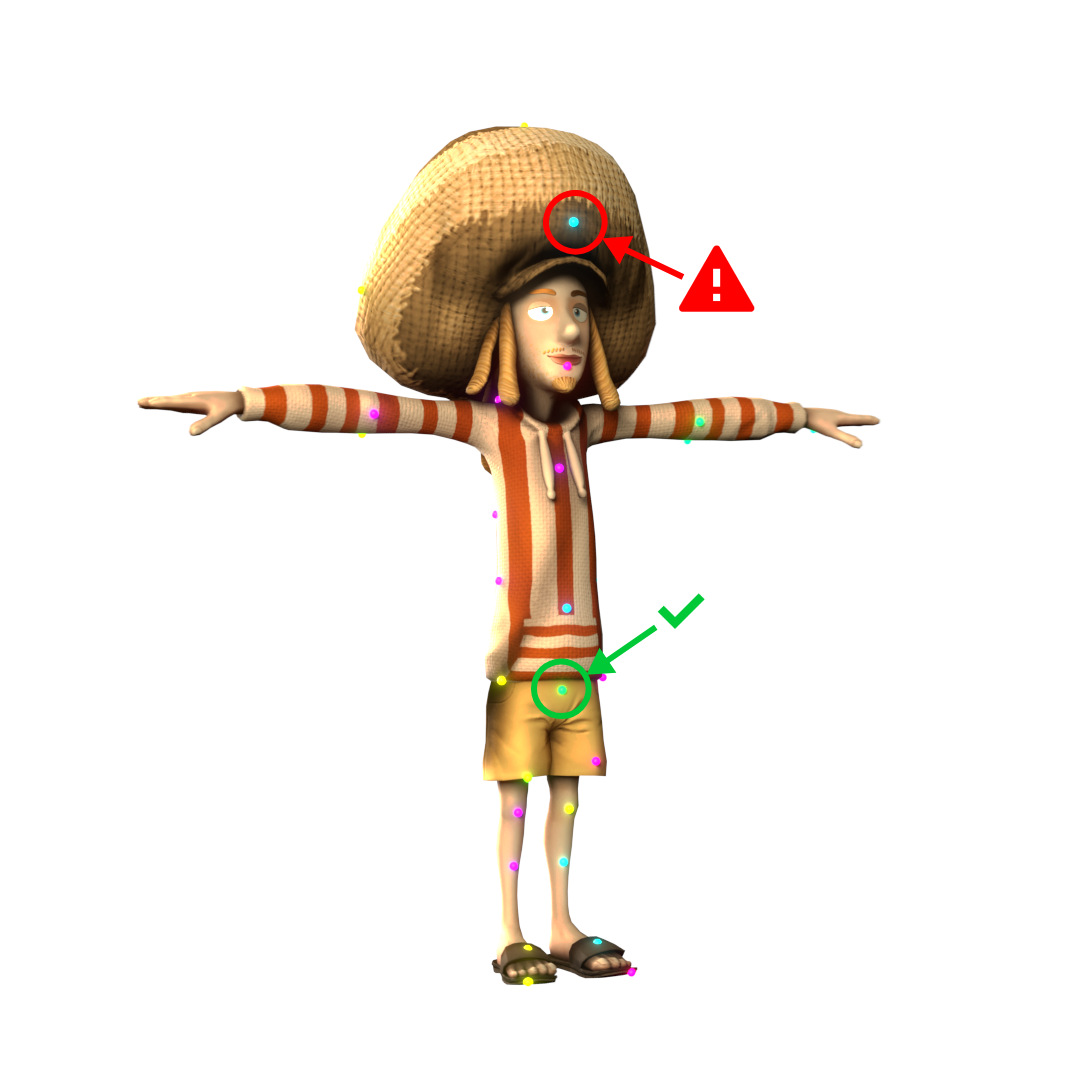}
            \caption{Nearly-symmetric character}
            \label{fig:symmetric}
        \end{subfigure}%
    \end{subfigure}

    \begin{subfigure}[b]{\columnwidth}
        \begin{subfigure}[b]{.45\columnwidth}
            \centering
            \includegraphics[clip,width=\textwidth]{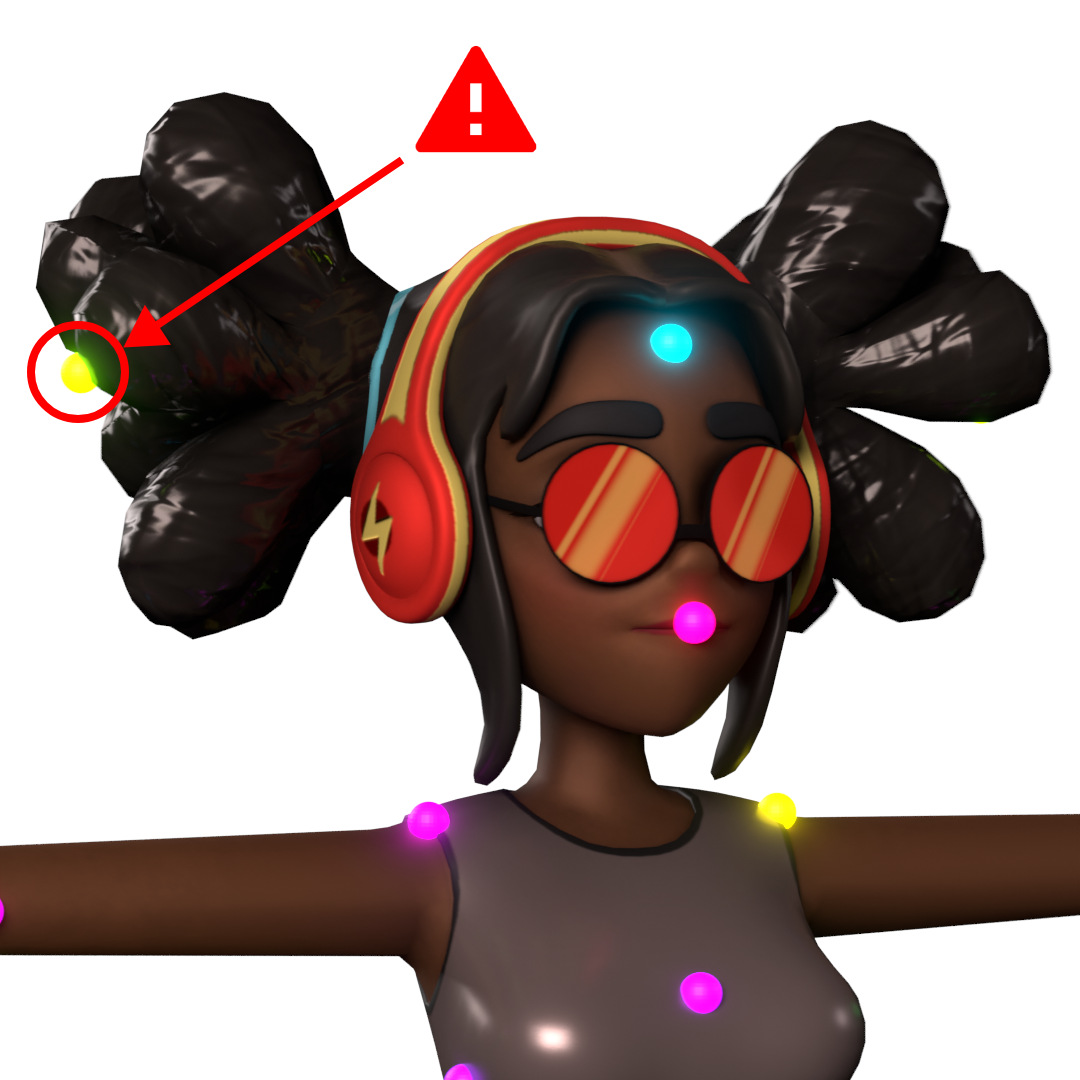}
            \caption{Large haircuts}
            \label{fig:michelle}
        \end{subfigure}%
        \hfill%
        \begin{subfigure}[b]{.45\columnwidth}
            \centering
            \includegraphics[clip,width=\textwidth]{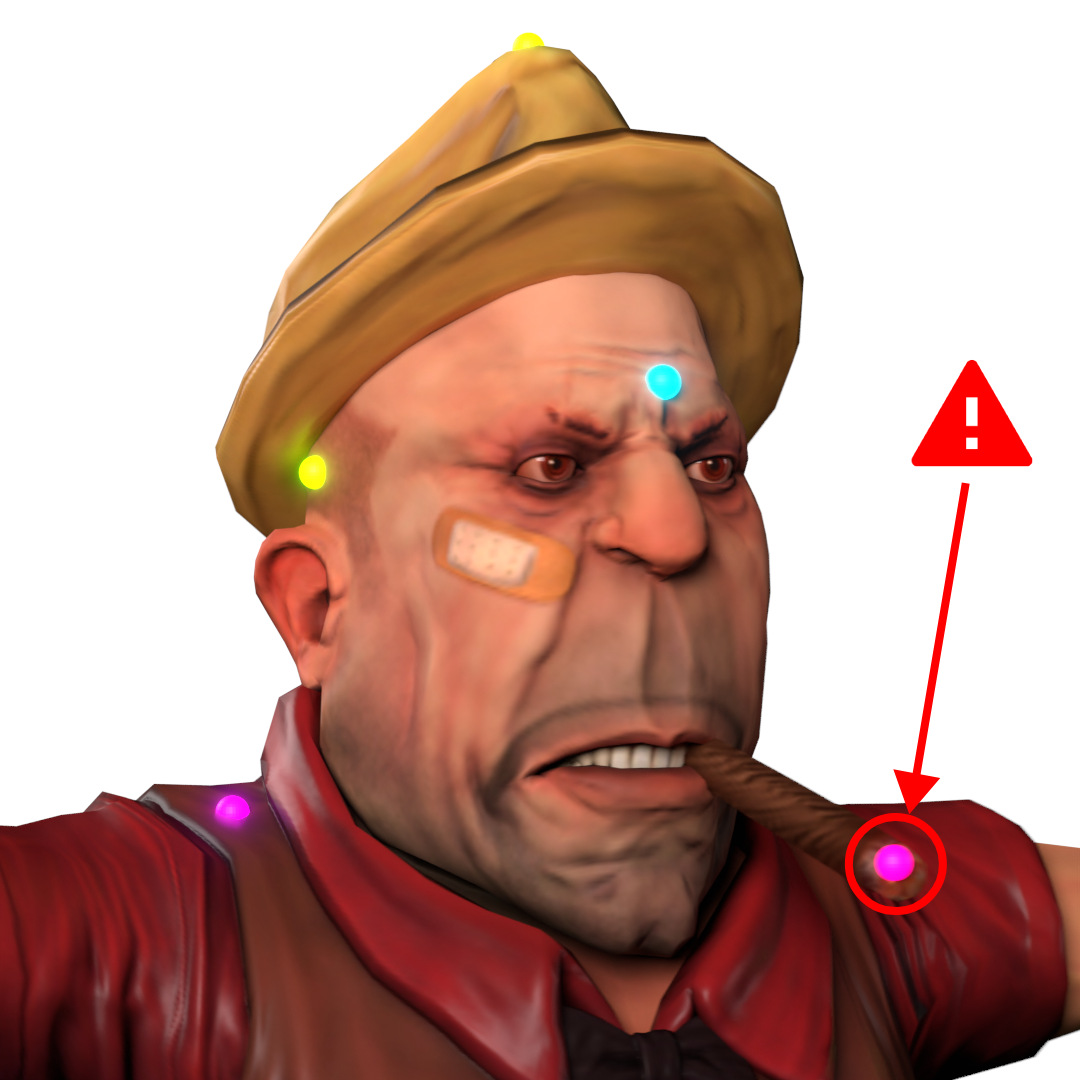}
            \caption{Accessories}
            \label{fig:theboss}
        \end{subfigure}%
    \end{subfigure}
    \caption{Examples of failure cases of the key-vertices transfer algorithm}
    \label{fig:failure-cases-kptransfer}

\end{figure}

The transfer of key vertices can yield suboptimal results if the target mesh contains too few vertices, such as the blocky characters from Minecraft~\cite{minecraft} depicted in Figure~\ref{fig:minecraft-low}. In that case, there are not enough vertices to accurately model the overall shape of the body parts, which prevents the method from placing key vertices in relevant places. A simple solution consists in subdividing the mesh to increase the number of vertices. Additionally, this issue could be automatically detected based on simple rules like face area and number of vertices.

We notice that the key vertices on the higher-resolution mesh are better located, as seen on Figure~\ref{fig:minecraft-high}, but there are still some errors : for instance, the chin vertex and eye vertex are too high compared to the location of the chin and eyes in the texture. This cannot be taken into account based on geometry alone.

\subsubsection{Asymmetry}


Figure~\ref{fig:asymmetric} shows the impact of an asymmetric mesh. Due to the presence of a satchel on the side of the character, the key-vertices are pulled in to the right, which could deteriorate the pertinence of the semantic loss, and lead to contacts in the wrong places. In particular, the key-vertex ``hips-front'' (green) ends up closer to the right thigh. This can be corrected by manually removing the satchel that makes the character asymmetric, as shown in Figure~\ref{fig:symmetric}.

\subsubsection{Accessories/Props}


The key-vertices transfer algorithm also tends to be affected by the presence of additional props in the target character, even when they are symmetrical. For instance, some cartoon characters come with a large hat/ haircut, or accessories, like those seen in Figures~\ref{fig:symmetric}, ~\ref{fig:michelle} and~\ref{fig:theboss}. In that case, the key-vertices of the head are likely to be placed on the most extreme points. As an example, on Figure~\ref{fig:michelle}, the ``ear'' vertices are misplaced, on Figure~\ref{fig:symmetric}, the ``eye'' vertex is misplaced, and on Figure~\ref{fig:theboss}, the ``chin'' vertex is misplaced. This may have a positive impact on some motions, as the optimization will prevent collisions with the hair or cigar. Nevertheless, it can also lead to a loss of semantics because contact might occur at the wrong position on the mesh in some motions.

\subsubsection{Overlapping meshes}

Some characters have overlapping meshes (for instance, the top of their pants being hidden behind their top). In that case, the transfer algorithm might place the vertices on ``hidden'' vertices, causing small amounts of self-penetration as the outer mesh is neglected. To alleviate this issue, a simple filtering of those vertices from the mathematical formulation would be sufficient.

\subsection{Retargeting}


\subsubsection{Complex poses}
\begin{figure}[!t]
    \centering
    \begin{subfigure}[b]{\columnwidth}
        \centering
        \includegraphics[clip,trim=600 150 400 500,width=\linewidth]{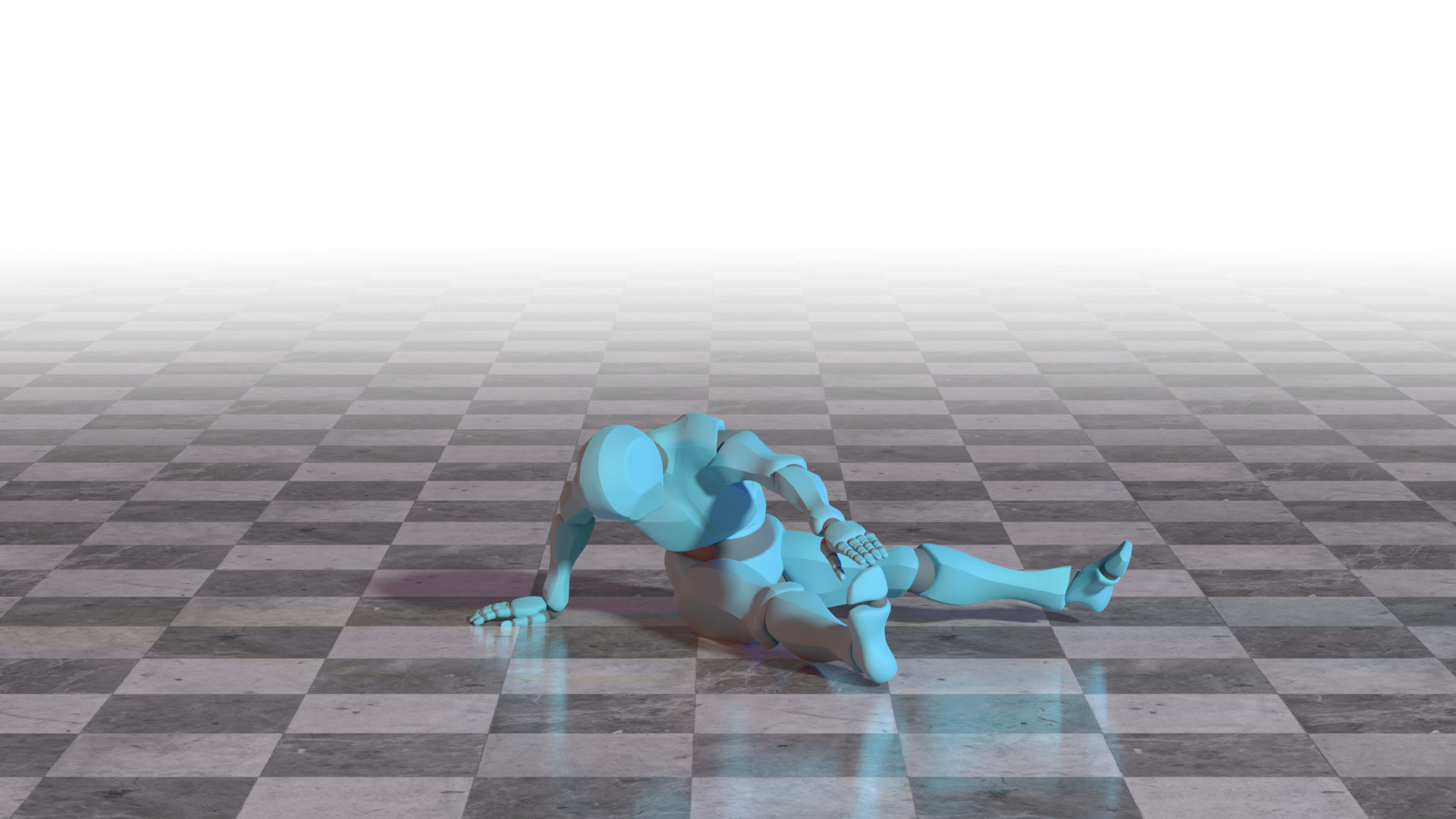}
        \caption{Source pose}
        \label{fig:failure-cases-1.1}
    \end{subfigure}
    \begin{subfigure}[b]{\columnwidth}
        \centering
        \includegraphics[clip,trim=600 150 400 500,width=\linewidth]{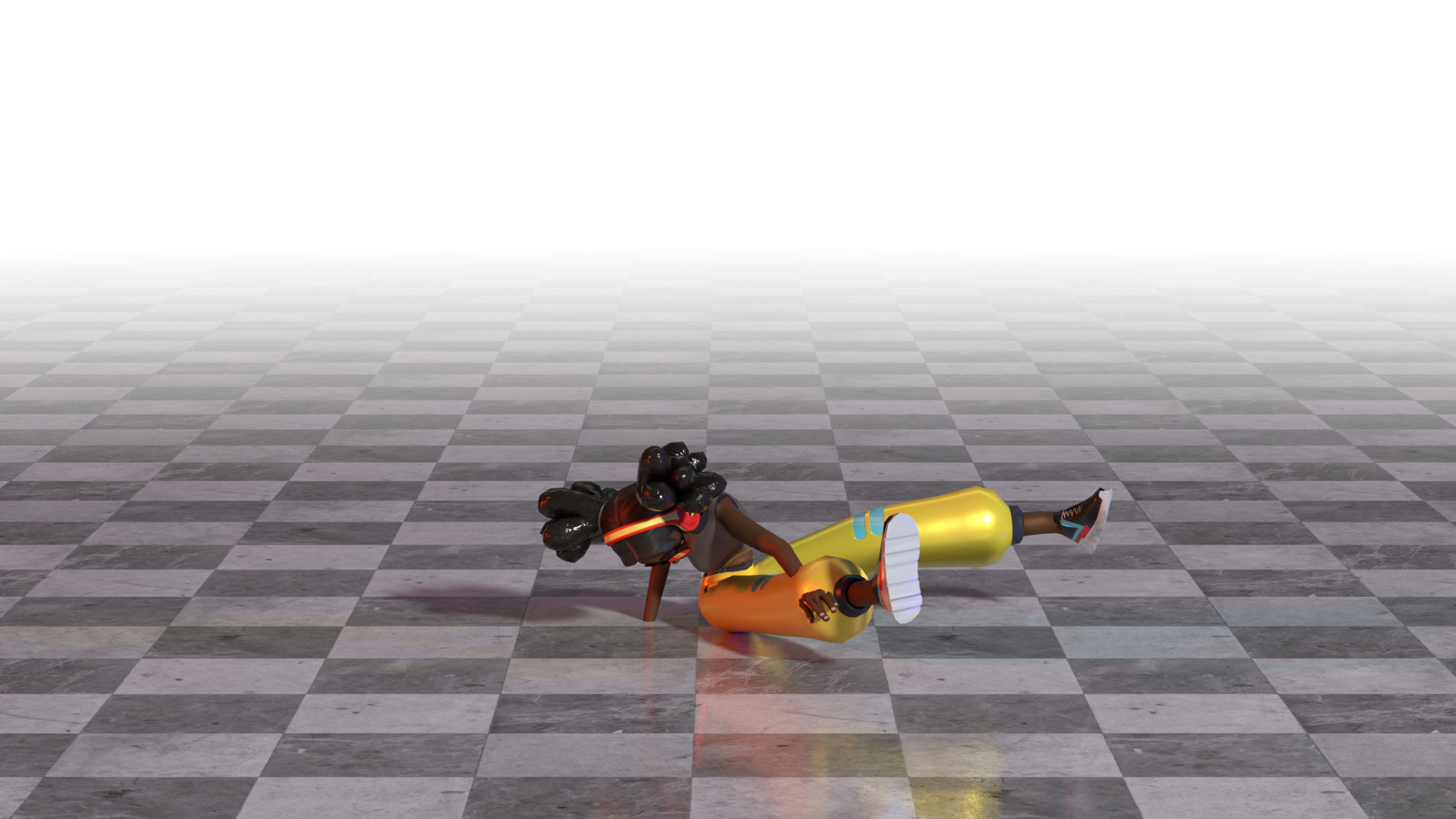}
        \caption{``Copy Rotations'' (for reference)}
        \label{fig:failure-cases-1.2}
    \end{subfigure}
    \begin{subfigure}[b]{\columnwidth}
        \centering
        \includegraphics[clip,trim=600 150 400 500,width=\linewidth]{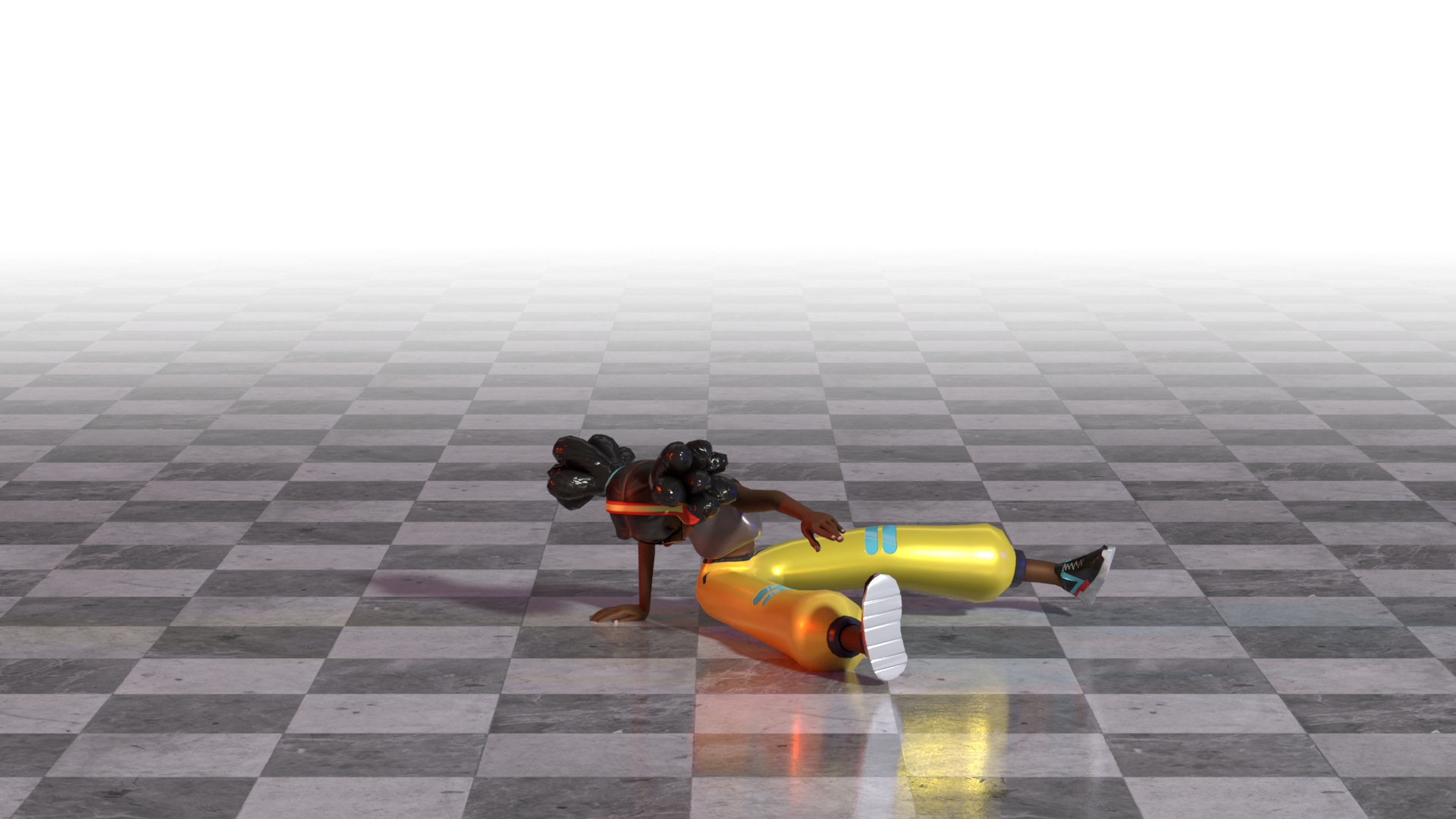}
        \caption{Output of our method}
        \label{fig:failure-cases-1.3}
    \end{subfigure}
\caption{Failure case on a complex pose with several contacts.}
\label{fig:failure-cases-1}
\end{figure}

In very complex poses, with several points of contact, it becomes quite difficult for the optimization to find a compromise between the various semantic aspects of the motion. 
Figure~\ref{fig:failure-cases-1} shows an examples of a complex failure case : the floor contacts (left hand, both feet), that were lost by ``copy rotations'' (Figure~\ref{fig:failure-cases-1.2}) are then restored by our method (Figure~\ref{fig:failure-cases-1.3}), but the foot-hand contact is lost. 

\begin{figure}[!b]
    \centering
    \includegraphics[trim={10 120 10 160},clip,width=0.8\columnwidth]{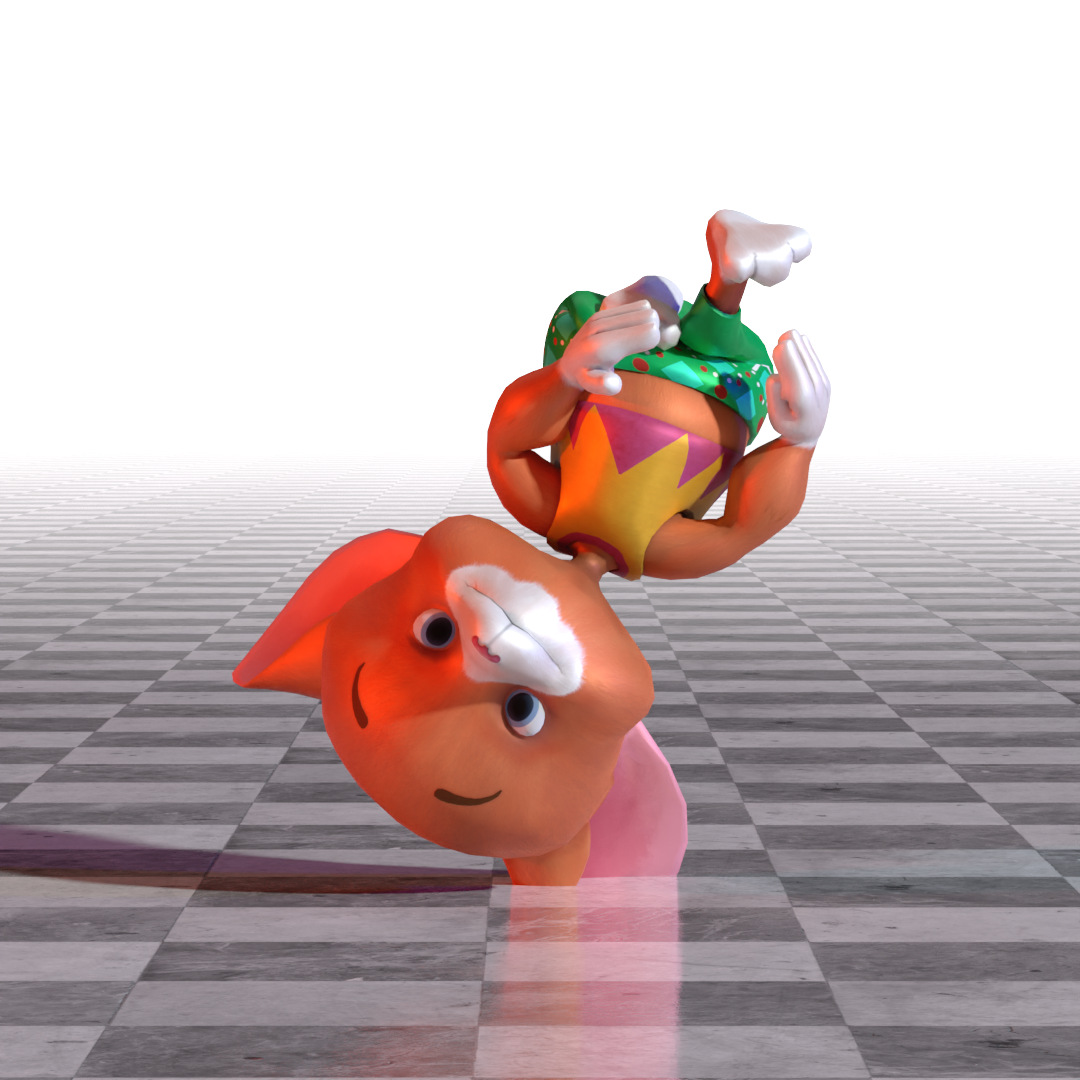}
    \caption{Collisions can appear on parts of the mesh where no key-vertices are placed (motion : Backflip, character : Mousey)}
    \label{fig:mousey-failure}
\end{figure}

\subsubsection{Unrealistic motion trajectories}
\begin{figure*}
   \begin{subfigure}[b]{.65\textwidth}
        \vspace*{1\baselineskip}
        \begin{subfigure}[b]{\textwidth}
            \centering
            \scalebox{-1}[1]{\includegraphics[trim=430 100 0 0, clip, width=\linewidth]{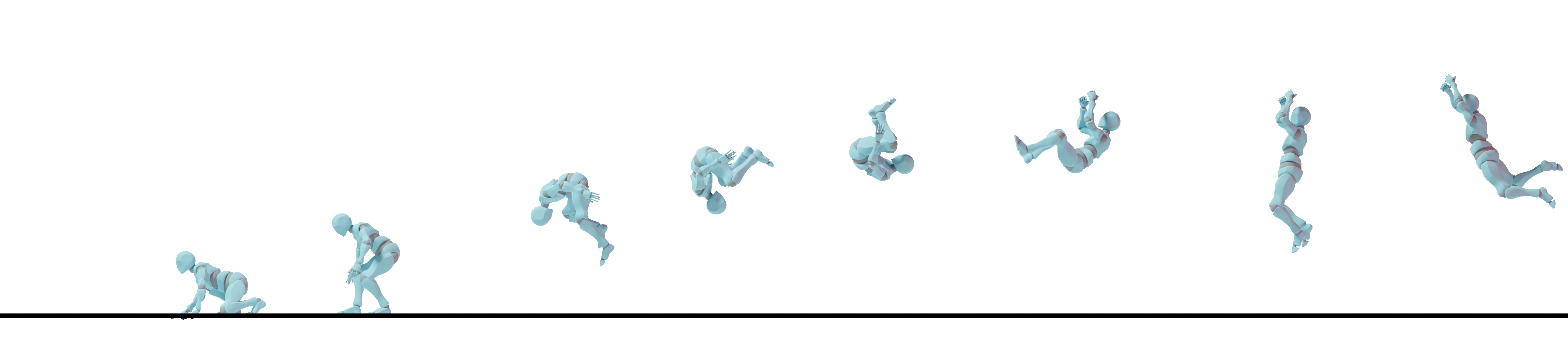}}
            \vspace*{-1.1\baselineskip}
            \caption{Chronophotography of the source motion}
            \label{fig:chronophotography_source}
        \end{subfigure}\\
        \vspace*{2\baselineskip}
        \begin{subfigure}[b]{\textwidth}
            \centering
            \scalebox{-1}[1]{\includegraphics[trim=430 100 0 0, clip, width=\linewidth]{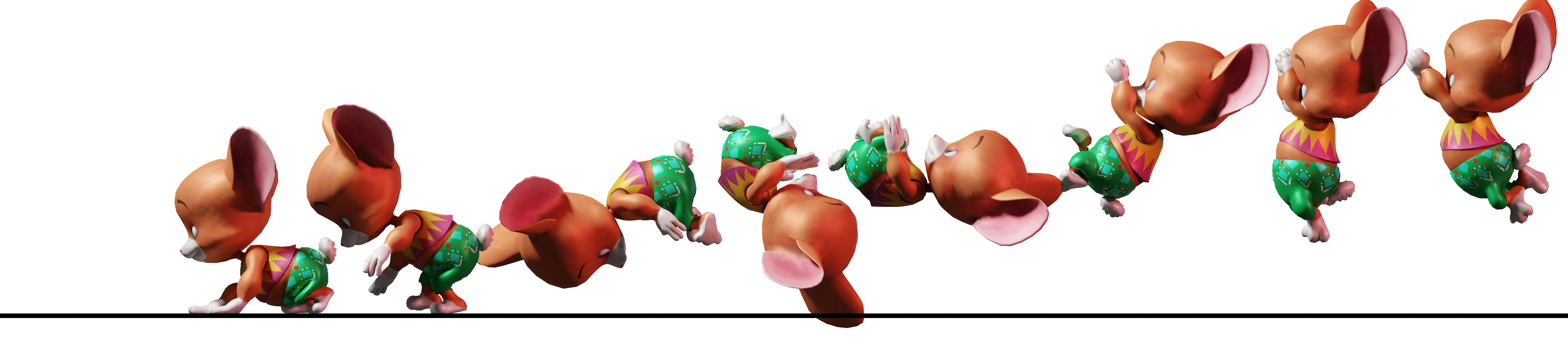}}
            \vspace*{-1.1\baselineskip}
            \caption{Chronophotography of the retargeted motion}
            \label{fig:chronophotography_target}
        \end{subfigure}
        \vspace*{1\baselineskip}

    \end{subfigure}%
    \hfill%
    \begin{subfigure}[b]{.32\textwidth}
        \centering
        \includegraphics[trim=0 0 0 0, clip, width=\linewidth]{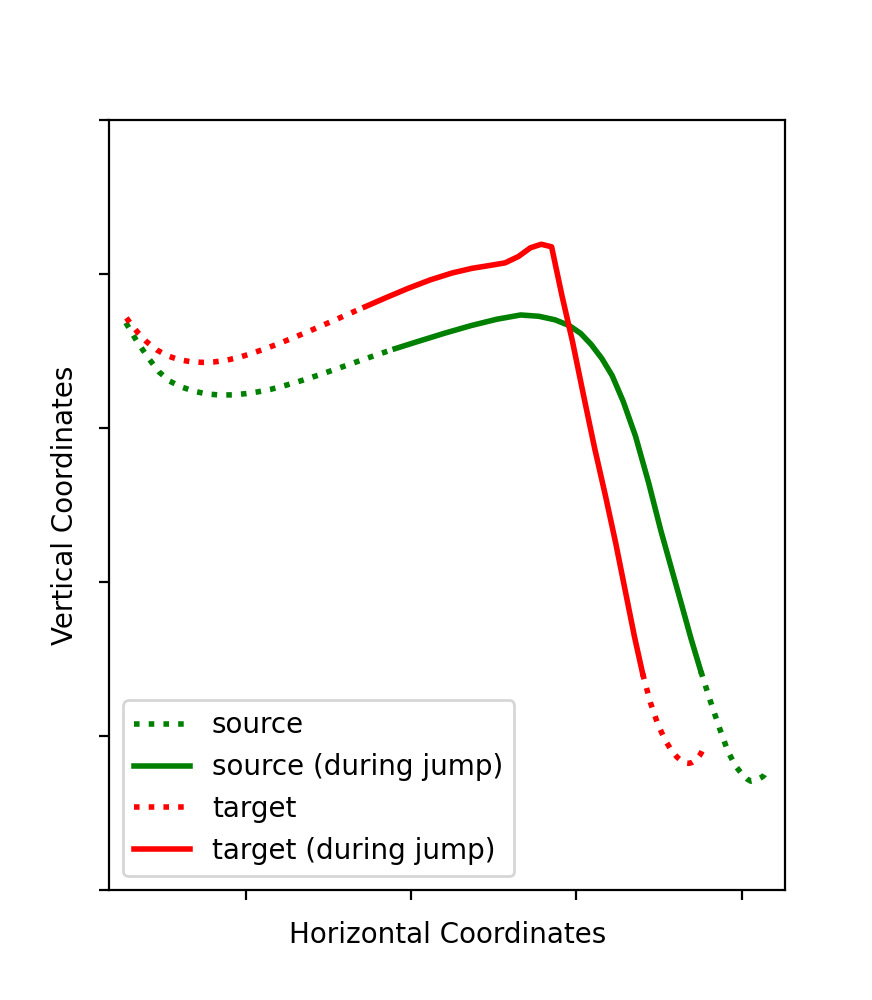}
        \caption{Trajectory of the pelvis position for both characters}
        \label{fig:chronophotography}
    \end{subfigure}

    \caption{Failure case on an unrealistic trajectory (motion : Swing To Land, character : Mousey)}
    \label{fig:trajectory}
    
\end{figure*}

Our optimization considers all frames almost independently, apart from $\mathcal{L}_\text{smooth}$. When the retargeting imposes a considerable change of position on a portion of the animation, there are no guarantee that it will respect the laws of physics. This is shown on Figure~\ref{fig:trajectory} : to prevent the head from colliding with the floor during a flip, the character's position is shifted vertically, making the trajectory look less like a parabola. In cases like this, the floor penetration loss leads to a decrease in motion plausibility. An additional term taking into account the global trajectory of the center of mass could be used to alleviate this issue.

\subsubsection{Mesh over-simplification}
In cases where the key-vertices are not sufficient to capture important parts of the mesh, collisions might still occur. This can be seen on Figure~\ref{fig:mousey-failure}, where the upper part of the ear penetrates the ground freely as there are no key-vertices encoding this part of the head. Increasing the number of key-vertices is likely to solves this kind of issues, as this allows for a finer encoding of the mesh.

\subsection{Additional characters}
We demonstrate that, unlike previous methods, ReConForM can adapt to a variety of characters, regardless of how different their skeleton structure or morphology might be. For this, we show results on five characters used in animated short films, graciously provided by Blender Studio \cite{blenderstudio} under CC-BY license. The characters are: Rain, Rex, Gabby, Sprite and Phil. The result of key-vertices transfer can be seen in the main paper (for $N=41$) and in Figure~\ref{fig:keyvertices_transfer_big} (for $N=96$). Moreover, some results of several motions retargeted on these characters are provided in the main paper.

\section{User study}

\subsection{Details}
We selected 8 characters from Mixamo for the user study, and we retargeted each of the 45 animations composing our evaluation set on a randomly chosen character, as described in Table~\ref{table:dataset}. We rendered 45 videos containing the source animation on the left, and ours / R$^2$ET's \cite{r2et} animations on the right (in a random order). 

We used Prolific \cite{prolific} to collect answers from 116 English-speaking users (users were compensated on average £10.39 per hour), and we also asked 17 animation students and professionals to participate in our survey. Each participant was asked to fill a form containing 12 questions, with an additional question to test their attention and filter out inattentive users. In total, we collected a total of 1596 individual answers. Amongst all respondents:
\begin{itemize}
    \item 1 was under 18
    \item 88 were aged 18-30
    \item 38 were aged 30-50
    \item 6 were over 50
\end{itemize}
    
We also asked them to report their familiarity with the task at hand (multiple answers allowed): 
\begin{itemize}
    \item 85 of them declared that they were ``familiar with video games or interactive experiences including 3D characters''
    \item 17 of them declared ``having used 3D animation software before''
    \item 7 of them declared being ``a professional / student in the field of character animation''
    \item 36 declared having ``no experience in particular'' in the field of 3D animation
\end{itemize}

For each of the 12 videos, the users were tasked to answer two questions: 
\begin{itemize}
    \item ``In the example above, with regards to fidelity to the source animation:''
    \item ``In the example above, which output is the most visually pleasing (regardless of the source):''
\end{itemize}
The possible answers were:
\begin{itemize}
    \item A is better
    \item B is better
    \item A and B are tied (both good)
    \item A and B are tied (could both be improved)
    \item Not sure
\end{itemize}

\subsection{Filtering}

Upon reviewing the results obtained with R$^2$ET's pre-trained weights, we observed a lot of jitter and flicker. After exchanging with the authors about this issue, we decided to use a OneEuro filter \cite{1eurofilter} with a cutoff frequency of $f_c = 0.02$ Hz and $\beta = 0.5$, which resulted from a manual tuning in order to obtain the best results. We used this filter on R$^2$ET's results for our user study, and (when explicitly mentioned) for our quantitative comparison.

\clearpage

\begin{figure*}
    \centering
    \begin{subfigure}[b]{\textwidth}
        \centering
        \input{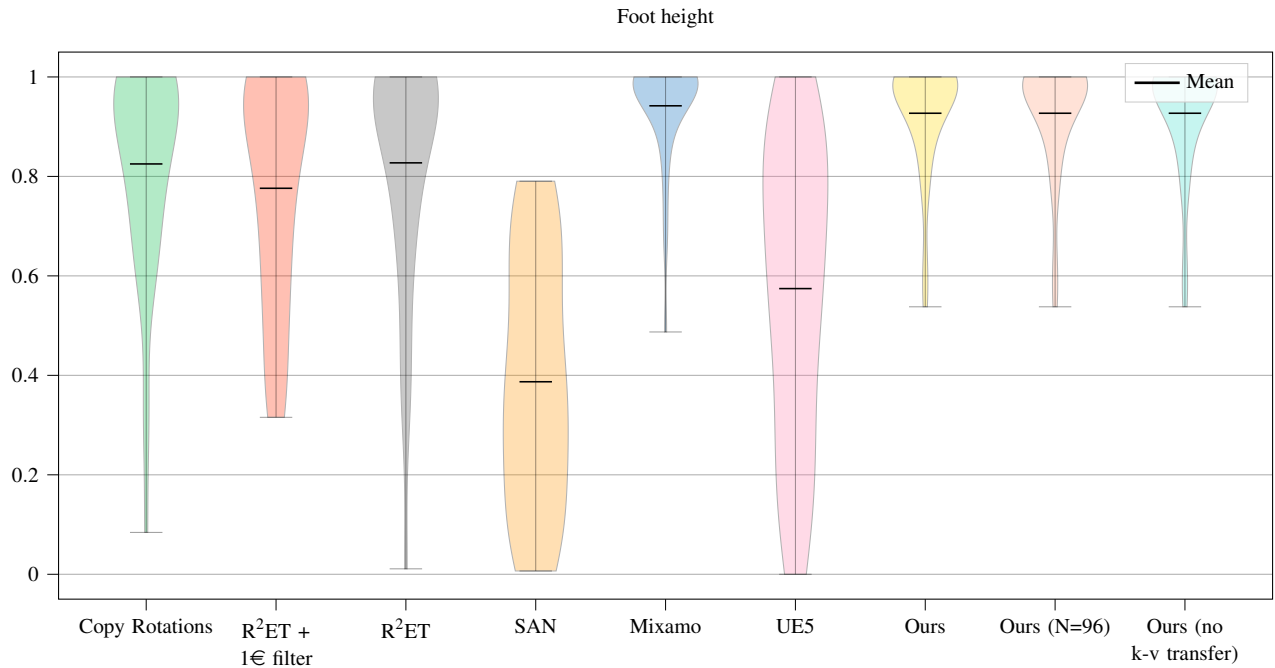}    
        \caption{Comparison of the foot-ground contacts F-1 score (source taken as ground truth)  ($\uparrow$)}
    \end{subfigure}%
    \newline
    \begin{subfigure}[b]{\textwidth}
        \centering
        \input{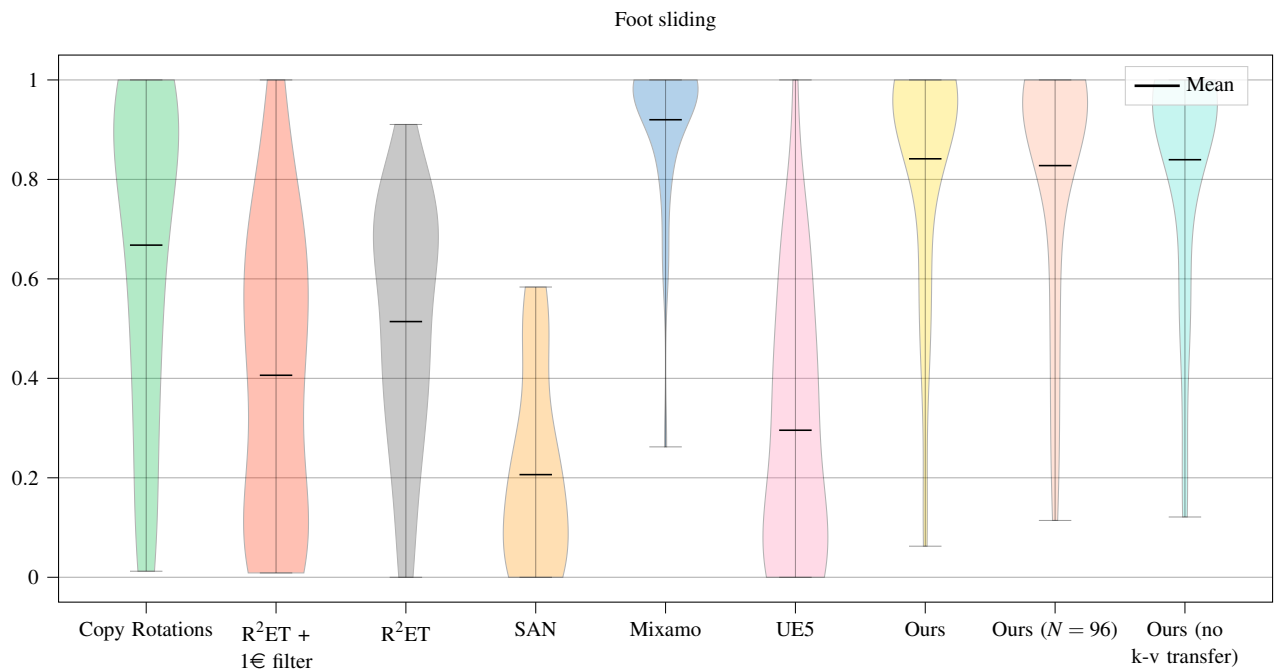}    
        \caption{Comparison of the foot sliding F-1 score (source taken as ground truth) ($\uparrow$)}
    \end{subfigure}%
    \caption{Distribution of the metrics across all 45 animations of the validation dataset. For each method, the minimum, maximum and mean values across all animations are indicated by horizontal lines.}
\end{figure*}%
\begin{figure*}\ContinuedFloat
    \centering
        \begin{subfigure}[b]{\textwidth}
        \centering
        \input{contents/graphs/violin_plot_metrics2/pen-violinplot}    
        \caption{Comparison of the limb penetration volume ($\downarrow$)}
    \end{subfigure}%
    \newline
    \begin{subfigure}[b]{\textwidth}
        \centering
        \input{contents/graphs/violin_plot_metrics2/floorpen-violinplot}    
        \caption{Comparison of the floor penetration volume ($\downarrow$)}
    \end{subfigure}%
    \caption{Distribution of the metrics across all 45 animations of the validation dataset. For each method, the minimum, maximum and mean values across all animations are indicated by horizontal lines. When relevant, the metric is computed for the source animation as a reference.}
    \label{fig:violinplots}
\end{figure*}%
\begin{figure*}\ContinuedFloat
    \centering
        \begin{subfigure}[b]{\textwidth}
        \centering
        \input{contents/graphs/violin_plot_metrics2/jerk-violinplot}    
        \caption{Comparison of the jerk values ($\downarrow$)}
    \end{subfigure}
    \caption{Distribution of the metrics across all 45 animations of the validation dataset. For each method, the minimum, maximum and mean values across all animations are indicated by horizontal lines. When relevant, the metric is computed for the source animation as a reference.}
    \label{fig:violinplots-2}
\end{figure*}


\begin{table*}[!ht]
\vspace{2\baselineskip}
    \centering
    \begin{tabular}{|l|l|l||l|}
        \hline
        \textbf{Animation name} & \textbf{Source} & \textbf{Difficulty}  & \textbf{Target character in our user study} \\ \hline\hline     
        Angry & XBot & high & Michelle \\ \hline
        Arm Stretching & YBot & high & TheBoss \\ \hline
        Backflip & XBot & low & Mousey \\ \hline
        Breakdance Foot to Idle & XBot & high & Ortiz \\ \hline
        Brooklyn Uprock & XBot & high & BigVegas \\ \hline
        Butterfly Twirl & XBot & low & SportyGranny \\ \hline
        Chapeau de Couro & YBot & high & Michelle \\ \hline
        Charge & YBot & low & Mousey \\ \hline
        Combo Punch & YBot & low & TheBoss \\ \hline
        Crazy Gesture & XBot & low & Ortiz \\ \hline
        Cross Jumps & XBot & low & SportyGranny \\ \hline
        Cross Jumps Rotation & YBot & low & Mousey \\ \hline
        Crying & XBot & high & Ortiz \\ \hline
        Dive roll & YBot & high & Michelle \\ \hline
        Ducking & YBot & high & TheBoss \\ \hline
        Evading A Threat & YBot & low & BigVegas \\ \hline
        Football Stance & YBot & low & Ortiz \\ \hline
        Forward Jump & YBot & low & Sophie \\ \hline
        Gangnam Style & YBot & low & BigVegas \\ \hline
        Golf Pre-Putt & XBot & high & Mousey \\ \hline
        Golf Putt & YBot & low & Kaya \\ \hline
        Golf Tee Up & YBot & low & SportyGranny \\ \hline
        Hit On Side Of The Head & YBot & low & Sophie \\ \hline
        Insult & XBot & low & SportyGranny \\ \hline
        Kneeling & YBot & high & TheBoss \\ \hline
        Knocked Out & XBot & high & Sophie \\ \hline
        Looking & YBot & high & Mousey \\ \hline
        Low Crawl & XBot & high & Michelle \\ \hline
        PikeWalk & YBot & high & Ortiz \\ \hline
        Plotting & XBot & low & Kaya \\ \hline
        Push Up & XBot & high & Sophie \\ \hline
        Shooting & YBot & low & Michelle \\ \hline
        Shoulder Rubbing & XBot & high & Mousey \\ \hline
        Silly Dancing & YBot & high & TheBoss \\ \hline
        Situps & XBot & high & Michelle \\ \hline
        Standing Arguing & XBot & low & Sophie \\ \hline
        Standing Up & YBot & high & Kaya \\ \hline
        Stumble Backwards & YBot & high & BigVegas \\ \hline
        Stunned & YBot & high & SportyGranny \\ \hline
        Swagger Walk & XBot & low & Ortiz \\ \hline
        Talking on a Cell Phone & YBot & high & Kaya \\ \hline
        Thankful & XBot & low & SportyGranny \\ \hline
        Threatening & XBot & high & TheBoss \\ \hline
        Tripping & YBot & high & Michelle \\ \hline
        Waving & XBot & low & Kaya \\ \hline
    \end{tabular}
    \caption{Composition of our dataset and user study}
    \label{table:dataset}
    
\vspace{2\baselineskip}
\end{table*}




\clearpage

\section{``Copy Rotations" Algorithm}

We noticed that the ``copy rotations" baseline, despite being commonly used in several works, does not appear to have a formal definition. Intuitively, it can be understood as a way to copy the local rotation of the bones form the source to the target. However, in practice, the formulation is more complex, as it needs to take into account the bones' initial rotation in a common pose (in our case, a T-pose).

As such, for reproducibility and completeness purposes, we propose the following formalization of the ``copy rotations" algorithm, and we explain how it is implemented in ReConForM.

\noindent\textbf{Definition:}
A \textbf{skeleton} is a tuple $\mathcal{S} = \left( \mathcal{M}, \mathcal{H}\right)$, where:
\begin{itemize}
    \item $\mathcal{M} \in \mathbb{R}^{N\times 4 \times 4}$ is the world-space TRS (translation, rotation, scale) matrix of each of the $N$ bones, in a special pose (in our case, a T-pose). In our case, the scales are unitary, as we are only interested in rotations and positions.\\
    Thus, the rotation matrices have the following structure: \\
    $\mathcal{M}^i = 
    \left(\begin{array}{@{}c|c@{}}
        \begin{matrix}& & \\&\mathcal{R}^i & \\ & &\end{matrix}& \begin{matrix}\ \\ \mathcal{P}^i\\ \ \end{matrix}\\\hline
        \begin{matrix} 0 & 0 & 0 \end{matrix}& 1
    \end{array}\right)$
    \item $\mathcal{H} : \llbracket 1, N-1\rrbracket \rightarrow \llbracket 0, N-1\rrbracket$ is the ``hierarchy" function, taking as input an index of a bone, and returning the index of its parent bone. The first bone is said to be the ``root" bone, and has no parent ; it is the only one whose position is animated.
\end{itemize}

\noindent\textbf{Definition:}
A \textbf{pose} is a tuple $\left(p, \mathbf{q}\right)$, where:
\begin{itemize}
    \item $p \in \mathbb{R}^{3}$ is the world-space position of the root bone.
    \item $\mathbf{q} \in \mathbb{R}^{N\time 3\times 3}$ is the rotation of each bone, in local space (i.e., relative to its parent's coordinate frame)
\end{itemize}

\noindent\textbf{Definition:}
A \textbf{mapping} from a source skeleton $\mathcal{S}_S$ to a target skeleton $\mathcal{S}_T$, denoted as $\mathcal{M}_{S\to T}$, is a set of several pairs $\left(b_S, b_T\right)$, where:
\begin{itemize}
    \item $b_S \in \llbracket 0, N_S \rrbracket$ is the index of a bone in the source skeleton, which has $N_S$ bones
    \item $b_T \in \llbracket 0, N_T \rrbracket$ is the index of the corresponding bone in the target skeleton, which has $N_T$ bones
\end{itemize}

\subsection{Simple formulation}
\label{section:retargeting}
Given a source skeleton $\mathcal{S}_S := (\mathcal{M}_S, \mathcal{H}_S)$ and a target skeleton $\mathcal{S}_T := (\mathcal{M}_T, \mathcal{H}_T)$, our goal is to transfer a pose $(p_S, \mathbf{q_S})$ from the source to the target.

\subsubsection{Pelvis position}
The height (in T-pose) of the root bone of the source is noted $\mathcal{M}_S^{0}[3, 4] := \begin{smallmatrix}(0& 0& 0& 1)\end{smallmatrix} \cdot \mathcal{M}_S^{0} \cdot \begin{smallmatrix}(0& 0& 1& 0)\end{smallmatrix}^T$, as our coordinate system has the Z coordinate pointing upwards.

As such, we use the ratio between the source and target's heights to define $p_T = \displaystyle\frac{\mathcal{M}_T^{0}[3, 4]}{\mathcal{M}_S^{0}[3, 4]}p_S$

\subsubsection{Bone rotations}

To transfer the bone rotations, it is necessary to proceed in a specific older, so that every bone is processed after its parents, and before any of its children. This is because rotating a bone will update the world matrices of all of its successors.
In order to do this, a topological sort of the target skeleton is necessary.

During the execution of our retargeting process, the rotations of the bones are updated, starting from the root bone and moving in topological order. Let us denote by $\mathcal{M}_T^\text{curr} \in \mathbb{R}^{N_T \times 4 \times 4}$ the current world matrices of the target skeleton, during the execution.
Also, let us denote by $\mathcal{M}_S^\text{curr} \in \mathbb{R}^{N_S \times 4 \times 4}$ the world matrices of the source skeleton when the local pose $(p_S, \mathbf{q}_S)$ is applied.

When considering the next pair of bones to retarget, $(b_S, b_T) \in \mathcal{M}$, let us define $b_S' = \mathcal{H}_S(b_S)$ and $b_T' := \mathcal{H}_T(b_T)$, their respective parents. We have the following:
\begin{itemize}
    \item $q_S^{b_S}$ is the rotation we want to transfer, expressed in local space (i.e., with respect to the parent of $b_S$).
    \item $\left(\mathcal{R}_S^{b_S'}\right)^{-1} \mathcal{R}_S^{b_S}$ is the rotation of $b_S$ in T-pose, expressed in local space (with respect to its parent).
\end{itemize}
Knowing that, the aim is to transfer the relative rotation with respect to the T-pose, which is: \\
$L_S^{b_S} := \left(\left(\mathcal{R}_S^{b_S'}\right)^{-1} \mathcal{R}_S^{b_S}\right)^{-1}q_S^{b_S}$\\
or, more simply, $L_S^{b_S} := \left(\mathcal{R}_S^{b_S}\right)^{-1} \mathcal{R}_S^{b_S'} q_S^{b_S}$

To transfer this rotation to the target, we must first express it in world space: \\
$W^{b_S} := \mathcal{R}_S^{\text{curr}, b_S'} L_S^{b_S} \left(\mathcal{R}_S^{\text{curr}, b_S'}\right)^{-1}$

This gives us the local rotation to apply to the target bone:\\
$L_T^{b_T} := \left(\mathcal{R}_T^{\text{curr}, b_T'}\right)^{-1} W^{b_S} \mathcal{R}_T^{\text{curr}, b_T'}$.

``Applying" this rotation to the bone in its current state means that, \textit{in fine}, $\mathbf{q}_T^{b_T} = \mathcal{R}_T^{b_T} L_T^{b_T}$.

\subsection{Complex cases}
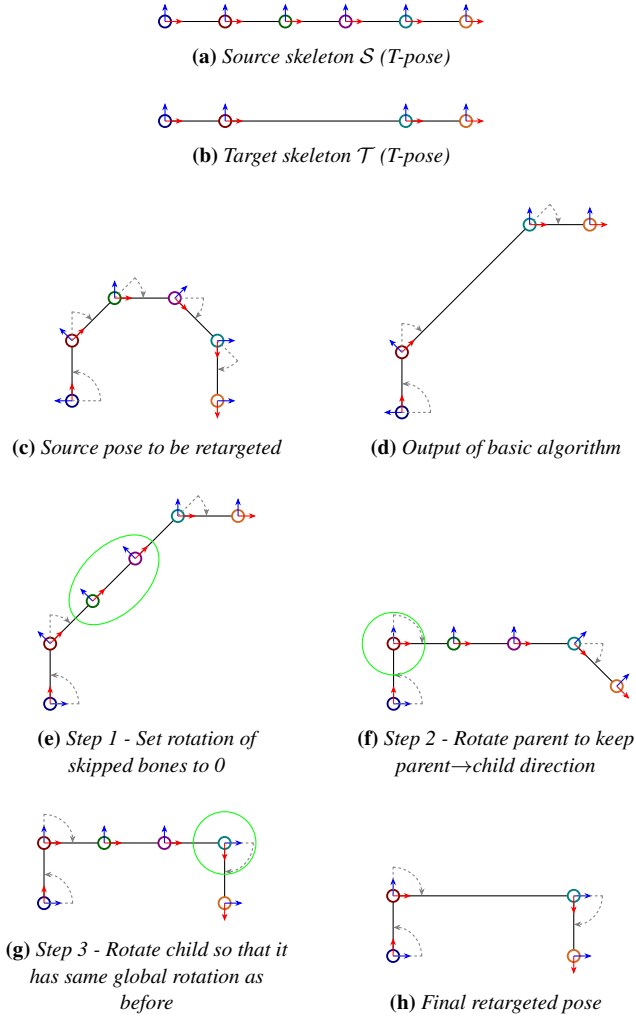
\begin{figure}[ht]
    \centering
    \begin{subfigure}[b]{0.65\columnwidth}
        \centering
        \begin{tikzpicture}[scale=0.8]
            \foreach \x in {0,1,2,3,4}
              \draw[black] (\x,0) -- (\x+1,0);
            
            \draw[line width=0.25mm, draw=lightblue, fill=white] (0,0) circle (0.1);
            \draw[line width=0.25mm, draw=darkyellow, fill=white] (1,0) circle (0.1);
            \draw[line width=0.25mm, draw=teal, fill=white] (2,0) circle (0.1);
            \draw[line width=0.25mm, draw=darkred, fill=white] (3,0) circle (0.1);
            \draw[line width=0.25mm, draw=darkpink, fill=white] (4,0) circle (0.1);
            \draw[line width=0.25mm, draw=olive, fill=white] (5,0) circle (0.1);

            \foreach \x in {0,1,2,3,4,5}
              \draw[-{Stealth[length=1mm]}, red] (\x,0) -- (\x+0.3,0);
            
            \foreach \x in {0,1,2,3,4,5}
              \draw[-{Stealth[length=1mm]}, blue] (\x,0) -- (\x,0.3);
        \end{tikzpicture}
        \caption{Source skeleton $\mathcal S$ (T-pose)}
        \label{fig:fewer_bones1}
    \end{subfigure}
    
    \vskip\baselineskip

    \begin{subfigure}[b]{0.65\columnwidth}
        \centering
        \begin{tikzpicture}[scale=0.8]
            \draw[black] (0,0) -- (1,0);
            \draw[black] (1,0) -- (4,0);
            \draw[black] (4,0) -- (5,0);
            
            \draw[line width=0.25mm, draw=lightblue, fill=white] (0,0) circle (0.1);
            \draw[line width=0.25mm, draw=darkyellow, fill=white] (1,0) circle (0.1);
            \draw[line width=0.25mm, draw=darkpink, fill=white] (4,0) circle (0.1);
            \draw[line width=0.25mm, draw=olive, fill=white] (5,0) circle (0.1);

            \foreach \x in {0,1,4,5}
              \draw[-{Stealth[length=1mm]}, red] (\x,0) -- (\x+0.3,0);
            
            \foreach \x in {0,1,4,5}
              \draw[-{Stealth[length=1mm]}, blue] (\x,0) -- (\x,0.3);
        \end{tikzpicture}
        \caption{Target skeleton $\mathcal T$ (T-pose)}
        \label{fig:fewer_bones2}
    \end{subfigure}
    
    \vskip\baselineskip
    
    \begin{subfigure}[b]{0.45\columnwidth}
        \centering
        \begin{tikzpicture}[scale=0.8]

            \def\sq{0.70710678} 
            
            \draw[black] (0, 0) -- (0, 1);
            \draw[black] (0, 1) -- (\sq, 1+\sq);
            \draw[black] (\sq, 1+\sq) -- (1+\sq, 1+\sq);
            \draw[black] (1+\sq, 1+\sq) -- (1+2*\sq, 1);
            \draw[black] (1+2*\sq, 1) -- (1+2*\sq, 0);
            
            \draw[line width=0.25mm, draw=lightblue, fill=white] (0,0) circle (0.1);
            \draw[line width=0.25mm, draw=darkyellow, fill=white] (0, 1) circle (0.1);
            \draw[line width=0.25mm, draw=teal, fill=white] (\sq, 1+\sq) circle (0.1);
            \draw[line width=0.25mm, draw=darkred, fill=white] (1+\sq, 1+\sq) circle (0.1);
            \draw[line width=0.25mm, draw=darkpink, fill=white] (1+2*\sq, 1) circle (0.1);
            \draw[line width=0.25mm, draw=olive, fill=white] (1+2*\sq, 0) circle (0.1);
            
            \draw[-{Stealth[length=1mm]}, red] (0, 0) -- (0, 0.3);
            \draw[-{Stealth[length=1mm]}, red] (0, 1) -- (0.3*\sq, 1+0.3*\sq);
            \draw[-{Stealth[length=1mm]}, red] (\sq, 1+\sq) -- (\sq+0.3, 1+\sq);
            \draw[-{Stealth[length=1mm]}, red] (1+\sq, 1+\sq) -- (1+\sq+0.3*\sq, 1+\sq-0.3*\sq);
            \draw[-{Stealth[length=1mm]}, red] (1+2*\sq, 1) -- (1+2*\sq, 1-0.3);
            \draw[-{Stealth[length=1mm]}, red] (1+2*\sq, 0) -- (1+2*\sq, -0.3);
            
            \draw[-{Stealth[length=1mm]}, blue] (0, 0) -- (-0.3, 0);
            \draw[-{Stealth[length=1mm]}, blue] (0, 1) -- (-0.3*\sq, 1+0.3*\sq);
            \draw[-{Stealth[length=1mm]}, blue] (\sq, 1+\sq) -- (\sq, 1+\sq+0.3);
            \draw[-{Stealth[length=1mm]}, blue] (1+\sq, 1+\sq) -- (1+\sq+0.3*\sq, 1+\sq+0.3*\sq);
            \draw[-{Stealth[length=1mm]}, blue] (1+2*\sq, 1) -- (1+2*\sq+0.3, 1);
            \draw[-{Stealth[length=1mm]}, blue] (1+2*\sq, 0) -- (1+2*\sq+0.3, 0);

            \draw (0,0) pic[gray, dash pattern=on 1pt off 1pt, -{Stealth[length=1mm]}]{carc=0:90:.38};
            \draw (0,1) pic[gray, dash pattern=on 1pt off 1pt, -{Stealth[length=1mm]}]{carc=90:45:.38};
            \draw (\sq, 1+\sq) pic[gray, dash pattern=on 1pt off 1pt, -{Stealth[length=1mm]}]{carc=45:0:.38};
            \draw (1+\sq, 1+\sq) pic[gray, dash pattern=on 1pt off 1pt, -{Stealth[length=1mm]}]{carc=0:-45:.38};
            \draw (1+2*\sq, 1) pic[gray, dash pattern=on 1pt off 1pt, -{Stealth[length=1mm]}]{carc=-45:-90:.38};
            
            \draw[gray, dash pattern=on 1pt off 1pt] (0,0) -- (0.52, 0);
            \draw[gray, dash pattern=on 1pt off 1pt] (0,1) -- (0, 1+0.52);
            \draw[gray, dash pattern=on 1pt off 1pt] (\sq, 1+\sq) -- (\sq + 0.52*\sq, 1+\sq+0.52*\sq);
            \draw[gray, dash pattern=on 1pt off 1pt] (1+\sq, 1+\sq) -- (1+\sq+0.52, 1+\sq);
            \draw[gray, dash pattern=on 1pt off 1pt] (1+2*\sq, 1) -- (1+2*\sq+0.52*\sq, 1-0.52*\sq);
            
        \end{tikzpicture}
        \caption{Source pose to be retargeted}
        \label{fig:fewer_bones3}
    \end{subfigure}
    \hfill
    \begin{subfigure}[b]{0.45\columnwidth}
        \centering
        \begin{tikzpicture}[scale=0.8]
            \def\sq{0.70710678} 
            
            \draw[black] (0, 0) -- (0, 1);
            \draw[black] (0, 1) -- (3*\sq, 1+3*\sq);
            \draw[black] (3*\sq, 1+3*\sq) -- (1+3*\sq, 1+3*\sq);
            
            \draw[line width=0.25mm, draw=lightblue, fill=white] (0,0) circle (0.1);
            \draw[line width=0.25mm, draw=darkyellow, fill=white] (0, 1) circle (0.1);
            \draw[line width=0.25mm, draw=darkpink, fill=white] (3*\sq, 1+3*\sq) circle (0.1);
            \draw[line width=0.25mm, draw=olive, fill=white] (1+3*\sq, 1+3*\sq) circle (0.1);
            
            \draw[-{Stealth[length=1mm]}, red] (0, 0) -- (0, 0.3);
            \draw[-{Stealth[length=1mm]}, red] (0, 1) -- (0.3*\sq, 1+0.3*\sq);
            \draw[-{Stealth[length=1mm]}, red] (3*\sq, 1+3*\sq) -- (3*\sq+.3, 1+3*\sq);
            \draw[-{Stealth[length=1mm]}, red] (1+3*\sq, 1+3*\sq) -- (1+3*\sq+.3, 1+3*\sq);
            
            \draw[-{Stealth[length=1mm]}, blue] (0, 0) -- (-0.3, 0);
            \draw[-{Stealth[length=1mm]}, blue] (0, 1) -- (-0.3*\sq, 1+0.3*\sq);
            \draw[-{Stealth[length=1mm]}, blue] (3*\sq, 1+3*\sq) -- (3*\sq, 1+3*\sq+.3);
            \draw[-{Stealth[length=1mm]}, blue] (1+3*\sq, 1+3*\sq) -- (1+3*\sq, 1+3*\sq+.3);

            \draw (0,0) pic[gray, dash pattern=on 1pt off 1pt, -{Stealth[length=1mm]}]{carc=0:90:.38};
            \draw (0,1) pic[gray, dash pattern=on 1pt off 1pt, -{Stealth[length=1mm]}]{carc=90:45:.38};
            \draw (3*\sq, 1+3*\sq) pic[gray, dash pattern=on 1pt off 1pt, -{Stealth[length=1mm]}]{carc=45:0:.38};

            \draw[gray, dash pattern=on 1pt off 1pt] (0,0) -- (0.52, 0);
            \draw[gray, dash pattern=on 1pt off 1pt] (0,1) -- (0, 1+0.52);
            \draw[gray, dash pattern=on 1pt off 1pt] (3*\sq, 1+3*\sq) -- (3*\sq + 0.52*\sq, 1+3*\sq+0.52*\sq);
            
        \end{tikzpicture}
        \caption{Output of basic algorithm}
        \label{fig:fewer_bones4}
    \end{subfigure}

    \vskip\baselineskip
    
    \begin{subfigure}[b]{0.45\columnwidth}
        \centering
        \begin{tikzpicture}[scale=0.8]

            \def\sq{0.70710678} 
            
            \draw[black] (0, 0) -- (0, 1);
            \draw[black] (0, 1) -- (\sq, 1+\sq);
            \draw[black] (\sq, 1+\sq) -- (2*\sq, 1+2*\sq);
            \draw[black] (2*\sq, 1+2*\sq) -- (3*\sq, 1+3*\sq);
            \draw[black] (3*\sq, 1+3*\sq) -- (1+3*\sq, 1+3*\sq);
            
            \draw[line width=0.25mm, draw=lightblue, fill=white] (0,0) circle (0.1);
            \draw[line width=0.25mm, draw=darkyellow, fill=white] (0, 1) circle (0.1);
            \draw[line width=0.25mm, draw=teal, fill=white] (\sq, 1+\sq) circle (0.1);
            \draw[line width=0.25mm, draw=darkred, fill=white] (2*\sq, 1+2*\sq) circle (0.1);
            \draw[line width=0.25mm, draw=darkpink, fill=white] (3*\sq, 1+3*\sq) circle (0.1);
            \draw[line width=0.25mm, draw=olive, fill=white] (1+3*\sq, 1+3*\sq) circle (0.1);
            
            \draw[-{Stealth[length=1mm]}, red] (0, 0) -- (0, 0.3);
            \draw[-{Stealth[length=1mm]}, red] (0, 1) -- (0.3*\sq, 1+0.3*\sq);
            \draw[-{Stealth[length=1mm]}, red] (\sq, 1+\sq) -- (\sq+0.3*\sq, 1+\sq+0.3*\sq);
            \draw[-{Stealth[length=1mm]}, red] (2*\sq, 1+2*\sq) -- (2*\sq+0.3*\sq, 1+2*\sq+0.3*\sq);
            \draw[-{Stealth[length=1mm]}, red] (3*\sq, 1+3*\sq) -- (3*\sq+0.3, 1+3*\sq);
            \draw[-{Stealth[length=1mm]}, red] (1+3*\sq, 1+3*\sq) -- (1+3*\sq+0.3, 1+3*\sq);
            
            \draw[-{Stealth[length=1mm]}, blue] (0, 0) -- (0.3, 0);
            \draw[-{Stealth[length=1mm]}, blue] (0, 1) -- (-0.3*\sq, 1+0.3*\sq);
            \draw[-{Stealth[length=1mm]}, blue] (\sq, 1+\sq) -- (\sq-0.3*\sq, 1+\sq+0.3*\sq);
            \draw[-{Stealth[length=1mm]}, blue] (2*\sq, 1+2*\sq) -- (2*\sq-0.3*\sq, 1+2*\sq+0.3*\sq);
            \draw[-{Stealth[length=1mm]}, blue] (3*\sq, 1+3*\sq) -- (3*\sq, 1+3*\sq+0.3);
            \draw[-{Stealth[length=1mm]}, blue] (1+3*\sq, 1+3*\sq) -- (1+3*\sq, 1+3*\sq+0.3);
            
            \draw (0,0) pic[gray, dash pattern=on 1pt off 1pt, -{Stealth[length=1mm]}]{carc=0:90:.38};
            \draw (0,1) pic[gray, dash pattern=on 1pt off 1pt, -{Stealth[length=1mm]}]{carc=90:45:.38};
            \draw (3*\sq, 1+3*\sq) pic[gray, dash pattern=on 1pt off 1pt, -{Stealth[length=1mm]}]{carc=45:0:.38};
            
            \draw[gray, dash pattern=on 1pt off 1pt] (0,0) -- (0.52, 0);
            \draw[gray, dash pattern=on 1pt off 1pt] (0,1) -- (0, 1+0.52);
            \draw[gray, dash pattern=on 1pt off 1pt] (3*\sq, 1+3*\sq) -- (3*\sq+0.52*\sq, 1+3*\sq+0.52*\sq);

            \draw[rotate around={45:(1.5*\sq,1+1.5*\sq)}, green] 
            (1.5*\sq,1+1.5*\sq) ellipse [x radius=0.9, y radius=0.55];
        
        \end{tikzpicture}
        \caption{Step 1 - Set rotation of skipped bones to 0}
        \label{fig:fewer_bones5}
    \end{subfigure}
    \hfill
    \begin{subfigure}[b]{0.45\columnwidth}
        \centering
        \begin{tikzpicture}[scale=0.8]

            \def\sq{0.70710678} 
            
            \draw[black] (0, 0) -- (0, 1);
            \draw[black] (0, 1) -- (1, 1);
            \draw[black] (1, 1) -- (2, 1);
            \draw[black] (2, 1) -- (3, 1);
            \draw[black] (3, 1) -- (3+\sq, 1-\sq);
            
            \draw[line width=0.25mm, draw=lightblue, fill=white] (0,0) circle (0.1);
            \draw[line width=0.25mm, draw=darkyellow, fill=white] (0, 1) circle (0.1);
            \draw[line width=0.25mm, draw=teal, fill=white] (1, 1) circle (0.1);
            \draw[line width=0.25mm, draw=darkred, fill=white] (2, 1) circle (0.1);
            \draw[line width=0.25mm, draw=darkpink, fill=white] (3, 1) circle (0.1);
            \draw[line width=0.25mm, draw=olive, fill=white] (3+\sq, 1-\sq) circle (0.1);
            
            \draw[-{Stealth[length=1mm]}, red] (0, 0) -- (0, 0.3);
            \draw[-{Stealth[length=1mm]}, red] (0, 1) -- (0.3, 1);
            \draw[-{Stealth[length=1mm]}, red] (1, 1) -- (1+0.3, 1);
            \draw[-{Stealth[length=1mm]}, red] (2, 1) -- (2+0.3, 1);
            \draw[-{Stealth[length=1mm]}, red] (3, 1) -- (3+\sq*0.3, 1-\sq*0.3);
            \draw[-{Stealth[length=1mm]}, red] (3+\sq, 1-\sq) -- (3+\sq+0.3*\sq, 1-\sq-0.3*\sq);

            \draw[-{Stealth[length=1mm]}, blue] (0, 0) -- (0.3, 0);
            \draw[-{Stealth[length=1mm]}, blue] (0, 1) -- (0, 1.3);
            \draw[-{Stealth[length=1mm]}, blue] (1, 1) -- (1, 1+0.3);
            \draw[-{Stealth[length=1mm]}, blue] (2, 1) -- (2, 1+0.3);
            \draw[-{Stealth[length=1mm]}, blue] (3, 1) -- (3+\sq*0.3, 1+\sq*0.3);
            \draw[-{Stealth[length=1mm]}, blue] (3+\sq, 1-\sq) -- (3+\sq+0.3*\sq, 1-\sq+0.3*\sq);
            
            \draw (0,0) pic[gray, dash pattern=on 1pt off 1pt, -{Stealth[length=1mm]}]{carc=0:90:.38};
            \draw (0,1) pic[gray, dash pattern=on 1pt off 1pt, -{Stealth[length=1mm]}]{carc=90:0:.38};
            \draw (3, 1) pic[gray, dash pattern=on 1pt off 1pt, -{Stealth[length=1mm]}]{carc=0:-45:.38};
            
            \draw[gray, dash pattern=on 1pt off 1pt] (0,0) -- (0.52, 0);
            \draw[gray, dash pattern=on 1pt off 1pt] (0,1) -- (0, 1+0.52);
            \draw[gray, dash pattern=on 1pt off 1pt] (3, 1) -- (3+0.52, 1);

            \draw[green] 
            (0, 1) ellipse [x radius=0.52, y radius=0.52];
            
        \end{tikzpicture}
        \caption{Step 2 - Rotate parent to keep parent$\to$child direction}
        \label{fig:fewer_bones6}
    \end{subfigure}

    \vskip\baselineskip
    
    \begin{subfigure}[b]{0.45\columnwidth}
        \centering
        \begin{tikzpicture}[scale=0.8]

            \def\sq{0.70710678} 
            
            \draw[black] (0, 0) -- (0, 1);
            \draw[black] (0, 1) -- (1, 1);
            \draw[black] (1, 1) -- (2, 1);
            \draw[black] (2, 1) -- (3, 1);
            \draw[black] (3, 1) -- (3, 0);
            
            \draw[line width=0.25mm, draw=lightblue, fill=white] (0,0) circle (0.1);
            \draw[line width=0.25mm, draw=darkyellow, fill=white] (0, 1) circle (0.1);
            \draw[line width=0.25mm, draw=teal, fill=white] (1, 1) circle (0.1);
            \draw[line width=0.25mm, draw=darkred, fill=white] (2, 1) circle (0.1);
            \draw[line width=0.25mm, draw=darkpink, fill=white] (3, 1) circle (0.1);
            \draw[line width=0.25mm, draw=olive, fill=white] (3, 0) circle (0.1);
            
            \draw[-{Stealth[length=1mm]}, red] (0, 0) -- (0, 0.3);
            \draw[-{Stealth[length=1mm]}, red] (0, 1) -- (0.3, 1);
            \draw[-{Stealth[length=1mm]}, red] (1, 1) -- (1+0.3, 1);
            \draw[-{Stealth[length=1mm]}, red] (2, 1) -- (2+0.3, 1);
            \draw[-{Stealth[length=1mm]}, red] (3, 1) -- (3, 1-0.3);
            \draw[-{Stealth[length=1mm]}, red] (3, 0) -- (3, -0.3);

            \draw[-{Stealth[length=1mm]}, blue] (0, 0) -- (0.3, 0);
            \draw[-{Stealth[length=1mm]}, blue] (0, 1) -- (0, 1.3);
            \draw[-{Stealth[length=1mm]}, blue] (1, 1) -- (1, 1+0.3);
            \draw[-{Stealth[length=1mm]}, blue] (2, 1) -- (2, 1+0.3);
            \draw[-{Stealth[length=1mm]}, blue] (3, 1) -- (3+0.3, 1);
            \draw[-{Stealth[length=1mm]}, blue] (3, 0) -- (3+0.3, 0);
            
            \draw (0,0) pic[gray, dash pattern=on 1pt off 1pt, -{Stealth[length=1mm]}]{carc=0:90:.38};
            \draw (0,1) pic[gray, dash pattern=on 1pt off 1pt, -{Stealth[length=1mm]}]{carc=90:0:.38};
            \draw (3, 1) pic[gray, dash pattern=on 1pt off 1pt, -{Stealth[length=1mm]}]{carc=0:-90:.38};
            
            \draw[gray, dash pattern=on 1pt off 1pt] (0,0) -- (0.52, 0);
            \draw[gray, dash pattern=on 1pt off 1pt] (0,1) -- (0, 1+0.52);
            \draw[gray, dash pattern=on 1pt off 1pt] (3, 1) -- (3+0.52, 1);
            
            \draw[green] 
            (3, 1) ellipse [x radius=0.52, y radius=0.52];
            
        \end{tikzpicture}
        \caption{Step 3 - Rotate child so that it has same global rotation as before}
        \label{fig:fewer_bones7}
    \end{subfigure}
    \hfill
    \begin{subfigure}[b]{0.45\columnwidth}
        \centering
        \begin{tikzpicture}[scale=0.8]

            \def\sq{0.70710678} 
            
            \draw[black] (0, 0) -- (0, 1);
            \draw[black] (0, 1) -- (3, 1);
            \draw[black] (3, 1) -- (3, 0);
            
            \draw[line width=0.25mm, draw=lightblue, fill=white] (0,0) circle (0.1);
            \draw[line width=0.25mm, draw=darkyellow, fill=white] (0, 1) circle (0.1);
            \draw[line width=0.25mm, draw=darkpink, fill=white] (3, 1) circle (0.1);
            \draw[line width=0.25mm, draw=olive, fill=white] (3, 0) circle (0.1);
            
            \draw[-{Stealth[length=1mm]}, red] (0, 0) -- (0, 0.3);
            \draw[-{Stealth[length=1mm]}, red] (0, 1) -- (0.3, 1);
            \draw[-{Stealth[length=1mm]}, red] (3, 1) -- (3, 1-0.3);
            \draw[-{Stealth[length=1mm]}, red] (3, 0) -- (3, -0.3);

            \draw[-{Stealth[length=1mm]}, blue] (0, 0) -- (0.3, 0);
            \draw[-{Stealth[length=1mm]}, blue] (0, 1) -- (0, 1.3);
            \draw[-{Stealth[length=1mm]}, blue] (3, 1) -- (3+0.3, 1);
            \draw[-{Stealth[length=1mm]}, blue] (3, 0) -- (3+0.3, 0);
            
            \draw (0,0) pic[gray, dash pattern=on 1pt off 1pt, -{Stealth[length=1mm]}]{carc=0:90:.38};
            \draw (0,1) pic[gray, dash pattern=on 1pt off 1pt, -{Stealth[length=1mm]}]{carc=90:0:.38};
            \draw (3, 1) pic[gray, dash pattern=on 1pt off 1pt, -{Stealth[length=1mm]}]{carc=0:-90:.38};
            
            \draw[gray, dash pattern=on 1pt off 1pt] (0,0) -- (0.52, 0);
            \draw[gray, dash pattern=on 1pt off 1pt] (0,1) -- (0, 1+0.52);
            \draw[gray, dash pattern=on 1pt off 1pt] (3, 1) -- (3+0.52, 1);
            
        \end{tikzpicture}
        \caption{Final retargeted pose}
        \label{fig:fewer_bones8}
    \end{subfigure}
    
    \caption{Explanation of ``Copy Rotations'' algorithm when the target has fewer bones than the source}
    \label{fig:fewer_bones}
\end{figure}

The previous formulation needs to be altered when both $b_S$ and $b_T$ have a single child (respectively $\overline{b_S}$ and $\overline{b_T}$ and that $ (\overline{b_S}, \overline{b_T}) \notin \mathcal{M}$. We can distinguish two specific cases:
\begin{itemize}
    \item The case where several source bones correspond to a single target bones, as illustrated in Figure~\ref{fig:fewer_bones}
    \item The case where a single source bone corresponds to multiple target bones, as illustrated in Figure~\ref{fig:more_bones}
\end{itemize}

\subsubsection{Missing bones on the target}
In the first case, transferring only the rotation of bones in the mapping means that some rotations are lost, which has a negative impact on the target pose, especially if the amount of rotation is important.

This is shown in Figure~\ref{fig:fewer_bones}. Let's suppose that the source skeleton  corresponds to the one on Figure~\ref{fig:fewer_bones1}, shown in its T-pose, and that the target skeleton corresponds to the one on Figure~\ref{fig:fewer_bones2}, with identical joint color defining the mapping between the source and target. Applying the algorithm mentioned in Section~\ref{section:retargeting} to the pose shown on Figure~\ref{fig:fewer_bones3} yields the result on Figure~\ref{fig:fewer_bones4}: we can see that the pose does not look similar at all.

To fix this problem, we modify the source pose so that all non-mapped bones keep their T-pose rotation (Figure~\ref{fig:fewer_bones5}). Then, we rotate the previous bone $B_1$ (which appears in the mapping) to align the child of the last non-mapped bone $B_2$ (which appears in the mapping), so that the vector $\overrightarrow{B_1B_2}$ keeps the same direction as before (albeit with a different norm), as shown on Figure~\ref{fig:fewer_bones6}. Finally, we rotate $B_2$ so that it has the same global orientation as it had at first, in order to keep the rest of the kinematic chain unchanged, as shown on Figure~\ref{fig:fewer_bones7}.

After modifying the source pose this way, the retargeting algorithm described in Section~\ref{section:retargeting} yields the output shown in Figure~\ref{fig:fewer_bones8}, which is a much better output than that of Figure~\ref{fig:fewer_bones4}.

\subsubsection{Additional bones on the target}
In the second case, if we only transfer the rotation of the bones which appear in $\mathcal{M}$, we lose the benefit of the additional bones present on the target.

\begin{figure}[ht]
    \centering
        \begin{subfigure}[b]{0.65\columnwidth}
        \centering
        \begin{tikzpicture}[scale=0.8]
            \draw[black] (0,0) -- (2,0);
            \draw[black] (2,0) -- (4,0);
            \draw[black] (4,0) -- (6,0);
            
            \draw[line width=0.25mm, draw=lightblue, fill=white] (0,0) circle (0.1);
            \draw[line width=0.25mm, draw=teal, fill=white] (2,0) circle (0.1);
            \draw[line width=0.25mm, draw=darkpink, fill=white] (4,0) circle (0.1);
            \draw[line width=0.25mm, draw=last, fill=white] (6,0) circle (0.1);

            \foreach \x in {0,2,4,6}
              \draw[-{Stealth[length=1mm]}, red] (\x,0) -- (\x+0.3,0);
            
            \foreach \x in {0,2,4,6}
              \draw[-{Stealth[length=1mm]}, blue] (\x,0) -- (\x,0.3);
        \end{tikzpicture}
        \caption{Source skeleton $\mathcal S$ (T-pose)}
        \label{fig:more_bones1}
    \end{subfigure}

    \vskip\baselineskip
    
    \begin{subfigure}[b]{0.65\columnwidth}
        \centering
        \begin{tikzpicture}[scale=0.8]
            \foreach \x in {0,1,2,3,4,5}
              \draw[black] (\x,0) -- (\x+1,0);
            
            \draw[line width=0.25mm, draw=lightblue, fill=white] (0,0) circle (0.1);
            \draw[line width=0.25mm, draw=darkyellow, fill=white] (1,0) circle (0.1);
            \draw[line width=0.25mm, draw=teal, fill=white] (2,0) circle (0.1);
            \draw[line width=0.25mm, draw=darkred, fill=white] (3,0) circle (0.1);
            \draw[line width=0.25mm, draw=darkpink, fill=white] (4,0) circle (0.1);
            \draw[line width=0.25mm, draw=olive, fill=white] (5,0) circle (0.1);
            \draw[line width=0.25mm, draw=last, fill=white] (6,0) circle (0.1);

            \foreach \x in {0,1,2,3,4,5,6}
              \draw[-{Stealth[length=1mm]}, red] (\x,0) -- (\x+0.3,0);
            
            \foreach \x in {0,1,2,3,4,5,6}
              \draw[-{Stealth[length=1mm]}, blue] (\x,0) -- (\x,0.3);
        \end{tikzpicture}
        \caption{Target skeleton $\mathcal T$ (T-pose)}
        \label{fig:more_bones2}
    \end{subfigure}
    
    \vskip\baselineskip
    
    \begin{subfigure}[b]{0.45\columnwidth}
        \centering
        \begin{tikzpicture}[scale=0.8]
            \def\sq{0.70710678} 
            \def\sqq{0.86602540} 
            
            \draw[black] (0, 0) -- (2*\sqq, 1);
            \draw[black] (2*\sqq, 1) -- (2*\sqq+1, 2*\sqq+1);
            \draw[black] (2*\sqq+1, 2*\sqq+1) -- (2*\sqq+3, 2*\sqq+1);
            
            \draw[line width=0.25mm, draw=lightblue, fill=white] (0,0) circle (0.1);
            \draw[line width=0.25mm, draw=teal, fill=white] (2*\sqq, 1) circle (0.1);
            \draw[line width=0.25mm, draw=darkpink, fill=white] (2*\sqq+1, 2*\sqq+1) circle (0.1);
            \draw[line width=0.25mm, draw=last, fill=white] (2*\sqq+3, 2*\sqq+1) circle (0.1);
            
            \draw[-{Stealth[length=1mm]}, red] (0, 0) -- (0.3*\sqq, 0.3*.5);
            \draw[-{Stealth[length=1mm]}, red] (2*\sqq, 1) -- (2*\sqq+0.3*.5, 1+0.3*\sqq);
            \draw[-{Stealth[length=1mm]}, red] (2*\sqq+1, 2*\sqq+1) -- (2*\sqq+1+0.3, 2*\sqq+1);
            \draw[-{Stealth[length=1mm]}, red] (2*\sqq+3, 2*\sqq+1) -- (2*\sqq+3+.3, 2*\sqq+1);

            \draw[-{Stealth[length=1mm]}, blue] (0, 0) -- (-0.3*.5, 0.3*\sqq);
            \draw[-{Stealth[length=1mm]}, blue] (2*\sqq, 1) -- (2*\sqq-0.3*\sqq, 1+0.3*0.5);
            \draw[-{Stealth[length=1mm]}, blue] (2*\sqq+1, 2*\sqq+1) -- (2*\sqq+1, 2*\sqq+1+0.3);
            \draw[-{Stealth[length=1mm]}, blue] (2*\sqq+3, 2*\sqq+1) -- (2*\sqq+3, 2*\sqq+1+0.3);

            \draw (0,0) pic[gray, dash pattern=on 1pt off 1pt, -{Stealth[length=1mm]}]{carc=0:30:.38};
            \draw (2*\sqq, 1) pic[gray, dash pattern=on 1pt off 1pt, -{Stealth[length=1mm]}]{carc=30:60:.38};
            \draw (2*\sqq+1, 2*\sqq+1) pic[gray, dash pattern=on 1pt off 1pt, -{Stealth[length=1mm]}]{carc=60:0:.38};

            \draw[gray, dash pattern=on 1pt off 1pt] (0,0) -- (0.52, 0);
            \draw[gray, dash pattern=on 1pt off 1pt] (2*\sqq, 1) -- (2*\sqq+0.52*\sqq, 1+0.52*0.5);
            \draw[gray, dash pattern=on 1pt off 1pt] (2*\sqq+1, 2*\sqq+1) -- (2*\sqq+1 + 0.52*0.5, 2*\sqq+1 +0.52*\sqq);

        \end{tikzpicture}
        \caption{Source pose to be retargeted}
        \label{fig:more_bones3}
    \end{subfigure}
    \hfill
    \begin{subfigure}[b]{0.45\columnwidth}
        \centering
        \begin{tikzpicture}[scale=0.8]
            \def\sq{0.70710678} 
            \def\sqq{0.86602540} 
            
            \draw[black] (0, 0) -- (2*\sqq, 1);
            \draw[black] (2*\sqq, 1) -- (2*\sqq+1, 2*\sqq+1);
            \draw[black] (2*\sqq+1, 2*\sqq+1) -- (2*\sqq+3, 2*\sqq+1);
            
            \draw[line width=0.25mm, draw=lightblue, fill=white] (0,0) circle (0.1);
            \draw[line width=0.25mm, draw=darkyellow, fill=white] (\sqq, .5) circle (0.1);
            \draw[line width=0.25mm, draw=teal, fill=white] (2*\sqq, 1) circle (0.1);
            \draw[line width=0.25mm, draw=darkred, fill=white] (2*\sqq+.5, \sqq+1) circle (0.1);
            \draw[line width=0.25mm, draw=darkpink, fill=white] (2*\sqq+1, 2*\sqq+1) circle (0.1);
            \draw[line width=0.25mm, draw=olive, fill=white] (2*\sqq+2, 2*\sqq+1) circle (0.1);
            \draw[line width=0.25mm, draw=last, fill=white] (2*\sqq+3, 2*\sqq+1) circle (0.1);
            
            \draw[-{Stealth[length=1mm]}, red] (0, 0) -- (0.3*\sqq, 0.3*.5);
            \draw[-{Stealth[length=1mm]}, red] (\sqq, .5) -- (\sqq+0.3*\sqq, 0.5+0.3*.5);
            \draw[-{Stealth[length=1mm]}, red] (2*\sqq, 1) -- (2*\sqq+0.3*.5, 1+0.3*\sqq);
            \draw[-{Stealth[length=1mm]}, red] (2*\sqq+0.5, 1+\sqq) -- (2*\sqq+0.5+0.3*.5, \sqq+1+0.3*\sqq);
            \draw[-{Stealth[length=1mm]}, red] (2*\sqq+1, 2*\sqq+1) -- (2*\sqq+1+0.3, 2*\sqq+1);
            \draw[-{Stealth[length=1mm]}, red] (2*\sqq+2, 2*\sqq+1) -- (2*\sqq+2+0.3, 2*\sqq+1);
            \draw[-{Stealth[length=1mm]}, red] (2*\sqq+3, 2*\sqq+1) -- (2*\sqq+3+.3, 2*\sqq+1);

            \draw[-{Stealth[length=1mm]}, blue] (0, 0) -- (-0.3*.5, 0.3*\sqq);
            \draw[-{Stealth[length=1mm]}, blue] (\sqq, .5) -- (\sqq-0.3*.5, .5+0.3*\sqq);
            \draw[-{Stealth[length=1mm]}, blue] (2*\sqq, 1) -- (2*\sqq-0.3*\sqq, 1+0.3*0.5);
            \draw[-{Stealth[length=1mm]}, blue] (2*\sqq+.5, 1+\sqq) -- (2*\sqq+.5-0.3*\sqq, \sqq+1+0.3*0.5);
            \draw[-{Stealth[length=1mm]}, blue] (2*\sqq+1, 2*\sqq+1) -- (2*\sqq+1, 2*\sqq+1+0.3);
            \draw[-{Stealth[length=1mm]}, blue] (2*\sqq+2, 2*\sqq+1) -- (2*\sqq+2, 2*\sqq+1+0.3);
            \draw[-{Stealth[length=1mm]}, blue] (2*\sqq+3, 2*\sqq+1) -- (2*\sqq+3, 2*\sqq+1+0.3);

            \draw (0,0) pic[gray, dash pattern=on 1pt off 1pt, -{Stealth[length=1mm]}]{carc=0:30:.38};
            \draw (2*\sqq, 1) pic[gray, dash pattern=on 1pt off 1pt, -{Stealth[length=1mm]}]{carc=30:60:.38};
            \draw (2*\sqq+1, 2*\sqq+1) pic[gray, dash pattern=on 1pt off 1pt, -{Stealth[length=1mm]}]{carc=60:0:.38};

            \draw[gray, dash pattern=on 1pt off 1pt] (0,0) -- (0.52, 0);
            \draw[gray, dash pattern=on 1pt off 1pt] (2*\sqq, 1) -- (2*\sqq+0.52*\sqq, 1+0.52*0.5);
            \draw[gray, dash pattern=on 1pt off 1pt] (2*\sqq+1, 2*\sqq+1) -- (2*\sqq+1 + 0.52*0.5, 2*\sqq+1 +0.52*\sqq);
            
        \end{tikzpicture}
        \caption{Output of basic algorithm}
        \label{fig:more_bones4}
    \end{subfigure}

    \vskip\baselineskip
    
    \begin{subfigure}[b]{0.45\columnwidth}
        \centering
        \begin{tikzpicture}[scale=0.8]

            \def\sq{0.70710678} 
            \def\sqq{0.86602540} 
            \def\cf{0.96592582628} 
            \def\sf{0.2588190451} 
            
            \draw[black] (0, 0) -- (\cf, \sf);
            \draw[black] (\cf, \sf) -- (\cf+\sqq, \sf+.5);
            \draw[black] (\cf+\sqq, \sf+.5) -- (\cf+\sqq+\sq, \sf+.5+\sq);
            \draw[black] (\cf+\sqq+\sq, \sf+.5+\sq) -- (\cf+\sqq+\sq+.5, \sf+.5+\sq+\sqq);
            \draw[black] (\cf+\sqq+\sq+.5, \sf+.5+\sq+\sqq) -- (\cf+\sqq+\sq+.5+\sqq, \sf+.5+\sq+\sqq+.5);
            \draw[black] (\cf+\sqq+\sq+.5+\sqq, \sf+.5+\sq+\sqq+.5) -- (\cf+\sqq+\sq+.5+\sqq+1, \sf+.5+\sq+\sqq+.5);
            
            \draw[line width=0.25mm, draw=lightblue, fill=white] (0,0) circle (0.1);
            \draw[line width=0.25mm, draw=darkyellow, fill=white] (\cf, \sf) circle (0.1);
            \draw[line width=0.25mm, draw=teal, fill=white] (\cf+\sqq, \sf+.5) circle (0.1);
            \draw[line width=0.25mm, draw=darkred, fill=white] (\cf+\sqq+\sq, \sf+.5+\sq) circle (0.1);
            \draw[line width=0.25mm, draw=darkpink, fill=white] (\cf+\sqq+\sq+.5, \sf+.5+\sq+\sqq) circle (0.1);
            \draw[line width=0.25mm, draw=olive, fill=white] (\cf+\sqq+\sq+.5+\sqq, \sf+.5+\sq+\sqq+.5) circle (0.1);
            \draw[line width=0.25mm, draw=last, fill=white] (\cf+\sqq+\sq+.5+\sqq+1, \sf+.5+\sq+\sqq+.5) circle (0.1);
            
            \draw[-{Stealth[length=1mm]}, red] (0, 0) -- (0.3*\cf, 0.3*\sf);
            \draw[-{Stealth[length=1mm]}, red] (\cf, \sf) -- (\cf+0.3*\sqq, \sf+0.3*0.5);
            \draw[-{Stealth[length=1mm]}, red] (\cf+\sqq, \sf+.5) -- (\cf+\sqq+0.3*\sq, \sf+.5+.3*\sq);
            \draw[-{Stealth[length=1mm]}, red] (\cf+\sqq+\sq, \sf+.5+\sq) -- (\cf+\sqq+\sq+0.5*0.3, \sf+.5+\sq+\sqq*0.3);
            \draw[-{Stealth[length=1mm]}, red] (\cf+\sqq+\sq+.5, \sf+.5+\sq+\sqq) -- (\cf+\sqq+\sq+.5+.3*\sqq, \sf+.5+\sq+\sqq+.3*.5);
            \draw[-{Stealth[length=1mm]}, red] (\cf+\sqq+\sq+.5+\sqq, \sf+.5+\sq+\sqq+.5) -- (\cf+\sqq+\sq+.5+\sqq+.3, \sf+.5+\sq+\sqq+.5);
            \draw[-{Stealth[length=1mm]}, red] (\cf+\sqq+\sq+.5+\sqq+1, \sf+.5+\sq+\sqq+.5) -- (\cf+\sqq+\sq+.5+\sqq+1+.3, \sf+.5+\sq+\sqq+.5);

            \draw[-{Stealth[length=1mm]}, blue] (0, 0) -- (-0.3*\sf, 0.3*\cf);
            \draw[-{Stealth[length=1mm]}, blue] (\cf, \sf) -- (\cf-0.3*0.5, \sf+0.3*\sqq);
            \draw[-{Stealth[length=1mm]}, blue] (\cf+\sqq, \sf+.5) -- (\cf+\sqq-.3*\sq, \sf+.5+.3*\sq);
            \draw[-{Stealth[length=1mm]}, blue] (\cf+\sqq+\sq, \sf+.5+\sq) -- (\cf+\sqq+\sq-0.3*\sqq, \sf+.5+\sq+0.5*0.3);
            \draw[-{Stealth[length=1mm]}, blue] (\cf+\sqq+\sq+.5, \sf+.5+\sq+\sqq) -- (\cf+\sqq+\sq+.5-0.5*0.3, \sf+.5+\sq+\sqq+.3*\sqq);
            \draw[-{Stealth[length=1mm]}, blue] (\cf+\sqq+\sq+.5+\sqq, \sf+.5+\sq+\sqq+.5) -- (\cf+\sqq+\sq+.5+\sqq, \sf+.5+\sq+\sqq+.5+.3);
            \draw[-{Stealth[length=1mm]}, blue] (\cf+\sqq+\sq+.5+\sqq+1, \sf+.5+\sq+\sqq+.5) -- (\cf+\sqq+\sq+.5+\sqq+1, \sf+.5+\sq+\sqq+.5+.3);

            \draw (0,0) pic[gray, dash pattern=on 1pt off 1pt, -{Stealth[length=1mm]}]{carc=0:15:.38};
            \draw (\cf, \sf) pic[gray, dash pattern=on 1pt off 1pt, -{Stealth[length=1mm]}]{carc=15:30:.38};
            \draw (\cf+\sqq, \sf+.5) pic[gray, dash pattern=on 1pt off 1pt, -{Stealth[length=1mm]}]{carc=30:45:.38};
            \draw (\cf+\sqq+\sq, \sf+.5+\sq) pic[gray, dash pattern=on 1pt off 1pt, -{Stealth[length=1mm]}]{carc=45:60:.38};
            \draw (\cf+\sqq+\sq+.5, \sf+.5+\sq+\sqq) pic[gray, dash pattern=on 1pt off 1pt, -{Stealth[length=1mm]}]{carc=60:30:.38};
            \draw (\cf+\sqq+\sq+.5+\sqq, \sf+.5+\sq+\sqq+.5) pic[gray, dash pattern=on 1pt off 1pt, -{Stealth[length=1mm]}]{carc=30:0:.38};

            \draw[gray, dash pattern=on 1pt off 1pt] (0,0) -- (0.52, 0);
            \draw[gray, dash pattern=on 1pt off 1pt] (\cf, \sf) -- (\cf+0.52*\cf, \sf+0.52*\sf);
            \draw[gray, dash pattern=on 1pt off 1pt] (\cf+\sqq, \sf+.5) -- (\cf+\sqq+0.52*\sqq, \sf+.5+.5*.52);
            \draw[gray, dash pattern=on 1pt off 1pt] (\cf+\sqq+\sq, \sf+.5+\sq) -- (\cf+\sqq+\sq+.52*\sq, \sf+.5+\sq+.52*\sq);
            \draw[gray, dash pattern=on 1pt off 1pt] (\cf+\sqq+\sq+.5, \sf+.5+\sq+\sqq) -- (\cf+\sqq+\sq+.5+0.52*.5, \sf+.5+\sq+\sqq+0.52*\sqq);
            \draw[gray, dash pattern=on 1pt off 1pt] (\cf+\sqq+\sq+.5+\sqq, \sf+.5+\sq+\sqq+.5) -- (\cf+\sqq+\sq+.5+\sqq+0.52*\sqq, \sf+.5+\sq+\sqq+.5+0.52*0.5);
            
        \end{tikzpicture}
        \caption{Final retargeted pose}
        \label{fig:more_bones5}
    \end{subfigure}
    
    \caption{Explanation of ``Copy Rotations'' algorithm when the target has more bones than the source}
    \label{fig:more_bones}
\end{figure}
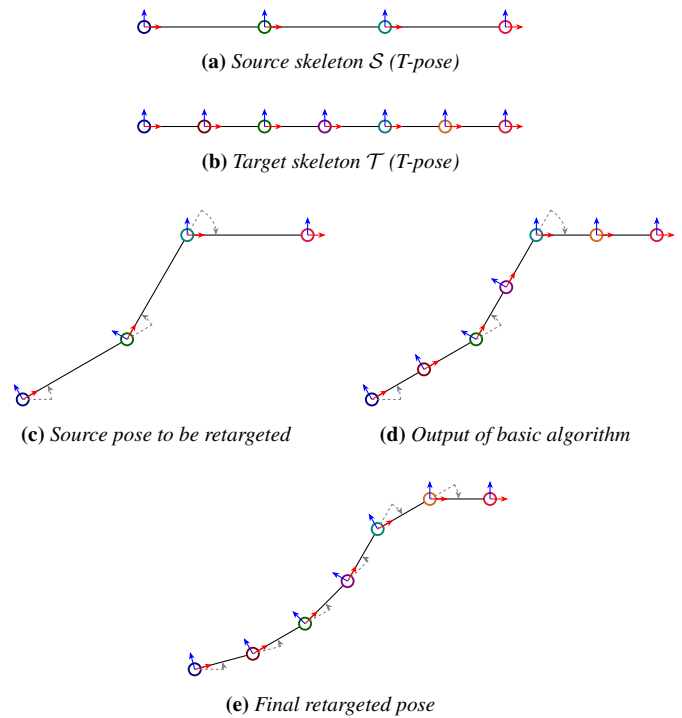

This is visible in Figure~\ref{fig:more_bones}. Once again, let's suppose that the source skeleton (in T-pose) corresponds to the one on Figure~\ref{fig:more_bones1}, and that the target skeleton corresponds to the one on Figure~\ref{fig:more_bones2}. If we simply transfer the rotations of the mapped bones, for a given pose on Figure~\ref{fig:more_bones3}, we obtain the result show on Figure~\ref{fig:more_bones4}.

To mitigate this, we propose to ``split" the rotation of the source bone, and to spread it across all of the additional bones. This gives the result shown on Figure~\ref{fig:more_bones5}, which makes use of all bones equally and gives a more ``flexible" appearance to the character.

\printbibliography